\documentclass[a4paper,12pt]{these}  
 	
\usepackage{these}          
\usepackage{thesetitle}     
\usepackage{thesefonts}     
\usepackage{icomma,natbib}
\usepackage{bm}
\usepackage{lscape}
\usepackage{supertabular}
\usepackage{boites}
\usepackage{rotating}
\titre{Observations de Pulsars avec le Fermi Gamma-ray Space Telescope}

\nom{Parent}
\prenom{Damien}
\ed{des Sciences Physiques et de l'Ing\'enieur}              			   
\specialite{Astrophysique, Plasmas, et Corpuscules}       
\date{13 novembre 2009}                                  
\annee{2009}                                           
\labo{CENTRE D'\'ETUDES NUCL\'EAIRES DE BORDEAUX GRADIGNAN} 
\logolabo{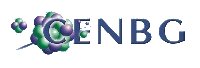}                       
\rapporteuronename{M. X}                          
\rapporteuronefonc{XXXX}
\rapporteuroneuniv{Universit� de XXXX}
\rapporteurtwoname{M. Y}
\rapporteurtwofonc{XXXXX}		
\rapporteurtwouniv{Labo XXXX} 
\presidentname{M. Z}
\presidentfonc{XXXX}
\presidentuniv{XXXX}
\directeurname{M. D.~A.~SMITH}
\directeurfonc{Directeur de Recherche CNRS}
\directeuruniv{Universit� XXXX}

\juryonename{xxx}
\juryonefonc{xxx}
\juryoneuniv{xxx} 
\jurytwoname{xxx}
\jurytwofonc{xxx}
\jurytwouniv{xxx}

\ordre{3890}
\makeindex    

\dominitoc    

\begin{document}

\maketitle                      

\newpage
\thispagestyle{empty2}
\null
\newpage

\thispagestyle{empty2}

\dedicace{A Tous Ceux Qui Ont Participé.}

\newpage
\thispagestyle{empty2}
\null
\newpage



\vspace*{10cm}

\epigraphe{``Se réveiller, c'est se mettre à la recherche du monde.''}{Alain (\'Emile-Auguste Chartier),\\ \normalfont\itshape Vigiles de l'esprit (1942)}

\thispagestyle{empty2}

\thispagestyle{empty2} 
\chapter*{Remerciements}

Merci à David Smith et Thierry Reposeur, mes directeurs, pour m'avoir permis de faire cette thèse, guidé durant ces trois années, et
pour avoir supporté la magnifique senteur du thé aux fruits rouges (enfin surtout pour un). Je leur en suis sincèrement reconnaissant
car c'est un peu grâce à eux que je suis dans cette superbe ville de D.C. avec un lot de pans. \\

Je tiens à remercier également les autres membres de l'équipe Astroparticules du CENBG pour leur précieuse aide et les discussions passionnantes : Marianne Lemoine-Goumard et sa compère Marie-Hélène Grondin ou "mais euh", Denis Dumora, Lucas Guillemot, Benoît Lott et Lise Escande. N'oublions pas Funix ! Merci également à Bernard Haas pour m'avoir accueilli et apporté son soutien au sein du CENBG, et je tire mon chapeau à l'ensemble du personnel pour son efficacité et sa sympathie. J'ai une pensée pour mes autres (anciens) collègues du CENBG : Jérémie Messud, Rémi Huguet, Hao Tran Viet Nhan et Julien Le Bloas, notamment pour leurs discussions footballistiques, sur les docs de sociétés, et à savoir si un chercheur français gagne bien sa vie. \\

Je fais également part de ma gratitude à Pascal Vincent et Pierre Jean pour avoir accepté d'être les rapporteurs de ma thèse, ainsi
qu'à l'ensemble du jury : Yves Gallant, Franck Gobet, David et Thierry. Je remercie encore ces derniers pour leur soutien dans l'écriture de ce
manuscrit et la préparation de la soutenance. \\

N'oublions pas que Fermi n'est pas qu'un instrument mais aussi une collaboration. Je tiens donc à témoigner ma reconnaissance aux collègues gamma et radio pour leur contribution à mes travaux : Ismaël Cognard, Gilles Theureau, Jean-Marc Casandjian, Philippe Bruel, Dave Thompson, Roger Romani, Alice Harding, Max Razzano, Tyrel Johnson, Matthew Kerr, Patrick Weltevrede, Fernando Camilo, Kyle Watters, Eric Grove et le reste de la collaboration. \\

Je terminerai par exprimer ma sincère gratitude à mes proches, à mes parents et à ma soeur qui m'ont apporté leur appui toutes
ces années, et à ma très chère épouse pour son sourire et ses bons petits plats. Enfin, je ne remercierai pas mes amis pour leur intérêt à l'égard
de mon travail mais pour leur amitié indestructible : Toons, French, Balland, Rocco, Renaud, Jérémy, Heather, Rodolphe, Olive et
Greg ! \\

J'espère n'avoir oublié personne ...
 
\tableofcontents                

\begin{fmffile}{diagram}        

\partie*{\textit{AVANT-PROPOS}} 
\vspace*{40pt}

Prédit par Baade et Zwicky en 1934, l'existence d'étoiles à neutrons a été confirmée par la découverte d'un pulsar radio en 1967. Cette découverte fut un tournant important en astrophysique, tant par la confirmation d'objets macroscopiques de densité nucléaire dans l'univers, que par l'enjeu scientifique de comprendre la complexité d'une étoile à neutrons fortement magnétisée en rotation rapide, et par delà, d'étudier la physique sous des conditions extrêmes. Les pulsars sont formés au sein des supernovae ou par le transfert de moment angulaire d'une étoile compagnon vers une étoile à neutrons. Par ces deux processus de formation, on distingue deux types principaux de pulsars: les pulsars normaux (sujet principal de cette thèse) et les pulsars ``milliseconde''. 


Ces objets sont utilisés pour étudier la physique des étoiles à neutrons, la relativité générale, le champ magnétique galactique, la physique des plasmas, la distribution de densité d'électrons dans la Galaxie, ainsi que la magnétohydrodynamique relativiste des vents de pulsars interagissant avec le milieu ambiant. Cependant, même après plus de 40 années de recherches intenses, utilisant des instruments de plus en plus performants, il reste des questions fondamentales qui ne sont pas résolues:
\begin{itemize}
\item Combien de pulsars se trouvent dans la Galaxie et quel est leur taux de formation ?
\item Quelle est la relation entre la supernova d'une étoile massive et les propriétés intrinsèques des pulsars ?
\item Quelle est la limite supérieure de la masse d'une étoile à neutrons ?
\item Est-ce que le vent de particules chargées injecté par le pulsar dans le mileu interstellaire peut alimenter une partie du spectre des rayons cosmiques ? 
\item Quelle est la composition de la magnétosphère d'un pulsar et comment interagit-elle avec le champ magnétique intense ?
\item Où et comment l'émission radio et l'émission à haute énergie (de l'infrarouge au rayonnement $\gamma$) sont-elles produites dans la magnétosphère du pulsar ? 
\item Est-ce que les processus physiques sont les mêmes pour tous les pulsars ?
\item Quel est le rapport entre les pulsars radio émetteurs $\gamma$ tels que le Crabe ou Vela, et les pulsars $\gamma$ non-émetteurs en radio tels que Geminga et CTA1 ?
\end{itemize}
L'étude de ces questions va bien au-delà du sujet de cette thèse, mais montrent l'étendue de la recherche sur les étoiles à neutrons et les pulsars. \\

La période de rotation des pulsars est relativement stable mais elle n'est pas constante. Tous ralentissent à cause de leur rayonnement magnétique dipolaire (ondes électromagnétiques à la fréquence de rotation du pulsar), et de leur vent de particules relativistes qui peut éventuellement créer en contact avec le milieu interstellaire une nébuleuse à vent de pulsar. Une fraction de ces particules accélérées le long des lignes de champ magnétique produit un faisceau d'ondes électromagnétiques qui balaye le ciel comme un phare, générant pour l'observateur une pulsation par période de rotation. La base de données ``ATNF Pulsar Catalogue'' répertorie 1827 pulsars répartis dans notre Galaxie et dans les nuages de Magellan, dont la pulsation a été mise en évidence, principalement en radio (une cinquantaine émet également à haute énergie, de l'optique au rayonnement $\gamma$). Théoriquement, les ondes électromagnétiques de la radio jusqu'aux hautes énergies sont produites dans différentes régions de la magnétosphère du pulsar. Les ondes radio polarisées sont émises depuis les pôles magnétiques de l'étoile par émission synchrotron. Les photons $\gamma$ quant à eux sont produits principalement par le rayonnement de courbure et la diffusion inverse Compton des électrons (et positons) accélérés le long des lignes de champ magnétique. Cette propriété est importante puisque les particules chargées accélérées rayonnent leur énergie très vite et sont liées aux sites d'accélération. C'est dans ce cadre que se place ce travail de thèse, qui ambitionne d'analyser les pulsars $\gamma$ détectés par le télescope spatial \textit{Fermi Gamma-ray Space Telescope} et d'essayer de contraindre la géométrie et les mécanismes des zones d'accélération, en étudiant en détail les courbes de lumière et les spectres résolus en phase d'une large population de pulsars $\gamma$. \\

La première partie de ce mémoire présente les propriétés intrinsèques des pulsars, ainsi que les modèles d'émission $\gamma$ basés sur les observations de l'expérience EGRET. On dressera ensuite pour la mission \textit{Fermi} une liste de pulsars candidats $\gamma$ établie sur leur énergie de ralentissement et de leur distance.

Le \textit{Large Area Telescope}, instrument principal de \textit{Fermi} dédié à l'observation du ciel des hautes énergies ($>$ 20\,MeV), sera brièvement décrit dans la deuxième partie de cette thèse. La mesure des paramètres spectraux des pulsars (i.e. le flux intégré, l'indice spectral, et l'énergie de coupure), nécessaires pour discriminer les modèles d'émission $\gamma$ des pulsars, dépend des fonctions de réponse de l'instrument. Nous présenterons une méthode développée pour la validation de la surface efficace en utilisant le pulsar de Vela, qui compare l'efficacité des coupures entre les vraies données et les données simulées à chaque niveau de la classification des événements du LAT. Nous finirons cette partie par la description des analyses temporelle et spectrale des pulsars.

Enfin, la troisième et dernière partie présentera les résultats obtenus à partir des méthodes d'analyses pour trois pulsars individuels, PSRs J0205+6449, J2229+6114, et J1048$-$5832, ainsi que du premier catalogue de pulsars du télescope \textit{Fermi}. Une interprétation de la mesure des paramètres spectraux et du profil des pulsations $\gamma$ des pulsars détectés, sera présentée. \\

Cette thèse se base sur les papiers suivants : \\
\begin{itemize}
\item ``The First \textit{Fermi} Large Area Telescope Catalog of Gamma-ray Pulsars'', \\
Abdo, A. A. et al. 2009, submitted to Astrophysical Journal, \\
Contact authors : A. Caliandro, E. Ferrara, \textbf{D. Parent}, R. Romani \\

\item ``Discovery of Pulsations from the Pulsar J0205+6449 with the Fermi Gamma-ray Space Telescope'', 
Abdo, A. A. et al. 2009, Astrophysical Journal Letters, 699, L102, \\
Contact author : \textbf{Damien Parent} \\

\item ``Fermi LAT detection of pulsed gamma-rays from the Vela-like pulsars PSR J1048-5832 and PSR J2229+6114'', 
Abdo, A. A. et al. 2009, Astrophysical Journal, 706, 1331, \\
Contact authors : \textbf{Damien Parent}, Alice K. Harding, Massimiliano Razzano \\ \\

\item ``The Radio Polarization of Six Gamma-ray Pulsars seen with the \textit{Fermi} LAT'', \\
Weltevrede, P. et al. 2009, submitted to Astrophysical Journal \\

\item ``PSR J1410-6132: a young, energetic pulsar associated with the EGRET source 3EG J1410-6147'', 
O'Brien ... \textbf{Parent, D.}, ... et al. 2008, Monthly Notices of the Royal Astronomical Society, 388, L1 \\

\item Signataire de 6 autres publications \textit{Science} et \textit{ApJ} sur les pulsars radio émetteurs $\gamma$, 6 autres sont en cours de publication.
\end{itemize}


\partie{INTRODUCTION}          

%

\chapter[Les Pulsars]{Les Pulsars}\label{chap:pulsar}


\minitoc

\section{Découverte}

Depuis la fin de la seconde guerre mondiale, l'astronomie radio connaît un développement considérable. Les nouvelles techniques d'observations ouvrent de nouveaux champs de recherches et mènent à la découverte inattendue des quasars \citep{schmidt63}. Suite à cela, Antony Hewish et son étudiante Jocelyn Bell de l'université de Cambridge espèrent différencier les quasars des radio-galaxies en utilisant la scintillation interplanétaire, réfraction des ondes radio due à l'atmosphère terrestre et au gaz ionisé interstellaire. Ils construisent pour cela un radiotélescope suffisamment sensible aux sources faibles et variables dans le temps. Au cours du mois de juillet 1967, J. Bell découvre une source qui émet de façon constante et régulière des impulsions radio qui diffèrent de la scintillation. Elles apparaissent avec un décalage de 4 minutes chaque jour, caractéristique d'un objet d'origine céleste. Un enregistreur avec une réponse en temps plus rapide est ensuite utilisé. La  nouvelle prise de données confirme un signal extrêmement stable dont la périodicité est de 1.337 secondes (Figure \ref{fig:psrdiscovery}). L'annonce de cette découverte est publiée dans la revue \textit{Nature} en février 68 \citep{hewish68}, et soumet l'idée que la rotation ou l'oscillation d'un astre, tel que les naines blanches ou les étoiles à neutrons, peut être à l'origine de ce rayonnement électromagnétique. Notons qu'aujourd'hui la source radio d'Hewish et Bell est connue sous le nom de PSR~B1919+21 ou PSR~J1921+2153\footnote{PSR signifie ''Pulsating Source of Radio''. $B$ et $J$ correspondent respectivement aux équinoxes standards utilisées pour les coordonnées du pulsar (B1950 et J2000), tandis que 1919 (ou 1921) et +21 (ou +2153) correspondent aux coordonnées de la source (ascension droite de 19h19' et déclinaison de +21\degr).}. 

\fig[Pulsations de CP1919 (PSR~B1919+21)]{scale=1.0}{c1_psr_discovery.jpg}{Enregistrement des pulsations du premier pulsar connu, CP1919 (PSR~B1919+21), le 28 novembre 1967. Les signaux observés ont une période de 1.337\,s.}{fig:psrdiscovery}

C'est seulement à la fin de l'année 1968, après la découverte de signaux périodiques de 88 ms et de 33 ms, respectivement dans les restes de supernova de Vela \citep{large68} et du Crabe \citep{staelin68}, que l'on attribue avec certitude l'origine de ce phénomène, non pas aux naines blanches relativement connues, mais à une étoile à neutrons en rotation. La période minimale de rotation d'une étoile avec une densité uniforme $\rho$ est donnée par $P_{min} =$ ($3\pi/G\rho$)$^{1/2}$. Pour les naines blanches dont la densité moyenne est de 10$^{8}$\,g\,cm$^{-3}$, la période minimale de rotation est de l'ordre de la seconde. Quant à leurs oscillations, \citet{melzer66} montrent également des périodes minimales voisines de la seconde. En conséquence, les signaux ne pouvaient venir que d'objets encore plus compacts, notamment les hypothétiques étoiles à neutrons. Le nom de pulsar est ainsi associé à cet objet, association d'autant plus pertinente qu'un ralentissement dans la période du Crabe est observé \citep{richards69}, ce que prévoyait les modèles de pulsar de \citet{gold68} et \citet{pacini68}.

Cette remarquable découverte confirmera les prédictions faites en 1934 par \citet{baade34} de l'existence d'étoiles à neutrons au sein des supernovae (pour une revue récente, voir \citet{Green2009}), stade ultime de l'évolution stellaire, et récompensera Hewish du prix Nobel de physique en 1974.

\section{\'Evolution de la Rotation et Propriétés Intrinsèques}\label{sec:propneutronstar}

Le lien entre étoiles à neutrons et pulsars étant établi par l'observation de leur pulsation, intéressons-nous à présent à la formation et aux propriétés intrinsèques des pulsars. \\

L'évolution d'une étoile est liée à sa masse initiale. Plus l'étoile est massive, moins elle passe de temps dans la séquence principale à synthétiser des noyaux d'hélium à partir des protons et des électrons. On pourra se reporter au diagramme d'Hertzsprung et Russell pour parcourir l'évolution stellaire \citep{russell21}. Quand les noyaux d'hydrogène viennent à manquer, la pression radiative qui compense l'attraction gravitationnelle chute. Le coeur de l'étoile se contracte et augmente en température, permettant la combustion de l'hélium. Ce phénomène s'accompagne de l'expansion des couches externes. Pour les étoiles de types O et B, supérieures à 8 -- 10 masses solaires, la fusion thermonucléaire d'éléments de plus en plus lourds va continuer jusqu'au fer. La nucléosynthèse de tout autre élément plus lourd ne produit pas d'énergie, au contraire elle en consomme. A ce stade la pression radiative créée par l'énergie nucléaire s'arrête. Le coeur de l'étoile s'effondre créant une onde de choc qui expulse les couches supérieures de l'étoile dans le milieu interstellaire. Ce phénomène est appelé \textit{supernova} \citep{baade34} de type Ib/c, ou II \citep{cappellaro00}. Si le coeur restant excède la masse critique de Chandrasekhar \citep{chandrasekhar35}, masse à laquelle la pression de dégénérescence des électrons s'oppose à la force gravitationnelle, les électrons et les protons vont se combiner pour former des neutrons. L'effondrement s'arrêtera quand la pression des neutrons dégénérés compensera la pression gravitationnelle, formant ainsi une étoile à neutrons. 

\subsection{\'Etoiles à Neutrons}\label{sec:neutronstar}

Les étoiles à neutrons sont l'un des objets les plus compacts que l'on connaisse dans notre univers (après les trous noirs). Leur structure (composition interne, masse, et rayon) est encore mal connue et dépend de l'équation d'état de la matière. Cette discipline constitue un sujet d'études important en physique nucléaire. L'analyse théorique de la structure des étoiles à neutrons est basée sur la chronométrie des pulsars et les observations optiques et $X$ des étoiles elles-mêmes. 

La figure \ref{fig:neutronstar} (gauche) présente un des modèles d'étoile à neutrons pour une masse de 1.4 masse solaire et un rayon de 15 km. Entre la surface de l'étoile et son centre, la densité volumique varie de 10$^{6}$ à 10$^{15}$\,g\,cm$^{-3}$. L'étoile est composée d'une croûte surfacique rigide d'une épaisseur de 1 km formée essentiellement de noyaux de fer et d'un noyau de neutrons superfluides. La séparation entre ces deux parties se situe près de $\rho = 4.3 \times 10^{11}$\,g\,cm$^{-3}$. La plupart des modèles de matière dense prédisent un déconfinement des quarks et une possible apparition de matière exotique (quark étrange) au-dessus d'une densité de $\sim 5 \times 10^{14}$\,g\,cm$^{-3}$, la densité de la matière nucléaire froide. Cette structure composée de différents états de la matière pourrait expliquer à la fois l'observation de perturbations continues de la rotation du pulsar appelées ``timing noise'', ainsi que l'accélération brusque de la période de rotation des pulsars que l'on nomme ``glitch'' \citep{anderson75,janssen07}. Ces événements sont probablement la conséquence du transfert du moment angulaire de la croûte solide, qui tourne à la période mesurée du pulsar, jusqu'à la matière  au centre de l'étoile tournant plus vite. Les \textit{glitch} sont suivis par un rétablissement de la fréquence de rotation, qui est la conséquence d'un nouvel équilibre entre les deux composants. 

La figure \ref{fig:neutronstar} (droite) présente la masse des étoiles à neutrons mesurée à partir de pulsars présents dans des systèmes binaires, révélant une masse moyenne autour de 1.4 masse solaire (masse de Chandrasekhar). L'intervalle de masse prédit par les modèles théoriques d'étoiles à neutrons s'étend de 0.2 à 3.0 masses solaires, et dépend des modèles d'équation d'état de la matière nucléaire. Par la suite, nous prendrons une masse d'étoile à neutrons $M_{ns}$ de 1.4 $M_{\odot}$.

Pour plus de détails sur les étoiles à neutrons, on pourra se reporter à l'article de \citet{lattimer07} et au cours en ligne de \citet{gourgoulhon05}.

\sfig[Etoiles à neutrons.]{neutron_star.tex}{\textbf{A:}~Structure possible d'une étoile à neutrons. \textbf{B:}~Masses mesurées d'étoiles à neutrons localisées dans des systèmes binaires. Pour chaque région, la moyenne des masses est indiquée en pointillé, tandis que la moyenne pondérée est représentée par la ligne discontinue. La figure est extraite de \citet{lattimer07}.}{fig:neutronstar}

\subsection{Rotation}

Pendant la phase de supernova, le système `coeur de l'étoile massive / étoile à neutrons' passe en quelques secondes de $R_{0} \sim 10^{5}$ km (rayon moyen) à $R_{ns} \sim 10$ km. Si l'on assimile l'étoile à une sphère, son moment d'inertie $I$ est égal à $2/5 \times M_{ns} R_{ns}^{2}$\,, soit $\sim 10^{45}$\,g\,cm$^{2}$. En supposant que le moment angulaire $I\Omega$ avec $\Omega = 2\pi / P$ et la masse du système soient en partie conservés pendant l'effondrement, la période initiale de rotation de l'étoile à neutrons $P_{ns}$ sera déterminée par la période de l'étoile initiale $P_{0}$, avec :
\begin{equation}
P_{ns} \simeq P_{0} (\frac{R_{ns}}{R_{0}})^{2}
\end{equation}
Ainsi, pour une étoile initiale de période $P_{0} \sim 10^{6}$\,s (notre soleil tourne sur lui-même en $2 \times 10^{6}$\,s $\sim$ 27 jours), la période de rotation de l'étoile à neutrons est de l'ordre de 10\,ms. Ce chiffre est en accord avec la période observée des jeunes pulsars tels que le Crabe (33\,ms), PSR~J0205+6449 (64\,ms), ou PSR~J1833-1034 (61.9\,ms). \\

D'autre part, comme l'ont observé \citet{richards69} pour le Crabe, la période de rotation des pulsars augmente au cours du temps. En d'autres termes, ils	ralentissent, indiquant un transfert de leur énergie cinétique de rotation. Si l'on considère le pulsar comme un rotateur rigide avec un moment d'inertie $I$, l'énergie cinétique disponible est donnée par $\frac{1}{2}I\Omega^{2}$, avec $\Omega$ la vitesse angulaire. En conséquence, l'énergie de rotation perdue par seconde est :
\begin{equation}\label{eq:energie_rotation}
\dot{E}_{rot} = - \frac{d}{dt}(\frac{1}{2} I \Omega^{2}) = - I \Omega \dot{\Omega} = 4 \pi^{2} I (\dot{P} / P^{3}) \ {\rm erg\,s^{-1}}.
\end{equation}
Avec un ralentissement $\dot{P}$ de $4.2 \times 10^{-13}$\,s\,s$^{-1}$ \citep{atnf} et adoptant un moment d'inertie $I$ de $\sim 10^{45}$\,g\,cm$^{2}$, le taux d'énergie de rotation perdue par le pulsar du Crabe est de $4.6 \times 10^{38}$\,erg\,s$^{-1}$ \footnote{	
1 erg $= 6.24150974 \times 10^{11}$ électron-volt}. Les lignes pour les différentes valeurs de $\dot{E}$ en fonction de ($P$,$\dot{P}$) sont dessinées sur la figure \ref{fig:ppdot}.


\subsection{Champ Magnétique}\label{sec:chpmag}

Durant l'effondrement gravitationnel le flux magnétique $\phi \propto B R^{2}$ est également conservé à l'intérieur de l'étoile. Le champ magnétique polaire à la surface de l'étoile à neutrons $B_{ns}$ est ainsi déterminé par le champ magnétique de l'étoile initiale $B_{0}$, par :
\begin{equation}
B_{ns} \simeq B_{0} (\frac{R_{0}}{R_{ns}})^{2} \ .
\end{equation}
Utilisant les valeurs standards précédentes pour le rayon et adoptant un champ magnétique de 100\,G (10$^{-2}$ Tesla) pour l'étoile initiale, le champ magnétique résultant est de l'ordre de $10^{12}$\,G. \citet{ostriker69} montrent qu'une étoile à neutrons de rayon $R_{ns}$ tournant à la vitesse angulaire $\Omega$ dans le vide, avec un champ magnétique dipolaire non parallèle à l'axe de rotation rayonne de l'énergie. La puissance rayonnée est donnée par:
\begin{equation}\label{eq:energie_dipole}
\frac{dW}{dt} = \frac{\Omega^{4}}{3c^{3}} R^{6}_{ns} B^{2}_{ns} sin^{2}\alpha
\end{equation}
où $\alpha$ est l'angle entre l'axe du dipôle et l'axe de rotation (voir la figure \ref{fig:standardmodele}) et $c$ est la vitesse de la lumière. En reliant les équations \ref{eq:energie_rotation} et \ref{eq:energie_dipole}, on peut estimer, avec les seules observables ($P$ la période de rotation et $\dot{P}$ sa dérivée), le champ magnétique à la surface de l'étoile :
\begin{equation}\label{eq:magfieldsurface}
B_{ns} = 3.2 \times 10^{19} (P \dot{P})^{1/2} \ {\rm G}.
\end{equation}
Pour le pulsar du Crabe, le champ magnétique de surface obtenu est de l'ordre de $4 \times 10^{12}$\,G, en accord avec le champ moyen estimé après l'effondrement. Les lignes pour les différentes valeurs de $B_{ns}$ en fonction de ($P$,$\dot{P}$) sont dessinées sur la figure \ref{fig:ppdot}. Le champ magnétique au niveau du cylindre de lumière (voir la section \ref{sec:modstandard}) est quant à lui décrit par:
\begin{equation}\label{eq:magfieldcylinder}
B_{LC} = (\frac{3 I \ 8\pi^{4} \ \dot{P}}{c^3 P^5})^{1/2} \approx 2.943 \times 10^8 (\dot{P} P^{-5})^{1/2} \ {\rm G}.
\end{equation}

\subsection{Indices de Freinage et Age du Pulsar}\label{indicefreinage}

Rappelons que les résultats précédents sont vérifiés dans l'approximation dipolaire du pulsar. Nous allons voir que la loi qui détermine le ralentissement de la rotation est légèrement différente de celle observée dans le cas d'un dipôle magnétique. A partir des équations \ref{eq:energie_rotation} et \ref{eq:energie_dipole}, la variation de la vitesse angulaire du pulsar est donnée par :
\begin{equation}
\dot{\Omega} = - \frac{B^{2}_{ns} R^{6}_{ns} sin^{2}\alpha}{6 I c^{3}} \Omega^{3} = - k \Omega^{3}.
\end{equation}
Plus généralement, si l'on suppose que le pulsar est formé avec une grande vitesse angulaire, on admet que la vitesse angulaire suit une simple loi de puissance d'indice $n$ de la forme:
\begin{equation}\label{eq:omega}
\dot{\Omega} = -k \Omega^{n},
\end{equation}
où k est une constante et $n$ est défini comme l'indice de freinage. En dérivant l'expression précedente, on peut décrire l'indice $n$ à partir de la période de rotation du pulsar et ses dérivées :
\begin{equation}
n = \frac{\Omega \ddot{\Omega}}{\dot{\Omega}^{2}} = 2 - \frac{P\ddot{P}}{\dot{P}^{2}}.
\end{equation}
Notons qu'il est très difficile de mesurer l'indice de freinage. Cela demande de connaître parfaitement la seconde dérivée, laquelle est dominée par les instabilités du pulsar. Quelques valeurs ont été mesurées pour les pulsars du Crabe ($n = 2.525 \pm 0.005$), de PSR~B1509$-$58 ($n = 2.8 \pm 0.2$) et de PSR~B0540$-$69 ($n = 2.01 \pm 0.02$). On remarquera l'écart entre les indices de freinages mesurés et la valeur théorique $n = 3$ dans le cas unique de radiation par un dipôle magnétique. Cependant, plusieurs effets n'ont pas été pris en compte dans ce modèle très simplifié:~la dissipation de l'énergie par l'émission d'ondes gravitationnelles, la déformation du champ magnétique $\vec{B}$ par la rotation du pulsar, les effets de \textit{glitch} et de \textit{timing noise}, ainsi que le vent d'électrons relativistes créé par le pulsar qui heurte le milieu interstellaire (voir \ref{sec:modstandard}). \\

En intégrant l'équation \ref{eq:omega}, et en supposant que la période $P$ est grande devant la période initiale $P_{0}$, on peut déduire l'âge caractéristique du pulsar par :
\begin{equation}
\tau = \frac{P^{n-1}}{(2\pi)^{n-1}K} = \frac{P}{(n-1) \dot{P}}
\end{equation}
Toujours dans l'approximation du dipôle magnétique ($n$ = 3), l'âge caractéristique peut être décrit très simplement par $\tau = P/2\dot{P}$. Dans le cas du pulsar du Crabe créé par la supernova en 1054 (955 ans) avec une période $P$ de 33\,ms et une dérivée $\dot{P}$ de $4.2 \times 10^{-13}$, l'âge caractéristique obtenu est de $\sim$ 1260 ans, ce qui est légèrement différent (mais raisonnable) si on utilise la valeur expérimentale $n=2.525$ (1600\,ans). Dans le cas du pulsar PSR~J0205+6449, nous verrons au chapitre \ref{chap:j0205} que l'âge estimé de cette façon est en désaccord avec la supernova associée à cet objet.

\section{Premières Observations de l'\'Emission Pulsée $\gamma$}\label{sec:premier_obsgamma}

Un an après la découverte du premier pulsar en radio par \citet{hewish68}, les astronomes entamèrent une recherche dans les autres domaines du spectre électromagnétique. En 1969, \citet{cocke69} publient la découverte de pulsations optiques du Crabe, tandis que \citet{fritz69} et \citet{bradt69} annoncent sa découverte en X établie grâce à une fusée placée dans la haute atmosphère. Le rayonnement électromagnétique de haute énergie tels que le rayonnement X ou $\gamma$, ainsi que certaines longueurs d'ondes en infra-rouge sont absorbés par l'atmosphère terrestre et n'atteignent pas le sol. Il est donc indispensable, pour observer ces domaines d'énergie, d'utiliser des instruments situés dans la haute atmosphère ou au-delà.

\fig[Carte du troisième catalogue d'EGRET]{scale=0.8}{c1_egret_cat_bw.pdf}{Cette carte indique la position des sources du troisième catalogue d'EGRET \citep{hartman99}. Les sources sont placées suivant leurs coordonnées galactiques et sont identifiées suivant un code de forme. Cette figure est extraite de \citet{thompson08}.}{fig:egretsky}

L'astronomie gamma\footnote{Le domaine \gam est défini au-delà de 100\,keV (des photons jusqu'à la centaine de TeV sont détectés). Etant donné le vaste domaine, les méthodes de détection sont différentes suivant l'énergie du photon \gam. On distingue le domaine de basse énergie entre 100\,keV et 30\,MeV (détection du $\gamma$ par effet photoélectrique), le domaine de haute énergie entre 30\,MeV et 30\,GeV (détection du $\gamma$ par conversion $e^{\pm}$), et le domaine de très haute énergie au-dessus de 30\,GeV (détection de la lumière Tcherenkov émise par la cascade $e^{\pm}$ créée par le photon $\gamma$ primaire).} débute en 1972 avec le Small Astronomy Satellite (SAS-2), qui met en évidence trois sources ponctuelles le long du plan galactique, toutes trois identifiées comme des pulsars. Deux d'entre elles sont mises en évidence par leur coïncidence spatiale et la pulsation de leur émission avec les pulsars radio du Crabe et de Vela \citep{kniffen74,thompson75}, tandis que la troisième source nommée Geminga n'est identifiée qu'en 1992 en tant que pulsar, par la découverte d'une périodicité de 237 ms dans les photons X \citep{halpern92} et les photons $\gamma$ \citep{bignami92}. Trois ans plus tard, l'Agence Spatiale Européenne met en orbite le Cosmic Ray Satellite (COS-B). L'instrument va révéler la présence de nombreuses sources ponctuelles dans le plan galactique \citep{hermsen77}, et cartographie pour la première fois le fond diffus \gam qui résulte de la désintégration de $\pi^{0}$ produits par l'interaction des rayons cosmiques avec le gaz interstellaire.

Cependant, c'est en 1991 que l'astronomie \gam connaît un essor important. La NASA lance le Compton Gamma Ray Observatory (CGRO) qui compte à son bord quatre instruments : BATSE (Burst And Transient Source Experiment) dédié à l'observation des sursauts $\gamma$ entre 20 et 100\,keV, OSSE (Oriented Scintillation Spectrometer Experiment) utilisé pour l'étude des raies de désintégration des noyaux radioactifs dans les vestiges de supernovae, COMPTEL (imaging COMPton TELescope) dédié à l'étude des sources entre 1 et 30\,MeV, et EGRET (Energetic Gamma Ray Experiment Telescope) pour l'étude des sources entre 30\,MeV et 30\,GeV. Ce dernier, précurseur du télescope Fermi, présente une meilleure résolution angulaire et une plus grande surface de collection que ses prédécesseurs, et dresse ainsi le premier catalogue des sources $\gamma$ de hautes énergies \citep{hartman99}. La figure \ref{fig:egretsky} présente la position dans le ciel des 271 sources du troisième catalogue d'EGRET, dont 94 associations avec des noyaux actifs de galaxie, un sursaut solaire, le nuage de Magellan, et 5 (+1) pulsars\footnote{Le pulsar PSR~B1951+32 n'est pas inclut dans le 3$^{ème}$ catalogue d'EGRET à cause de son flux trop faible, mais est représenté sur la carte.}. 
Il est important de souligner que 170 sources n'ont pas été identifiées en raison de l'incertitude sur leur position, notamment au niveau du plan galactique. Ce sujet constitue l'un des enjeux majeurs de l'expérience Fermi.


\subsection{Pulsars CGRO 1991\ --\ 1999}\label{sec:egretpsr}

Les télescopes à bord du satellite CGRO ont identifié au total 7 pulsars avec un haut niveau de confiance. Le télescope EGRET a détecté l'émission $\gamma$ pour 6 pulsars (9 possibles), comprenant 5 pulsars radio et le pulsar Geminga dont la pulsation radio n'a jamais été observée. Quant au septième pulsar PSR~B1509$-$58, il a été détecté en dessous de 5\,MeV par le détecteur COMPTEL \citep{matz94}. Le tableau \ref{tab:cgropsr} référence les principales propriétés des pulsars détectés par les deux instruments COMPTEL et EGRET, ainsi que 3 pulsars détectés par EGRET avec une significativité plus faible, PSR~J1048$-$5832 (chapitre \ref{chap:j1048}), PSR~J0659+1414 (chapitre \ref{chap:catalog}), et le pulsar milliseconde J0218+4232 (\citealt{kuiper00}, chapitre \ref{chap:catalog}). $L_{\gamma}$ est la luminosité $\gamma$, donnée par:
\begin{equation}\label{eq:luminosite}
L_{\gamma}= 4\pi \ f_{\Omega} (\alpha,\zeta_E) \ F_{E,obs} \ d^{2}
\end{equation}
où $d$ est la distance du pulsar depuis la Terre (voir la section \ref{sec:distance}), $F_{E,obs}$ est le flux énergétique observé depuis la Terre et moyenné sur toute la phase, et $f_{\Omega}$($\alpha$,$\zeta_E$) est le facteur de correction qui prend en compte la géométrie du faisceau donné par \citep{watters09}:
\begin{equation}\label{eq:fbeam}
f_{\Omega}(\alpha,\zeta_{E}) = \frac{\int F_{\gamma}(\alpha;\zeta,\phi)sin(\zeta)d\zeta d\phi}{2 \int F_{\gamma}(\alpha;\zeta_{E},\phi) d\phi}
\end{equation}
où $F_{\gamma}(\alpha;\zeta,\phi)$ est le flux rayonné en fonction de l'angle d'observation $\zeta$ par rapport à l'axe de rotation (ligne de visée) et la phase du pulsar ($\phi)$, tandis que $f_{\Omega}$ est le rapport entre l'émission $\gamma$ globale sur tout le ciel et le flux prévu moyenné sur toute la phase pour une courbe de lumière observée depuis la Terre. Enfin, l'efficacité $\gamma$ notée $\eta$ est le rapport entre la luminosité (équation \ref{eq:luminosite}) et le ralentissement $\dot{E}$ (équation \ref{eq:energie_rotation}).

\tab[Paramètres des pulsars observés par CGRO]{cgropsr}{Paramètres intrinsèques et mesurés des pulsars et des possibles candidats détectés par les télescopes COMPTEL et EGRET à bord du satellite CGRO. $P$ est la période de rotation. $\tau$ est l'âge caractéristique (voir \ref{indicefreinage}). $\dot{E}$ est le taux de perte d'énergie. $F_{E}$ est le flux en énergie observé à haute énergie (E$>$1\,MeV). $d$ est la distance entre le pulsar et la Terre	. $L_{\gamma}$ est la luminosité $\gamma$, tandis qu'$\eta$ est l'efficacité de conversion entre l'énergie de rotation $\dot{E}$ et la luminosité.}{tab:cgropsr}

Une grande partie de l'information que l'on connaît sur les pulsars dérive des propriétés temporelles. Pourtant, les paramètres $\dot{E}$, $\vec{B}_{ns}$, $\tau$, ne répondent pas à la question de l'origine du rayonnement électromagnétique. C'est à partir des profils pulsés et des spectres en énergie résolus en phase que l'on va pouvoir identifier les processus d'émission qui produisent le rayonnement pulsé (voir \ref{sec:predicmodele}). Les sous-sections suivantes présentent les profils pulsés et les spectres en énergies des pulsars détectés par la mission CGRO.

\subsection{Courbes de Lumière}

La figure \ref{fig:cgropulsar} montrent les courbes de lumière\footnote{La courbe de lumière ou profil pulsé est l'histogramme du nombre de coups en fonction du temps, modulo la période du pulsar.} des 7 pulsars détectés par la mission CGRO pour 5 intervalles en énergies~: de haut en bas, radio, optique, rayonnement $X$ mou ($<$1\,keV), $X$ dur/$\gamma$ mou (10\,keV - 1\,MeV), et rayonnement $\gamma$ (au-dessus de 100\,MeV). Les particularités de ces courbes de lumières sont, d'une part leurs profils différents en fonction de la longueur d'onde observée, ce qui illustre une géométrie et des mécanismes d'émission qui dépendent de l'énergie, émission radio cohérente, rayonnement X thermique, et rayonnement $\gamma$ non thermique. D'autre part, pour la plupart des pulsars, les courbes de lumières affichent une structure en double pics couvrant environ 50\% de la rotation, ce qui pourrait signifier un large cône d'émission $\gamma$. PSRs~J1709$-$4429 (B1706$-$44) et J1057$-$5226 (B1055$-$52) semblent avoir un profile $\gamma$ plus complexe. Ainsi, comprendre la différence entre les profils radio (la plupart sont formés d'un pic) et les profils $\gamma$ constitue un défi pour la mission \textit{Fermi}. Notons également que PSR~B1509$-$58 qui a le champ magnétique le plus intense des 7 pulsars, n'a pas été observé par EGRET. 

\fig[Phasogrammes des 7 pulsars de CGRO]{scale=0.3}{c1_phasosEGRET.pdf}{Courbes de lumières pour 5 intervalles en énergie des 7 pulsars détectés par la mission CGRO~: radio, optique, rayonnement $X$ mou ($<$1\,keV), $X$ dur / $\gamma$ mou (10\,keV - 1\,MeV), et rayonnement $\gamma$ ($>$100\,MeV). Les pulsars sont rangés de gauche à droite par âge croissant. Chaque case montre une rotation de l'étoile à neutrons. Cette figure est extrait de la revue de \citet{thompson08}. }{fig:cgropulsar}

\subsection{Spectres Multi-longueurs d'Onde}

La figure \ref{fig:spectracgropsr} présente les spectres multi-longueurs d'onde, de la radio au rayonnement $\gamma$ pour les 7 pulsars détectés par la mission CGRO. La caractéristique commune est une luminosité maximale dans l'intervalle d'énergie $\gamma$, ainsi qu'une réelle différence entre l'émission radio (processus cohérent) et l'émission de haute énergie (accélération de particules chargées), notamment pour les pulsars du Crabe et de Vela. Pour les pulsars de Geminga, B1055$-$52, et Vela, le spectre $X$ indique une émission thermique qui provient de la surface de l'étoile à neutrons, tandis que le spectre $\gamma$ pour les 7 pulsars est relativement plat (indice spectral autour de 2), et montre au moins pour les pulsars EGRET une coupure autour du GeV. Seul PSR~B1951+32 n'en possède pas. Rappelons également qu'une détection pulsée du pulsar du Crabe, au-dessus de 25\,GeV, a été observée par l'expérience Cherenkov MAGIC\footnote{Major Atmospheric Gamma-ray Imaging Cherenkov telescope: http:$//$magic.mppmu.mpg.de/} \citep{magic08}, soulignant l'importance de ces coupures dans les modèles des pulsars (voir la section \ref{sec:predicmodele}). Une limite supérieure de l'émission pulsée avait été déterminée quelques années auparavant par la collaboration CELESTE indiquant une coupure dans le spectre $\gamma$ d'EGRET \citep{denaurois02}.

\fig[Spectres multi-longueurs d'onde des 7 pulsars de CGRO]{scale=0.9}{c1_spectra_cgropsr.pdf}{Spectres multi-longueurs d'onde des 7 pulsars détectés par la mission CGRO \citet{thompson08}.}{fig:spectracgropsr}

\section{Conséquences et Modèles d'\'Emission Gamma}

Les astronomes comprirent rapidement que les pulsars ne pulsaient pas, mais que l'effet de pulsation provenait de la rotation d'un faisceau d'onde électromagnétique émis dans l'environnement du pulsar. Cependant, comment expliquer une émission pulsée sur tout le domaine d'énergie et notamment à haute énergie, comme pour les pulsars du Crabe et de Vela ? Quels sont les processus physiques mis en jeu pour produire des photons jusqu'au GeV ? Où et comment les particules chargées sont-elles accélérées dans la magnétosphère du pulsar ? Est-ce que les processus sont les mêmes pour tous les pulsars ? Quel est le rapport entre les pulsars radio émetteurs $\gamma$ tels que le Crabe ou Vela, et les pulsars $\gamma$ silencieux en radio tel que Geminga\footnote{Dans la littérature, le terme radio-loud est employé pour définir les pulsars émetteurs radio et $\gamma$, tandis que le terme radio-faint est employé pour les pulsars $\gamma$ dont l'émission radio n'a pas été mis en évidence.} ? \\

Seule une petite partie du faisceau électromagnétique balaye notre ligne de visée. La conséquence est qu'il n'est pas simple de construire la géométrie complète de l'émission à partir des observations (voir la figure \ref{fig:cgropulsar}). Théoriquement, l'idée générale est que les émissions de la radio jusqu'aux hautes énergies sont produites dans différentes régions de la magnétosphère du pulsar. Les ondes radio polarisées sont émises depuis la calotte polaire par émission synchrotron, zone au-dessus des pôles magnétiques de l'étoile. Quant aux photons de haute énergie (optique, rayonnement X, et rayonnement $\gamma$), ils sont produits principalement par le rayonnement de courbure et la diffusion inverse Compton des électrons (et positons) accélérées le long des lignes de champ magnétique. Cette propriété est un réel avantage, puisque les particules accélérées rayonnent leur énergie très vite et sont liées aux sites d'accélération. De plus, la luminosité des 6 pulsars détectés par EGRET atteint son maximum dans le domaine des $\gamma$ (voir la figure \ref{fig:spectracgropsr}), laquelle est une fraction de 0.1\% à 10\% de l'énergie cinétique de rotation ($\sim$ $10^{-4}-10^{-5}$ pour la radio). Dans ce sens, l'étude des mécanismes d'accélération des particules passe principalement par l'observation du rayonnement $\gamma$ qui en résulte. Ainsi, l'un des enjeux du télescope Fermi est de contraindre la géométrie et les mécanismes des zones d'accélération, en étudiant en détails les spectres résolus en phase d'une large population de pulsars $\gamma$. \\

Les trois sections suivantes présentent la théorie d'émission standard des pulsars et les trois modèles d'émission $\gamma$, lesquelles se différencient par l'emplacement de l'origine du faisceau, le modèle de la calotte polaire ou \textit{Polar Cap} (PC), le modèle \textit{Slot Gap} (SG) ou \textit{Two Pole Caustic} (TPC) et le modèle de la cavité externe ou \textit{Outer Gap} (OG).

\subsection{Modèle Standard}\label{sec:modstandard}

Comme nous l'avons vu précédemment, il est parfaitement raisonnable de penser que les pulsars sont des étoiles à neutrons en rotation rapide, fortement magnétisées, formées dans l'explosion d'une supernova. Ceci est le principe de base des théories d'émission des pulsars. Nous introduisons ici certaines propriétés importantes sur la magnétosphère des pulsars, où le champ magnétique et le moment angulaire vont jouer un rôle prépondérant dans sa structure. Notons que ces propriétés sont décrites dans le cas relativement simple où l'axe de rotation du pulsar est aligné avec son axe magnétique (modèle du pulsar aligné).

\fig[Illustration des modèles d'émission gamma]{scale=0.6}{c1_ModelesPulsar.pdf}{Illustration des modèles Polar Cap, Slot Gap et Outer Gap. $\Omega$ représente l'axe de rotation, tandis que $B$ représente l'axe magnétique. L'inclinaison magnétique $\alpha$ est l'angle entre $\Omega$ et $B$. Les zones colorées représentent les différents sites d'accélération et d'émission $\gamma$, avec les surfaces de densité de charge nulle ($\rho_{GJ} = 0$) représentées par les lignes bleues.}{fig:standardmodele}

Si l'on considère que l'étoile à neutrons est un parfait conducteur, le champ magnétique dipolaire $\vec{B}$ tournant avec une vitesse angulaire $\vec{\Omega}$ induit près de l'étoile un champ électrique $\vec{E}$ :
\begin{equation}\label{eq:electricfield}
\vec{E} + \frac{1}{c} (\vec{\Omega} \times \vec{r}) \times \vec{B} = 0.
\end{equation}
\citet{goldreich69} montrent dans le cas où l'axe de rotation est aligné à l'axe magnétique, que les forces électromagnétiques en dehors de l'étoile dominent complètement les processus physiques, et sont suffisamment intenses pour arracher les particules de la surface de l'étoile et créer une magnétosphère composée d'un plasma chargé. La densité de charge, appelée également la densité de charge de Goldreich-Julian $\rho_{GJ}$, est définie par :
\begin{equation}\label{eq:chargedensity}
\rho = \frac{1}{4\pi} \vec{\nabla} \cdot \vec{E} \approx - \frac{\vec{\Omega} \cdot \vec{B}}{2 \pi c} \equiv \rho_{GJ} ,
\end{equation}
tandis que la densité de particules est donnée par :
\begin{equation}
n = 7 \times 10^{-2} \ B_{z} \ P^{-1} \ \ {\rm particules\ cm^{-3}},
\end{equation}
avec $B_{z}$ la composante axiale du champ magnétique en gauss, et $P$ la période de rotation. \\

Le champ électrique intense créé influence la région comprise entre le pulsar et le cylindre de lumière $r_{c} = c/\Omega$, distance où les lignes de champ magnétique et les particules en corotation avec l'étoile à neutrons atteignent la vitesse de la lumière. La figure \ref{fig:standardmodele} montre la géométrie de la magnétosphère et la zone d'émission des modèles $\gamma$ dans le cas où l'axe de rotation et l'axe magnétique ne sont pas alignés. La zone fermée (partie grisée) contient le plasma en corotation, tandis que la zone ouverte permet au plasma de s'écouler le long des lignes de champ, et de s'échapper dans le milieu interstellaire créant un ``vent'' relativiste\footnote{Ce vent en contact avec le milieu interstellaire peut créer une source : une nébuleuse à vent de pulsar (PWN: Pulsar Wind Nebula).}. Les zones au niveau de la surface de l'étoile d'où proviennent les lignes ouvertes sont définies comme les calottes polaires, lesquelles sont délimitées par la dernière ligne de champ fermée, soit par un cône de demi-angle d'ouverture $\theta_{cp}$ :
\begin{equation}\label{eq:cone}
sin (\theta_{cp}) = (\frac{\Omega R_{ns}}{c})^{1/2},
\end{equation}
avec $R_{ns}$ le rayon de l'étoile à neutrons. En supposant que le rayon du cylindre de lumière $r_{c}$ est très supérieur à celui de l'étoile à neutrons, le rayon de la calotte polaire $r_{cp}$ à la surface de l'étoile est donnée par:
\begin{equation}\label{eq:rayoncp}
r_{cp} = R_{ns} \theta_{cp} \simeq (\frac{\Omega R^{3}_{ns}}{c})^{1/2}.
\end{equation}
Le flux de particules chargées accélérées le long des lignes de champs ouvertes au niveau de la surface de l'étoile est donné par :
\begin{equation}
\dot{N}_{cp} \simeq (\pi r^{2}_{cp}) (\frac{\rho}{e}) \simeq \frac{\Omega^{2} R^{3}_{ns} B_{s}}{2 e c},
\end{equation}
avec $B_{s}$ le champ magnétique à la surface de l'étoile. La différence de potentiel entre le pôle de l'étoile et la dernière ligne de champ fermée est proportionnelle à la perte de l'énergie de rotation \ref{eq:energie_rotation}, et peut être exprimée à partir des équations \ref{eq:electricfield} et \ref{eq:cone} par:
\begin{equation}\label{eq:diffpot}
\Delta \Phi = \int E \cdot ds \simeq \frac{\Omega^{2} R^{3}_{ns} B_{s}}{2 c^{2}} \propto \dot{E}^{1/2}.
\end{equation}

Les grandeurs données aux équations \ref{eq:cone}, \ref{eq:rayoncp}, et \ref{eq:diffpot} définissent les principales caractéristiques du modèle standard des pulsars. En appliquant ces équations au pulsar du Crabe, le demi-angle d'ouverture de la calotte polaire obtenu est de $\theta_{cp} = 4.8$\,$^{\circ}$ avec un rayon $r_{cp} = 0.84$\,km, tandis que la différence de potentiel atteinte entre le centre et le bord de la calotte est de $\Delta\Phi \simeq 2 \times 10^{16}$\,V.

\subsection{Les Modèles d'\'Emission Gamma}\label{sec:modgam}

D'après \citet{goldreich69}, le champ électrique induit par le champ magnétique dipolaire de l'étoile à neutrons est si intense que la magnétosphère se remplit de particules chargées, arrachées à la surface de l'étoile. L'équation \ref{eq:chargedensity} implique à l'intérieur de la magnétosphère une densité de charge en corotation avec le champ magnétique et qui écrante la composante du champ électrique parallèle au champ magnétique ($\vec{E} \cdot \vec{B} = 0$). Cependant, il existe des zones définies par $\vec{\Omega} \cdot \vec{B} = 0$ où cette densité est nulle, et où le champ électrique n'est pas écranté ($\vec{E} \cdot \vec{B} \neq 0$) (voir la figure \ref{fig:standardmodele}). Cette propriété joue un rôle essentiel puisque à l'intérieur de ces cavités les particules chargées sont accélérées à des énergies ultra-relativistes. Les processus physiques mis en jeu pour créer des photons de haute énergie sont le rayonnement de courbure, l'émission synchrotron, et la diffusion inverse Compton. Pour plus détails sur ces processus, on se reportera à l'article de \citet{rudak02}.

Les tentatives pour modéliser les accélérateurs de particules dans la magnétosphère du pulsar sont réparties en deux scénarios. Les modèles  \textit{Polar Cap} et \textit{Slot Gap} placent la source d'émission au-dessus des pôles magnétiques jusqu'à quelques rayons de l'étoile à neutrons, tandis que le modèle \textit{Outer gap} propose une émission près du cylindre de lumière. \\

\subsubsection{Polar Cap et Slot Gap}\label{sec:polarcap}

Le modèle Polar Cap fut proposé par \citet{sturrock71} pour expliquer l'émission radio. Il suppose que les particules arrachées à la surface de l'étoile sont accélérées dans une petite zone au-dessus du pôle magnétique par le champ électrique (voir la figure \ref{fig:goldjulmodel}). A partir de l'approximation $\rho = (1/4 \pi)\nabla^{2} \Phi$ et de l'équation \ref{eq:chargedensity}, le potentiel responsable du flux de particules chargées est donné par $\Phi \sim (2\Omega B_{s}/c)h^{2}$, avec $h \sim r_{cp}$, la hauteur de cette région. \citet{sturrock71} démontre que ce potentiel est suffisant pour accélérer les particules à des vitesses relativistes ($\gamma \leq 10^{7}$) le long des lignes de champ magnétique. Ces particules primaires (e.g. électrons) rayonnent des photons $\gamma$ par rayonnement de courbure \citep{ruderman75}, dont l'énergie caractéristique est :
\begin{equation}
\epsilon_{\gamma} = \frac{3}{2} c \hbar \frac{\gamma^{3}}{\rho_{cr}} \ {\rm eV},
\end{equation}
avec $\gamma$ le facteur de Lorentz et $\rho_{cr}$ le rayon de courbure de la ligne de champ. Lorsque ces photons se déplacent à travers le champ magnétique intense, ils créent des paires électron-positon ($\gamma + B \rightarrow e^{+} e^{-}$) \footnote{Un photon d'énergie $E_{\gamma}$ se propageant selon un angle $\psi$ par rapport à la direction du champ magnétique, peut se matérialiser en paire e$^{\pm}$ si $E_{\gamma} sin(\psi) \geq 2 m_{e} c^{2}$.}. Ces nouvelles particules chargées sont elles-mêmes accélérées et produisent à leur tour des $\gamma$ qui se matérialisent en paires e$^{\pm}$ et ainsi de suite. Ces cascades électromagnétiques s'échappant le long des lignes de champ ouvertes, alimentent le plasma de la magnétosphère et écrantent le champ accélérateur $\vec{E}_{\parallel}$, limitant la zone accélératrice. Les paires secondaires (voire tertiaires) produites le long de l'axe magnétique, à basse altitude, sont à l'origine de l'émission radio observée. 

\fig[Schéma de la magnétosphère d'un pulsar]{scale=0.25}{c1_PSRs_magnetosphere.png}{Schéma de la magnétosphère d'un pulsar dans le cas du modèle de Goldreich-Julian. La figure de droite illustre les particules chargées arrachées à la surface de l'étoile. Les particules accélérées par un champ électrique intense émettent des photons $\gamma$ par rayonnement de courbure.}{fig:goldjulmodel}

L'émission observée à haute énergie est due, quant à elle, à la diffusion inverse Compton des paires e$^{\pm}$ sur les photons infra-rouges \citep{daugherty86}. Ce mécanisme est très efficace dans le perte d'énergie des particules chargées, produisant des photons $\gamma$ plus énergétiques que dans le processus de rayonnement de courbure. Seulement, étant donné que l'émission est située au-dessus des pôles, le champ magnétique est intense. Le modèle prédit ainsi une coupure de type super-exponentielle dans le spectre pulsé du pulsar en raison de l'absorption des photons $\gamma$ par le processus de création de paires. Notons finalement que ce modèle reproduit correctement le spectre résolu en phase observé pour le pulsar de Vela, mais sous certaines conditions, telles qu'une zone d'accélération à trois rayons de l'étoile. \\

Pour atténuer les effets du champ magnétique sur les photons $\gamma$, \citet{arons83} a examiné un autre site d'accélération compris entre la calotte polaire et le cylindre de lumière. Le site est relativement fin et confiné le long de la dernière ligne de champ ouverte (zone rose sur la figure \ref{fig:standardmodele}). Ce type d'accélérateur est appelé ``Slot Gap'' (SG) ou ``Two-Pole Caustic'' (TPC) \citep{dyks03,harding06}. Il reproduit correctement les propriétés de polarisation, les profils pulsés et les spectres résolus en phase pour le pulsar du Crabe entre 0.3\,keV et 10\,GeV, ainsi que le spectre à haute énergie du pulsar milliseconde PSR~J0218+4232 \citep{kuiper00}.

\subsubsection{Modèle de la Cavité Externe: Outer Gap}\label{sec:outergap}

Dans le modèle \textit{Outer Gap} \citep{cheng86,romani95}, l'émission $\gamma$ se produit près du cylindre de lumière, entre la ligne de densité de charge nulle ($\vec{\Omega} \cdot \vec{B} = 0$), la dernière ligne de champ fermée, et le cylindre de lumière (section \ref{sec:modstandard}). La figure \ref{fig:standardmodele} illustre la zone d'émission (zone bleue) dans le cas du modèle OG. \citet{holloway73} appuie l'idée que les charges qui s'échappent de cette région ne sont pas remplacées, et laissent une cavité vide séparée par des densités de charges opposées. La conséquence est l'apparition d'un champ électrique parallèle aux lignes de champ magnétique. 

Les électrons et les positons se propageant le long des lignes de champs près de cette cavité émettent des photons par rayonnement de courbure ou par diffusion inverse Compton sur des photons de plus basse énergie. Ces photons créés vont pénétrer dans la cavité, et à leur tour se matérialiser en paires $e^{\pm}$. La différence de potentiel présente dans la cavité va accélérer les particules chargées, lesquelles vont émettre des photons de haute énergie. Comme dans le modèle PC, ce flux de particules chargées va délimiter la cavité accélératrice et maintenir un intense courant chargé à l'origine de l'émission $\gamma$. Généralement, ces cavités externes ne sont pas prévues pour être associées à l'émission radio, bien que le Crabe soit peut-être une exception. De plus, le champ magnétique local étant beaucoup plus faible qu'aux pôles, l'atténuation spectrale à haute énergie est moins importante que dans le modèle PC.

Selon ce modèle, on distingue deux types de pulsars, associés au Crabe et à Vela. Pour les pulsars de type Crabe relativement jeunes, la cavité produit des photons au GeV par rayonnement de courbure, puis produit des électrons et des positons, lesquels rayonnent par émission synchrotron et par diffusion inverse Compton. Dans le cas des pulsars de type Vela, les paires $e^{\pm}$ sont créées dans la cavité, et accélérées à des énergies relativistes dans le sens opposé au vent. Ces particules chargées primaires vont produire des photons $\gamma$ par diffusion inverse Compton sur des photons infrarouge (IR). Les photons $\gamma$ suffisamment énergétiques vont à leur tour se matérialiser en paires $e^{\pm}$, par interaction avec la composante IR. Suite à cela, l'émission synchrotron de cette deuxième population de particules chargées va donner un faisceau de photons $X$ et $\gamma$, qui vont interagir pour produire une troisième population de paires $e^{\pm}$ moins énergétique. Les photons IR seraient créés par l'émission synchroton de ces paires tertiaires. Notons également que ce modèle prédit une diffusion inverse Compton des photons IR sur les particules primaires; cette diffusion serait par ailleurs à l'origine d'une émission au TeV qui n'a pas été encore observée.

Finalement, les deux modèles sont assez différents. Dans le cas des pulsars de type Crabe, les paires $e^{\pm}$ primaires perdent la quasi totalité de leur énergie par le rayonnement de courbure, tandis que dans le cas des pulsars de type Vela, les paires perdent leur énergie par émission synchrotron. La conséquence est que la luminosité $\gamma$ prédite dans les deux modèles est différente. Pour les pulsars de type Crabe, elle est donnée par :
\begin{equation}
L_{\gamma} \cong 1.5 \times 10^{37} \ P \ \ {\rm erg \ s^{-1}},
\end{equation}
avec $P$ la période de rotation. Pour les pulsars de type Vela, elle est donnée par :
\begin{equation}
L_{\gamma} \propto P^{-4} \ B^{2}_{S} \ \ {\rm erg \ s^{-1}},
\end{equation}
avec $P$ la période de rotation et $B_{S}$ le champ magnétique de surface.

\subsection{Prédictions des Modèles}\label{sec:predicmodele}

\subsubsection{Courbes de Lumière}

La figure \ref{fig:phasovela} présente le profil pulsé $\gamma$ du pulsar de Vela observé par EGRET (en haut à gauche), ainsi que les courbes de lumières prédites par les modèles PC, SG, et OG. Les trois modèles arrivent à produire un profil pulsé avec deux pics. Cependant, dans le cas du modèle PC, l'observation d'un tel profil demande un angle d'inclinaison très petit ($\leq 10$\,$^{\circ}$). Les deux autres modèles génèrent des caustiques\footnote{L'effet de caustique est dû aux photons émis à des altitudes différentes de l'étoile le long des lignes de champ magnétique. Ceci est d'autant plus vrai que l'inclinaison magnétique $\alpha$ est grande.} \citep{morini83} en raison de l'émission $\gamma$ à plus haute altitude dans la magnétosphère. L'émission n'a pas la forme d'un faisceau comme en radio, mais peut être vue en forme d'``éventail''. Pour le modèle SG, on observe l'émission de la caustique des 2 pôles, tandis que pour le modèle OG, seule l'émission d'un des deux pôles est visible.

\fig[Profil pulsé du pulsar de Vela vu par EGRET]{scale=0.8}{c1_phaso_vela_alice.pdf}{\textbf{En haut à gauche:}~Profil pulsé du pulsar de Vela, observé par le télescope EGRET au-dessus de 30 MeV. \textbf{Autres figures:}~Profils pulsés prévus par les modèles Polar Cap, Slot Gap, et Outer Gap. L'encadré du haut représente la phase en fonction de l'angle d'observation ($\zeta$) à partir de l'axe de rotation du pulsar, tandis que l'encadré du bas représente le nombre de coups en fonction de la phase $\phi$ pour une inclinaison donnée $\alpha$. Les chiffres 1 et 2 définissent les 2 pôles de l'étoile. La figure est extraite de \citep{harding07b}.}{fig:phasovela}

\subsubsection{Spectres en \'Energie}

Le modèle Polar Cap prédit une coupure rapide (de type super-exponentielle) dans le spectre pulsé à haute énergie, due à la fois à la limite énergétique des processus d'émission $\gamma$ et à l'atténuation des photons par la création de paires $e^{\pm}$ avec le champ magnétique. Les modèles Slot Gap et Outer Gap annoncent, quant à eux, une coupure moins brutale, car leur émission est à plus haute altitude. L'énergie de ces coupures est estimée entre 1 et 20\,GeV et dépend principalement du champ magnétique à la surface de l'étoile, de la période du pulsar, et de l'altitude d'émission. Le spectre en énergie des pulsars peut être décrit par une loi de puissance multipliée par une coupure exponentielle ou super-exponentielle :
\begin{equation}\label{eq:powerlawexpcutoff}
\frac{dF}{dE} = K \ E^{-\Gamma} e^{(E/E_{Cutoff})^{\beta}},
\end{equation}
avec $K$ un facteur de normalisation, $\Gamma$ l'indice spectral, $E_{Cutoff}$ la coupure en énergie, et $\beta$ l'indice d'atténuation ($\simeq 1$ pour les modèles OG et SG, et $\simeq 2$ pour le modèle PC). Cette interprétation proposée par \citet{nel95} est la conséquence des différents processus d'émission cités précédemment. La figure \ref{fig:spectrevela} présente la simulation du spectre de Vela dans le cas des modèles PC et OG. \\

Finalement, l'un des enjeux de la mission Fermi et de cette thèse est de mesurer la coupure de ces spectres pour pouvoir discriminer l'un des modèles, voire de les rejeter. La partie \textit{Résultats} de cette thèse présente, entre autres, l'analyse spectrale des pulsars détectés par le \textit{Large Area Telescope}. Cependant, la détermination des paramètres spectraux dépend d'une bonne calibration de l'appareil, et notamment de la surface efficace (Chapitre \ref{chap:irfs}). 

\fig[Spectre à haute énergie du pulsar de Vela]{scale=0.9}{c1_spectrevela.pdf}{Spectre à haute énergie du pulsar de Vela. Les croix sont les données du télescope EGRET. La ligne pointillée représente les prédictions du modèle Outer Gap, tandis que la ligne discontinue représente les prédictions du modèle Polar Cap. Les barres d'erreurs sont estimées pour un an d'observation. La figure est extraite de \citep{mcenery04}.}{fig:spectrevela}                 
\chapter[Les Pulsars Candidats Gamma]{Les Pulsars Candidats Gamma}\label{chap:candidat}

\minitoc

\section{Population Galactique}

\subsection{Distribution Galactique}\label{sec:distribpsr}

Plus de quarante années après la découverte du premier pulsar radio \citep{hewish68}, la population de pulsars connus à ce jour excède 1800 objets. La majorité provient des observations du ciel par les radiotélescopes de Jodrell Bank (Angleterre, \citealt{hobbs04}), Green Bank Telescope (Etats-Unis,  \citealt{kaplan05}), Nançay (France, \citealt{theureau05}), Arecibo (Porto Rico, \citealt{dowd00}), et Parkes (Australie, \citealt{manchester08}). En fait, la moitié des découvertes résulte de la campagne multi-faisceaux du télescope de Parkes durant ces 8 dernières années \citep{manchester01}. La figure \ref{fig:distribpulsar} présente la répartition en coordonnées galactiques des 1827 pulsars\footnote{Ce chiffre provient de la base de données $ATNF$ (Australia Telescope National Facility). Parmi ces détections, 8 pulsars ont été observés dans les nuages de Magellan, et environ une centaine dans les amas globulaires.} découverts dans notre Galaxie, et les nuages de Magellan. Il faut noter que cette distribution est limitée par la sensibilité des instruments, et biaisée par la sélection de la région observée. Par exemple, la concentration de pulsars vers la longitude 50\degr\ provient d'une recherche approfondie du télescope d'Arecibo. D'autre part, les pulsars observés sont fortement sélectionnés par leur flux apparent $F \propto L / d^{2}$. En effet, les pulsars dont la luminosité $L$ est faible, ne sont vus qu'à des petites distances $d$. De plus, les effets du milieu interstellaire sur la propagation du signal radio (voir \ref{sec:dispersion}) sont très importants dans le plan galactique, d'autant plus qu'on se rapproche du centre. La conséquence de ces nombreux effets est que l'échantillon actuel représente juste une petite part de la population totale des pulsars actifs dans la galaxie. Celle-ci est estimée entre $10^{5}$ et $10^{6}$ \citep{psrastro}.

\fig[Distribution des pulsars du catalogue ATNF]{scale=0.65}{c2_sky_map_pulsar.jpg}{Distribution des pulsars connus dans le ciel en coordonnées galactiques, en projection de Hammer-Aitoff (source:ATNF).}{fig:distribpulsar}

La figure \ref{fig:distribpulsar} indique que les pulsars se répartissent essentiellement dans le plan galactique comme les étoiles massives O et B, sur une épaisseur d'environ 1 kiloparsec (kpc) pour un rayon de 10 kpc autour du centre galactique. Cette observation conforte l'hypothèse que les étoiles à neutrons naissent de l'effondrement gravitationnel des étoiles massives. \citet{kaspi98} associe une vingtaine de pulsars radio à des restes de supernova. Il apparaît également que de nombreux pulsars sont en dehors du plan ($\mid b \mid$ grand). Ceci peut être dû à la proximité du pulsar par rapport à la Terre, ou il est possible que les explosions violentes de supernova ne soient pas totalement symétriques, donnant au pulsar une impulsion à leur naissance \citep{janka05}. La mesure des mouvements propres d'un échantillon de 233 pulsars montre une vitesse moyenne de propagation de $246 \pm 22$\,km\,s$^{-1}$ \citep{hobbs06b}. Les plus rapides peuvent atteindre 1000\,km\,s$^{-1}$. La conséquence est qu'une fraction importante des pulsars s'échappe de leur reste de supernova, et subséquemment du plan galactique.

\fig[Diagramme $P-\dot{P}$ (version EGRET)]{scale=0.6}{c2_ppdot_distrib_v1.pdf}{Diagramme $P-\dot{P}$, distribution des pulsars en fonction de leur période de rotation et de leur dérivée. Les lignes vertes représentent l'âge caractéristique. Les lignes bleues représentent le champ magnétique à la surface de l'étoile, tandis que les lignes rouges représentent le taux de perte d'énergie de rotation $\dot{E}$. Les points noirs sont les pulsars listés dans le catalogue ATNF. Les étoiles bleues sont les pulsars détectés par la mission EGRET avec un bon niveau de confiance, tandis que les triangles rouges sont ceux détectés avec un faible niveau de confiance. Les principales caractéristiques des 9 pulsars EGRET sont énumérées dans le tableau \ref{tab:cgropsr}.}{fig:ppdot}

\subsection{Diagramme $P$-$\dot{P}$}

Dans le chapitre \ref{chap:pulsar}, nous avons vu que seules les mesures de la période de rotation $P$ et du taux de freinage $\dot{P}$ pouvaient nous  renseigner sur l'évolution de la rotation et les propriétés intrinsèques des pulsars. La figure \ref{fig:ppdot} appelée le diagramme $P-\dot{P}$, montre la distribution des pulsars en fonction de la période de rotation et de sa dérivée. Deux groupes sont clairement distincts. On y retrouve en haut à droite les pulsars normaux ($P \sim 0.5$\,s et $\dot{P} \sim 10^{-15}$\,s\,s$^{-1}$, section \ref{sec:psrnorm}), tandis que le petit groupe en bas à gauche représente les pulsars milliseconde ($P \sim 3$\,ms et $\dot{P} \sim 10^{-20}$\,s\,s$^{-1}$, section \ref{sec:msp}). La différence dans $P$ et $\dot{P}$ implique fondamentalement des différences d'âge et de champ magnétique pour ces deux populations. Le champ magnétique $B \propto (P \dot{P})^{0.5}$, l'âge caractéristique $\tau \propto P / \dot{P}$ sont indiqués par différentes lignes. Le taux de perte d'énergie de rotation $\dot{E} \propto \dot{P} / P^{3}$ est également indiqué. Les pulsars naissent avec une période relativement petite et subissent un ralentissement. L'évolution naturelle pour un pulsar normal est ainsi de migrer des petites périodes de rotation (région en haut à gauche) vers les plus grandes périodes dans un intervalle de temps de 10$^{5\ -\ 6}$ années. Au delà de 10$^{7}$ années, le flux du pulsar est certainement trop faible pour être détectable. Remarquons que le pulsar du Crabe et de Vela font partie de la classe de pulsars la plus jeune, la plus énergétique, avec un champ magnétique intense.

Les étoiles bleues représentent les pulsars $\gamma$ détectés par la mission EGRET, tandis que les triangles rouges représentent les détections marginales (\ref{sec:egretpsr}). Ces pulsars sont relativement jeunes et sont concentrés (à l'exception du pulsar milliseconde candidat) dans la région où le champ magnétique est intense. Tous ont également un taux de perte d'énergie de rotation $\dot{E}$ élevé. Pour le Crabe, il est de $4.5 \times 10^{38}$\,erg\,s$^{-1}$, tandis que pour Vela, il est de $7 \times 10^{36}$\,erg\,s$^{-1}$. On constate ainsi que les pulsars observés à haute énergie ont un profil $P$-$\dot{P}$ similaire, ce qui va permettre de sélectionner nos pulsars candidats $\gamma$.

\subsection{Pulsars Normaux}\label{sec:psrnorm}

La première classe définit les pulsars ``normaux'' que l'on observent après la phase de supernova, tels que PSR~B1919+21 (le premier pulsar détecté), et les pulsars du Crabe et de Vela. Ils ralentissent à cause du rayonnement dipolaire et s'éteignent au bout d'une dizaine de millions d'années (voir \ref{sec:chpmag}). Ils sont concentrés le long du plan galactique et représente 90\% des pulsars observés. Signalons que dans cette thèse, nous analyserons essentiellement des pulsars normaux (chapitres 6 à 9).

\subsection{Pulsars Milliseconde}\label{sec:msp}

La seconde classe identifie les pulsars ``milliseconde''. 80\% d'entre eux sont présents dans des systèmes binaires (moins de 1\% pour les pulsars normaux). Initialement, le système est composé de deux étoiles de la séquence principale. L'étoile la plus massive va évoluer plus vite et s'effondrer pour former une étoile à neutrons. Dans moins de 10\% des cas, malgré la violence de l'explosion de la supernova, l'étoile reste liée gravitationnellement à son compagnon. Si les paramètres orbitaux sont favorables, le champ gravitationnel intense de l'étoile à neutrons attire de la matière de l'étoile compagnon et forme un disque d'accrétion. Le résultat de ce transfert de masse et de moment angulaire, est un rayonnement X intense du système, ainsi qu'une accélération de la rotation de l'étoile à neutrons jusqu'à des périodes de l'ordre de la milliseconde ($P<30$\,ms). Les particules chargées dans l'environnement du nouveau pulsar vont être accélérées et produire un faisceau d'onde électromagnétique. 

L'âge caractéristique des pulsars milliseconde est compris entre 10$^{8}$ et 10$^{10}$ années. Leur taux de freinage est relativement petit ($\dot{P}<10^{-17}$\,s\,s$^{-1}$) comparé à celui des pulsars normaux. Quant au champ magnétique de surface, il est possible qu'il se dissipe lors de la phase d'accrétion \citep{bisnovatyi74}. Il est 10$^{4}$ plus petit que pour les pulsars normaux. Finalement, leur vitesse de propagation est de $87 \pm 13$\,km\,s$^{-1}$ en moyenne.

Le premier pulsar milliseconde a été découvert en 1982 à Arecibo avec une période de 1.6\,ms \citep{backer82}. A ce jour, 168 pulsars sont listés dans le catalogue $\textit{ATNF}$ avec une période inférieure à 30\,ms et un freinage  $\dot{P}<10^{-17}$. Parmi eux, 96 se trouvent dans les amas globulaires \citep{camilo05}, et 8 dans les nuages de Magellan. Pour plus de détails, on se reportera à la thèse intitulée ``Détections de pulsars milliseconde avec le FERMI Large Area Telescope'' \citep{guillemot09}.

\section{Distance des pulsars}\label{sec:distance}

Dans cette section, nous présentons les différentes méthodes utilisées pour estimer la distance des pulsars. Cette quantité difficile à mesurer de façon directe (\ref{sec:parallaxe}) est essentielle à la fois pour les études statistiques des régions de formation des étoiles massives, mais également pour comprendre les propriétés physiques des pulsars, notamment la luminosité. A noter qu'aucune relation ne permet encore d'évaluer la distance à partir des paramètres des pulsars tels que leur luminosité $\gamma$. Cependant, la situation pourrait changer, étant donné le nombre grandissant de pulsars $\gamma$ détectés avec la mission $Fermi$. La distance est également primordiale pour sélectionner nos candidats $\gamma$. Rappelons que le flux observé est une fonction de la luminosité du pulsar, mais aussi de la distance au carré ($F \propto L / d^{2}$). \\

Il existe plusieurs méthodes pour évaluer la distance des pulsars, soit par la mesure de leur parallaxe (\ref{sec:parallaxe}), soit par l'absorption d'hydrogène neutre et l'association d'objet (\ref{sec:absHI}), ou encore par la mesure de la dispersion du signal radio couplé à un modèle de la distribution de densité des électrons; les deux les plus utilisés sont le modèle de \citet{taylor93} et le NE~2001 \citep{cordes02} (\ref{sec:dispersion}). Les 2 premières méthodes pour estimer la distance permettent de contraindre les modèles de densité des électrons. Le tableau \ref{tab:psrcatdistance} présente la distance des pulsars $\gamma$ du premier catalogue de pulsars du \textit{Large Area Telescope} (Chapitre \ref{chap:catalog}).

\subsection{Parallaxe}\label{sec:parallaxe}

Les distances des pulsars les plus proches, situés à moins de 1\,kpc du Soleil, peuvent être obtenues à partir de la mesure de leur parallaxe. La méthode consiste à mesurer la courbure des fronts d'onde radio provenant du pulsar au long d'une année complète. La courbure est évaluée par la mesure très précise du temps d'arrivée des pulsations à différents endroits de l'orbite terrestre. 

\subsection{Absorption d'Hydrogène et Association d'Objet}\label{sec:absHI}

Le spectre des pulsars situés à des latitudes galactiques faibles, à l'intérieur des bras spiraux de la Galaxie, peuvent présenter une raie d'absorption de l'hydrogène neutre. L'hydrogène présent sous forme de nuage absorbe le rayonnement à la longueur d'onde de 21\,cm (1420\,MHz). Cette raie H$_{I}$ peut montrer un décalage spectral dû à l'effet Doppler causé par le déplacement du gaz dans les bras spiraux. A partir des modèles dynamiques de la rotation de la Galaxie, le décalage spectral de l'hydrogène peut être relié à une distance \citep{fich89}. Cette technique peut être également utilisée pour estimer la distance d'un objet associé au pulsar (ex: restes de supernova, nuages de Magellan, amas globulaires), par l'absorption ou l'émission de l'hydrogène neutre. Dans le cas du pulsar PSR~J0205+6449, la distance est déterminée à partir du PWN découvert avant le pulsar \citep{roberts93}. 

Les mesures d'absorption de l'hydrogène neutre (H$_{I}$) à 21\,cm sont une importante estimation des diverses propriétés du milieu interstellaire, et permettent de contraindre les modèles de densité des électrons.

\subsection{Mesure de Dispersion}\label{sec:dispersion}

Avant d'arriver sur Terre, la large bande d'émission radio émise par le pulsar traverse un plasma composé principalement d'électrons. Celui-ci provient du milieu interstellaire. La conséquence est que les ondes radio de basse fréquence $\nu_{1}$ arrivent avec un retard par rapport aux ondes radio de haute fréquence $\nu_{2}$. Celle-ci peut en principe être corrigée avec exactitude. Le retard en temps relatif entre deux signaux pulsés pour deux fréquences différentes ($\nu_{1}$ et $\nu_{2}$) peut être énoncé par :
\begin{equation}\label{eq:tpsrela}
\Delta t = \frac{1}{K} (\frac{1}{\nu^{2}_{1}} - \frac{1}{\nu^{2}_{2}}) DM, 
\end{equation}
où $K = 2.410331 \times 10^{-4}$\,MHz$^{-2}$\,cm$^{-3}$\,s$^{-1}$\,pc est la constante de dispersion \citep{backer93} et
\begin{equation}
DM = \int_{0}^{L} n_{e} dl,
\end{equation}
est la mesure de dispersion (pc\,cm$^{-3}$) avec $L$ la distance entre la Terre et le pulsar en parsec et $n_{e}$ la densité d'électrons en cm$^{-3}$. L'acronyme DM provient du terme anglais ``Dispersion Measure''. Généralement, la dispersion est déterminée durant la découverte du pulsar, à partir du temps d'arrivée des photons et de deux fréquences différentes (voir l'équation \ref{eq:tpsrela}). La figure \ref{fig:dispersion} montre le signal reçu pour chaque canal de fréquence en fonction de la phase du pulsar pour PSR~J0437$-$4714. Signalons que pour les photons X ou $\gamma$, ce plasma n'a aucun effet. \\

\fig[Effet de la dispersion du signal]{scale=0.5}{c2_dispersion.jpg}{\textbf{Gauche:}~Signal radio reçu pour chaque canal de fréquence en fonction de la phase du pulsar pour PSR~J0437$-$4714. En l'absence de correction de dispersion du signal, les basses fréquences radio arrivent après les hautes fréquences. \textbf{Droite:}~Profil radio après l'application de la mesure de dispersion. Les données proviennent du télescope radio de Parkes en Australie. La figure est extraite de \citet{han09}}{fig:dispersion}

La mesure de dispersion est un excellent outil pour non seulement identifier l'émission radio des pulsars, mais aussi pour modéliser la structure du gaz ionisé dans le milieu interstellaire. \citet{taylor93} et \citet{cordes02} ont construit des modèles quantitatifs sur la distribution des électrons libres dans la galaxie. Ces modèles sont contraints à partir des mesures de dispersion des pulsars et de leurs distances obtenues par des mesures indépendantes (voir \ref{sec:parallaxe} et \ref{sec:absHI}). Ils sont modélisés suivant la forme des bras spiraux, qui dérivent d'observations radio et optiques des régions $H_{II}$ . Généralement, on suppose des densités uniformes d'électrons dans les bras spiraux et des transitions entre les bras. La conséquence est qu'à partir d'un modèle de densité d'électrons et de la mesure de la dispersion, on peut estimer la distance du pulsar \footnote{http:$//$rsd-www.nrl.navy.mil/7213/lazio/ne$\_$model/}. Cependant, l'incertitude de ces distances dérivées peut excéder 50\% pour certains pulsars. N'oublions pas que le milieu interstellaire est irrégulier. Le signal radio peut donc être détérioré en fonction du temps et de la fréquence observée.

\tab[Estimation de la Distance des Pulsars]{DistanceTable.tex}{Estimation de la distance des pulsars du catalogue (chapitre \ref{chap:catalog}). \\
HI: distance évaluée par l'observation d'HI; P: mesure de la parallaxe; DM: mesure de dispersion \citep{cordes02}; A: autres mesures. La référence des distances pour chaque pulsar est indiquée dans le papier \citet{abdo09psrcat} 
}{tab:psrcatdistance}

\section{Les Candidats Pour Fermi}\label{sec:candidat}

\subsection{La Campagne pour les \'Ephémérides}\label{sec:campephem}

Parmi les 1827 pulsars connus \citep{atnf}, seulement sept pulsars ont été détectés comme émetteurs $\gamma$ par la mission CGRO. Avant le lancement du Large Area Telescope (Chapitre \ref{chap:lat}), les prévisions sur les découvertes de pulsars $\gamma$ par le télescope tablaient entre une dizaine et plusieurs centaines de détections selon les modèles \citep{harding07,jiang06}. Les deux méthodes pour la détection de pulsations $\gamma$ sont: d'une part la recherche de pulsations en aveugle en utilisant uniquement les données du LAT \citep{atwood06,abdo09blindsearch}, et d'autre part la recherche de pulsations pour des pulsars déjà connus dans une autre longueur d'onde.

Nous avons fait le choix à Bordeaux d'étudier les pulsars préalablement observés à une autre fréquence, le plus souvent en radio. La collaboration $Fermi$ a ainsi démarré, à partir de 2007, une campagne de suivi des pulsars, avec la coopération des radiotélescopes énumérés à la section \ref{sec:distribpsr}, et des télescopes $XMM$ et $RXTE$ \citep{smith08}. Les communautés radio et $X$ se sont engagées à fournir les éphémérides des pulsars (voir la section \ref{sec:ephem}), c'est-à-dire, les paramètres de rotation qui permettent en contrôlant les variations de la rotation du pulsar, d'intégrer à la bonne phase rotationnelle les photons $\gamma$ qui nous parviennent. Seulement, le LAT re\c{c}oit du pulsar de Vela, qui est la source la plus émissive du ciel $\gamma$, en moyenne un photon tous les 4 minutes. Cela est très peu comparé aux autres longueurs d'onde. Pour les pulsars les plus faibles, le temps entre deux évènements peut être de quelques jours. Il est donc indispensable pour une détection efficace des pulsars $\gamma$ et une analyse rigoureuse, d'avoir des éphémérides de qualité. Les pulsars montrent également des irrégularités dans leur rotation (\textit{timing noise} et \textit{glitch}). La conséquence est qu'il est très difficile d'extrapoler la période de rotation et ses dérivées sur une très grande période. Un exemple concret est la découverte des pulsars avec un fort $\dot{E}$, PSR~J2229+6114 et PSR~J1410$-$6132, après la fin de la mission CGRO. Une recherche de pulsations dans les archives des données EGRET a été effectuée, mais a échoué pour trouver un signal pulsé car les photons étaient trop peu nombreux et trop vieux comparés aux éphémérides \citep{thompson02,obrien08}. L'observation des pulsars par les télescopes radio et $X$ doit être régulière, mais représente un coût, ce qui justifie de déterminer une liste bien précise de pulsars candidats. La section suivante présente la sélection de nos pulsars candidats $\gamma$.



\subsection{La Sélection}\label{sec:selection}

\citet{arons96} montre qu'au-dessus d'une certaine valeur du potentiel au niveau des lignes de champs ouvertes $V$, les cascades de paires $e^{\pm}$ qui génèrent les photons $\gamma$ dans la magnétosphère vont se produire. Ce potentiel proportionnel à $\dot{E}$ peut être décrit en fonction de la période de rotation du pulsar et de sa dérivée par $V = 4 \times 10^{20} \ P^{-3/2} \ \dot{P}^{1/2} \simeq 3.18 \times 10^{-3} \sqrt{\dot{E}}$ (équation \ref{eq:diffpot} en fonction de $P$ et $\dot{P}$). De plus, l'efficacité $\eta_{\gamma} = L_{\gamma}/\dot{E}$ augmente quand ce potentiel au niveau des lignes de champs ouvertes décroît, soit $\eta_{\gamma} \sim 1/V \sim 1 / \sqrt{\dot{E}}$. La conséquence est qu'il existe une limite de production du rayonnement $\gamma$. Basée sur les 7 pulsars détectés à haute énergie, cette limite semble être proche de $\dot{E} \simeq 3 \times 10^{34}$\,erg\,s$^{-1}$. Suivant ces observations empiriques, une liste de 224 pulsars candidats $\gamma$ a été sélectionnée avec $\dot{E} > 1 \times 10^{34}$\,erg\,s$^{-1}$. Les sources ont été classées selon la quantité $\sqrt{\dot{E}} / d^{2}$, qui correspond au flux du pulsar incident sur Terre, et souligne l'importance sur l'estimation de la distance $d$. Ces critères de sélection ainsi que la liste complète des sources sont référencés dans le papier \citet{smith08}.

Le tableau \ref{tab:psrcandidat} présente parmi la liste des 224 objets, une sous-sélection des pulsars normaux associés à une source EGRET non identifiée \citep{hartman99}, ou/et à un reste de supernova / nébuleuse à vent de pulsars \citep{kaspi98}, ou/et à une source détectée au TeV par l'expérience Cherenkov HESS\footnote{High Energy Stereoscopic System : http:$//$www.mpi-hd.mpg.de/hfm/HESS/}. Le but de ce tableau est de pré-définir des objets potentiellement intéressants pour la suite de la mission (et de la thèse), et de signaler que l'observation $\gamma$ n'est pas une finalité, mais que la compréhension de ces sources passe également par les autres longueurs d'onde et les sources associées aux pulsars. Signalons au lecteur que ce tableau est basé sur des observations ou des références de l'année 2007 (ou antérieures). Les pulsars sont rangés selon $\sqrt{\dot{E}} / d^{2}$, où $d$ est la distance obtenue dans le catalogue ATNF (``Dist''). Ce tri ne prend pas en compte la géométrie du faisceau par rapport à la ligne de visée terrestre. De plus, la colonne ``EGRET events'' présente les photons $\gamma$ détectés par EGRET au-dessus de 10\,GeV dans un rayon de 1$^{\circ}$ autour de la position du pulsar, sur toute la durée de la mission. Sur les 1506 photons détectés au-dessus de 10\,GeV sur tout le ciel, 35 sont localisés à moins de 1 degré de la position des 6 pulsars EGRET. Cette étude est basée sur les travaux de \citep{thompson05}. Le tableau résume également l'émission pulsée détectée dans les différentes longueurs d'onde, de la radio au rayonnement $\gamma$ ($O$: optique, $X_{s}$: X mou, $X_{h}$: X dur, $G_{s}$: gamma mou, $G_{h}$: gamma dur). Les deux dernières colonnes présentent respectivement la densité du flux moyen radio pour une observation à 1400\,MHz, et l'initiale du radiotélescope observant le pulsar (P:~Parkes, N:~Nançay, J:~Jodrell). \\

Précisons finalement que la collaboration avec les astronomes des radiotélescopes est un véritable succès. Les groupes basés à Manchester (Jodrell Bank) et à Nançay nous ont fourni en 2008, suite à nos nombreuses découvertes, plus de 500 éphémérides des pulsars qu'ils observent régulièrement. La plupart d'entre eux présente un ralentissement $\dot{E}$ inférieur à 10$^{34}$\,erg\,s$^{-1}$. Cela nous permet de couvrir un plus grand espace de phase $P$-$\dot{P}$ et de ne pas imposer une limite théorique sur $\dot{E}$ pour la production de photons $\gamma$. Les résultats sont résumés au chapitre \ref{chap:catalog} (catalogue de pulsars).

\clearpage
\newpage
{

\addtolength{\oddsidemargin}{-0.cm} 
\addtolength{\textheight}{0.5cm}

\begin{landscape}

\begin{table}
\begin{center}
\begin{tiny}

\begin{tabular}{ccccc|ccccc|cc}
\hline
\\
Rang & Nom & Associations 3EG & Edot & Evts EGRET & Associations  & Associations  & Emission pulsée détectée & Commentaires & S1400 & Obs. \\
$\sqrt{Edot}/D^{2}$ &	    &     & ergs\,s$^{-1}$ & $>$10GeV  & PWN/SNR &  TeV & R \ O \ Xs \ Xh \ Gs \ Gh &      & mJy   &	   \\	
\\
\hline
\\
1   & B0833-45	  & 3EG J0834-4511   & 6.92e+36 & 5	& SNR:Vela		& HESSJ0835-456   & $P \ P \ P \ P \ P \ P$ & EGRET pulsar	& 1100	&P	\\
2   & J0633+1746  & 3EG J0633+1751   & 3.25e+34 & 10	& GRS:Geminga		&		  & $? \ P \ P \ - \ - \ P$ & 	EGRET pulsar	& *	&	\\ 
3   & B0531+21    & 3EG J0534+2200   & 4.61e+38 & 10	& SNR:Crab		& HESSJ0534+220	  & $P \ P \ P \ P \ P \ P$ &	EGRET pulsar	& 14	&N/J \\ 
5   & B0656+14    &		     & 3.81e+34 & 	& SNR:Monogem Ring	&		  & $P \ P \ P \ - \ - \ ?$ &  $\gamma$ detection with $3.6 \sigma$	& 3.7 &N/J \\
6   & B1706-44    & 3EG J1710-4439   & 3.41e+36 & 8	& SNR:G343.1-2.3	&		  & $P \ - \ P \ - \ - \ P$ &	EGRET pulsar	& 7.3	&Parkes	\\
\\
7   & J0205+6449  &		     & 2.70e+37 & 3	& PWN:3C58,SN1181	&		  & $P \ - \ P \ - \ - \ -$ &	  & 0.045 &N/J  \\
9   & J1833-1034  &		     & 3.37e+37 & 1	& PWN		 	&		  & $P \ - \ - \ - \ - \ -$ &	  & 0.071 &N	\\
10  & B1951+32    &		     & 3.74e+36 & 2	& PWN:CTB 80		&		  & $P \ - \ - \ - \ - \ P$ & EGRET pulsar  & 1.0   &A/N/J \\ 
11  & J1740+1000  &		     & 2.32e+35 &	&			&		  & $P \ - \ - \ - \ - \ -$ & Gb=20.4	  & 9.2   &	 \\
12  & J1747-2958  & 3EG J1744-3011   & 2.51e+36 & 5	& PWN:Mouse		& HESS J1745-303  & $P \ - \ - \ - \ - \ -$ & Parkes 2005   & 0.25  &N	\\
\\
14  & B1509-58    &		     & 1.77e+37 & 	& SNR:G320.4-1.2	& HESS J1514-591  & $P \ - \ P \ P \ P \ -$ &	COMPTEL pulsar& 0.94  &P	 \\
20  & B1046-58    & 3EG J1048-5840   & 2.01e+36 &	& PWN:puppy		&		  & $P \ - \ - \ - \ - \ ?$ &	 & 6.5   &P	 \\
21  & J1930+1852  &		     & 1.16e+37 &	& SNR:G54.1+0.3   	&		  & $P \ - \ P \ - \ - \ -$ &	 & 0.06  &A/N/J  \\
22  & J1811-1925  &		     & 6.42e+36 & 3	& SNR/PWN:G11.2-0.3	&		  & $- \ - \ P \ - \ - \ -$ &	 & *	  &RXTE   \\
24  & J1124-5916  &		     & 1.19e+37 & 	& SNR/PWN:G292.0+1.8	&		  & $P \ - \ P \ - \ - \ -$ &	 & 0.08  &P	\\
\\
26  & J1357-6429  &		     & 3.10e+36 &	& PWN			&       	  & $P \ - \ P \ - \ - \ -$ &	 & 0.44  &P	  \\
27  & B0740-28    &		     & 1.43e+35 &	& PWN			&		  & $P \ - \ - \ - \ - \ -$ &   & 15.0  &P/J    \\
28  & J0538+2817  & 3EG J0542+2610   & 4.94e+34	& 5	& SNR:S147		&		  & $P \ - \ P \ - \ - \ -$ &	 & 1.9   &N/J   \\
31  & B1823-13    & 3EG J1826-1302   & 2.84e+36 & 4	& PWN:Turkey		& HESS J1825-137  & $P \ - \ - \ - \ - \ -$ &  & 2.1   &P/J   \\
32  & J1809-1917  &  		     & 1.78e+36 & 1     & PWN		        & HESS J1809-193  & $P \ - \ - \ - \ - \ -$ &	 & 2.5   &P	 \\
\\
33  & J1617-5055  &		     & 1.60e+37 & 	& SNR/GRS:J1616-508 	& HESS J1616-508  & $P \ - \ P \ - \ - \ -$ &	  & *	  &P	\\
34  & B1800-21    & 		     & 2.22e+36 & 1 \ 4	& SNR/PWN/GRS:G8.7-0.1  & HESS J1804-216  & $P \ - \ - \ - \ - \ -$ &	   & 7.6   &P/J  \\
37  & B1754-24    &		     & 2.59e+36 &	& PWN:Duck		&		  & $P \ - \ - \ - \ - \ -$ &  & 3.9   &P/J    \\
39  & B1055-52    & 3EG J1058-5234   & 3.01e+34 & 	&			&		  & $P \ - \ P \ - \ - \ P$ &	 EGRET pulsar  & *	  &P	  \\ 
42  & J1718-3825  & 3EG J1714-3857   & 1.25e+36 & 3	& PWN			& HESSJ1718-385	  & $P \ - \ - \ - \ - \ -$ &	  & 1.3   &P/N   \\
\\
45  & B1727-33    & 		     & 1.23e+36 & 9     &			&		  & $P \ - \ - \ - \ - \ -$ &	  & 3.2   &P/N/J  \\
46  & B1853+01    & 3EG J1856+0114   & 4.30e+35 & 1	& PWN:W44,3C392	        &		  & $P \ - \ - \ - \ - \ -$  &	   & 0.19  &A/J	 \\
48  & J1420-6048  & 3EG J1420-6038   & 1.04e+37 & 2	& PWN:kookaburra	&		  & $P \ - \ P \ - \ - \ -$ &	  & 0.9   &P	 \\
53  & B1259-63    &		     & 8.25e+35 &	&			& HESS J1303-638  & $P \ - \ - \ - \ - \ -$ &	 OB asso.	  & 1.70  &P?	 \\
56  & J1015-5719  & 3EG J1014-5705   & 8.27e+35 &	&			&		  & $P \ - \ - \ - \ - \ -$ &		  & 0.90  &P?	  \\
\\
57  & J1837-0604  & 3EG J1837-0606   & 2.00e+36 & 	&			&		  & $P \ - \ - \ - \ - \ -$ &	 	  & 0.7   &P/J   \\
59  & J2229+6114  & 3EG J2227+6122   & 2.25e+37 &	& G106.6+2.9 		&                 & $P \ - \ P \ - \ - \ -$ &	   & 0.25  &N/J   \\
60  & J1105-6107  & 3EG J1102-6103   & 2.48e+36 & 	& MSH 11-62(SNR)?/PWN	&		  & $P \ - \ - \ - \ - \ -$ &	 	  & 0.75  &P 	 \\
66  & B1737-30	  & 3EG J1744-3011   & 8.24e+34 & 	&			&		  & $P \ - \ - \ - \ - \ -$ &	   & 6.4   &P/J/N  \\
71  & J1637-4642  & 3EG J1639-4702   & 6.40e+35 &	&		  	&   		  & $P \ - \ - \ - \ - \ -$ &	 	  & 0.78  &P?	  \\
\\
73  & B1830-08    &           	     & 5.84e+35	& 6 	&		        &		  & $P \ - \ - \ - \ - \ -$ &	    & 3.6   &P/J/N  \\
76  & J1702-4128  &		     & 3.42e+35 & 3     &			&		  & $P \ - \ - \ - \ - \ -$ &     & 1.1   &P     \\
79  & J1928+1746  & 3EG J1928+1733   & 1.61e+36 & 	&			&		  & $P \ - \ - \ - \ - \ -$ &     & 0.25  &A/N   \\
81  & J1016-5857  & 3EG J1013-5915   & 2.58e+36 &	& SNR/PWN:G248.3-1.8(?) &		  & $P \ - \ - \ - \ - \ -$ &   & 0.46  &P	 \\
82  & J1705-3950  & 		     & 7.37e+34 & 2	& 			&		  & $P \ - \ - \ - \ - \ -$ &	 & 1.5   &P	  \\
\\
87  & B1338-62    &		     & 1.38e+36 & 1	& SNR/PWN:G308.8-0.1	&		  & $P \ - \ - \ - \ - \ -$ &	   & 1.9   &P	 \\
96  & J1648-4611  &		     & 2.09e+35 & 5	& 			&		  & $P \ - \ - \ - \ - \ -$ &	   & 0.58  &	  \\
97  & J1853+0056  &                  & 4.03e+34 & - \ 3 &			&		  & $P \ - \ - \ - \ - \ -$ &	   & 0.21  &A/N/J  \\
100 & B1643-43    &		     & 3.58e+35 &	& SNR/PWN:G341.2+0.9(?) &		  & $P \ - \ - \ - \ - \ -$ &   & 0.98  &P	 \\
106 & B0611+22    & 3EG J0617+2238   & 6.24e+34	& 1	&			&		  & $P \ - \ - \ - \ - \ -$ &	  & 2.2   &N/J   \\
\\
124 & J1846-0258  &		     & 8.09e+36 & 1	& SNR/PWN:Kes 75	&		  & $- \ - \ P \ - \ - \ -$ &	 	  & *	  &RXTE   \\
126 & J1815-1738  & 		     & 3.93e+35 & 1	&			&		  & $P \ - \ - \ - \ - \ -$ &	    & 0.25  &P?	 \\
135 & J1413-6141  & 3EG J1410-6147   & 5.65e+35 & - \ 3 &			&		  & $P \ - \ - \ - \ - \ -$ &	  & 0.61  &	  \\
140 & B1832-06    &                  & 5.58e+34 & 4 \ 6 &    			&		  & $P \ - \ - \ - \ - \ -$ &    & 0.17  &P/N  \\
149 & B0540-69    & 3EG J0533-6916   & 1.48e+38 &	& PWN:N158A		&		  & $P \ P \ P \ P \ P \ -$ &	  & 0.024 &RXTE   \\
\\
151 & J2021+3651  & 3EG J2021+3716   & 3.38e+36 & 	& PWN			&		  & $P \ - \ - \ - \ - \ -$ &	  & 0.1   &A/N/J  \\
165 & J1412-6145  & 3EG J1410-6147   & 1.24e+35 & - \ 3	&			&		  & $P \ - \ - \ - \ - \ -$ &	    & 0.47  &P	 \\
169 & J1755-2534  &                  & 3.47e+34 & 3	&			&		  & $P \ - \ - \ - \ - \ -$ &	   & 1.3	  &N	  \\	
191 & J1820-1529  & 		     & 4.04e+34	& 3	&		        &		  & $P \ - \ - \ - \ - \ -$ &	  & 0.61  &P	 \\
199 & J1726-3530  &		     & 3.51e+34 & 4	& SNR:G352.2-0.1(?)	&	  	  & $P \ - \ - \ - \ - \ -$ &   	  & 0.3	  &P/N/J  \\
\hline

\end{tabular}
\label{tab:candidate}
\end{tiny}
\end{center}
\caption{Liste de pulsars candidats gamma du LAT.\label{tab:psrcandidat}}
\end{table}

\end{landscape}
}

\partie{L'INSTRUMENT}          

\chapter[Le Large Area Telescope]{Le Large Area Telescope}\label{chap:lat}


\minitoc

\section{Le Télescope Spatial Fermi}

Le télescope spatial \textit{Fermi Gamma-ray Space Telescope} (figure \ref{fig:fermi}) est une mission scientifique internationale dédiée à l'observation du ciel des hautes énergies. Il est composé de deux instruments réalisés par la collaboration de divers instituts, agences spatiales et universités d'Allemagne, des \'Etats-Unis, de France, d'Italie, du Japon et de Suède. Cinq équipes françaises de l'IN2P3/CNRS, de l'INSU/CNRS et de l'IRFU/CEA contribuent à ce projet. Le satellite de 4,2 tonnes a été lancé le 11 juin 2008 par la NASA à Cap Canaveral en Floride par une fusée Delta II Boeing. Il suit actuellement une orbite quasi circulaire autour de la Terre, de période 95 minutes, à une altitude d'environ 550 km. La durée de la mission est de 5 ans, prolongeable jusqu'à 10 ans. Avant le décollage et la phase de vérification, le télescope portait le nom de GLAST pour \textsl{Gamma-ray Large Area Space Telescope}. La NASA l'a rebaptisé \textit{Fermi}, en l'honneur du célèbre scientifique italien Enrico Fermi, pionnier des mécanismes d'accélération des rayons cosmiques. \\

L'objectif de la mission \textit{Fermi} est d'étudier l'univers à travers le rayonnement $\gamma$ de haute énergie. Cela inclut: (1) la compréhension des mécanismes d'accélération des particules dans les noyaux actifs de galaxie, les pulsars, les nébuleuses à vent de pulsar, et les restes de supernova, (2) l'identification des sources non-identifiées et l'origine de l'émission diffuse relevée par EGRET, (3) de comprendre le comportement des sursauts \gam, des sources transitoires, et des sursauts solaires, et finalement (4) d'utiliser l'observation du rayonnement \gam comme une sonde de la matière noire.

Sa construction est justifiée par le fait qu'aucun détecteur au sol n'observe de photons dans sa gamme d'énergie. Les détecteurs Cherenkov comme HESS et MAGIC, ont un seuil en énergie  supérieur à environ 30\,GeV malgré leur grande surface collective. Ils observent les photons créés par effet Cherenkov, issus des gerbes électromagnétiques produites dans l'atmosphère. Cependant, en dessous de cette limite, les gerbes prennent place trop haut dans l'atmosphère pour que les télescopes actuels les observent, et le bruit de fond des particules cosmiques chargées noit le signal. L'instrument permet également de réaliser des campagnes multi-longueur d'onde avec d'autres télescopes. \\

\fig[Illustration du satellite \textit{Fermi}]{scale=0.8}{c3_fermi_satellite.jpg}{Illustration du satellite \textit{Fermi Gamma-ray Space Telescope}.}{fig:fermi}

Le télescope est composé de deux instruments (figure \ref{fig:fermi_lat_bgm}):
\begin{ablist} 
\item le Large Area Telescope (LAT), l'instrument principal du télescope \citep{atwood09}. Il mesure la direction, l'énergie entre 0.02\ et 300\,GeV\footnote{Au-delà de 300\,GeV, la résolution en énergie est supérieure à 20\%, et la reconstruction des événements devient très difficile.}, et le temps d'arrivée des photons ainsi que des particules chargées. Il a la particularité d'explorer l'ensemble du ciel en trois heures grâce à son très grand champ de vue (20\% du ciel à tout moment). De nombreuses sources de rayons gamma étant variables, cette surveillance continuelle du ciel permet d'alerter la communauté scientifique en cas d'éruptions. Signalons que pour l'étude des pulsars qui est le sujet principal de cette thèse, seules les données et les performances provenant du LAT seront utilisées.
\item le Gamma-ray Burst Monitor (GBM) qui vient compléter le LAT, est dédié à l'étude des sursauts gamma entre 10\,keV et 30\,MeV. Il est le successeur de l'instrument BATSE sur CGRO. Il est composé de 12 détecteurs scintillateurs NaI et de 2 détecteurs scintillateurs BGO, ceci avec un angle solide de 8 sr pour une précision de 10 arcmin. Il sert notamment de déclencheur pour le repointage du LAT en cas de détection d'un sursaut $\gamma$. Pour plus d'informations, on pourra se reporter à l'article de \citet{meegan09}.
\end{ablist} 

Notons que la contrainte due à l'environnement spatial impose un coût et une conception bien particulière, lesquels jouent sur les performances de l'instrument telles que la surface du détecteur et donc la sensibilité de l'appareil. La puissance électrique est également limitée à 700\,W, ce qui contraint le traitement des données à bord du satellite et les technologies (par exemple, la limitation des photomultiplicateurs). Finalement, le satellite doit pouvoir résister aux conditions spatiales. Une série de tests tels que la résistance aux vibrations, et la résistance aux basses et hautes températures, a été effectuée avant le décollage. \\


\sfig[Le LAT et le GBM]{fermi_lat_gbm.tex}{Le Large Area Telescope (LAT) et le Gamma-ray Burst Monitor (GBM)}{fig:fermi_lat_bgm}

\section{Principes de Détection des $\gamma$ par le LAT}

Intéressons-nous à présent aux principes d'interactions du LAT. Nous nous plaçons dans le cadre de l'électromagnétisme. Les processus d'interactions $\gamma$-matière varient en fonction de l'énergie de l'onde électromagnétique/photon. Aux énergies du LAT (au-dessus du MeV), les principales interactions sont l'ionisation, le \textit{bremsstrahlung} ou rayonnement de freinage, et la création de paires $e^{\pm}$.

Au-dessus du MeV, le photon $\gamma$ qui entre dans le trajectographe (section \ref{sec:tracker}), a une chance d'interagir avec le champ électromagnétique des atomes du milieu, et de créer une paire électron-positon $e^{\pm}$. Ce mécanisme est illustré par la figure \ref{fig:fermi_lat_bgm}. La probabilité d'interaction (ou section efficace) d'un photon dépend de la densité du matériau qu'il traverse. La paire $e^{\pm}$ créée va ensuite interagir avec le milieu. Selon les équations de Maxwell, toute charge dont la vitesse varie en valeur absolue ou en direction rayonne d'autant plus facilement que sa masse est petite. Ce rayonnement s'appelle le rayonnement de freinage, décrit par $e^{-} + A \rightarrow e^{-} + \gamma$, avec A le champ électromagnétique du milieu. Les électrons et positons vont ainsi émettre des $\gamma$ qui vont à leur tour créer des paires $e^{\pm}$, et ainsi de suite. Les premières populations de photons et de paires $e^{\pm}$ vont se dérouler dans le trajectographe, tandis que le reste de la cascade électromagnétique va se former dans le calorimètre (section \ref{sec:cal}), à cause de sa plus grande densité. A chaque itération l'énergie de départ est divisée, et les photons $\gamma$ sortent peu à peu du régime de création de paires. Ensuite, chaque électron de la gerbe dépose de l'énergie par ionisation et rayonnement de freinage dans le CsI du calorimètre. Le dépôt d'énergie par ionisation est décrit par la formule de Bethe-Bloch.

Notons que les sections efficaces de création de paires $e^{\pm}$ et de rayonnement de freinage peuvent être décrites à l'aide de la longueur de radiation X$_{0}$. Celle-ci dépend du matériau traversé, soit X$_{0} \sim 716$\,g\,cm$^{-2} A / Z^{2}$. La conséquence est que la résolution en énergie qui dérive de la reconstruction de la gerbe électromagnétique, dépend de la profondeur du calorimètre du LAT. Au delà d'une énergie initiale de 30\,GeV, le calorimètre mesure moins de la moitié de la cascade. Il est donc très difficile ensuite de reconstruire l'énergie de la particule incidente. Le deuxième facteur limitant dans la reconstruction des particules est la diffusion multiple. Quand un électron traverse un matériau, il est dévié par le champ des atomes, d'autant plus que sa vitesse est petite. Si l'électron a une trop grande déviation, le trajectographe ne pourra pas identifier correctement la direction initiale de la particule. 

Ces processus physiques vont limiter les performances de l'instrument, et ainsi contraindre la surface efficace, la résolution en énergie, et la résolution angulaire. Le chapitre suivant décrit ces caractéristiques et présente la validation de la surface efficace.

\section{Fonctionnement de l'Instrument}\label{LAT}

Le Large Area Telescope (LAT) est composé de 16 tours formant un carré 4 $\times$ 4 enveloppé par un bouclier d'anti-coïncidence pour l'identification des particules chargées provenant du rayonnement cosmique et de l'albédo terrestre. Chaque tour est elle-même composée d'un trajectographe, d'un calorimètre, et d'un système d'acquisition. Ces différentes parties seront décrites par la suite. Le tableau \ref{tab:LATperform} énumère les principaux paramètres et capacités du LAT. 

\tab[Paramètres et capacités du LAT]{LAT_performance}{Paramètres et capacités du LAT pour les fonctions de réponse de l'instrument (IRFs) appelées Pass6\_v3.}{tab:LATperform}

\subsection{Trajectographe}\label{sec:tracker}

Le trajectographe est l'un des sous-détecteurs utilisé dans la détermination de la direction des particules incidentes. Cette direction est reconstruite à partir d'algorithmes utilisant les coordonnées ($x_{i}^{TKR}$, $y_{i}^{TKR}$, $z_{i}^{TKR}$) des pistes silicium $i$ activées par les paires $e^{\pm}$ créées par le photon incident. Le trajectographe est également la voie principale de déclenchement de l'acquisition du LAT. 

\fig[Module trajectographique du LAT]{scale=1.0}{c3_silicon_glast_diag.jpg}{Module trajectographique du LAT.}{fig:tracker}

Un module du système est composé de 18 plans x\ --\ y, sensibles au passage d'une particule chargée (figure \ref{fig:tracker}). Chaque plan x et y est composé de 16 détecteurs à piste de silicium (Silicon Strip Detector, SSD), qui permettent de visualiser les traces laissées par les particules chargées. La dimension z est obtenue en empilant les plans. La taille d'un détecteur est 8.95 $\times$ 8.95\,cm pour une épaisseur de 400\,${\mu}$m, et compte 384 pistes de silicium (jonction PN). Leur efficacité de déclenchement est supérieure à 99\%, et le bruit électronique est de 45$\pm$8 pistes déclenchées lors du passage d'un événement, ceci pour l'ensemble des tours. Notons qu'un événement allume quelques dizaines de pistes concentrées sur un cylindre de quelques centimètres de rayon. Cette technologie apporte également un temps morts beaucoup plus court, par rapport à la chambre à ionisation utilisée par le détecteur EGRET ($\sim$ 5\,ms). 

Entre ces plans sont intercalés 16 plans convertisseurs en tungstène permettant de convertir le photon incident en une paire $e^{\pm}$. Etant donné que la probabilité d'interaction du photon dépend de la densité du matériau traversé, le trajectographe a été conçu pour amorcer la cascade électromagnétique qui se développera pour le reste dans le calorimètre. Ces plans convertisseurs sont maintenus par une structure en carbone composite, et ont des épaisseurs différentes. Les 12 premiers plans (section mince) ont une épaisseur de 0.03 longueur de radiation (X$_{0}$), pour éviter que les électrons et les positons n'interagissent trop par rayonnement de freinage ou par diffusion multiple. Cela permet d'augmenter la qualité des traces, mais diminue la probabilité de conversion en paire $e^{\pm}$ pour le photon $\gamma$ entrant. Les 4 plans suivants (section épaisse) ont une épaisseur de 0.18 longueur de radiation, permettant un taux de conversion des photons de 65\%. Cette caractéristique donne une plus grande sensibilité aux sources faibles, mais dégrade considérablement la reconstruction des photons de basse énergie ($<$100\,MeV). Pour ces photons, la gerbe électromagnétique ne se développe pas jusqu'au calorimètre. Ainsi, les 2 derniers plans ne possèdent pas de plans convertisseurs pour éviter d'accentuer la conversion. 

La reconstruction du vertex créé par la conversion en paire $e^{\pm}$ du photon incident se fait au sol, et est déterminée par des algorithmes de reconnaissance de formes, tels que le \textit{track finding} et le \textit{track fitting}. Les traces sont reconstruites en tenant compte du nombre de pistes touchées $i$, et de la diffusion multiple qui donne une indication sur l'énergie des paires $e^{\pm}$. Notons qu'un autre algorithme utilise le procédé appelé filtre de Kalman \citep{jones98}, dont le principe est de propager un vecteur d'état d'une couche vers la couche inférieure pour reconstruire la trace. Finalement, une seconde analyse basée sur des ``Classification Trees'' (voir la section \ref{sec:anactb}) sélectionne le meilleur algorithme pour reconstruire la direction de la particule incidente notée $\theta$\footnote{$\theta$ est défini comme l'angle d'incidence, soit l'angle entre l'axe normal au plan du LAT et la direction du photon incident.}. \\

Les principales caractéristiques du trajectographe sont détaillées dans le tableau \ref{tab:tracker}, ainsi que dans \citet{atwood07}. Ce sous-sytème joue un rôle principal dans la résolution angulaire du LAT. Cette performance sera ainsi limitée par la trace, et donc par l'imprécision sur la direction du photon incident.

\tab[Caractéristiques du trajectographe du LAT]{tracker_performance}{Caractéristiques du trajectographe du LAT.}{tab:tracker}

\subsection{Calorimètre}\label{sec:cal}

Le calorimètre du LAT a pour but principal de mesurer l'énergie $E_{\gamma}$ des photons incidents au-dessus de 20\,MeV. Il est également utilisé pour la trajectographie des événements et l'identification des rayons cosmiques. La figure \ref{fig:calorimetre} présente un des modules du calorimètre et expose sa structure en couche des faces latérales : plaque de fermeture en aluminium, circuit électronique de lecture, puis protection thermique et électromagnétique en aluminium. Sa masse totale, qui est limitée par la nécessité d'une mise en orbite du satellite, est de 1800\,kg. \\

Le principe du calorimètre est de créer une cascade électromagnétique grâce à sa petite longueur de radiation (X$_{0}(CsI)=1.85$\,cm), et de mesurer l'énergie ($e_{i}^{CAL}$) et la position ($x_{i}^{CAL}$,$y_{i}^{CAL}$,$z_{i}^{CAL}$) des dépôts d'énergie. Les photons et les $e^{\pm}$ ionisent les cristaux, qui émettent à la suite de la lumière visible créée par la désexcitation de leurs atomes. La lumière est ensuite récupérée par des photodiodes et convertie en signal analogique. 

\fig[Module du calorimètre du LAT]{scale=0.5}{c3_calorimeter_LAT.jpg}{Module du calorimètre du LAT \citep{bregeon05}.}{fig:calorimetre}

Chacun des 16 modules du calorimètre contient 8 couches alternativement croisées de 12 barreaux de Iodure de Césium dopé au thalium CsI(Tl), lesquelles sont soutenues par une structure en carbone composite, développée et construite au laboratoire Leprince-Ringuet de l'\'Ecole Polytechnique. Chaque couche dont l'épaisseur est de 19.9\,mm, soit 1.08\,X$_{0}$, est orthogonale à la précédente, formant une structure hodoscopique. Les barreaux qui mesurent 326\,mm $\times$ 26.7\,mm $\times$ 19.9\,mm, ont à leurs extrêmités une double photodiode, laquelle est composée d'une grande diode (LE: Low energy), et d'une diode six fois plus petite (HE: High Energy). De plus, la sortie de chaque diode est connectée à deux circuits électroniques permettant d'amplifier le signal. Ce système permet ainsi une lecture du dépôt d'énergie sur quatre gammes, soit un intervalle s'étendant de 2\,MeV à 102.4\,GeV, avec un recouvrement des gammes deux à deux (voir le schéma \ref{fig:cal_fonc}). 

Le système permet également d'estimer la position moyenne du dépôt d'énergie des photons et des $e^{\pm}$. La mesure se fait grâce à l'atténuation de la lumière lors de sa propagation à l'intérieur d'un barreau, dont la surface est dépolie. Le signal est lu aux deux extrémités du barreau. Son amplitude est asymétrique en fonction de la position longitudinale du passage des particules. En supposant que l'atténuation de la lumière soit linéaire dans tout le cristal, la position du dépôt d'énergie à une distance $x$ de son centre est donnée par:
\begin{equation}
x = \frac{1}{\beta} \frac{L-R}{L+R},
\end{equation}
avec $\beta$ le coefficient d'atténuation de la lumière par unité de longueur, et $L$ et $R$ respectivement la lumière collectée à l'extrémité gauche et droite du barreau. La résolution sur la position de la gerbe est de l'ordre de quelques millimètres. Ces caractéristiques permettent ainsi de visualiser la gerbe électromagnétique en trois dimensions, et notamment d'aider à la reconstrution de la direction de la particule incidente. 

\fig[Schéma de fonctionnement du calorimètre]{scale=0.35}{c3_cal_fonctionnement.jpg}{Schéma de fonctionnement du calorimètre. La particule qui traverse le cristal dépose de l'énergie, par excitation des atomes du cristal. L'énergie déposée se transforme en lumière par scintillation. Ensuite, le signal lumineux se propage jusqu'aux diodes où il est converti en signal analogique. Cette figure est extraite de \citep{davezac06}.}{fig:cal_fonc}

La reconstruction de l'énergie totale de la gerbe est, quant à elle, obtenue par trois méthodes distinctes, ayant des spécificités pour chaque population de particules (photons, particules chargées, basses et hautes énergies, petits et grands angles d'incidence). Signalons que seuls les événements reconstruits par le trajectographe sont traités, et que l'on corrige l'énergie déposée dans les parties inactives du calorimètre. La première méthode prend en compte la profondeur du maximum de la cascade électromagnétique confinée dans deux barreaux, ainsi que l'énergie du photon incident. La seconde s'appuie sur le profil longitudinal de la cascade mesuré par les 12 couches, qui est ensuite ajusté à la forme analytique. Celle-ci ne sera utilisé qu'à haute énergie. Enfin, la troisième méthode utilise une simulation de gerbe qui est ensuite comparée au profil observé. Ces méthodes sont présentées dans la thèse de \citet{davezac06}. Finalement, une des trois méthodes sera sélectionnée par un ``Classification Tree'' (voir la section \ref{sec:anactb}) pour reconstruire l'énergie de la particule incidente $E$.

Comme dans le cas du trajectographe, un ensemble de ces variables est présent dans les fichiers de pré-analyse. Elles nous servent entre autres à valider l'instrument en orbite et à identifier les photons.

\tab[Caractéristiques du calorimètre du LAT]{cal_performance}{Caractéristiques du calorimètre du LAT.}{tab:calor}

\subsection{Détecteur Anti-Coïncidence}

Le but du détecteur anti-coïncidence (ACD) est d'identifier les particules chargées, d'une part pour la rejection du fond de rayonnement cosmique (il joue le rôle de veto), et d'autre part pour l'étalonnage en énergie du calorimètre avec les ions lourds du rayonnement cosmique.

Le détecteur est composé de 89 détecteurs scintillateurs plastiques, appelés tuiles (figure \ref{fig:tuile}), et couvre 8.3\,m$^{2}$ du détecteur. Les tuiles ont une taille d'environ 30\,cm sur 30\,cm, qui dépend de leur emplacement. Elles ne recouvrent pas les faces du calorimètre, pour éviter des vétos dus à la rétro-diffusion. Quand une particule chargée traverse l'ACD, elle dépose de l'énergie. La lumière produite dans les tuiles est récupérée par des fibres optiques à décalage spectral. Cette caractéristique modifie la longueur d'onde de la lumière pour une meilleure adéquation avec les photomultiplicateurs, qui mesurent la lumière. Pour alimenter les 18 photomultiplicateurs, le LAT dispose de châssis électroniques et d'alimentations haute tension ajustables, pour prendre en compte le vieillissement des photomultiplicateurs.

\fig[Schéma d'un détecteur scintillateur plastique]{scale=0.3}{c3_tuileACD.jpg}{Schéma d'un détecteur scintillateur plastique. Extrait de \citet{johnson02}}{fig:tuile}

Le taux de rayons cosmiques qui interagit avec le LAT est d'environ 10\,kHz, comparé aux quelques Hz de photons $\gamma$. Le cahier des charges du LAT impose un bruit de fond résiduel chargé inférieur à 10\% de l'intensité du fond gamma diffus. Cette condition impose une suppression d'un facteur 10$^{6}$ pour les protons et 10$^{4}$ pour les électrons au niveau de la totalité de l'instrument. La conséquence est que l'efficacité de détection des particules chargées par l'ACD doit être de 0.9997, sans pour autant réduire le nombre de photons détectés. La topologie des événements dans le trajectographe et la forme des cascades hadroniques dans le cas des protons vont aider à rejeter les protons et les électrons non détectés par l'ACD. Notons que le seuil de déclenchement est de 0.3\,MIP (``Minimum Ionizing Particle'') sur chaque tuile. Finalement, la rejection du fond évite de stocker trop d'événements compte tenu des contraintes télémétriques.

Une cascade électromagnétique qui se produit dans le calorimètre peut produire des photons jusqu'à des angles de 180$^{\circ}$ par rapport au sens de développement de la cascade (voir la figure \ref{fig:backsplash}). Ces photons de seconde génération qui ont une énergie de quelques keV vont interagir par effet Compton avec des électrons, lequels vont générer un signal dans l'ACD. Ce phénomène s'appelle la rétro-diffusion ou ``backsplash'' en anglais. Le LAT, contrairement à EGRET, est composé d'un détecteur d'anti-coïncidence ségmenté, qui permet de corréler l'alignement de la trace induite par la particule incidente avec la position les tuiles activées. Ce système limite les effets de self-veto dus à l'émission vers l'arrière et préserve la surface efficace, notamment à haute énergie.

\fig[Illustration d'un phénomène de rétro-diffusion]{scale=0.3}{c3_backsplash.jpg}{Illustration d'un phénomène de rétro-diffusion (``backsplash'') à l'intérieur du LAT. Extrait de \citep{johnson02}}{fig:backsplash}

\subsection{Déclenchement}\label{sec:declenchement}

L'instrument est composé de plusieurs niveaux de déclenchement pour l'acquisition. L'ensemble est géré par le \textit{Global trigger Electronics Module} (GEM). Il utilise pour cela les performances du trajectographe, du calorimètre et du bouclier anti-coïncidence ; le but étant d'avoir d'une part, un rapport signal-sur-bruit maximum en raison de la limitation de la bande passante télémétrique, et d'autre part, un temps mort minimum entre deux prises de données. Le satellite a la capacité d'enregistrer jusqu'à 36 heures de données. Il transmet ses données 2 fois par jour à un débit de 3 $\cdot$ 10$^{5}$\,bits\,s$^{-1}$.

Le premier niveau (essentiellement déclenché par les particules chargées) est composé de 8 triggers individuels, et détermine si la particule a les conditions requises pour être analysée. La fenêtre de déclenchement est ouverte dès qu'un des triggers est activé. Elle est ajustée en temps (entre 50 et 1550\,ns) pour permettre aux (éventuels) autres triggers individuels d'arriver, mais suffisamment étroite pour éviter que d'autres événements interagissent avec le LAT. Nous verrons qu'il peut y avoir des accidents (voir la section \ref{sec:highlevel}). Notons que chaque trigger est neutralisé si le flux de données produit par un événement est trop grand (par exemple, un trop grand nombre de pistes de silicium touchées), ou parce que le taux de comptage est trop important. Pour cela, des échelles sont associées à chaque trigger individuel pour diviser le nombre d'entrées des événements à fort taux de comptage. La liste suivante présente ces 8 triggers individuels utilisés dans le LAT au niveau du hardware. 

\begin{ablist} 
\item \textbf{TKR:}~Ce déclenchement, dont la fréquence moyenne est de 4\,kHz, est donné par le trajectographe, si 3 plans x\ --\ y consécutifs sont touchés. 
\item \textbf{CAL-LO:}~Obtenu par le calorimètre, le système déclenche la lecture des voies si un barreau de CsI mesure un dépôt de 100\,MeV. Il réduit le taux de déclenchement TKR dans le cas où un photon convertit toute son énergie dans le trajectographe.
\item \textbf{CAL-HI:}~Identique au déclenchement de type CAL-LO, mais avec un seuil de 1\,GeV.
\item \textbf{ROI:}~Certains événements $\gamma$ peuvent activer les tuiles de l'ACD à cause de la rétro-diffusion. Pour éviter de rejeter tous les événements à forte émission vers l'arrière, un algorithme analyse le déclenchement de type TKR et les tuiles touchées.
\item \textbf{CNO:}~Activé quand $\sim$ 20\,MIP sont déposés dans une tuile de l'ACD. Il permet la calibration du calorimètre avec des ions lourds. CNO:~Carbone, Azote, Oxygène.
\item \textbf{Periodic:}~Ce déclencheur ouvre toutes les voies du détecteur. Il permet de vérifier les piédestaux de l'ACD et du calorimètre, et le taux d'occupation du trajectographe en vol.
\item \textbf{Solicited:}~Utilisé pour le software du satellite.
\item \textbf{External:}~Rien n'est connecté. Ne permet pas l'ouverture d'une fenêtre. Historiquement utilisé pour les tests au sol \citep{bregeon05}.
\end{ablist} 

A la fermeture de cette fenêtre, on enregistre l'état de ces 8 conditions. Le second niveau de déclenchement, le \textit{Global LAT Trigger} (GLT), utilise la combinaison de ces triggers individuels que l'on appelle ``trigger engine''. Il contient les instructions de lecture et les configurations relatives au type d'événement que l'on veut enregistrer (photons, ions). Par exemple, la sélection des photons $\gamma$ demande que la condition $TKR$ soit vraie, tandis que les autres doivent être fausses. Le dernier niveau que l'on nomme le \textit{On Board Filter} (OBF) est un filtre software, intégré au système embarqué d'acquisition des données. Son but est de rejeter un nombre suffisant d'événements du fond, sans pour autant éliminer les photons $\gamma$. Le grand avantage de ce filtre est qu'il est reprogrammable comparé aux deux premiers niveaux. Le taux de particules après l'OBF est d'environ 300\,Hz. 

Finalement, la combinaison de ces trois niveaux triggers réduit le fond par un facteur de presque 10$^{2}$, tandis qu'il préserve 75\% des photons qui ont interagit avec le LAT. La suite de l'analyse se fait au sol et discrimine les photons et les particules chargées (voir la section \ref{sec:anactb}).

\subsection{\'Etalonnage en \'Energie du Calorimètre}

L'étalonnage en énergie du LAT est un enjeu majeur, d'une part pour nos mesures de la coupure en énergie des spectres des pulsars (discrimination des modèles), et d'autre part pour la nouvelle physique telle que la détection des signaux des particules de la matière noire. Un premier étalonnage en énergie a été effectué au sol en utilisant les particules chargées et des injections de charges. La méthode consiste à calibrer la gamme d'énergie la plus basse (LEX8\footnote{LEX8:Low Energy, grande diode et amplificateur à fort gain. LEX1: Low Energy, grande diode et amplificateur à gain faible.}) avec des muons cosmiques, qui laissent un dépôt d'énergie d'environ 11.2\,MeV par barreau, proche du minimum d'ionisation. Plusieurs millions d'événements déclenchés par le calorimètre sont ainsi enregistrés, et ensuite analysés. L'étalonnage au sol des autres voies se fait par inter-étalonnage grâce aux injections de charge.

L'étalonnage en vol utilise les rayons cosmiques. Un ion dépose environ $Z^{2} \times 11.2 {\rm cos} \theta$\,MeV au minimum d'ionisation dans chaque barreau traversé, avec $\theta$ l'angle d'incidence. Ainsi, un proton permet d'étalonner la gamme d'énergie la plus basse en déposant 11.2\,MeV, tandis qu'un noyau de fer déposant environ 7.6\,GeV permet d'étalonner la voie enregistrant les dépôts les plus énergétiques (HEX1\footnote{HEX8:High Energy, petite diode et amplificateur à fort gain. HEX1: High Energy, petite diode et amplificateur à gain faible.}). Notons également que pour un même dépôt d'énergie la quantité de lumière produite dépend du type de particule. Cette non-linéarité ou effet de ``quenching'' doit être prise en compte. Une campagne de tests sous faisceau au GSI\footnote{http:$//$www.gsi.de/index.html} a permis de mesurer ces non-linéarités pour l'ensemble des ions et la gamme en énergie utilisée pour l'étalonnage en vol \citep{lott06}. En résumé, du proton jusqu'au fer, le spectre des rayons cosmiques couvre parfaitement les quatres gammes en énergie, ce qui évite d'utiliser les injections de charges. Un algorithme (``GCRCalib'') permet ensuite de garder les événements bien reconstruits et rejette ceux qui ont interagit par réaction nucléaire. Finalement, comme le rayonnement est relativement intense, une journée est suffisante pour obtenir une bonne statistique sur les ions légers, soit un taux de $\sim$ 1.8\,Hz pour les événements CNO, et $\sim$ 10$^{-2}$\,Hz pour les événements de fer. De plus, étant donné que les 4 gammes en énergie ont un recouvrement deux à deux, il est possible d'inter-étalonner en vol les voies touchées par un même événement.

%
\chapter[Validation de la Surface Efficace du Large Area Telescope]{Validation de la Surface Efficace du Large Area Telescope}\label{chap:irfs}

\minitoc

\section{Introduction}

Le télescope LAT est conçu pour répondre à des objectifs scientifiques tels que la compréhension des mécanismes d'accélération des particules dans la magnétosphère des pulsars. Ce travail passe par l'étude des spectres incluant les paramètres énumérés à l'équation \ref{eq:powerlawexpcutoff} et le flux intégré. Pour accéder à ces quantités physiques, on utilise les fonctions de réponse de l'instrument (IRFs, Instrument Response Functions), qui permettent de traduire une quantité reconstruite (énergie $E$ et direction $\theta$ des photons) en une quantité physique vraie. Ces fonctions basées sur une simulation Monte Carlo (voir la section \ref{sec:simuLAT}) sont gouvernées par la conception du LAT, des algorithmes de reconstruction des événements (voir les sections \ref{sec:tracker} et \ref{sec:cal}), et l'identification des particules (voir la section \ref{sec:anactb}). Ce chapitre présente la validation de l'une des trois IRFs, la surface efficace.

\section{Simulation de l'Instrument}\label{sec:simuLAT}

La simulation détaillée du LAT et des événements est faite par Gleam (GLAST LAT Event Analysis Machine), qui se base sur l'outil Monte Carlo GEANT4 \footnote{http:$//$geant4.web.cern.ch/geant4/} \citep{agostinelli03}, logiciel de simulation des interactions entre les particules et la matière. Cela inclut la diffusion multiple, la production de paires et la diffusion Compton pour les photons, le rayonnement de freinage pour les $e^{\pm}$, aussi bien que les interactions hadroniques. L'information sur la géométrie du détecteur, la structure du satellite, et les matériaux utilisés (pistes silicium du trajectographe, cristaux de CsI et les diodes du calorimètre, tuiles du bouclier anti-coïncidence) est stockée dans des fichiers XML. Notons que ces fichiers sont également utilisés pour les algorithmes de reconstruction de l'énergie et de la direction des événements. La simulation est utilisée pour étudier les algorithmes de reconstruction et la classification des événements, ainsi que les fonctions de réponse de l'instrument telles que la résolution angulaire, la résolution en énergie, et la surface efficace.

Avant le lancement du satellite, une caractérisation expérimentale de l'instrument a été effectuée, pour vérifier que les réponses de l'instrument étaient en accord avec les prédictions de la simulation Monte Carlo. Il est essentiel d'obtenir une bonne reproductibilité entre les quantités mesurées directes (énergie déposée $e_{i}^{CAL}$, activation des pistes $xyz_{i}^{TKR}$, etc) et l'analyse haut-niveau (reconstruction de l'énergie $E$ et de la direction $\theta$). De plus, une mesure directe des IRFs est requise pour certaines configurations critiques, telles que les photons de basse énergie à cause de la diffusion multiple. Pour ces raisons, une campagne de tests sous faisceau a été réalisée au près des lignes du PS\footnote{PS:~Faisceaux de photons ``étiquetés'', d'énergie comprise entre 50 MeV et 3 GeV, et d'électrons et de protons entre 300 MeV/c et 15 GeV/c.} et du SPS\footnote{SPS:~Faisceaux d'électrons et de protons dans un domaine entre 10 GeV/c et 300 GeV/c.} au CERN en été 2006 \citep{baldini07}. Une première calibration avait également été effectuée en utilisant des muons produits par le rayonnement cosmique dans l'atmosphère. Notons que toutes les configurations n'ont pas été testées, en raison de l'espace de phase important du LAT en terme d'angle, de position, et d'énergie.

\section{Reconstruction et Classification des \'Evénements}\label{sec:anactb}

La figure \ref{fig:analysis} présente les différentes étapes de ``l'analyse'' des événements du LAT, incluant la reconstruction et la classification des événements. A leur arrivée sur Terre ou en sortie de Gleam, en d'autres termes qu'ils soient simulés ou vrais, les événements sont reconstruits et classés par la même chaîne d'analyse, ce qui rend les choses comparables ! 

\fig[\'Etapes de la reconstruction des événements du LAT]{scale=0.55}{c4_schema_analysis.png}{Organigramme représentant les différentes étapes de la reconstruction des événements du LAT. \textit{fits} (Flexible Image Transport System) est un format de fichiers utilisé en astronomie : http:$//$fits.gsfc.nasa.gov}{fig:analysis}

La première étape notée (1) est la reconstruction de la direction et de l'énergie de la particule incidente à partir des données brutes des sous-sytèmes du LAT, comprenant le trajectographe ($xyz_{i}^{TKR}$) et le calorimètre ($e_{i}^{CAL}$ et $xyz_{i}^{CAL}$). Les sections \ref{sec:tracker} et \ref{sec:cal} respectivement décrivent brièvement les algorithmes de reconstruction de la direction et de l'énergie d'une particule incidente. On trouvera plus de détails dans la thèse de \citet{davezac06}. L'analyse produit un fichier appelé ``recon'' avec une multitude de variables utilisées pour l'analyse suivante. Notons que le développement de la reconstruction repose fortement sur la simulation Monte Carlo des événements. 

La deuxième étape notée (2) associe à l'événement une probabilité caractérisant le degré de confiance de la quantité mesurée. Cette analyse est basée sur les outils statistiques d'aide à la décision et à l'exploration des données, que l'on nomme les arbres de décisions (Classification Trees (CT), \citealt{breiman84}). La construction de ces arbres a été réalisée à l'aide de la simulation Monte Carlo Gleam. Les particules qui passent à travers l'analyse sont réparties en groupes selon un ensemble de variables, tels que la direction de la particule dans le trajectographe, l'emplacement de la trace, ou l'énergie déposée dans le calorimètre. Cette analyse se déroule en trois étapes. Premièrement, l'énergie finale pour chaque événement est assignée à l'énergie déterminée par la méthode de reconstruction la plus appropriée (méthode paramétrique, corrélation avec la dernière couche du calorimètre, estimation du profil de la gerbe). Celle-ci est sélectionnée par un CT, que l'on nomme ``CTBBestEnergy''. Ensuite, un second CT permet d'estimer la probabilité que l'énergie soit bien reconstruite (``CTBBestEnergyProb''). La deuxième étape est une analyse similaire qui attribue à la particule la meilleure direction disponible (``CTBest[XYZ]Dir''), ainsi qu'une probabilité correspondante d'être dans le coeur de la distribution de la PSF (``CTBCORE''). Finalement, quand l'énergie et la direction de la particule ont été évaluées, la probabilité que l'événement soit un photon est examinée. La rejection du fond est une analyse difficile mais essentielle dans la reconstruction des particules, en raison de l'espace de phase important du LAT et du faible rapport signal-sur-bruit ($\sim$1/300) des données entrantes. Cette dernière analyse est divisée en trois niveaux, chacun associé aux trois sous-sytèmes du LAT (ACD, trajectographe, et calorimètre). La probabilité que l'événement soit un photon est évaluée pour chaque partie. Ces variables sont nommées ``CTBCPFGamProb'', ``CTBCALGamProb'', ``CTBTKRGamProb''\footnote{Cette page liste l'ensemble des variables $CTB$ présent dans l'analyse:
 \scriptsize{http:$//$www.slac.stanford.edu/exp/glast/ground/software/status/documentation/GlastRelease/latest/GlastClassify/latest/merittuple.html}}. 

Finalement, trois classes d'événements sont définies à partir de l'analyse décrite ci-dessus. Elles sont basées également sur le fond de particules chargées prévu en orbite et la connaissance du ciel $\gamma$. Ces classes se différencient par leur niveau de discrimination des photons, l'efficacité, et la résolution angulaire et en énergie. L'utilisateur peut également définir ses propres catégories d'événements selon le but de l'analyse, comme par exemple maximiser la résolution angulaire pour l'identification des sources, mais au prix de réduire la surface efficace. Les trois classes standards sont :

\begin{ablist} 
\item la classe \textit{transient}: elle est définie avec les coupures les plus lâches et présente un taux de fond résiduel d'environ 2\,Hz. Ceci équivaut en moyenne à un photon du fond toutes les 5 secondes à l'intérieur d'une 10$^{\circ}$ autour de la source. La surface efficace est maximisée notamment à basse énergie et donc est appropriée pour l'étude des sources transitoires tels que les sursauts $\gamma$. 
\item la classe \textit{source}: elle a été élaborée pour que la contamination du fond résiduel soit similaire au flux extragalactique $\gamma$ estimé par EGRET, soit un taux de photons d'environ 0.4\,Hz. Cette classe est appropriée pour l'étude des sources ponctuelles.
\item la classe \textit{diffuse}: elle a la meilleure rejection de fond, soit un taux de photons d'environ 0.1\,Hz. Elle a été élaborée pour étudier les sources diffuses les plus faibles. Notons que pour l'instant cette classe est utilisée pour l'analyse actuelle des pulsars $\gamma$.
\end{ablist} 

Ces trois classes d'analyse définies pour la classification appelée ``Pass6'' sont documentées dans \citet{atwood09}. Pour l'analyse des sources de cette thèse, nous utilisons la version 3 de cette classification, $Pass6\_v3$. Contrairement à la version 1 utilisée pendant les premiers mois de la mission, la version 3 prend en compte les effets de traces fantômes observés pendant la phase de calibration de l'instrument (voir la section \ref{sec:highlevel} et la figure \ref{fig:ghosttrack}). Un nouveau schéma appelé ``Pass7'' prévu pour la fin de l'année 2009 inclut une analyse spécifique des particules du fond, telle que les électrons, les protons, et les ions lourds.

\section{Fonctions de Réponse de l'Instrument}\label{sec:irfs}

Les fonctions de réponse de l'instrument (IRFs) sont le lien entre le flux de photon vrai et les événements détectés. Comme nous l'avons vu précédemment, la détection dépend des caractéristiques de l'instrument, mais également de l'analyse qui assigne une probabilité que l'événement soit un photon. Les IRFs peuvent être formulées comme la surface multipliée par la probabilité, qu'un photon avec une énergie vraie $E'$ et une direction vraie $\theta'$ soit détecté aux quantités reconstruites $E$ et $\theta$. En négligeant les corrélations entre l'énergie et la direction de la particule et en supposant que la PSF est circulaire autour de la position vraie de la source, les IRFs peuvent être décrites par:
\begin{equation}\label{eq:irfs}
R(E,\theta;E',\theta') = A_{eff}(E',\theta') \ P_{PSF}(\theta;E',\theta') \ P_{E}(E;E',\theta')
\end{equation}

Le terme $A_{eff}(E',\theta')$ représente la surface efficace différentielle, soit l'efficacité de détection multipliée par la surface géométrique en fonction de l'énergie et la direction vraie du photon incident. La dispersion de l'énergie $P_{E}$ est la densité de probabilité qu'un photon avec l'énergie $E'$ et la direction $\theta'$ soit détecté avec l'énergie reconstruite $E$. Tandis que $P_{PSF}$ désigne la fonction d'étalement du point (le terme en anglais est ``Point Spread Function''), c'est-à-dire, la densité de probabilité qu'un photon d'énergie $E'$ et de direction $\theta'$ soit détecté avec la direction reconstruite $\theta$.

Les IRFs sont basées sur la simulation Monte Carlo Gleam décrite dans la section \ref{sec:simuLAT}. Un ensemble de photons et particules chargées est généré isotropiquement sous tous les angles d'inclinaisons possibles et sur toute la gamme d'énergie du LAT (de 20\,MeV à 300\,GeV), et sont analysés séparément par la procédure décrite dans la section \ref{sec:anactb}. La comparaison entre les paramètres reconstruits de l'événement et le photon entrant, dont on connaît exactement la direction et l'énergie, donne les IRFs. La simulation des particules chargées permet, quant à elle, d'estimer la contamination des différentes classes de photons. Les trois fonctions de réponse de l'instrument (résolution angulaire, résolution en énergie, et surface efficace) sont présentées dans les sections suivantes.

\subsection{Résolution Angulaire}\label{sec:psf}

La résolution angulaire du LAT est déterminée par sa PSF, traduisant le pouvoir de localisation d'une source. Il s'agit de la précision avec laquelle une source d'émission $\gamma$ va être reconstruite par le détecteur. La PSF est dominée par la diffusion multiple des premières paires $e^{\pm}$ dans le trajectographe, principalement à basse énergie (E$<$100\,MeV). A haute énergie, la PSF est dominée par la résolution spatiale des plans de positionnement. Ceci est défini par le terme $\delta$ dans l'équation suivante. A ces incertitudes s'ajoute celle sur le pointé du satellite ($<$40\,arcmin). De façon générale, la PSF peut être décrite par :
\begin{equation}
PSF(E_{\gamma}) = \frac{\alpha}{\sqrt{E_{\gamma}}} \oplus \frac{\beta}{\sqrt{E_{\gamma}}} \oplus \delta \ , 
\end{equation}
avec $E_{\gamma}$ l'énergie du photon incident. Le terme $\alpha$ définit la dépendance en énergie de la diffusion multiple, tandis que $\beta$ résulte de la distance entre l'entrée du photon dans le LAT et la première conversion $e^{\pm}$. La figure \ref{fig:angularresolution} gauche illustre pour la classe \textit{diffuse}, la PSF du LAT à 68\% et 95\% de contenance en fonction de l'énergie pour les différentes sections de conversions dans le trajectographe et pour une incidence normale, tandis que la figure droite montre la PSF en fonction de l'angle d'incidence pour des photons de 10\,GeV. La PSF peut être décrite, en se basant sur la correction de la formule de Molière pour la diffusion multiple, par: 
\begin{equation}
PSF(E_{\gamma}) = a \times E_{\gamma}^{-b},
\end{equation}
avec $E_{\gamma}$ en GeV. Dans le cas de l'analyse ``Pass6\_v3'' que l'on utilisera pour l'analyse des pulsars des chapitres 6 à 9, les termes $a$ et $b$ sont respectivement égaux à 0.8 et 0.75.

\fig[Résolution angulaire du LAT pour la classe \textit{diffuse}]{scale=0.82}{c4_panelpsf.pdf}{PSF du LAT pour la classe d'événements \textit{diffuse}, $Pass6\_v3$. \textbf{Gauche:}~Angles à 68\% (ligne continue) et 95\% (ligne pointillée) de contenance de la direction reconstruite du photon entrant en fonction de l'énergie, pour une incidence normale ($cos~\theta>0.9$), et pour des conversions dans la section mince (front), épaisse (back), mince et épaisse (total) du trajectographe. \textbf{Droite:}~Angles à 68\% (ligne continue) et 95\% (ligne pointillée) de contenance de la direction reconstruite du photon entrant en fonction de l'angle d'incidence pour des photons de 10\,GeV.}{fig:angularresolution}

\subsection{Résolution en \'Energie}

L'énergie reconstruite est le résultat des algorithmes de reconstruction du trajectographe et principalement du calorimètre. La figure \ref{fig:energyresolution} gauche montre pour la classe \textit{diffuse}, la résolution en énergie du LAT en fonction de l'énergie pour une incidence normale. Une part de la cascade se développe dans le trajectographe limitant la reconstruction par le calorimètre à toutes les énergies, tandis qu'à haute énergie (E~$>$~30\,GeV) seule la moitié de la gerbe électromagnétique ($\sim$ 50\%) est présente dans le calorimètre (fuites vers l'arrière). Remarquons également que suivant l'endroit de la conversion $e^{\pm}$ (couche mince ou épaisse) et l'énergie de la particule incidente, l'énergie reconstruite est plus ou moins bien connue. Le calorimètre contient seulement la fin de la cascade électromagnétique pour un photon de basse énergie converti dans les couches minces. La figure \ref{fig:energyresolution} droite représente la résolution en énergie en fonction de l'angle d'incidence pour des photons de 10\,GeV.

\fig[Résolution en énergie du LAT pour la classe \textit{diffuse}]{scale=0.82}{c4_edispanelpub.pdf}{\textbf{Gauche:}~Résolution en énergie du LAT pour les événements de la classe \textit{diffuse}, classification $Pass6\_v3$, en fonction de l'énergie pour une incidence normale ($cos~\theta>0.9$), et pour des conversions dans la section mince (front), épaisse (back), mince et épaisse (total) du trajectographe. \textbf{Droite:}~Résolution en énergie en fonction de l'angle d'incidence pour des photons de 10\,GeV.}{fig:energyresolution}

\subsection{Surface Efficace}
La surface efficace est la surface vraie acceptant les événements bien reconstruits. Intégrée en fonction de l'énergie $E$ et de l'angle d'incidence $\theta$, elle est le rapport entre le nombre de photons $\gamma$ détectés et acceptés par l'analyse, et le flux incident $\gamma$ de la source pendant un temps d'observation. Ainsi de façon simplifiée, le flux intégré d'une source peut être calculé par:
\begin{equation}
F_{\gamma} = \frac{N_{\gamma}}{\Delta t \times \int A_{eff}(E,\theta) dE d\theta} \ {\rm ph \ cm^{-2} s^{-1}}
\end{equation}

La surface efficace dépend de la surface géométrique du détecteur, de la reconstruction, et de la sélection des événements suivants leur classification. La figure \ref{fig:effectivearea} gauche représente pour les photons de la classe \textit{diffuse}, la surface efficace en fonction de l'énergie pour une incidence normale. Au-dessus de 100\,MeV, en utilisant les photons convertis à la fois dans les sections minces et épaisses (total), la surface efficace est compris entre 7000 et 8000\,cm$^{2}$. La figure droite représente la surface efficace en fonction de l'angle d'incidence pour des photons de 10\,GeV.

\fig[Surface efficace du LAT pour la classe \textit{diffuse}]{scale=0.82}{c4_aeffpanel.pdf}{\textbf{Gauche:}~Surface efficace du LAT pour la classe \textit{diffuse} en fonction de l'énergie pour une incidence normale, et pour une conversion dans la section mince (front), épaisse (back), mince et épaisse (total) du trajectographe. \textbf{Droite:}~Surface efficace en fonction de l'angle d'incidence pour des photons de 10\,GeV.}{fig:effectivearea}

\section{Validation de la Surface Efficace du LAT}

\subsection{Méthode d'Analyse}

La méthode de validation de la surface efficace a été développée au SLAC\footnote{SLAC National Accelerator Laboratory, Stanford, Californie.} pendant les simulations de la réception des premières données du LAT, entre octobre 2007 et mars 2008. Elle a été ensuite appliquée aux vraies données à partir du 30 juin 2008 pendant la phase de calibration et de vérification de l'instrument. Notons qu'elle est toujours utilisée par la collaboration pour valider les nouvelles versions de l'analyse de reconstruction des particules.

Conceptuellement, la surface efficace des photons peut être décrite comme :
\begin{equation}
A^{\gamma}_{eff} (E,\theta) = A_{geo} \times \prod_{i=1}^{M} \epsilon_{i}^{\gamma} (E,\theta),
\end{equation}
avec $A_{geo}$ la surface géométrique du détecteur ($\sim$22500\,cm$^{2}$), et $M$ le nombre de niveaux dans l'analyse des événements décrite en section \ref{sec:anactb}. $\epsilon_{i}^{\gamma} (E,\theta)$ est l'efficacité de la coupure ($i$) en fonction de l'énergie $E$ de la particule et de $\theta$ l'angle entre l'axe normal au plan du LAT ($z$) et la direction du photon incident. On définit l'efficacité par~:
\begin{equation}\label{eq:efficiency}
\epsilon_{i}^{\gamma} (E,\theta) = \frac{N_{\gamma}(i+1)}{N_{\gamma}(i)},
\end{equation}
avec $N_{\gamma}(i)$ le nombre d'événements restant après la coupure ($i$).

Pour valider la surface efficace $A^{\gamma}_{eff}$ des photons, nous proposons de comparer pour les données simulées et les données observées par le LAT, l'efficacité des coupures à chaque niveau ($i$) de la rejection du fond de particules chargées. Si les estimations ne sont pas en accord, nous déterminons de la même manière quelles sont les variables responsables, telles que les CT associés au trajectographe, au calorimètre, et au détecteur anti-coïncidence (section \ref{sec:lowlevel}). La problématique avant de comparer l'efficacité des coupures est la sélection des photons. Comment obtenir un échantillon de photons purs provenant d'une source observée par l'instrument ? Comment distinguer un photon d'une particule chargée observée par le LAT à chaque niveau de l'analyse ? Le choix des sources est ainsi limité par la capacité à sélectionner des photons. Il est également nécessaire d'avoir une source assez brillante pour l'observer avant la rejection du fond hadronique. Les pulsars permettent de sélectionner un échantillon pur de photons $\gamma$ en utilisant leur information temporelle. De plus, le pulsar de Vela est la source la plus brillante du ciel $\gamma$.

La figure \ref{fig:velaphase} présente le phasogramme (ou courbe de lumière) du pulsar de Vela pour deux coupures (rejection du fond) différentes. Rappelons que la courbe de lumière d'un pulsar est le nombre d'événements en fonction du temps, modulo la période. Nous reviendrons sur cette notion dans le chapitre \ref{chap:analyse}. La figure gauche montre la courbe de lumière observée après le filtre OBF et l'application d'une coupure minimale pour une région d'intégration (ROI) de 10$^{\circ}$ autour de la position du pulsar, tandis que la figure droite montre pour la même ROI la courbe de lumière de ce même pulsar pour la classe d'événements \textit{diffuse}. Les coupures appliquées sont énumérées dans la table \ref{tab:highlevelcut}. Ainsi, le nombre de photons peut être estimé par la soustraction du nombre d'événements de la région OFF normalisée (0.64 $< \phi < 1.0$), moins le nombre d'événements de la région d'émission des pics (région ON). Le premier pic est défini entre $0.08 < \phi < 0.18$, tandis que le second pic est défini entre $0.5 < \phi < 0.62$. Cette méthode s'appelle la technique ON -- OFF. Le nombre de photons $N_{\gamma}$ est donc égal à $N_{p1}/0.1 + N_{p2}/0.12 - N_{off}/0.36$, avec $N_{p1}$ et $N_{p2}$ le nombre d'événements dans les pics et $N_{off}$ le nombre d'événements dans la région OFF. 

Notons que l'information temporelle du pulsar (sa phase) n'est pas inclus dans les fichiers de données. Pour ce travail, nous avons ainsi créé un outil nommé ``addPhase'' qui permet d'ajouter dans les fichiers bas-niveau (fichiers $merit$) et haut-niveau (fichiers $fits$) la phase du pulsar en se basant sur ses paramètres de rotation (voir le chapitre \ref{chap:analyse}).

\fig[Phasogrammes du pulsar Vela, avant et après coupures]{scale=0.5,angle=90}{c4_vela_phase.pdf}{\textbf{Gauche:}~Courbe de lumière pour le pulsar de Vela après le filtre OBF et l'application d'une coupure minimale (voir la table \ref{tab:highlevelcut}). La partie grisée correspond aux deux pics du pulsar (région $ON$). La ligne pointillée indique le début de la région $OFF$ ($0.64 < \phi < 1.0$). \textbf{Droite:}~Courbe de lumière du pulsar de Vela pour les événements ``diffuse'' (coupure finale). Notons que le fond résiduel passe d'environ 5000 coups par intervalle de phase à une vingtaine de coups en appliquant les coupures standards. Dans le même temps, le signal passe d'une significativité ($signal/\sqrt{fond}$) d'environ 5$\sigma$ à 20$\sigma$.}{fig:velaphase}

\subsection{Validation à Haut Niveau}\label{sec:highlevel}

La validation à haut niveau correspond à l'analyse des efficacités pour les classes d'événements \textit{transient}, \textit{source}, et \textit{diffuse}, définies dans la section \ref{sec:anactb}. Pour illustrer ce travail, nous présenterons dans cette thèse uniquement la validation de la classe ``Diffuse'' utilisée pour l'analyse des pulsars. L'efficacité des coupures $\epsilon_{i}^{\gamma} (E,\theta)$ pour la classe \textit{diffuse} est: le nombre de photons estimé par la techinique $ON$ -- $OFF$ et passant l'analyse \textit{diffuse} (rejection du fond et reconstruction des événements), sur le nombre de photons après la coupure minimale (voir l'équation \ref{eq:efficiency}). Etant donné l'important espace de phase du LAT, l'efficacité de coupures sur les données LAT et Monte Carlo est évaluée suivant un nombre maximum d'intervalles en énergie et en angle, lequels sont limités par la statistique des données LAT.

\tab[Définition des coupures standards (Transient, Source, Diffuse)]{highlevelcut}{Définition des coupures standards pour l'analyse des événements du LAT (Transient, Source, Diffuse) et de la coupure minimale permettant d'avoir une reconstruction d'événement. \\ \\
\textit{Notes:} \\
-- $TkrNumTracks>0$ demande au moins une trace dans le trajectographe. \\
-- $CalCsIRLn>4$ impose que la trace intercepte le calorimètre ($>$4\,X$_{0}$). \\
-- $CalEnergyRaw>5$ impose au moins 5\,MeV d'énergie déposée dans le calorimètre. \\
-- $CTBBestEnergyRatio$ est le rapport entre la meilleure énergie estimée (``CTBBestEnergy'') et l'énergie déposée dans le calorimètre. \\
-- $CTBCLassLevel$ est une variable qui classe les événements selon le degré de confiance qu'il soit un photon (0=fond hadronique, 1=\textit{transient}, 2=\textit{source}, 3=\textit{diffuse}). \\
-- $FT1Dec$ et $FT1Ra$ sont les coordonnées de l'événement. Les coordonnées du pulsar Vela utilisées sont (ra,dec) = (-45.176354$^{\circ}$,128.835881$^{\circ}$).  
}{tab:highlevelcut}

\fig[Efficacité des coupures pour la classe \textit{diffuse}]{scale=0.86,angle=180}{c4_efficiency_energy.pdf}{Efficacité des coupures pour la classe \textit{diffuse} en fonction de l'énergie (``CTBBestLogEnergy'') et de l'angle d'incidence du photon. $CTBBestZDir$ correspond à $-cos~\theta$, avec $\theta$ l'angle entre l'axe normale au plan du LAT et la direction incidente du photon. Les points noirs correspondent aux données du LAT (incertitude statistique), tandis que les courbes rouge et bleues sont des données simulées. (Crédit: P.Bruel).}{fig:efficiencydiffuse}

La figure \ref{fig:efficiencydiffuse} montre pour 8 intervalles en angle défini par ``CTBBestZDir''\footnote{Cette variable correspond à $-cos~\theta$, avec $\theta$ l'angle entre l'axe normale au plan du LAT et la direction incidente du photon.}, l'efficacité $\gamma$ pour la classe d'événement \textit{diffuse} en fonction de l'énergie définie par la variable ``CTBBestLogEnergy'' pour la classe \textit{diffuse}. Les points noirs correspondent aux données du LAT, tandis que les courbes rouges et bleues correspondent à deux échantillons de photons simulés. La différence entre ces deux simulations est la prise en compte pour la simulation v17r7 (courbe rouge) des traces fantômes (figure \ref{fig:ghosttrack}). On notera que pour le deuxième intervalle en angle (-0.9$<$ CTBBestZDir $<$-0.8), l'efficacité à partir de quelques GeV se dégradent. L'effet est d'environ 2\,$\sigma$, et nous n'en connaissons pas la raison exacte. Dans la sélection \textit{diffuse}, il se peut qu'il y ait une dépendance en angle que l'on ne prenne pas en compte dans l'analyse. \\

Dans les données Monte Carlo, chaque événement est généré indépendamment, et les interactions entre ces événements ne sont pas pris en compte. Plusieurs cas de ce qu'on appelle les traces fantômes ont été observées dans les données réelles pendant la période de mise en route. La plupart des événements du fond hadronique sont facilement identifiés en tant que tels, et les triggers individuels du LAT ne sont pas activés pour enregistrer ce type d'événement. Cependant, il arrive qu'un photon interagissant avec le LAT ouvre la fenêtre d'acquisition des données, et au même moment une particule chargée entre dans le détecteur. Il est également possible que le signal d'un ion soit encore présent lors du déclenchement du détecteur par un photon. Les traces laissées par la particule chargée sont ainsi enregistrées. La figure \ref{fig:ghosttrack} montre un exemple de trace fantôme. La conséquence est qu'un certain nombre de photons bien reconstruits sont identifiés comme des mauvais événements lors de l'analyse de la reconstruction au sol. Ces événements sont rejetés et conduisent à une dégradation des performances du LAT et notamment de la surface efficace. Ainsi, pour être en accord avec les données réelles, la simulation v17r7 inclut aléatoirement des faux événements pour prendre en compte ces problèmes de coïncidence.

\fig[Exemple d'une trace fantôme dans le LAT]{scale=0.9}{c4_ghost_track.pdf}{Exemple d'un événement fantôme dans le LAT. Sur la droite, un photon interagit dans le trajectographe et dans le calorimètre. Sur la gauche, il apparaît une trace fantôme dans le trajectographe résultant d'une coïncidence temporelle accidentelle d'une particule chargée. Dans le trajectographe et le calorimètre, seules les parties actives sont montrées. Les traces candidates sont également dessinées.}{fig:ghosttrack}

La figure \ref{fig:efficiencyresidual} présente le rapport entre les vraies données et les données simulées pour les mêmes intervalles en énergie et en angle que la figure \ref{fig:efficiencydiffuse}. Les points bleus résultent de la comparaison entre les vraies données et la simulation v15r51p1 qui n'inclut pas les effets d'événements fantômes, tandis que les points noirs utilisent les données simulées prenant en compte les traces fantômes. On remarque clairement que la dernière simulation reproduit mieux les données observées par le LAT. A partir de ces résultats, une enveloppe correspondant aux incertitudes systématiques induite par la surface efficace a été estimée. Les résultats sont résumés dans la section \ref{errorsys}. 


\subsection{Validation des Variables Individuelles}\label{sec:lowlevel}

Pour vérifier la procédure de validation à haut-niveau, les principales variables de la classification sont également analysées. Elles permettent d'affiner la validation et notamment de vérifier si il n'y a pas un défaut sous-jacent aux trois classes de photons. Nous ne présenterons ici que la variable $CTBCALGamProb$ qui représente la probabilité estimée par le calorimètre que la particule soit un photon. 

\fig[Efficacité de la variable CTBCALGamProb]{scale=0.38,angle=0}{c4_CTBCALGamProb.png}{\textbf{Gauche:}~Efficacité de la variable $CTBCALGamProb$ par intervalle de 0.1. Les points bleus représentent l'efficacité des données Monte Carlo v15r21. Cette version de la simulation ne prend pas en compte les effets fantômes. Les points rouges représentent les données LAT. \textbf{Droite:}~Déviation de la variable $CTBCALGamProb$.}{fig:ctbcalgamprob}

La figure \ref{fig:ctbcalgamprob} (gauche) montre l'efficacité de la variable $CTBCALGamProb$ par intervalle de 0.1 entre les données du LAT et les données simulées v15r21. Les points bleus représentent l'efficacité des données Monte Carlo, tandis que les points rouges représentent les données LAT avec la prise en compte de l'incertitude statistique. A droite, la déviation de la variable est représentée. Malgré l'absence de prise en compte des effets fantômes, elle est inférieure à 20\%. La majorité des points (excepté le point 0 -- 0.1) sont en accord aux incertitudes près avec une déviation égale à 0. Ces variables sont évaluées à chaque mise à jour de l'analyse de la classification des événements. Rappelons néanmoins que le but est d'avoir une incertitude systématique suffisamment petite sur les quantités mesurées, tels que le flux et l'indice spectral des sources. 

\subsection{Estimation des Incertitudes Systématiques}\label{errorsys}

La figure \ref{fig:efficiencyresidual} montre à partir du rapport des efficacités, l'estimation des incertitudes systématiques induite par la surface efficace. L'enveloppe supérieure représentée en noire associée à la classification $Pass6\_v1$ montre une erreur de 20\% en dessous de 100\,MeV, environ 10\% en dessous de 1\,GeV, et de 30\% au-dessus de 10\,GeV. L'enveloppe rouge (inférieure) associée à la classification $Pass6\_v3$, qui prend en compte les effets de traces fantômes, montre une erreur de 10\% en dessous de 100\,MeV, environ 5\% en dessous de 1\,GeV, et de 20\% au-dessus de 10\,GeV. A partir de ces résultats, deux échantillons d'IRFs ont été créés pour estimer les effets maximaux de la variation de la surface efficace. Ils ne prennent pas en compte les erreurs sur la PSF et sur la résolution en énergie. Ces deux échantillons sont ensuite utilisés pour évaluer les erreurs systématiques sur le flux $\gamma$ intégré, l'indice spectral $\Gamma$, la coupure en énergie $E_{C}$, et l'indice d'atténuation $\beta$ des pulsars (voir l'équation \ref{eq:powerlawexpcutoff}). \\

Nous verrons aux chapitres 6, 7, et 8 (résultats sur les pulsars individuels), comment est évaluée l'incertitude systématique sur les paramètres spectraux et notamment l'indice d'atténuation $\beta$, paramètre qui permet de discrimer les différents modèles d'émission $\gamma$ des pulsars. Brièvement, la méthode consiste à ajuster les paramètres aux données en utilisant premièrement l'IRF standard, et ensuite les deux IRFs modifièes pour estimer un intervalle de confiance.

\fig[Rapport des efficacité des coupures pour la classe \textit{diffuse}]{scale=0.55,angle=-90}{c4_showcompeff2_v17r7.png}{Rapport des efficacités entre les vraies données et les données Monte Carlo. Les points bleus résultent de la comparaison entre les vraies données et la simulation v15r51p1 n'incluant pas les effets d'événements fantômes. Les points noirs résultent de la comparaison entre les vraies données et la simulation v17r7 prenant en compte les effets de trace fantôme. Les courbes rouges et noirs représentent l'estimation des systématiques pour la version initiale de la procédure initiale d'analyse des événements (P6\_v1) et la version améliorée (P6\_v3) prenant en compte les effets de trace fantôme. (Crédit:~Ph.Bruel).}{fig:efficiencyresidual}
%

\chapter[Analyse des Données]{Analyse des Données}\label{chap:analyse}


\minitoc

\section{Introduction}

Ce chapitre décrit les analyses temporelles et spectrales des pulsars $\gamma$ à partir des données du LAT\footnote{Les données sont publiques depuis le 19 août 2009 et peuvent être téléchargées ici: http:$//$fermi.gsfc.nasa.gov/cgi-bin/ssc/LAT/LATDataQuery.cgi}, des outils standards développés par la collaboration \textit{Fermi} (les $Science Tools$), ainsi que du programme d'analyse des pulsars radio $TEMPO2$ \citep{hobbs06}. Une première partie présente la construction d'une courbe de lumière (ou phasogramme, histogramme à n intervalles en phase) à l'aide d'une éphéméride de pulsar, et une seconde partie présente l'analyse spectrale résolue en phase d'un pulsar en prenant en compte les sources avoisinantes et le fond d'émission diffus. \\

Notes sur les $Science Tools$ (ST): \\
Les Sciences Tools sont un ensemble d'outils utilisés pour l'analyse des données haut niveau de \textit{Fermi}, données stockées dans des fichiers au format $fits$ (voir le schéma \ref{fig:analysis}). Les outils sont publiques, développés en langage C++ et Python, et se répartissent en sous-groupes chacun associé à différents types d'analyse:~l'extraction et la sélection des données, l'analyse spectrale des sources ponctuelles, les sursauts $\gamma$, l'analyse des pulsars, la simulation, etc. Le lecteur désireux de débuter sa première analyse trouvera une documentation complète et mise à jour des ST sur: http:$//$glast-ground.slac.stanford.edu/workbook/sciTools\_Home.htm et http:$//$fermi.gsfc.nasa.gov/ssc/ .

\section{Analyse Temporelle}\label{sec:anatemp}

Signalons au préalable, que le temps d'arrivée des événements du LAT doit être correct et précis afin que cette étude soit menée à bien. Des tests sur les horloges du LAT ont été effectués avant et après la mise en orbite du satellite pour s'assurer de leur bon fonctionnement \citep{abdo09calib}. Au sol, les tests ont consisté à mesurer la différence des temps d'arrivée des muons entre les systèmes GPS du LAT et un système GPS de référence, tandis qu'en orbite nous avons utilisé la position des pics $\gamma$ des pulsars les plus brillants. \\

Pour construire la courbe de lumière d'un pulsar (ou phasogramme, voir la figure \ref{fig:cgropulsar}), il faut déterminer à quel stade de la rotation de l'étoile à neutrons le photon détecté par le télescope a été émis. En d'autres termes, il faut attribuer à chaque photon détecté une phase rotationnelle. Signalons que l'abcisse d'un phasogramme est très souvent définie en tant que phase du pulsar. Il s'agit en fait de la phase normalisée à $2\pi$, c'est-à-dire la fraction de tour de l'étoile à neutrons que l'on représente entre 0 et 1. 

La première étape est de remettre en ordre le temps des événements, en transférant les temps d'arrivée du photon observé par le LAT vers le barycentre du système solaire (section \ref{sec:bary}), ceci pour s'affranchir des mouvements propres de l'observatoire et du pulsar. Ensuite, à l'aide de ces temps barycentrés et des paramètres de rotation du pulsar décrits par une éphéméride (section \ref{sec:ephem}), une phase est assignée à chaque photon. Dans cette thèse et dans une grande partie des publications scientifiques de la collaboration \textit{Fermi}, la construction des phases pour les pulsars est obtenue avec le programme d'analyse $TEMPO2$ et le ST ``gtpphase''. Notons cependant que l'outil gtpphase ne peut prendre en compte que les deux premières dérivées de la fréquence pour calculer la phase (voir l'équation \ref{eq:phase}). Etant donné que les pulsars que l'on étudie sont relativement jeunes et présentent une fréquence de rotation instable (\textit{timing noise} et \textit{glitch}), il est souvent nécessaire de recourir à la troisième voire la quatrième dérivée de la fréquence. Cette situation m'a amené à utiliser uniquement l'outil $TEMPO2$ pour éviter toute confusion dans l'analyse.

\subsection{Barycentrisation des \'Evénements}\label{sec:bary}

Pour analyser le temps des événements émis par le pulsar, ou en d'autres termes suivre sa rotation de façon précise, il est nécessaire idéalement de se situer dans un référentiel au repos par rapport à l'objet. Cependant, notre référentiel d'observation n'est pas inertiel vis-à-vis du pulsar, que ce soit pour des télescopes situés sur Terre en mouvement autour du soleil, ou pour le satellite \textit{Fermi} en orbite autour de la Terre. Le meilleur référentiel est le barycentre du système solaire (SSB, \textbf{S}olar \textbf{S}ystem \textbf{B}arycenter), point proche du soleil. La ``barycentrisation'' consiste ainsi à transférer des temps d'arrivée topocentriques (enregistrés par un observatoire) vers le centre de masse du système solaire qui varie suivant la position des astres dans le système solaire. Ceci revient à dater un événement supposé observé au barycentre du sytème solaire, et permet d'avoir, pour tous les événements, une échelle de temps homogène notée TDB, pour ``\textbf{T}emps \textbf{D}ynamique \textbf{B}arycentrique''. Ce temps barycentré $t_{SSB}$ est calculé par $TEMPO2$ en fonction du temps $t_{MET\_TT}$ de l'événement enregistré par le LAT. Les temps $t_{MET\_TT}$ sont exprimés en \textbf{M}ission \textbf{E}lapsed \textbf{T}ime (MET) TT (\textbf{T}emps \textbf{T}errestre)\footnote{Le temps TT est le temps que mesurerait une horloge parfaite, située au niveau de la géoïde: surface équipotentielle coïncidant avec le niveau de la mer en première approximation.}. Le temps $t_{SSB}$ est calculé de la façon suivante:
\begin{equation}
t_{SSB} = t_{MET\_TT} + \Delta_{Einstein} + \Delta_{Roemer} + \Delta_{Shapiro} \ {\rm .}
\end{equation}

\begin{ablist}
\item $\Delta_{Einstein}$:~cette échelle regroupe la dilatation du temps introduite par le mouvement de la Terre dans le potentiel gravitationnel du système solaire, ainsi que le décalage vers le rouge gravitationnel induit par les corps du système solaire, et permet de passer du référentiel géocentrique (temps en TT) au référentiel barycentrique (temps TDB). Les positions du soleil, de la Lune, et des 8 planètes de notre système solaire sont définies dans des éphémérides du système solaire DE200 ou DE405 publiées par le JPL\footnote{Jet Propulsion Laboratory} \citep{standish98}.
\item $\Delta_{Roemer}$:~cet effet géométrique correspond au décalage séparant l'arrivée des photons du LAT par rapport au barycentre du système solaire.
\item $\Delta_{Shapiro}$:~cet effet est une correction temporelle induit par le passage proche des photons dans l'espace-temps courbé d'un objet massif, tel que le Soleil.
\end{ablist}

Pour plus de précisions sur la ``barycentrisation'' ainsi que sur les échelles de temps, on se reportera à l'excellent ouvrage de \citet{handbook05}.

\subsection{Les \'Ephémérides}\label{sec:ephem}

Les éphémérides des pulsars décrivent les paramètres de la dynamique de rotation des pulsars en fonction du temps (i.e. la fréquence de rotation et ses dérivées) dans le référentiel du barycentre du système solaire (TDB), ainsi que les paramètres astrométriques nécessaires pour la barycentrisation, tels que la position du pulsar, le mouvement du pulsar dans le plan du ciel, et les paramètres orbitaux pour les systèmes multiples. Rappelons que la section \ref{sec:campephem} décrit brièvement la campagne de chronométrie des pulsars qui fournit les éphémérides. En pratique, une éphéméride est construite en minimisant l'écart entre la phase observée, calculée à partir des temps d'arrivée des photons (TOAs, \textbf{T}imes \textbf{O}f \textbf{A}rrival) d'une observation donnée, et la phase calculée à partir d'un modèle de rotation du pulsar, à savoir les paramètres décrits précédemment (la position, la fréquence de rotation, etc). Le calcul de la phase d'un pulsar à partir du temps d'arrivée des photons est donné par l'équation \ref{eq:phase}. La plupart des paramètres de l'éphéméride sont établis pendant la découverte du pulsar, tandis que la fréquence et ses dérivées sont réajustées à mesure que les observations progressent. L'observation d'un pulsar pour un radiotélescope peut durer de quelques minutes à quelques heures, cela dépend de l'intensité du signal radio. 

La figure \ref{fig:residualphase} présente les résidus des phases (phase prédite moins phase observée) pour trois pulsars observés par GBT, dont PSR~J2229+6114 présenté au chapitre \ref{chap:j2229}. Remarquons que ces résidus sont dispersés autour de 0. Cette dispersion peut s'expliquer principalement par du \textit{Timing Noise}, par l'incertitude des observations radio, ou encore par des effets non pris en compte dans l'ajustement de la dynamique de la rotation du pulsar. Il est donc essentiel pour certains pulsars très bruités (i.e. un \textit{timing noise} important), notamment pour les pulsars jeunes tels que le pulsar du Crabe, de les observer régulièrement. 

\fig[Résidus des phases pour PSR~J1747-2958, PSR~J1930+1852, et PSR~2229+6114]{scale=0.6}{c5_ephem_residual.pdf}{Résidus des phases pour PSR~J1747-2958, PSR~J1930+1852, et PSR~2229+6114, observés par le Green Bank Telescope à 2\,GHz (Crédit:~F.Camilo).}{fig:residualphase}

La Table \ref{tab:ephem2229} présente les principaux paramètres de l'éphéméride pour le pulsar J2229+6114 basée sur les observations des télescopes GBT et Jodrell Bank. Cela inclut:~la position du pulsar fixée aux coordonnées des observations $X$ \citep{halpern01b}, la fréquence et ses dérivées, une époque T$_{0}$ de référence (point central de l'ajustement du développement de Taylor, voir l'équation \ref{eq:phase}), une date de référence (``TZRMJD'') pour laquelle la phase est égale à 0, l'intervalle de validité de l'ajustement, le nombre de TOAs utilisés pour construire l'éphéméride, la mesure de dispersion, et le modèle d'éphéméride du système solaire utilisé pour la ``barycentrisation''.

\begin{table}[ht!]
\begin{center}
\shadowbox{\begin{minipage}[b][1.05\totalheight][c]{16cm}
\begin{center} 
\begin{tabular}{ll}
\hline 
\hline
Paramètres & Valeurs \\
\hline
Ascension Droite (J2000)\dotfill & 22$^{h}$29$^{m}$05.28000084$^{s}$ \\
Déclinaison (J2000)\dotfill & +61${^\circ}$14'09.3000042'' \\
Fréquence (s$^{-1}$)\dotfill & 19.363953152(1) \\
Dérivée première de la fréquence (s$^{-2}$)\dotfill & -2.92006(5) $\times$10$^{-11}$ \\
Dérivée seconde de la fréquence (s$^{-3}$)\dotfill & 1.50(8)$\times$10$^{-21}$ \\
Epoque, T$_0$ (MJD[TDB])\dotfill & 54774.0 \\
Date de référence pour le $\phi_{0}$ (MJD[TDB])$^{a}$\dotfill & 54780.986659550693183 \\
Intervalle de validité en MJD \dotfill & 54634.286 -- 54913.622 \\
Nombre de TOAs\dotfill & 69 \\
Mesure de dispersion, DM(cm$^{-3}$\,pc)\dotfill & 204.97 $\pm$ 0.02 \\
Modèle d'éphéméride du S.S. \dotfill & DE405\\
\hline
\hline
\end{tabular}
\parbox{15cm}{\caption[Ephéméride du pulsar PSR~J2229+6114]{Ephéméride radio du pulsar PSR~J2229+6114 basée sur les observations des télescopes Green Bank Telescope (GBT) et Jodrell Bank. Les chiffres entre parenthèses représentent les incertitudes données par $TEMPO2$. \\
$^{a}$ Date à laquelle la phase $\phi = 0$ (TZRMJD).}\label{tab:ephem2229}}
\end{center}
\end{minipage}}\end{center}
\end{table}

Les éphémérides radio et $X$ utilisées pour la détection de pulsations $\gamma$ avec le LAT sont stockées sur les serveurs du CENBG Bordeaux-Gradignan\footnote{http:$//$www.cenbg.in2p3.fr/ephem/} et de la NASA\footnote{http:$//$fermi.gsfc.nasa.gov/ssc/data/access/lat/ephems/}.

\subsection{Construction d'une Courbe de Lumière}

Soit $\phi(t)$ la phase du pulsar à l'instant $t$. En utilisant les paramètres d'une éphéméride, la courbe de lumière est l'histogramme de la phase calculée à partir des temps d'arrivée des photons, préalablement transformés dans le référentiel du centre masse du système solaire suivant le développement de Taylor :
\begin{equation}\label{eq:phase}
\phi(t) = \phi_0 + \sum_{j = 0}^{j = N} \frac{f_{j} \times (t - T_0)^{j+1}}{(j+1)!}
\end{equation}
où $T_{0}$ est la date de référence de l'éphéméride du pulsar (voir la table \ref{tab:ephem2229}), $f_{j}$ est la dérivée de la fréquence de rotation à l'ordre $j$, et $\phi_0$ est la phase de référence. Cette dernière est donnée par la date de référence (``TZRMJD'') pour laquelle la phase est égale à 0 :
\begin{equation}\label{eq:phase0}
\phi_0 = - \sum_{j = 0}^{j = N} \frac{f_{j} \times (TZRMJD - T_0)^{j+1}}{(j+1)!}
\end{equation}

\subsection{Significativité}\label{sec:signi}

A partir des événements barycentrés et empilés suivant l'équation \ref{eq:phase}, la distribution de phases est ensuite analysée pour rechercher un éventuel signal pulsé à la fréquence radio attendue. Dans le cas du pulsar de Vela (voir la figure droite \ref{fig:velaphase}), la mise en évidence de la pulsation est claire. On peut simplement, pour estimer la significativité du signal, compter le nombre de coups pulsés au-dessus du fond que l'on compare au nombre de coups du fond ($signal/\sqrt{fond}$). Cependant dans certains cas, la fraction pulsée peut être beaucoup plus faible et le fond d'émission $\gamma$ plus important. Les tests statistiques suivants sont codés et utilisés pour rechercher un signal périodique à partir d'une courbe de lumière.

\subsubsection{Test Statistique $\chi^{2}$}

De façon générale, le test $\chi^{2}$ mesure la signification de l'écart entre des distributions observées et des distributions attendues, soit :
\begin{equation}
\chi^2 = \sum_{i=1}^{N} \frac{(x_i - \mu_i)^2}{\sigma_{i}^{2}} \ {\rm ,}
\end{equation}
où $N$ est le nombre d'intervalles, $x_i$ est le nombre d'événements dans un intervalle $i$, $\mu_i$ est la distribution attendue de coups, et $\sigma_{i}^{2}$ est la variance associée. Dans le cas d'un phasogramme, le test est utilisé pour tester l'hypothèse selon laquelle la distribution de phases est plate, c'est-à-dire, que les phases sont uniformément distribuées, soit $\mu_i = \langle x \rangle$, avec $\langle x \rangle$ la valeur moyenne des $x_i$. Pour un nombre suffisamment grand d'événements, la distribution des $x_i$ suit la loi de Poisson, ainsi $\sigma_{i}^{2}$ est équivalent à $\langle x \rangle$. Finalement, le test $\chi^{2}$ pour un phasogramme est défini par:
\begin{equation}
\chi^2 = \sum_{i=1}^{N} \frac{(x_i - \langle x \rangle)^2}{\langle x \rangle} \ {\rm ,}
\end{equation}
La conséquence est que pour un fort signal pulsé, la déviation par rapport à une distribution plate sera importante, ainsi que le résultat du $\chi^{2}$. On trouvera dans \citet{leahy83} la probabilité associé à un $\chi^{2}$ d'observer une distribution plate. Notons que ce test dépend du nombre d'intervalles utilisé pour construire la courbe de lumière.

\subsubsection{Test Statistique $Z^{2}_{m}$}

De la même façon que le test $\chi^{2}$, le $Z^{2}_{m}$ \citep{buccheri83} s'appuie sur la mesure de la différence entre une distribution plate et mesurée. Il se base sur le calcul du spectre de puissance de Fourier de la distribution de phases, en utilisant les m premières harmoniques. De plus, il est indépendant du choix du nombre d'intervalles du phasogramme construit. Ainsi, la déviation de la distribution de phase mesurée par rapport à une distribution plate est calculée à partir des $N_i$ valeurs de phases obtenues par l'équation \ref{eq:phase} (provenant des $N_i$ temps d'arrivée des photons), soit:
\begin{equation}
Z_m^2 = \frac{2}{N_i} \sum_{k=1}^{m} (\sum_{j=1}^{N_i} cos \ k \phi_j)^2 + (\sum_{j=1}^{N_i} sin \ k \phi_j)^2
\end{equation}
La probabilité de $Z^{2}_{m}$ dans le cas d'une absence de signal	est en fait une probabilité de $\chi^{2}$ avec 2m degrés de liberté. Notons que suivant le choix de l'harmonique m choisie, le test est plus ou moins performant selon la forme du pic. Pour les pics étroits une harmonique élevée est plus appropriée, tandis que pour des pics larges une harmonique petite est préférable. Il faut donc avoir une idée a priori de la forme de la courbe de lumière, ce qui n'est pas très satisfaisant.

\subsubsection{Test Statistique $H$-test}

Basé sur le $Z^{2}_{m}$, \citet{dejager89} proposent le test statistique H-test qui calcule la valeur optimale $m$ de l'harmonique. L'avantage de ce test est qu'il n'est pas biaisé par la forme du pic. L'harmonique optimale correspond à l'harmonique $m$ comprise entre 1 et 20, pour laquelle la valeur $H$ est maximisée, soit:
\begin{equation}
H = Z^{2}_{m} - 4m + 4 \ {\rm .}
\end{equation}
L'expression est applicable pour un nombre d'événements supérieur à 100. Finalement, la probabilité associée à $H$ d'obtenir une distribution plate est donnée par:
\begin{equation}
P(H) = \left\{
    \begin{array}{ll}
        0.9999755 e^{-0.39802 H} & \mbox{si } 0 < H < 23 \\
        1.210597 e^{(-0.45901 H + 0.0029H^2)} & \mbox{si } 23 < H < 50 \\
        < 4 \times 10^{-8} & \mbox{si } H > 50 \\
    \end{array}
\right.
\end{equation}


\section{Analyse Spectrale Résolue en Phase}\label{sec:anaspec}

\subsection{Introduction}

L'analyse spectrale des pulsars (sources ponctuelles) se base sur une liste de photons détectés par le LAT à l'intérieur d'une région d'intégration (ROI) qui est la région d'extraction des données en fonction de leur énergie, de leur direction, et de leur temps d'arrivée. Cette liste de photons provient de la reconstrution et de la classification des événements à partir des données mesurées du LAT (voir la section \ref{sec:anactb}). Cependant, le faible nombre de photons observés pour une source ponctuelle après l'application des coupures d'une classe de photons donnée signifie que les méthodes statistiques telles que les tests $\chi^2$ ne sont pas applicables pour évaluer les paramètres spectraux du pulsar. Il est nécessaire d'estimer ces paramètres par un maximum de vraisemblance en supposant une distribution discrète des événements (loi de Poisson). L'estimation a pour but de trouver les meilleurs valeurs possibles des paramètres pour que la fonction de distribution modélisée des événements s'ajuste le mieux aux données disponibles. Dans notre cas, les paramètres des sources sont le flux intégré et les paramètres spectraux du pulsar (voir l'équation \ref{eq:powerlawexpcutoff}). Finalement, les termes ``résolue en phase'' signifient que les photons sont sélectionnés dans un intervalle de phase défini. A condition que la statistique soit suffisante, ceci permet d'étudier, par l'estimation des paramètres spectraux, les différents composants d'une courbe de lumière, tels que les pics ou la région entre les pics. Avant de présenter la méthode, il est essentiel de rappeler quelques points importants spécifiques au télescope qui doivent être pris en compte dans l'analyse:

\begin{ablist} 
\item Le LAT ne fixe pas un point fixe du ciel comme le télescope EGRET et les télescopes $X$, il opère selon un mode de balayage lui permettant d'observer le ciel entier en deux orbites\footnote{Rappelons qu'une orbite dure environ 95 minutes.}. La conséquence est que les sources ne sont pas observées continûment en temps et avec le même angle d'incidence.
\item Le champ de vue du LAT est important (2.4\,sr). De nombreuses sources sont vues en même temps.
\item La PSF du LAT est relativement large (voir la figure \ref{fig:angularresolution}). De nombreuses sources, notamment au niveau du plan galactique, se recouvrent spatialement. De plus, le fond d'émission diffus du plan galactique est fortement structuré. Il est donc impossible, sauf éventuellement à haute latitude galactique, d'analyser une source ponctuelle isolée. L'analyse implique donc l'ajustement simultané de plusieurs sources ponctuelles et du fond d'émission diffus.
\end{ablist} 

Finalement, il n'est pas simple d'analyser une source ponctuelle dont les photons arrivent généralement discontinus dans le temps et avec un angle d'incidence différent. Ainsi, l'analyse suppose de connaître suffisamment bien les réponses de l'instrument (nécessité de les valider, voir le chapitre \ref{chap:irfs}), ainsi que la liste des sources avoisinantes et l'émission diffuse du fond (voir \ref{sec:catalogsrc}). L'analyse se compose ainsi en trois étapes:~(1) la sélection des données en tenant compte de la PSF, (2) la création d'un fichier contenant la liste des sources ponctuelles et diffuses de la région sélectionnée, (3) l'ajustement des paramètres spectraux par un maximum de vraisemblance. 

\subsection{Sélection des Données}\label{sec:selectiondata}

La première étape pour l'analyse spectrale d'un pulsar est la sélection des données. La PSF étant large à basse énergie (e.g. 68\% des photons d'une source ponctuelle sont contenus dans une région de 4.9 degrés à 100\,MeV centrée sur la source, voir la figure \ref{fig:angularresolution}), il est indispensable de prendre une région (ROI) suffisamment grande pour analyser l'ensemble des photons provenant de la source et ajuster correctement le fond diffus (environ 15$^{\circ}$ à l'intérieur du plan galactique et 10$^{\circ}$ à l'extérieur). La contrepartie est qu'à l'intérieur de cette région, les sources voisines contribuent à la liste de photons. Il faut donc inclure ces contributions dans l'analyse. La position et le spectre des sources ainsi que le fond diffus proviennent d'un catalogue (voir \ref{sec:catalogsrc}). Notons que la sélection des données stockées dans un fichier $fits$ (FT1) selon un intervalle de temps, un intervalle en énergie, ou une région spatiale, est effectuée par le ST $gtselect$. De plus, spécifiquement à l'étude des sources ponctuelles galactiques, seuls sont sélectionnés les événements de la classe \textit{diffuse} qui ont la plus grande probabilité d'être un photon. Enfin, les événements possédant un angle zénithal supérieur à 105$^{\circ}$ sont exclus de l'analyse. Au-delà de cette limite, une grande partie des photons $\gamma$ observés par le LAT résulte de l'interaction des rayons cosmiques avec l'atmosphère de la Terre produisant un fond brillant intense. Signalons que ces deux dernières coupures s'appliquent également à l'analyse temporelle, mais nous tâcherons de le rappeler dans les sections concernées. On peut également appliquer à cette étape une sélection en phase des données. Cela permet d'estimer les paramètres spectraux du pulsar en fonction de la phase, pour éventuellement identifier les processus physiques (i.e. rayonnement de courbure, synchrotron, et inverse Compton) qui correspondent aux différentes parties d'une courbe de lumière.

\subsection{Modèle des Sources}\label{sec:catalogsrc}

La deuxième étape dans l'analyse spectrale d'une source est la création d'un fichier contenant le modèle des sources contribuant à la ROI et au-delà. Ce modèle codé en format XML\footnote{XML:~eXtensible Markup Language.} pour être utilisé par les Science Tools inclut la position spatiale de la source à analyser, la position des sources avoisinantes, la forme fonctionnelle des spectres des sources, et la valeur des paramètres spectraux. Toutes ces informations sont extraites des catalogues de sources générés par la collaboration \textit{Fermi} (par exemple \citealt{abdo09bsl}). Rappelons brièvement que la figure \ref{fig:egretsky} présente la position dans le ciel des 271 sources du troisième catalogue d'EGRET \citep{hartman99}. L'analyse spectrale des sources étudiées dans cette thèse utilise le catalogue du LAT (6 mois de données), lequel comptabilise au total 778 sources ponctuelles. Le catalogue basé sur 11 mois contient plus de 1000 sources !

Le modèle inclut également les modèles d'émission diffuse galactique et extragalactique qui représente environ 90\% des événements du ciel $\gamma$. L'émission diffuse galactique est modélisée à partir de GALPROP\footnote{http:$//$galprop.stanford.edu/web\_galprop/galprop\_home.html} qui est un modèle numérique de propagation des rayons cosmiques dans la Galaxie, et est élaboré pour calculer l'émission $\gamma$ \citep{Strong04a,Strong04b}. Ce modèle utilise les distributions de gaz de la Galaxie basées sur des observations spectrales de $HI$ (21\,cm) et $CO$ (115\,GHz), ainsi qu'une modélisation détaillée de la distribution stellaire galactique et de la distribution de la poussière dans le milieu interstellaire. En résumé, l'émission diffuse $\gamma$ provient de la désintégration des pions produits par l'interaction des rayons cosmiques avec le gaz interstellaire, de l'interaction des rayons cosmiques avec une population de photons de basse énergie produits par les étoiles (processus inverse Compton, \citealt{porter08}), ainsi que du rayonnement de freinage et la diffusion Compton inverse des électrons. Notons que ce modèle est mis à jour régulièrement. Ainsi, la version de GALPROP utilisée pour chaque analyse spectrale des pulsars étudiés dans cette thèse sera indiquée dans des sections appropriées. L'émission diffuse extragalactique isotrope est, quant à elle, modélisée par une simple loi de puissance d'indice spectrale fixé à -2.25. Cela inclut également le fond résiduel instrumental. Il est possible que ce fond trouve son origine dans diverses classes d'objets extragalactiques non résolus comme les noyaux actifs de galaxie. 

En pratique, seuls les paramètres spectraux de la source étudiée, des sources avoisinantes à moins de 10$^{\circ}$ de la source, ainsi que de l'émission diffuse sont laissés libres pendant l'ajustement des paramètres de la source. Cela permet d'avoir une limitation des paramètres libres. Pour l'analyse des pulsars, tous les paramètres de l'équation \ref{eq:powerlawexpcutoff} peuvent être ajustés.

\subsection{Maximum de Vraisemblance}

L'analyse des données du LAT par une méthode de maximum de vraisemblance \citep{mattox96} est utilisée pour déterminer les paramètres spectraux de notre source. Cette méthode est implantée dans les ST sous le nom de $gtlike$. Pour la décrire, prenons le cas simplifié d'une source ponctuelle, sans fond diffus, dont les photons sont observés avec un angle d'incidence $\theta$ par le LAT. La source a un spectre en énergie $S(E',\theta')$ qui ne varie pas au cours du temps. La première étape est de créer une carte d'exposition. Les fonctions de réponse de l'instrument (voir \ref{sec:irfs}) sont une fonction de l'angle d'incidence $\theta$ et de l'énergie $E$ mesurés. Le nombre de photons observés d'une source dépend donc de la quantité de temps passé à observer la source pour un angle d'incidence donné. Ce temps dépend seulement de l'orientation du LAT pendant la période d'observation et non du modèle de la source. Habituellement, la carte d'exposition est essentiellement l'intégration de la surface efficace qui dépend de $E$ et $\theta$ sur le temps. Dans le cas du LAT, la carte d'exposition $\epsilon(E',\theta')$ est obtenue en intégrant les réponses de l'instrument (voir l'équation \ref{eq:irfs}) dans l'espace de phase des quantités mesurées:
\begin{equation}
\epsilon(E',\theta') = \int_{SR} dE \ d\theta \ dt \ R(E,\theta;E',\theta',t) \ {\rm ,}
\end{equation}
où $E$ et $\theta$ sont respectivement l'énergie et l'angle d'incidence mesurés. La ``Région des sources'' SR est une partie du ciel contenant toutes les sources qui contribue de manière significative à la ROI. Ceci est nécessaire pour s'assurer que les photons des sources à l'extérieur de la ROI sont pris en compte, en raison de la large PSF du LAT notamment à basse énergie. A partir de cette fonction d'exposition, le nombre prédit d'événements observés pour une source donnée est:
\begin{equation}
N_{pred} = \int_{ROI} dE' \ d\theta' \ S(E',\theta') \ \epsilon(E',\theta')
\end{equation}

En pratique, étant donné que le meilleur modèle a la plus grande probabilité de s'ajuster aux données, les paramètres de $S(E',\theta')$ sont ajustés pour maximiser le logarithme de la fonction de vraisemblance considérant une densité de probabilité discrète (loi de Poisson) des événements $i$:
\begin{equation}
ln(L) = \sum_i ln(M_i) - N_{pred} \ {\rm ,}
\end{equation}
où $M_i$ est la fonction de distribution des événements à la coordonnée de l'événement (i), soit:
\begin{equation}
M_i(E_i,\theta_i,t_i) = \int_{ROI} dE' \ d\theta' \ R(E_i,\theta_i,t_i;E_i',\theta_i') S(E_i',\theta_i')
\end{equation}
La quantité $ln(L)$ est, en d'autres termes, une fonction non-linéaire des paramètres spectraux de la source. On suppose que la vraisemblance $L$ est dérivable et que $L$ admet un maximum global pour une valeur précise des paramètres spectraux. Le maximum de $ln(L)$ s'obtient en recherchant pour quelle valeur de ces paramètres la dérivée de $L$ s'annule. On obtient en sortie de l'analyse la valeur avec l'incertitude des paramètres spectraux laissés libres dans le modèle des sources, le nombre de photons prédits provenant de la source, ainsi que sa significativité. Notons que le ST $gtlike$ propose plusieurs algorithmes pour maximiser le logarithme de la fonction de vraisemblance. Ils se différencient par leur vitesse d'exécution et leur précision.

\partie{R\'ESULTATS}

%

\chapter[PSR~J0205+6449]{PSR~J0205+6449}\label{chap:j0205}


\minitoc

\section{Introduction}

Ce chapitre rapporte la découverte de pulsations provenant du pulsar PSR~J0205+6449 dans la nébuleuse à vent de pulsar 3C~58 avec le télescope \textit{Fermi Gamma-Ray Space Telescope}. Dans la section \ref{sec:obsj0205}, nous décrivons les observations radio et $\gamma$ utilisées pour l'analyse, tandis que les sections \ref{sec:lcj0205} et \ref{sec:specj0205} incluent respectivement l'étude de la courbe de lumière $\gamma$ fournissant l'alignement avec l'émission radio et $X$, et l'étude spectrale de la source. Dans la section \ref{sec:discussionj0205}, nous discutons de la courbe de lumière, de la luminosité, de l'efficacité $\gamma$, de la géométrie de l'émission, et de l'âge du pulsar. Ce chapitre est tiré de l'article \citet{abdo09j0205} paru dans ApJL en 2009.


\section{Pulsations dans la Nébuleuse à Vent de Pulsar 3C~58}

La source radio 3C~58 a été reconnue en 1970 comme étant un reste de supernova (SNR~G130.7+3.1; \citealt{caswell70}), et plus tard classifiée comme une nébuleuse à vent de pulsar (PWN, également appelée plérion) par \citet{weiler78}. \citet{becker82} identifièrent une source ponctuelle dans le coeur de 3C~58 s'apparentant à un pulsar. La distance de la PWN est estimée à 3.2\,kpc \citep{roberts93}. Le pulsar PSR~J0205+6449 a finalement été découvert dans les données du \textit{Chandra X-ray Observatory} avec une période de 65.7\,ms, tandis que les archives du télescope \textit{Rossi X-ray Timing Explorer} ont permis une mesure de la dérivée de la période $\dot{P} = 1.93 \times 10^{-13}$ \citep{murray02}. Cela a été suivi par la découverte d'un très faible signal pulsé en radio, dont la densité de flux moyen est de $\sim$ 45\,$\mu$Jy \citep{camilo02}. Le pulsar a un taux d'énergie de rotation perdue de $\dot{E}$ de $2.7 \times 10^{37}$\,erg\,s$^{-1}$ qui en fait le troisième pulsar le plus énergétique de la galaxie, un champ magnétique de $3.6 \times 10^{12}$\,G, et un âge caractéristique $\tau$ de 5400 ans. On se reportera à la section \ref{sec:propneutronstar} pour le calcul de ces propriétés. Il présente également un haut niveau d'instabilités temporelles \citep{ransom04}. Deux \textit{glitchs} ont été observés depuis sa découverte \citep{livingstone08}. On trouvera dans le tableau \ref{tab:psrcatpar} les principales propriétés du pulsar.

L'association possible du système 3C~58/J0205+6449 et de la supernova historique SN 1181 (828 ans) est un véritable débat. Si la PWN 3C~58 coïncide spatialement et est énergétiquement compatible avec la supernova historique SN 1181 \citep{stephenson71,stephenson02}, de récents travaux sur les modèles de PWN \citep{chevalier05}, sur la vitesse d'expansion de la nébuleuse radio, et la vitesse des noeuds optiques impliquent un âge pour 3C~58 de plusieurs milliers d'années, plus proche de l'âge caractéristique du pulsar que de SN 1181. La nébuleuse 3C~58 est également très similaire à la nébuleuse du Crabe âgée de 955 ans (reste de supernova SN 1054), composée d'une nébuleuse radio avec un spectre plat, une émission $X$ étendue non thermique, et une source ponctuelle X due au pulsar central. Malgré cela, les deux objets diffèrent significativement en luminosité et en taille. La nébuleuse radio 3C~58, bien que deux fois plus grande, est moins lumineuse par un ordre de grandeur que la nébuleuse du Crabe \citep{ivanov04}, tandis que sa luminosité en $X$ est 2000 fois plus petite \citep{torii00}. Ces disparités pourraient être expliquées par un âge différent. 


\section{Observations}\label{sec:obsj0205}

\subsection{Observations Radio}

Le comportement rotationnel de PSR~J0205+6449 durant les 8 premiers mois de la mission \textit{Fermi} est décrit par la solution temporelle (éphéméride) présentée dans le tableau \ref{tab:ephemj0205}. Le pulsar a été observé à la fois par le NRAO Green Bank Telescope (GBT) et le Lovell télescope à Jodrell Bank. GBT a founi des données plus précises, tandis que des observations ont été réalisées plus fréquemment à Jodrell Bank. L'éphéméride utilisée pour empiler les photons $\gamma$ est basée sur les TOAs (voir la section \ref{sec:ephem}) obtenus par les deux télescopes entre le 17 juin 2008 et le 9 mars 2009. Dix-sept TOAs proviennent du radiotélescope GBT à la fréquence de 2.0\,GHz avec une incertitude moyenne de 0.27\,ms, chacun est basé sur 1 heure d'intégration. Quant aux 51 TOAs de Jodrell Bank dont l'incertitude moyenne est de 0.37\,ms, ils dérivent d'observations à 1.4\,GHz basées sur 2 heures d'intégration. Notons que la chronométrie de PSR~J0205+6449 est très bruitée. Dans le but de décrire correctement sa rotation pendant les 9 mois d'observation $\gamma$, l'éphéméride est composée des sept premières dérivées de la fréquence. Ce travail a été effectué avec le programme d'analyse TEMPO\footnote{TEMPO est l'ancienne version de TEMPO2:~http:$//$www.atnf.csiro.au/research/pulsar/tempo}. Un écart-type (RMS) de 0.4\,ms a été obtenue. La meilleure détermination de la mesure de dispersion (DM = 140.7 $\pm$ 0.3\,pc\,cm$^{-3}$) est citée dans \citet{camilo02}, laquelle est utilisée pour corriger les TOAs à une fréquence infinie, avec une incertitude de 0.3\,ms.

\tab[Ephéméride du pulsar PSR~J0205+6449]{ephemeris_j0205}{Paramètres rotationnels du pulsar PSR~J0205+6449 basés sur les observations des télescopes Green Bank Telescope (GBT) et Jodrell Bank. Les chiffres entre parenthèses représentent les incertitudes données par $TEMPO$. \\
$^{a}$ Date à laquelle la phase $\phi = 0$ (TZRMJD).}{tab:ephemj0205}

\subsection{Observations Gamma}

Les données collectées pour ce travail ont été obtenues par le \textit{Fermi} LAT dans deux modes d'observations différents, du 30 juin au 29 juillet 2008 durant la phase de vérification des instruments, et du 3 août 2008 au 9 mars 2009 quand le LAT opérait en mode de balayage. Nous avons utilisé les événements \textit{diffuse} ayant la probabilité la plus élevée d'être un photon et exclu ceux possédant un angle zénithal supérieur à 105$^{\circ}$ à cause de l'albédo terrestre. La source qui a une significativité statistique de 9.5 $\sigma$ ($<$ 10 $\sigma$) de 0.2 à 100 GeV pour les trois premiers mois de la mission (mode de balayage) n'apparaît pas dans le catalogue des sources brillantes \citep{abdo09bsl}. La figure \ref{fig:mapJ0205} montre la distribution des photons de la région entourant le pulsar. Notons la présence du micro-quasar LSI+61$^{\circ}$303 \citep{abdo09lsi61}, et de deux autres pulsars émetteurs $\gamma$, PSR~J0248+6021 \citep{cognard09} et PSR~J0007+7303 le pulsar du reste de supernova CTA1 \citep{abdo08}.

\fig[Distribution des photons pour PSR~J0205+6449]{scale=8.0}{c6_j0205_smooth.jpg}{Distribution des photons de la région qui entoure le pulsar PSR~J0205+6449. Les croix représentent les sources du catalogue pour 6 mois de données. Le nombre de degré par pixel est de 0.25.}{fig:mapJ0205}


\section{Analyse et Résultats}

\subsection{Courbe de Lumière}\label{sec:lcj0205}

Le pulsar est localisé dans le plan galactique où l'émission diffuse $\gamma$ est intense, et à 5.3$^{\circ}$ de la source brillante binaire LSI+61$^{\circ}$303 \citep{abdo09lsi61}. Pour l'analyse temporelle, un ensemble de photons avec une énergie supérieure à 100\,MeV a été sélectionné dans un cône dépendant en énergie de rayon $\theta_{68} \leqslant$ $0.8 \times E_{\rm GeV}^{-0.75}$ degrés, mais avec un rayon maximum de 1.5$^{\circ}$ centré sur la position $X$ du pulsar ($l=130.719^{\circ}$, $b=3.085^{\circ}$). Ce choix prend en compte les capacités de l'instrument et maximise le rapport signal-sur-bruit sur un large intervalle en énergie. Cela tronque la PSF à basse énergie et réduit le nombre d'événements du fond. Un total de 2922 photons restent après ces coupures. A partir du programme d'analyse TEMPO2, les temps d'arrivée des événements ont été transférés au barycentre du système solaire en utilisant l'éphéméride du système solaire JPL~DE405, et les événements ont été empilés suivant l'équation \ref{eq:phase} en utilisant les paramètres de l'éphéméride radio décrit dans le tableau \ref{tab:ephemj0205}.

La figure \ref{fig:lcJ0205} (cadre du haut) montre le phasogramme $\gamma$ au-dessus de 100\,MeV, sur lequel nous avons estimé la position et la largeur des pics. Le premier pic (P1) est décalé par rapport au pic radio (cadre du bas) par $0.08 \pm 0.01 \pm 0.01$ selon l'ajustement d'une lorentzienne de largeur à mi-hauteur (FWHM) de $0.15 \pm 0.01$. Pour la position, la première incertitude provient de l'ajustement du pic, et la seconde provient de l'incertitude de la mesure de dispersion en extrapolant les TOAs radio vers une fréquence infinie (voir l'équation \ref{eq:tpsrela}). Le second pic est asymétrique. Il a donc été évalué avec deux demi-lorentziennes pour prendre en compte les pentes différentes à gauche et à droite. L'ajustement place le pic à $0.57 \pm 0.01 \pm 0.01$, avec une largeur à mi-hauteur de $0.13 \pm 0.04$. Les deux pics sont séparés par $0.49 \pm 0.01$ en phase. Nous avons défini la région non pulsée (OFF) comme étant le minimum pulsé entre 0.65 et 1.0 en phase. La ligne pointillée représente le fond estimé à partir d'un anneau entre 2$^{\circ}$\ --\ 3$^{\circ}$ autour du pulsar. Un excès entre P1 et P2 ($\phi = 0.14 - 0.46$) apparaît avec une significativité (signal/$\sqrt{fond}$) de 5\,$\sigma$. Cependant, les données ne peuvent contraindre la dépendance en phase de cet excès. 

\fig[PSR~J0205+6449:~Courbes de lumière]{scale=0.62}{c6_J0205+6449_LC_multi_energy.pdf}{\textbf{Cadre du haut:}~Courbe de lumière de PSR~J0205+6449 au-dessus de 0.1\,GeV. Deux rotations sont représentées avec 50 intervalles en phase par période ($P=65.7$\,ms). La ligne pointillée montre le niveau de fond estimé à partir d'un anneau autour du pulsar (46 coups/intervalle). \textbf{Trois cadres suivants:}~Phasogrammes pour trois intervalles différents en énergie, chacun est représenté avec 50 intervalles en phase par période. \textbf{Second cadre à partir du bas:}~Nombre de coups dans la bande en énergie 2$-$60\,keV à partir des données \textit{RXTE} \citep{livingstone09}. \textbf{Cadre du bas:}~Profil pulsé radio basé sur 3.8 heures d'observations du télescope GBT à la fréquence de 2\,GHz avec 64 intervalles en phase. Cette figure est extraite de \citet{abdo09j0205}.}{fig:lcJ0205}

Pour examiner le profil $\gamma$ en fonction de l'énergie, des phasogrammes avec 50 intervalles en phase ont été tracés pour trois rangs en énergie, 0.1\ --\ 0.3 GeV, 0.3\ --\ 1 GeV, et $\ge$1 GeV (cadres du milieu). Entre 0.1 et 1\,GeV, deux pics et un possible épaulement après le premier pic sont distinctement présents, tandis qu'au-dessus d'1\,GeV seul le pic P2 est significatif. Nous observons également que le rapport P1/P2 (somme des coups) diminue quand l'énergie augmente, avec un rapport de $0.63\pm0.06$ entre 0.1\ --\ 0.3 GeV, $0.55\pm0.05$ entre 0.3\ --\ 1.0 et $0.24 \pm 0.06$ au-dessus de 1\,GeV. Les pulsars de Vela \citep{abdo09vela}, du Crabe \citep{abdo09crab}, de Geminga, et de PSR~B1951+32 montrent le même comportement \citep{thompson01}. Cette tendance générale suggère une dépendance spectrale en énergie de la courbe de lumière $\gamma$. Finalement, nous notons que le photon de plus haute énergie se situe dans le pic P2 avec une énergie de 8.6\,GeV.

La figure \ref{fig:lcJ0205} (derniers cadres) montre également la courbe de lumière $X$ entre 2 et 60 keV mesurée par le télescope $RXTE$ \citep{livingstone09}, ainsi que le profil radio à 2\,GHz provenant du télescope GBT. Le pic radio définit la phase de référence $\phi = 0$. \citet{livingstone09} montrent que le pic radio devance le premier pic $X$ par $\phi = 0.10 \pm 0.01$, et notent une séparation entre les deux pics étroits de $\Delta\phi = 0.5$. Le bon alignement entre les profils $X$ et $\gamma$ suggère une origine commune entre les deux composantes. Cette caractéristique est également observée dans les données LAT pour le pulsar de Vela, où le plus intense pic $X$ est aligné avec le premier pic $\gamma$ \citep{abdo09vela}.


\subsection{Spectre et Flux Résolus en Phase}\label{sec:specj0205}

Pour étudier le spectre résolu en phase de PSR~J0205$+$6449, une analyse spectrale par un maximum de vraisemblance a été effectuée en utilisant une carte $\gamma$ (ROI) de $20{^\circ}$ centrée sur la position du pulsar entre 0.1 et 200\,GeV. Rappelons que les erreurs systématiques sur la surface efficace estimées pour les IRFs ``Pass6\_v3'' sont $\le$ 5\% autour de 1\,GeV, 10\% en dessous de 0.1 GeV et 20\% au-dessus de 10\,GeV. L'émission diffuse pour le plan galactique a été modélisée en utilisant une carte basée sur le modèle GALPROP: version gll\_iem\_v01. Quant à l'émission extragalactique et au bruit de fond instrumental, ils ont été modélisés par une simple loi de puissance. Les sources ponctuelles à l'intérieur de la ROI et au-delà ($25{^\circ}$) ont été incluses (voir la figure \ref{fig:mapJ0205}). Pour le pulsar, la forme du spectre a été modélisée par une loi de puissance avec une coupure exponentielle (voir l'équation \ref{eq:powerlawexpcutoff}).

Premièrement, nous avons déterminé les composantes spectrales du fond diffus en sélectionnant la région OFF du pulsar. Ensuite, nous avons ajusté les données de la région ON ($\phi = 0.00 - 0.65$) du pulsar pour améliorer le rapport signal-sur-bruit. Le meilleur ajustement est décrit par :
\begin{equation}\label{eq:specj0205}
\frac{dF}{dE} = N_{0}\ E^{-\Gamma}\ e^{-E/E_{Cutoff}}
\ {\rm cm^{-2} s^{-1} GeV^{-1}}
\end{equation}
avec E en GeV, le terme $N_{0} = (1.4 \pm 0.1 \pm 0.1) \times 10^{-8}$\,cm$^{-2}$\,s$^{-1}$\,GeV$^{-1}$, l'indice spectral $\Gamma = 2.1 \pm 0.1 \pm 0.2$, et l'énergie de coupure $E_{Cutoff} = 3.0^{+1.1}_{-0.7} \pm 0.4$ GeV. Les erreurs sont respectivement les incertitudes statistiques et systématiques. Nous obtenons de cet ajustement entre 0.1\ --\ 200 GeV un flux de photon intégré de $(13.7 \pm 1.4 \pm 3.0) \times 10^{-8}$\,cm$^{-2}$\,s$^{-1}$ et un flux énergétique intégré de $F_{E,obs} = (6.7 \pm 0.5 \pm 1.0) \times 10^{-11}$\,ergs\,cm$^{-2}$\,s$^{-1}$. Les erreurs systématiques ont été obtenues avec les IRFs modifiées (voir la section \ref{errorsys}). La figure \ref{fig:sedJ0205} montre le spectre d'énergie multi-longueurs d'onde du pulsar. Les points \textit{Fermi} ont été obtenus en effectuant une analyse spectrale dans chaque intervalle en énergie en supposant une simple loi de puissance.

\fig[PSR~J0205+6449: Distribution d'\'Energie Spectrale]{scale=0.7,angle=90}{c6_multi_sed_j0205.pdf}{Spectre en énergie multi-longueurs d'onde pour le pulsar PSR~J0205+6449. Les références pour cette figure sont: radio: \citet{camilo02}; optique (WH Telescope): \citet{shearer08}; Chandra: \citet{murray02}; RXTE: Kuiper et Hermsen (données non publiés); Fermi: cette thèse; Veritas: \citep{aliu08}.}{fig:sedJ0205}

Pour vérifier l'hypothèse d'une coupure dans le spectre, nous avons ajusté le même lot de données avec une simple loi de puissance de la forme $dF/dE = N_{0}(E/1\rm GeV)^{-\Gamma}$. Le modèle spectral utilisant la coupure est mieux contraint avec une différence dans les logarithmes du maximum de vraisemblance de $\sim 4.5\,\sigma$, défavorisant l'hypothèse d'une simple loi de puissance. Nous avons également ajusté les données avec un indice d'atténuation $\beta$ libre\footnote{Rappelons que le modèle Polar Cap prédit un spectre avec une coupure en énergie de la forme Exp[-(E/E$_{C}$)$^{\beta}$], avec $\beta \sim 2$.}. L'ajustement donne $\beta = 0.8 \pm 0.2 \pm 0.3$, ce qui exclut le modèle avec une coupure super-exponentielle. 

Une recherche d'émission dans la région OFF du pulsar correspondant à une possible émission de la PWN 3C~58 a été effectuée. Nous avons ajusté les données de la région OFF par une loi de puissance et en supposant la nébuleuse ponctuelle. Aucun signal n'a été observé de la PWN. Finalement après une renormalisation du signal sur toute la phase, nous avons dérivé une limite supérieure à 95\% de confiance sur le flux intégré au-dessus de 200\,MeV de $1.7\times 10^{-8}$\,cm$^{-2}$\,s$^{-1}$.


\section{Discussion}\label{sec:discussionj0205}

\subsection{Courbe de Lumière}

Les courbes de lumières multi-longueurs d'onde sont essentielles pour localiser l'émission pulsée dans les régions des lignes de champs ouvertes et, par conséquent, pour comprendre les mécanismes d'accélération de particule. Le profil $\gamma$ de PSR~J0205+6449 couvre un large intervalle de phase et est similaire à la courbe de lumière de Vela \citep{abdo09vela}, incluant un alignement des pics $\gamma$ avec les pics $X$. Le décalage $\delta$ radio-$\gamma$ et la séparation $\Delta$ des deux pics $\gamma$ par $\sim$ 0.5 est en train de devenir un modèle prépondérant \citep{abdo09psrcat}. Les pulsars de Vela, B1951+32, et J2021+3651 \citep{halpern08,abdo09j2021} ont les mêmes caractéristiques, un premier pic $\gamma$ à 0.13, 0.16, et 0.17 respectivement par rapport au pic radio, et une séparation des pics $\gamma$ compris entre 0.4 et 0.5. Cela s'ajuste assez bien aux prédictions des modèles magnétosphériques externes, que ce soit le modèle traditionnel \textit{Outer Gap} (OG) ou le modèle plus récent \textit{Two-Pole Caustic} (TPC), et exclut donc le modèle \textit{Polar Cap}. Anticipons ! La figure \ref{fig:psrcat:distribution} présente la distribution $\delta - \Delta$ des pulsars $\gamma$ détectés par le LAT pour 6 mois de données.

\subsection{Luminosité, Efficacité, et Géométrie de l'\'Emission}

Pour connaître l'efficacité $\gamma$, il est essentiel de déterminer au préalable la luminosité totale $L_{\gamma}$ (voir l'équation \ref{eq:luminosite}). Cependant, une incertitude importante pour évaluer la luminosité est la détermination de la distance $D$. Des observations de l'absorption de l'hydrogène neutre (H~I) par \citet{roberts93} donnent une estimation de la vitesse radiale de la PWN 3C~58. Cette vitesse est convertie en une distance cinématique de 3.2\,kpc en supposant une courbure rotationnelle galactique plate \citep{fich89}. Les incertitudes sur la distance avec cette méthode sont de l'ordre de 25\%, rendant ce résultat compatible avec la distance de 2.6\,kpc précédemment reportée par \citet{green82}. La distance dérivée du DM est quant à elle de 4.5\,kpc selon le modèle NE~2001, mais avec une incertitude qui peut excéder 50\%. Pour ce travail de thèse, nous adopterons une distance de 3.2\,kpc étant donné l'incertitude de cette valeur.

A partir du flux énergétique intégré, nous estimons une luminosité :
\begin{equation}
L_{\gamma}= 8.3 \times 10^{34}\ (D/3.2\,{\rm kpc})^2\ f_{\Omega} \ \ {\rm erg\,s^{-1}}
\end{equation}
où $f_{\Omega}$ est le facteur de correction (voir l'équation \ref{eq:fbeam}), et déduisons pour un moment d'inertie de $10^{45}$\,g\,cm$^{2}$ une efficacité $\eta = L_{\gamma}/\dot{E} = 0.003\ f_{\Omega}\ (D/3.2\,{\rm kpc})^2$. 

A présent, il est nécessaire d'évaluer le facteur de correction qui prend en compte la géométrie de l'émission. En supposant les modèles magnétosphériques externes suite à la discussion sur la forme de la courbe de lumière et l'exclusion du facteur d'atténuation $\beta$ de la forme spectrale, nous pouvons déduire l'inclinaison magnétique $\alpha$ et l'angle d'observation $\zeta$ à partir de la séparation des pics $\gamma$ et en utilisant les courbes de lumières du papier ``Atlas'' de \citet{watters09}. Pour les modèles OG, nous estimons $f_{\Omega}\approx 0.9$--$1.0$ avec $\alpha \backsim 60^{\circ}$--90$^{\circ}$ et $\zeta \backsim 80^{\circ}$--85$^{\circ}$, tandis que pour les modèles TPC, $f_{\Omega} \approx 0.95$--1.25 avec à la fois le couple $\alpha \backsim 50^{\circ}$--$90^{\circ}$, $\zeta \backsim 85^{\circ}$--90$^{\circ}$ et le couple $\alpha \backsim 85^{\circ}$--$90^{\circ}$, $\zeta \backsim 45^{\circ}$--90$^{\circ}$. De plus, l'examen de la géométrie de la PWN 3C~58 utilisant les données {\em Chandra\/} \citep{ng04,ng08} mène à un angle d'observation de $\zeta = 91.6 \pm 0.2 \pm 2.5$ $^\circ$ (tore interne) basé sur l'angle d'inclinaison du tore au plan du ciel. Cette valeur est en accord avec les résultats du modèle OG et plus conforme avec le premier intervalle estimé pour le modèle TPC. Si on adopte une valeur moyenne de $f_{\Omega} = 1$ (compatible avec les deux modèles) et une distance de 3.2\,kpc, nous en déduisons une efficacité $\eta = 0.3\%$ pour convertir sa perte d'énergie rotationnelle en photons $\gamma$. Cette valeur de l'efficacité montre que PSR~J0205+6449 suit relativement bien la relation approximée par $\eta \varpropto \dot{E}^{-1/2} \varpropto 1/V$ (voir la figure \ref{fig:LvsEdot}).

\subsection{Age du Pulsar}

\citet{thompson99} soulignent que pour les pulsars EGRET, le maximum de luminosité pour les jeunes pulsars $\gamma$ tels que le Crabe ou PSR~B1509$-$58 est au niveau du rayonnement $X$, tandis que pour les pulsars moins jeunes tels que Vela ou Geminga le maximum de luminosité est à haute énergie, dans l'intervalle d'énergie du LAT (voir la figure \ref{fig:spectracgropsr}). Le tableau \ref{tab:luminosity} présente, pour les jeunes pulsars $\gamma$ connus\footnote{Nombre de pulsars $\gamma$ connus au moment de l'étude.}, la luminosité $X$ et $\gamma$, ainsi que le rapport L$_{X}$/L$_{\gamma}$ et l'âge caractéristique. Considérant une luminosité $X$ entre 0.5 et 8\,keV de $1.51 \times 10^{33}$\,ergs\,s$^{-1}$ \citep{kargaltsev08} et une luminosité L$_{\gamma}$ de $8.3 \times 10^{34}$\,ergs\,s$^{-1}$ avec $f_{\Omega} = 1$, PSR~J0205+6449 semble avoir un rapport $L_{X}/L{\gamma}$ plus proche de celui des pulsars tels que Vela et PSR~J2229+6114, que des très jeunes pulsars tels que le Crabe et  PSR~B1509$-$58. Par conséquent, cette étude suggère que l'association de la PWN 3C~58 qui est alimentée par le pulsar, avec la supernova historique SN 1181, est probablement incorrecte.

\tab[Luminosité $X$/$\gamma$]{luminosity}{Luminosités $L_X$ et L$_{\gamma}$ des jeunes pulsars connus. $\tau = P/2\dot{P}$ est l'âge caractéristique des pulsars. La luminosité est calculée en supposant le facteur de correction de l'émission $f_{\Omega} = 1$, à l'exception de (b). \\ \\
\textbf{Références:}\\
1.~\citep{kargaltsev08}; 2.~\citep{abdo09psrcat}; 3.~\citep{vanetten08}; 4.~\citep{abdo09vela}; 5.~\citep{abdo09j2021}; 6.~\citep{abdo09j0205}; 7.~\citep{abdo09j2229} \\ \\
\textbf{Notes:}\\
$^a$ Luminosité entre 0.5 et 8\,keV. \\
$^b$ Luminosité totale (optique et au-dessus) calculée en supposant $f_{\Omega}$ = 1/4$\pi$. \\
$^c$ Luminosité calculée en supposant une distance de 4\,kpc.
}{tab:luminosity}

\chapter[PSR~J2229+6114]{PSR~J2229+6114}\label{chap:j2229}

\minitoc

\section{Introduction}\label{sec:introj2229}

Ce chapitre rapporte la découverte de pulsations provenant du pulsar PSR~J2229+6114 avec le télescope \textit{Fermi Gamma-Ray Space Telescope}. Ce pulsar est sur beaucoup d'aspects similaire au pulsar de Vela\footnote{Ce pulsar est classé comme un pulsar ``vela-like''.}, tant par ses émissions pulsées radio, $X$, et $\gamma$, que par la présence d'une PWN autour du pulsar et une instabilité rotationnelle importante. \citet{kramer03} fournissent une discussion sur les pulsars de type Vela. Le pulsar PSR~J1048$-$5832 au chapitre suivant montre les mêmes caractéristiques. Dans la section \ref{sec:obsj2229}, nous décrivons les observations radio et $\gamma$ utilisées pour l'analyse, tandis que les sections \ref{sec:lcj2229} et \ref{sec:specj2229} incluent respectivement l'étude de la courbe de lumière $\gamma$ fournissant l'alignement avec l'émission radio et $X$, et l'étude spectrale de la source. Dans la section \ref{sec:discussionj2229}, nous discutons de la courbe de lumière, de la luminosité, de l'efficacité $\gamma$, et de la géométrie de l'émission. Ce chapitre est tiré de l'article \citet{abdo09j2229}, à paraître dans ApJ en 2009.

\section{Candidat pour la Source EGRET 3EG~J2227+6122}

PSR~J2229+6114 se trouve aux coordonnées galactiques (l,b)=(106$^{\circ}$,2.9$^{\circ}$) à l'intérieur de la boîte d'erreur de la source EGRET 3EG~J2227+6122 \citep{hartman99}. L'objet a premièrement été detecté dans les données $X$ de \textit{ROSAT} et \textit{ASCA} comme étant une source compacte, et ensuite identifié comme étant un pulsar grâce à la découverte de pulsations radio et $X$ de période 51.6\,ms \citep{halpern01b}. Le profil radio montre un simple pic étroit, tandis que la courbe de lumière $X$ entre 0.8 et 10\,keV se compose de deux pics, séparés par $\Delta\phi = 0.5$. La collaboration \textit{AGILE}\footnote{AGILE: Astro-rivelatore Gamma a Immagini LEggero} a récemment annoncé la découverte de pulsations $\gamma$ au-dessus de 100\,MeV \citep{pellizzoni09}. Le pulsar est aussi jeune que le pulsar de Vela (âge caractéristique $\tau = P/2\dot{P}$ = 10000 ans), et aussi énergétique ($\dot{E}=2.2 \times 10^{37}$\,erg\,s$^{-1}$). Il est probablement la source énergétique de la PWN ``Boomerang'' G106.65+2.96, qui appartient au reste de supernova G106.3+2.7 découvert par \citet{joncas90}. Notons que récemment la PWN a été détectée au TeV par le télescope \textit{MILAGRO} \citep{abdo09milagro}. Des études sur la vitesse radiale de l'hydrogène neutre et de l'environnement moléculaire placent le système à $\sim$\,800\,pc \citep{kothes01}, tandis que \citet{halpern01a} suggèrent une distance de 3\,kpc estimée à partir de l'absorption $X$. La distance déterminée par le DM en utilisant le modèle de la distribution d'électron dans la galaxie NE~2001 donne une distance de 7.5\,kpc, significativement au-dessus des autres estimations. Etant donné que les incertitudes sur cette distance sont grandes (jusqu'à 50\%), nous adopterons pour ce travail de thèse une distance de 3\,kpc. 

\section{Observations}\label{sec:obsj2229}

\subsection{Observations Radio}

PSR~J2229+6114 est observé par le télescope GBT et le télescope Lovell à Jodrell Bank. L'éphéméride utilisée ici pour le calcul des phases à partir des temps d'arrivée des photons $\gamma$ est basée sur des TOAs obtenus par les deux télescopes entre le 17 juin 2008 et le 23 mars 2009. Vingt-cinq TOAs proviennent du GBT avec une incertitude moyenne de 0.2\,ms, chacun représente 5 minutes d'observation à la fréquence centrale de 2.0\,GHz, tandis que quarante-quatre TOAs avec une incertitude moyenne de 0.3\,ms proviennent des observations de Jodrell Bank obtenus à partir de 30 minutes d'observations à la fréquence de 1.4\,GHz. Pour l'ajustement des paramètres de rotation avec TEMPO, la position du pulsar a été fixée à la position $X$ connue \citep{halpern01b}. PSR~J2229+6114 montre du \textit{timing noise} sur les 9 mois de données, et donc demande un ajustement de la fréquence et des deux premières dérivées. De plus, un petit glitch s'est produit le 12 novembre 2008 ($54782.6\pm0.5$~MJD) avec un pas de fréquence fractionnaire $\Delta F/F = (4.08\pm0.06)\times10^{-9}$ et $\Delta \dot F/{\dot F} = (2.0\pm0.4)\times10^{-4}$. Celui-ci a été pris en compte dans le modèle rotationnel. La rms de l'ajustement postérieur est de 0.24\,ms. La $\mbox{DM} = (204.97\pm0.02)$\,pc\,cm$^{-3}$ a été mesurée à partir d'observations de GBT pour trois fréquences différentes. Celle-ci est utilisée pour corriger le temps d'arrivée à 2\,GHz vers les fréquences infinies, pour la comparaison avec le profil $\gamma$.

\subsection{Observations Gamma}

Les données utilisées pour l'analyse spectrale ont été collectées à partir du 4 août jusqu'au 23 mars 2009 pendant que le télescope était en mode de balayage. Pour l'analyse temporelle, nous avons additionné les données de la phase de calibration de l'appareil, période s'étendant du 25 juin au 3 août 2008. Pendant cette phase, plusieurs configurations ont été testées qui ont affectées la résolution en énergie et la reconstruction des événements du LAT. Cependant, ces changements n'ont pas d'effets sur la chronométrie du LAT. De plus, seuls les photons marqués ``diffuse'' ont été selectionnées, et en addition nous avons exclu ceux possédant un angle zénithal supérieur à 105$^{\circ}$. La figure \ref{fig:mapJ2229} montre la distribution des photons de la région entourant le pulsar. Notons la présence du pulsar $\gamma$ PSR~J2238+59 (voir le chapitre \ref{chap:catalog}) à $\sim$2$^{\circ}$ de PSR~J2229+6114 \citep{abdo09blindsearch}.

\fig[Distribution des photons pour PSR~J2229+6114]{scale=8.0}{c7_J2229_smooth.jpg}{Distribution des photons de la région qui entoure le pulsar PSR~J2229+6114. Les croix représentent les sources du catalogue pour 6 mois de données. Le nombre de degré par pixel est de 0.25.}{fig:mapJ2229}

\section{Analyse et Résultats}

\subsection{Courbe de Lumière}\label{sec:lcj2229}

Pour l'analyse temporelle de PSR~J2229+6114, nous avons sélectionné les photons d'énergie supérieure à 100\,MeV dans un rayon de 1 degré autour de la position du pulsar radio. Ensuite, nous avons sélectionné les photons dans un cône dépendant en énergie de rayon $\theta_{68} \leqslant$ $0.8 \times E_{\rm GeV}^{-0.75}$ degrés, gardant tous les photons inclus dans un rayon de 0.35$^{\circ}$. Cette sélection se base sur la PSF du LAT et permet d'augmenter le rapport signal-sur-bruit sur une large bande en énergie. A partir du programme d'analyse TEMPO2, nous avons corrigé le temps d'arrivée des photons vers le barycentre du système solaire en utilisant l'éphéméride solaire JPL~DE405. La phase des événements $\gamma$ a ensuite été calculée à partir de l'éphéméride radio délivrée par les télescopes de GBT et de Jodrell Bank (voir le tableau \ref{tab:ephem2229}) suivant l'équation \ref{eq:phase}.

La figure \ref{fig:lcJ2229} (cadre du haut) montre l'histogramme résultant du nombre de coups en fonction de la phase au-dessus de 0.1\,GeV. En comparaison, le cadre du bas montre le profil radio pulsé aligné en phase provenant du télescope GBT. La courbe de lumière $\gamma$ montre un seul et asymétrique pic ($\phi = 0.15 - 0.65$). Nous avons ajusté le pic avec deux demi-lorentziennes pour prendre en compte les deux pentes. L'ajustement place le pic à $0.49 \pm 0.01 \pm 0.001$ avec une largeur à mi-hauteur de $0.23 \pm 0.03$. Nous avons estimé le niveau du fond à partir d'un anneau centré sur la position du pulsar entre 1 et 2$^{\circ}$ en sélectionnant le minimum pulsé de l'histogramme entre 0.65 et 0.15. Cela est représenté par la ligne pointillée (48 coups/intervalle) et est homogène avec la région OFF. A partir de ces résultats, le nombre total de photons pulsés est estimé à $1365 \pm 79$, avec une contribution du fond de $2431 \pm 49$ événements.

Pour examiner la forme du profil pulsé en fonction de l'énergie (voir la figure~\ref{fig:lcJ2229}, cadres du milieu), quatre phasogrammes ont été tracés suivant différents intervalles ($0.1-0.3$\,GeV, $0.3-1$\,GeV, $1-3$\,GeV et $>$3 GeV), montrant un possible déplacement du pic. Entre 0.1\ --\ 0.3 GeV, le pic $\gamma$ est décalé selon un ajustement avec deux demi-lorentziennes de $0.51 \pm 0.02$ en phase par rapport au pic radio, tandis que les décalages pour 0.3\ --\ 1\,GeV, 1\ --\ 3\,GeV et $>$ 3 GeV sont respectivement $0.48 \pm 0.01$, $0.49 \pm 0.01$, et $0.45 \pm 0.01$. Notons que la position du pic est relativement stable entre 0.1\ -- 3 GeV, mais il apparaît un léger désalignement au-dessus de 3\,GeV. Ainsi, les données ne suggère que légèrement la dépendance en fonction de l'énergie, même si le désalignement entre le profil $X$ et $\gamma$ est très prononcé. Finalement, notons que le photon de plus haute énergie est de 10.8\,GeV et se situe à la phase $\sim$ 0.30, où l'on observe un faible excès (au-dessus de 3\,GeV), qui pourrait correspondre à un second pic.

La figure~\ref{fig:lcJ2229} présente également le profil $X$ de PSR~J2229+6114 obtenu à partir d'une observation de {\em XMM\/} entre 1 et 10\,keV le 15 juin 2002 (52440~MJD) pour un temps d'exposition de 20\,ks. Les données ont été empilées en utilisant une éphéméride contemporaine basée sur des observations de Jodrell Bank et de GBT. Le pic le plus haut du profil $X$ a un retard de $\phi$ = $0.17\pm0.02$ par rapport au pic radio. L'analyse d'une observation du télescope {\em RXTE\/} à partir de 52250 MJD donne un décalage similaire. Il n'y a pas de dépendance en énergie de la forme pulsée $X$ dans l'intervalle 1--10\,keV, tandis que la finesse des pics aussi bien que leur forme spectrale indiquent que l'émission est majoritairement non-thermique.

\fig[PSR~J2229+6114:~Courbes de lumière]{scale=0.62}{c7_J2229+6114_LC_multi_energy.pdf}{\textbf{Cadre du haut:}~Courbe de lumière de PSR~J2229+6114 au-dessus de 0.1\,GeV. Deux rotations sont représentées avec 50 intervalles en phase par période ($P=51.6$\,ms). La ligne pointillée montre le niveau de fond estimé à partir d'un anneau autour du pulsar (48 coups/intervalle). \textbf{Quatre cadres suivants:}~Phasogrammes pour quatre intervalles différents en énergie, chacun est représenté avec 50 intervalles en phase par période. \textbf{Second cadre à partir du bas:}~Nombre de coups dans la bande en énergie 1$-$10\,keV à partir des données {\em XMM\/}. \textbf{Cadre du bas:}~Profil pulsé radio basé sur les observations du télescope GBT à la fréquence de 2\,GHz avec 128 intervalles en phase. Cette figure est extraite de \citet{abdo09j2229}.}{fig:lcJ2229}


\subsection{Spectre et Flux Résolus en Phase}\label{sec:specj2229}

Initialement, une région circulaire de 15\,$^{\circ}$ autour de la position du pulsar et le fond diffus galactique basé sur le modèle numérique GALPROP appelé 54\_59Xvarh7S\footnote{Modèle standard utilisé pour le catalogue 6 mois.} ont été utilisés pour ajuster le pulsar. Cependant au vu des résidus, certaines structures autour de l'objet n'ont pas été prises en compte dans le modèle galactique et pourraient surestimer ou sous-estimer le flux du pulsar. Finalement, nous avons adopté un modèle différent pour l'émission diffuse galactique basée à la fois sur six cartes ``anneaux'' galactocentriques de H~I\footnote{Les densités de colonnes de H~I proviennent des observations du L.A.B, http:$//$www.astro.uni-bonn.de/$\sim$webaiub/english/tools\_labsurvey.php} et de CO, et sur la distribution spatiale de l'intensité Compton inverse modélisée par le Galprop 54$\_$77. L'intensité aussi bien que les émissivités $\gamma$ par anneaux ont été ajustées pour maximiser l'accord avec les observations, ceci en tenant compte des sources ponctuelles détectées en $\gamma$. Cette approche est similaire à ce qui est décrit dans \citet{casandjian08} pour modéliser les données EGRET. Ce modèle est à présent utilisé par la collaboration LAT dans le catalogue 9 mois.

PSR~J2229+6114, référencé comme la source ponctuelle du LAT 0FGL~J2229.0+6114 dans le catalogue de 3 mois \citep{abdo09bsl}, a été modélisé par une loi de puissance avec une coupure exponentielle de la forme:
\begin{equation}\label{eq:specj2229}
\frac{dF}{dE} = N_{0}\ E^{-\Gamma}\ e^{-E/E_{c}}
\ {\rm cm^{-2} s^{-1} GeV^{-1}}
\end{equation}
avec E en GeV. L'ajustement a été fait dans la région ON entre 0.15 et 0.65 en phase. La figure \ref{fig:sedJ2229} (ligne continue) montre la distribution d'énergie spectrale pour la région ON à partir d'un ajustement par un maximum de vraisemblance. Cet ajustement donne un terme $N_{0}$ = $(5.2 \pm 0.4 \pm 0.1) \times 10^{-8}$\,cm$^{-2}$\,s$^{-1}$\,GeV$^{-1}$, un indice spectral $\Gamma$ = $1.85 \pm 0.06 \pm 0.10$ et une énergie de coupure $E_{c} = 3.6^{+0.9}_{-0.6} \pm 0.6 $\,GeV. A partir de ce résultat, nous avons estimé un flux de photon intégré de $(3.77 \pm 0.22 \pm 0.44) \times 10^{-7}$\,cm$^{-2}$\,s$^{-1}$. Cette valeur est 10\% inférieure au flux de $(4.13 \pm 0.61) \times 10^{-7}$\,cm$^{-2}$\,s$^{-1}$ de la source EGRET 3EG~J2227+6122 obtenue à partir de la loi de puissance avec un indice spectral de $2.24 \pm 0.14$, mais plus grande que le flux de $(2.6 \pm 0.4) \times 10^{-7}$\,cm$^{-2}$\,s$^{-1}$ mesuré par AGILE \citep{pellizzoni09}. Nous avons également ajusté les données avec une simple loi de puissance, menant à un flux de $(4.54 \pm 0.14) \times 10^{-7}$\,cm$^{-2}$\,s$^{-1}$ et à un indice spectral de $2.25 \pm 0.02$, compatible avec l'analyse EGRET. Notons que le modèle spectral avec une coupure exponentielle est mieux contraint avec une différence des logarithmes des maximums de vraisemblance de $\sim 9\sigma$, rejetant l'hypothèse d'une simple loi de puissance. Un modèle avec une coupure super-exponentielle de la forme Exp[-(E/E$_{C}$)$^{\beta}$] a été également testé. L'ajustement donne $\beta = 0.8 \pm 0.4 \pm 0.3$, ce qui est en accord avec une simple coupure exponentielle $\beta = 1$.

Une recherche d'une émission non-pulsée de la nébuleuse Boomerang a été effectuée à partir du même lot de données. Nous avons ajusté une source ponctuelle aux données de la région OFF à la position du pulsar radio entre 0.2 et 100\,GeV. Aucun signal n'a été observé de la PWN. Finalement, après une renormalisation du signal sur toute la phase, nous avons dérivé une limite supérieure à 95\% de confiance sur le flux intégré au-dessus de 200\,MeV de $4.0\times 10^{-8}$\,cm$^{-2}$\,s$^{-1}$.

\fig[PSR~J2229+6114:~Distribution d'\'Energie Spectrale]{scale=0.7,angle=90}{c7_sedJ2229.pdf}{Distribution d'énergie spectrale pour le pulsar PSR~J2229+6114. La ligne continue représente l'ajustement globale des données par un maximum de vraisemblance en supposant une loi de puissance avec une coupure exponentielle (voir l'équation \ref{eq:specj2229}). Les points spectraux sont le résultat de l'ajustement des données à partir d'une simple loi de puissance pour six intervalles en énergie séparés logarithmiquement.}{fig:sedJ2229}

\section{Discussion}\label{sec:discussionj2229}

\subsection{Courbe de Lumière}

La courbe de lumière $\gamma$ de PSR~J2229+6114 couvre (voir la figure \ref{fig:lcJ2229}) un large intervalle en phase, suggérant que le faisceau $\gamma$ émet selon un grand angle solide. Cette caractéristique semble en faveur des modèles d'émission magnétosphériques externe, en particulier le \textit{Outer Gap} (OG) et le \textit{Slot Gap/Two-Pole Caustic} (SG/TPC) (voir la section \ref{sec:modgam}).

\subsection{Luminosité, Efficacité, et Géométrie de l'Emission}

A partir de l'analyse spectrale du pulsar, le flux d'énergie intégré observé est évalué à $(23.7 \pm 0.7 \pm 2.5) \times 10^{-11}$\,erg\,cm$^{-2}$\,s$^{-1}$. La luminosité $\gamma$ résultante donne $L_{\gamma}$ = 2.6$\times$10$^{35}f_{\Omega}$($d$/3~kpc)$^{2}$ erg s$^{-1}$, tandis que l'efficacité donne $\eta_\gamma$ = 0.011 $f_{\Omega}$ ($d$/3~kpc)$^{2}$, qui est un facteur 1.6 plus grand que l'efficacité estimée par la relation $\eta \simeq (10^{33}/\dot{E})^{0.5} = 0.007$ \citep{watters09}. Notons qu'à partir de la distance estimée de la SNR (0.8\,kpc), l'efficacité diminue à 0.001 $f_{\Omega}$, tandis qu'avec la distance dérivée à partir du DM (7.5\,kpc) l'efficacité augmente à 0.07 $f_{\Omega}$, soulignant l'importance de déterminer correctement la distance.

A présent, il est nécessaire d'estimer le facteur de correction $f_{\Omega}$ qui prend en compte la géométrie de l'émission. Nous pouvons comparer les courbes de lumière à partir des modèles géométriques de \citet{watters09} pour dériver des contraintes sur la géométrie du faisceau. Les modèles magnétosphériques externes tels que TPC et OG avec $\eta_{\gamma} \sim 0.01$ peuvent décrire des profils $\gamma$ avec un simple pic. Les solutions pour une telle configuration apparaissent pour des inclinaisons magnétiques $\alpha \sim 20^{\circ} - 55^{\circ}$, et des angles d'observation $\zeta \sim 25^{\circ} - 50^{\circ}$ (géométrie TPC/SG) et pour $\alpha \sim 45^{\circ} - 80^{\circ}$, $\zeta \sim 35^{\circ} - 70^{\circ}$ et $f_{\Omega} \sim$ 0.47 - 1.08 (géométrie OG). Pour PSR~J2229+6114, nous avons une contrainte géométrique supplémentaire à partir de la modélisation des tores des PWN en $X$ entourant le pulsar, tels que mesurés par le télescope {\em Chandra}. \citet{ng04,ng08} ont déterminé à partir des données $X$ un angle d'observation $\zeta = 46^{\circ} \pm 2^{\circ} \pm 6^{\circ}$. Ceci donne un intervalle de recouvrement avec les solutions TPC et OG pour un simple pic. Finalement, bien qu'il y ait une certaine incertitude sur la détermination de l'inclinaison magnétique à partir du profil radio, le décalage du pic $\gamma$ par rapport au pic radio ($\phi \approx 0.50$) fournit une information supplémentaire. En examinant l'échantillon de courbe de lumière des figures 9 \& 10 de \citet{watters09}, nous trouvons que les profils à un seul pic pour cette phase sont déterminés seulement quand on peut identifier une composante P2, avec un pont (``bridge''), et un pic P1 très faible ou manquant. De telles courbes de lumières apparaissent seulement pour des efficacités $\eta < 0.03$ et seulement pour un intervalle en angle de $\alpha \sim 50^{\circ} - 70^{\circ}$, $\zeta \sim 40^{\circ} - 50^{\circ}$ (TPC/SG, avec P1 faible) ou $\alpha>45^{\circ}$, $\zeta = 45^{\circ} - 50^{\circ}$ (OG, avec P1 absent). Ceci est précisément l'intervalle $\zeta$ exigé par l'ajustement du tore de la PWN. En incluant la contrainte $\beta_{E}$ = $\alpha$ - $\zeta_{E}$ à partir de la détection de l'émission radio, nous voyons que les données radio, $X$, et $\gamma$, sont toutes conformes avec $\alpha = 55^{\circ} \pm 5^{\circ}$, $\zeta = 45 \pm 5^{\circ}$, impliquant $f_{\Omega} \approx 1$ pour les deux modèles. Ainsi, pour observer un tel profil multi-longueurs d'onde, cela nécessite une configuration géométrique de l'émission bien précise, une faible efficacité $\gamma$, et une émission à haute altitude.

%
\chapter[PSR~J1048$-$5832]{PSR~J1048$-$5832}\label{chap:j1048}

\minitoc

\section{Introduction}

Ce chapitre présente la découverte de pulsations provenant du pulsar PSR~J1048$-$5832 (PSR~B1046$-$58) avec le télescope \textit{Fermi Gamma-Ray Space Telescope}. Un signal $\gamma$ en provenance de ce pulsar avait été préalablement détecté avec un faible niveau de confiance par EGRET (voir le tableau \ref{tab:cgropsr}). De plus, au même titre que PSR~J2229+6114, PSR~J1048$-$5832 est un pulsar de type Vela (voir la section \ref{sec:introj2229}). Dans la section \ref{sec:obsj1048}, nous décrivons les observations radio et $\gamma$ utilisées pour l'analyse, tandis que les sections \ref{sec:lcj1048} et \ref{sec:specj1048} incluent respectivement l'étude de la courbe de lumière $\gamma$ fournissant l'alignement avec l'émission radio, et l'étude spectrale de la source. Dans la section \ref{sec:discussionj1048}, nous discutons de la courbe de lumière, de la luminosité, de l'efficacité $\gamma$, et de la géométrie de l'émission. Enfin, nous concluons ce chapitre par une comparaison des quantités mesurées entre ce pulsar et PSR~J2229+6114. Ce chapitre est tiré de l'article \citet{abdo09j2229}, à paraître dans ApJ en 2009.

\section{Candidat pour la Source EGRET 3EG~J1048$-$5840}

PSR~J1048-5832 (B1046$-$58) est localisé dans la région de la nébuleuse de Carina (NGC 3372) à basse latitude galactique ($l=287.42$\,${^\circ}$, $b=0.58$\,${^\circ}$). Il a été découvert durant une recherche de pulsations à 1.4\,GHz dans le plan galactique par le radiotélescope de Parkes, avec une période de rotation $P$ de 123.7\,ms \citep{johnston92}. Le pulsar a un ralentissement $\dot{E}$ de $2 \times 10^{36}$\,erg\,$^{-1}$ pour un moment d'inertie $I$ de $10^{45}$\,g\,cm$^{2}$, un champ magnétique dipolaire de surface de $3.5 \times 10^{12}$\,G, et un âge caractéristique $\tau_{c}$ de 20000 ans. Les observations à haute résolution d'une source ASCA\footnote{} coïncidant avec le pulsar par les télescopes \textit{Chandra} et \textit{XMM-Newton} ont révélé une nébuleuse à vent de pulsar asymétrique de $\sim6''\times11''$. Cependant, aucune pulsation à partir des données n'a été trouvée \citep{gonzalez06}. PSR~J1048-5832 a également été proposé comme étant la contrepartie de la source EGRET non-identifiée 3EG~J1048$-$5840 \citep{fierro95,pivovaroff00,nolan03}. Cette conclusion a été suggérée par la coïncidence de la position du pulsar avec la source EGRET et des propriétés spectrales et énergétiques de la source EGRET elle-même. Une analyse détaillée des données EGRET montre une possible pulsation $\gamma$ au-dessus de 400\,MeV, avec une courbe de lumière composée de deux pics séparés par $\sim$ 0.4 en phase (voir la figure \ref{fig:lcJ1048}). L'observation d'H~I dans l'environnement du pulsar donne une distance comprise entre 2.5 et 6.6\,kpc \citep{johnston96}. Le modèle NE~2001 assigne quant à lui une distance de 2.7\,kpc basée partiellement sur la détermination de la distance par l'observation H~I. Pour ce travail de thèse, nous adopterons une distance de 3\,kpc.

\section{Observations}\label{sec:obsj1048}

\subsection{Observations Radio}

Le pulsar PSR~J1048$-$5832 est observé mensuellement par le radiotélescope de 64\,m de Parkes en Australie depuis 2007 \citep{manchester08}. La durée d'une observation est de 2 minutes à la fréquence de 1.4\,GHz avec occasionnellement des observations à 0.7 et 3.1\,GHz. Des informations détaillées sur la procédure et sur l'analyse des données peuvent être trouvées dans \citet{weltevrede09}. Le pulsar est connu pour avoir une forte instabilité rotationnelle; un \textit{glitch} a été enregistré juste avant le décollage de \textit{Fermi} \citep{weltevrede09}. De plus, comme beaucoup de pulsars présentant un fort taux de ralentissement $\dot{E}$, PSR~J1048$-$5832 est hautement polarisé sur la large bande radio observée \citep{karastergiou05,johnston06}. L'éphéméride utilisée pour empiler les photons $\gamma$ à la bonne phase est basée sur vingt TOAs et dérive du programme d'analyse TEMPO2. La solution est ajustée pour la fréquence de rotation du pulsar et ses dérivées, et  lissée en utilisant les algorithmes nommés \textit{fitwaves} de TEMPO2. L'écart-type (RMS) résultant de 287$\mu$s présente une amélioration sur la solution pré-lissée. Notons que les paramètres \textit{fitwaves} ne sont pas pris en charge par le ST gtpphase. Des mesures du retard dû à la dispersion à 1369\,MHz donnent une mesure de la dispersion (DM) de 128.822$\pm$0.008\,pc\,cm$^{-3}$, sans indication que celle-ci varie au cours du temps. Cette mesure est utilisée pour corriger les TOAs vers les fréquences infinies pour calculer la phase des photons $\gamma$.

\tab[Ephéméride du pulsar PSR~J1048$-$5832]{ephemeris_j1048}{Paramètres rotationnels du pulsar PSR~J1048$-$5832 basés sur les observations du télescope Parkes. Les chiffres entre parenthèses représentent les incertitudes données par $TEMPO2$. \\
$^{a}$ Date à laquelle la phase $\phi = 0$ (TZRMJD).}{tab:ephemj1048}

\subsection{Observations Gamma}

Les données utilisées pour l'analyse spectrale ont été collectées à partir du 4 août 2008 jusqu'au 10 avril 2009 pendant que le télescope était en mode de balayage. Pour l'analyse temporelle, nous avons additionné à cet échantillon de photons les données de la phase de calibration de l'appareil, période s'étendant du 30 juin au 3 août 2008. Seuls les photons marqués \textit{diffuse} ont été selectionnés, et nous avons par ailleurs exclu ceux possédant un angle zénithal supérieur à 105$^{\circ}$. La figure \ref{fig:mapJ1048} montre la distribution des photons de la région entourant le pulsar. Notons la présence des pulsars $\gamma$ PSR~J1124$-$5916 \citep{abdo09psrcat}, PSR~J1028$-$5819 \citep{abdo09j1028}, et du pulsar EGRET PSR~J1057$-$5226 \citep{abdo09EGRETpsr}.

\fig[Distribution des photons pour PSR~J1048$-$5832]{scale=8.0}{c8_j1048_smooth.jpg}{Distribution des photons de la région qui entoure le pulsar PSR~J1048$-$5832. Les croix représentent les sources du catalogue 9 mois. Le nombre de degré par pixel est de 0.25.}{fig:mapJ1048}

\section{Analyse et Résultats}

\subsection{Courbe de Lumière}\label{sec:lcj1048}

Pour l'analyse temporelle de PSR~J1048$-$5832, nous avons sélectionné les photons avec une énergie supérieure à 100\,MeV dans un rayon de 1 degré autour de la position du pulsar radio. Ensuite, nous avons sélectionné les photons dans un cône dépendant en énergie de rayon $\theta_{68} \leqslant$ $0.8 \times E_{\rm GeV}^{-0.75}$ degrés, gardant tous les photons inclus dans un rayon de 0.35$^{\circ}$. Cette sélection se base sur la PSF du LAT et permet d'augmenter le rapport signal-sur-bruit dans une large bande en énergie. A partir du programme d'analyse TEMPO2, nous avons corrigé le temps d'arrivée des photons vers le barycentre du système solaire en utilisant l'éphéméride solaire JPL~DE405. La phase des événements $\gamma$ a ensuite été calculée à partir de l'éphéméride radio délivrée par le télescope de Parkes (voir le tableau \ref{tab:ephemj1048}). Le test de périodicité H-test (voir la section \ref{sec:signi}) sur cet échantillon de photon donne une probabilité $\leqslant 4 \times 10^{-8}$ que la pulsation soit une fluctuation. Cette valeur est par 4 ordres de grandeur plus petite que les précédents résultats d'EGRET \citep{kaspi00} et établit donc fermement la détection du pulsar.

La figure \ref{fig:lcJ1048} (cadre du haut) montre l'histogramme du nombre de photons $\gamma$ détectés en fonction de la phase au-dessus de 0.1\,GeV. Le profil radio à la fréquence de 1.4\,GHz est indiqué dans le cadre du bas pour la comparaison des phases. La courbe de lumière $\gamma$ est composée de deux pics, avec le pic P1 à la phase $0.15 \pm 0.01 \pm 0.0001$ et le pic P2 à la phase $0.57 \pm 0.01 \pm 0.0001$. La séparation en phase $\Delta\phi$ des deux pics est ainsi de $0.42 \pm 0.01 \pm 0.0001$. Les erreurs sont respectivement l'incertitude sur l'ajustement et l'incertitude sur le DM. Ces résultats sont en accord avec ceux trouvés par \citet{kaspi00}.

Les deux pics apparaissent asymétriques, avec un côté montant doux et un côté descendant prononcé. Nous avons ainsi ajusté chacun des pics avec deux demi-lorentziennes pour prendre en compte ces deux différences de pente. P1 ($\phi = 0.05 - 0.17$) est plus fin avec une largeur à mi-hauteur de $0.06 \pm 0.01$, tandis que P2 ($\phi = 0.45 - 0.65$) est plus large avec une largeur à mi-hauteur de $0.10\pm0.02$. Notons qu'une structure qui peut être associée à un épaulement ou à un ``bridge'' entre les pics est présente entre 0.17 et 0.30 en phase. Ce profil est très similaire à la courbe de lumière de Vela qui elle-même est composée de deux pics séparés par 0.43 en phase et d'un \textit{bridge} \citep{abdo09vela}. Nous avons également défini la région OFF comme le minimum de l'histogramme ($\phi = 0.7 - 1.05$). Pour vérifier cette hypothèse, nous avons estimé le fond représenté par une ligne pointillée (73 coups/intervalle) à partir d'un anneau centré sur le pulsar entre 1$^{\circ}$ et 2$^{\circ}$. Les sources proches ont été exclues, et nous avons renormalisé l'anneau au même espace de phase que notre sélection. Le résultat est en bon accord avec la région OFF. Ainsi, le nombre total de photons pulsés provenant du pulsar est estimé à $933 \pm 93$, avec une contribution du fond de $3654 \pm 60$ événements.

La figure \ref{fig:lcJ1048} (cadres du milieu) montre l'histogramme du nombre de photons $\gamma$ en fonction de la phase pour quatre intervalles en énergie différents (0.1\ --\ 0.3 GeV, 0.3\ --\ 1\,GeV, 1\ --\ 3\,GeV, $>$3\,GeV). L'évolution en énergie de la forme de la courbe de lumière est visible, bien que plus de données soient nécessaires pour contraindre la largeur des pics pour chaque intervalle. En dessous de 0.3\,GeV notamment, le premier pic semble plus large qu'à haute énergie. Cette caractéristique pourrait être expliquée par la contamination de la source $\gamma$ (EMS~0693) à moins de 1$^{\circ}$ du pulsar. Nous avons également regardé l'évolution du rapport P1/P2 en fonction de l'énergie, également observé par EGRET pour les pulsars de Vela, du Crabe, de Geminga, de PSR~B1951+32 \citep{fierro95} et par le LAT pour Vela \citep{abdo09vela} et PSR~J0205+6449 (voir le chapitre \ref{chap:j0205}). Dans chaque bande en énergie, nous avons calculé la hauteur du pic en respectant le niveau de fond estimé à partir de la région OFF. Le rapport des hauteurs montre qu'il n'y a aucune variation avec un niveau de confiance de 69\%. Il convient d'avoir plus de données pour être plus contraignant. Enfin, notons qu'entre 1 et 3\,GeV et au-dessus de 3 GeV l'épaulement après P1 ($\phi$ = 0.17\ --\ 0.33) ainsi que les 2 autres pics sont toujours significatifs, et que le photon de plus haute énergie est détecté dans P1 à 19\,GeV. 

\fig[PSR~J1048$-$5832:~Courbes de lumière]{scale=0.62}{c8_J1048-5832_LC_multi_energy.pdf}{\textbf{Cadre du haut:}~Courbe de lumière de PSR~J1048$-$5832 au-dessus de 0.1 GeV. Deux rotations sont représentées avec 50 intervalles en phase par période ($P=123.7$\,ms). La ligne pointillée montre le niveau de fond estimé avec un anneau centré sur la position du pulsar (73 coups/intervalle). \textbf{Quatre cadres suivants:}~Histogrammes du nombre de photons détectés en fonction de la phase pour quatre intervalles d'énergie. Chacun est représenté avec 50 intervalles en phase. \textbf{Second cadre à partir du bas:}~Courbe de lumière d'EGRET au-dessus de 400\,MeV \citep{kaspi00}. \textbf{Cadre du bas:}~Profil radio pulsé enregistré par le télescope de Parkes à la fréquence de 1.4\,GHz (1024 intervalles de phase).}{fig:lcJ1048}

\subsection{Spectre et Flux Résolus en Phase}\label{sec:specj1048}

Afin d'obtenir le spectre résolu en phase et le flux intégré de PSR~J1048$-$5832, une analyse spectrale par un maximum de vraisemblance a été effectuée en utilisant le ST ``gtlike'' (voir la section \ref{sec:anaspec}). Nous avons utilisé les IRFs ``Pass6\_v3'' lesquelles sont, rappelons-le, une mise à jour après le lancement du satellite qui corrige des effets de traces fantômes dans le LAT.

Une région circulaire de 15$^{\circ}$ centrée sur la position du pulsar a été modélisée en incluant les sources proches ainsi que l'émission diffuse. Les pulsars inclus dans la région ont été ajustés par une loi de puissance avec une coupure exponentielle (équation \ref{eq:specj1048}), tandis que les autres sources ont été ajustées par une simple loi de puissance. Le fond diffus galactique a été pris en compte en utilisant une carte basée sur le modèle numérique GALPROP appelé 54\_59Xvarh7S. Le fond isotropique a été quant à lui modélisé avec un spectre tabulé qui dérive d'un ajustement des données du LAT à haute latitude galactique. Nous avons ajusté le spectre de PSR~J1048$-$5832 par une loi de puissance avec une coupure exponentielle en énergie entre 0.1 et 0.7 en phase. Le spectre est décrit par l'équation:
\begin{equation}\label{eq:specj1048}
\frac{dF}{dE} = N_{0} \ E^{-\Gamma} e^{-E/E_{c}}
\ {\rm cm^{-2} s^{-1} GeV^{-1}}
\end{equation}
avec $E$ en GeV, le terme $N_{0} = (5.9 \pm 0.3 \pm 0.1) \times 10^{-8}$\,cm$^{-2}$\,s$^{-1}$\,GeV$^{-1}$, un indice spectral $\Gamma$ = $1.38 \pm 0.06 \pm 0.12$ et une coupure en énergie $E_{c} = 2.3^{+0.3}_{-0.4} \pm 0.3$\,GeV. La première incertitude est statistique, tandis que la seconde est systématique (utilisation des IRFs modifiées). A partir de ce résultat, nous avons obtenu un flux de photon intégré entre 0.1\ --\ 100 GeV de $(2.19 \pm 0.22 \pm 0.32) \times 10^{-7}$\,cm$^{-2}$\,s$^{-1}$, lequel correspond à environ un tiers du flux de la source EGRET 3EG~J1048$-$5840. La figure \ref{fig:sedJ1048} montre à la fois l'ajustement global entre 0.1 et 20 GeV (ligne continue), et les points spectraux obtenus par une analyse dans six intervalles en énergie espacés logarithmiquement, en supposant à chaque fois une simple loi de puissance. Pour vérifier l'hypothèse de la coupure en énergie dans le spectre, nous avons également modélisé le pulsar avec une simple loi de puissance de la forme $dF/dE = N_{0} E^{-\Gamma}$. Le modèle spectral utilisant la coupure est mieux contraint avec une différence dans les logarithmes du maximum de vraisemblance de $\sim 10\,\sigma$.

Le pulsar est répertorié dans le \textit{Bright Source List} du \textit{Fermi} LAT \citep{abdo09bsl} comme la source 0FGL~J1047.6$-$5834, qui est localisée à (RA,Dec) = (161.922$^{\circ}$, $-$58.577$^{\circ}$) avec une extension spatiale de rayon 0.138$^{\circ}$ à 95\% de niveau de confiance. Une autre source LAT est localisée à $\sim$1$^{\circ}$, 0FGL~J1045.6$-$5937. Nous avons modélisé cette source non-identifiée par une simple loi de puissance. Le meilleur ajustement donne un indice spectral de $2.2 \pm 0.1 \pm 0.1$ et un flux de photon intégré de $(4.49 \pm 0.40 \pm 0.80) \times 10^{-7}$\,cm$^{-2}$\,s$^{-1}$. La somme du flux de cette source et de PSR~J1048$-$5832 est $\sim$ 6.7 $\times$10$^{-7}$ cm$^{-2}$s$^{-1}$, identique au flux de la source EGRET 3EG~J1048$-$5840 de $(6.2 \pm 0.7) \times 10^{-7}$\,cm$^{-2}$\,s$^{-1}$. La source EGRET 3EG~J1048$-$5840 ainsi que la source COS--B 2CG 288--00 \citep{swan81} étaient apparemment constituées de ces deux sources, qui sont à présent résolues par le \textit{Fermi} LAT.

\fig[PSR~J1048$-$5832: Distribution d'\'Energie Spectrale]{scale=0.7,angle=90}{c8_sedJ1048.pdf}{Distribution d'énergie spectrale pour le pulsar PSR~J1048$-$5832. La ligne continue représente l'ajustement global par une loi de puissance avec une coupure exponentielle. Les points spectraux ont été obtenus par une analyse dans six intervalles en énergie espacés logarithmiquement, en supposant pour chaque intervalle une simple loi de puissance. Les erreurs sont uniquement statistiques.}{fig:sedJ1048}

\section{Discussion}\label{sec:discussionj1048}

\subsection{Courbe de Lumière}

La courbe de lumière $\gamma$ de PSR~J1048$-$5832 (voir la figure \ref{fig:lcJ1048}) couvre un large intervalle en phase, suggérant que le faisceau $\gamma$ couvre un grand angle solide. Le profil est très similaire à celui de Vela \citep{abdo09vela} ou même de PSR~J0205+6449 (voir le chapitre \ref{chap:j0205}). Cette caractéristique semble en faveur des modèles d'émission magnétosphériques externes, en particulier le \textit{Outer Gap} (OG) et le \textit{Slot Gap/Two-Pole Caustic} (SG/TPC) (voir la section \ref{sec:modgam}).

\subsection{Luminosité, Efficacité, et Géométrie de l'Emission}

Pour PSR~J1048$-$5832, le flux d'énergie intégré obtenu en intégrant l'équation \ref{eq:specj1048} sur l'énergie est $(19.4 \pm 1.0 \pm 3.1) \times 10^{-11}$\,erg\,cm$^{-2}$\,s$^{-1}$, menant à une luminosité $\gamma$ de $2.1 \times 10^{35}$ $f_{\Omega}$ ($d$/3~kpc)$^{2}$ et une efficacité $\eta$ = $L_{\gamma}/\dot{E}$ = 0.10 $f_{\Omega}$ ($d$/3~kpc)$^{2}$.

La différence de phase des deux pics $\gamma$ et l'efficacité sont utiles pour placer des contraintes sur les modèles d'émission magnétosphériques externes. En utilisant les courbes de lumière $\gamma$ de ``l'Atlas'' \citep{watters09}, nous avons estimé l'intervalle autorisé pour l'inclinaison magnétique $\alpha \sim$ 60$^{\circ}$ -- 85$^{\circ}$~, l'angle d'observation $\zeta \sim 70^{\circ}$ -- 80$^{\circ}$~, et le facteur de correction $f_{\Omega} \sim$ 0.7 -- 1.1 pour le modèle OG, et $\alpha \sim 50^{\circ}$ -- 75$^{\circ}$~, $\zeta \sim$ 50$^{\circ}$ -- 75$^{\circ}$~, et $f_{\Omega} \sim$ 1.1 pour le modèle SG/TPC. De plus, la variation sur l'angle de polarisation radio contraint l'angle $\beta_{E}$ = $\alpha$ - $\zeta_{E}$. En ajustant le modèle de vecteur rotationnel \citep{radhna69} pour les données radio de PSR~J1048$-$5832, une valeur de $\beta_{E}$ plus petite que 10$^{\circ}$ a été dérivée \citep{kara05,welte08a}. Les deux modèles théoriques ont ainsi un bon intervalle de solutions possibles, mais cela suppose une efficacité basée sur la relation $\eta \simeq (10^{33}/\dot{E})^{0.5}$ de \citet{watters09}, laquelle donne $\eta \simeq$ 0.02, soit un facteur cinq fois plus petit que l'efficacité dérivée à partir de la luminosité mesurée pour $f_{\Omega}=1$. En supposant notre quantité mesurée $\eta$ = 0.10, l'intervalle autorisé reste le même pour le modèle TPC et l'intervalle de $\alpha$ se déplace vers $\sim 70^{\circ}$\ --\ 90$^{\circ}$~pour le modèle OG. Le décalage de la phase entre le pic radio et le premier pic $\gamma$ est ainsi en bon accord avec les deux modèles théoriques.

\subsection{Comparaisons de PSR~J1048$-$5832 et PSR J2229+6114}

Bien que PSR~J1048$-$5832 et PSR~J2229+6114 soient deux pulsars de type Vela, leur profil $\gamma$ et la géométrie de leur émission sont assez différents. Le tableau \ref{tab:resume} résume les principales quantités mesurées pour les deux pulsars. La courbe de lumière de PSR~J1048$-$5832 est très similaire à celle de Vela, tandis que son efficacité est d'un facteur 10 $\times$ plus grand à celui-ci (en adoptant $f_{\Omega} = 1$ et $d = 3$\,kpc). Au contraire, l'efficacité de PSR~J2229+6114 est très similaire à celle de Vela, mais le pulsar ne présente qu'un seul et large pic, identique à PSR~J0357+32 découvert avec les données LAT \citep{abdo09blindsearch}.

\tab[Résultats temporelles et spectraux de PSR~J1048$-$5832 et PSR J2229+6114]{psr1048and2229.tex}{Ce tableau résume les résultats des analyses spectrale et temporelle des deux pulsars de type Vela, PSR~J1048$-$5832 et PSR J2229+6114. Les erreurs statistiques et systématiques sont données.}{tab:resume}

%
\chapter[Catalogue de Pulsars Gamma]{Catalogue de Pulsars Gamma}\label{chap:catalog}

\minitoc

\section{Introduction}

Les capacités du \textit{Large Area Telescope} avec sa sensibilité nettement améliorée comparé à ses prédécesseurs, nous donnent l'opportunité d'étudier une population nouvelle de pulsars émetteurs $\gamma$. Ce chapitre présente les 46 pulsars $\gamma$ détectés avec un haut niveau de confiance à partir des six premiers mois des données. Cela inclut les six pulsars observés clairement par le télescope EGRET, plus les trois pulsars détectés marginalement par EGRET (voir le tableau \ref{tab:cgropsr}). Nous présenterons essentiellement les propriétés des pulsars normaux détectés (non milliseconde). La section \ref{sec:obspsrcat} décrit brièvement les observations et les procédures d'analyse temporelle et spectrale, tandis que dans la section \ref{sec:psrcatdiscussion} nous discutons des résultats sur la luminosité, la population de pulsars, et les courbes de lumière. Signalons que ce chapitre est extrait du premier catalogue de pulsars avec le télescope \textit{Fermi Gamma-Ray Space Telescope} \citep{abdo09psrcat}. 

\tab[Paramètres intrinsèques des pulsars détectés par le LAT]{CharacParTable.tex}
{\scriptsize{Paramètres mesurés et intrinsèques des pulsars détectés par le LAT. Type: (r) pulsars normaux découverts en \textbf{r}adio, (g) pulsars découverts en \textbf{g}amma, (m) pulsars \textbf{m}illiseconde, (b) indiquent s'ils appartiennent à des systèmes binaires. $S_{1400}$: densité de flux radio à 1400\,MHz (source: ATNF) ou limite supérieure avec 1 \citep{Halpern2004} ; 2 \citep{camilo2009} ; 3 \citep{Roberts2002} ; 4 \citep{Halpern2007}. \\ 
\textbf{Références:} a \citep{abdo08} ; b \citep{abdo09blindsearch} ; c \citep{abdo09j0030} ; d \citep{abdo09msp} ; e \citep{abdo09j0205} ; f \citep{cognard09} ; h \citep{abdo09crab} ; i \citep{Weltevrede09b} ; j \citep{abdo09geminga} ; k \citep{abdo09vela} ; l \citep{abdo09j1028} ; m \citep{abdo09j2229} ; n \citep{abdo09EGRETpsr} ; o \citep{camilo09j1833} ; p \citep{abdo09j2021} ; q \citep{noutsos09} ; r \citep{abdo09j1907} ; s \citep{abdo09j2021p4026} ; t \citep{camilo2009} }
}{tab:psrcatpar}

\section{Observations et Analyses}\label{sec:obspsrcat}

Dans le but d'avoir une analyse uniforme, les données $\gamma$ utilisées pour l'étude de cette population de pulsars présentent le même intervalle de temps d'observation. Pour l'analyse spectrale, les données ont été collectées à partir du début de l'observation du télescope \textit{Fermi} en mode balayage (4 août 2008) jusqu'au 1 février 2009, tandis que l'analyse temporelle commence à partir du premier événement enregistré par le LAT (25 juin 2008) jusqu'au 1 février 2009. Comme dans les chapitres précédents, seuls les événements classés \textit{diffuse} ont été sélectionnés, et nous avons exclu ceux possédant un angle zénithal supérieur à 105$^{\circ}$. 


\subsection{Analyse Temporelle}

Deux méthodes pour la recherche de pulsations dans les données LAT ont été mises en place dès le début de la mission. 

La première méthode recherche des pulsations uniquement à partir du temps d'arrivée des photons en sélectionnant des candidats tels que des SNRs et des PWNs suspectées d'abriter un pulsar, et sur des sources du catalogue du LAT \citep{atwood06}. Au total après six mois d'observation, 16 nouveaux pulsars ont été découverts parmi lesquels le pulsar de la jeune supernova galactique CTA1 \citep{abdo08,abdo09blindsearch}. Soulignons deux points: (1)~les modèles théoriques d'émission prédisent une large émission $\gamma$ en éventail et un fin faisceau radio, cette méthode favorise donc les pulsars avec une grande inclinaison magnétique, si l'on suppose que l'émission radio vient du pôle magnétique; (2)~les pulsars de cette nouvelle population ne sont pas forcément des pulsars de type Geminga (\textit{radio-faint}). Des recherches de pulsations radio ont été effectuées sur certains de ces pulsars. Parmi eux, trois pulsars ont été observés avec un faible signal radio, PSRs J1741-2054 et J2032+4127 \citep{camilo2009}, PSR~J1907+0602 \citep{abdo09j1907}.

La deuxième méthode est la recherche de pulsations $\gamma$ en utilisant les éphémérides de pulsars déjà connus dans une autre longueur d'onde. Comme nous l'avons vu à la section \ref{sec:campephem}, il est essentiel pour pouvoir suivre et détecter des pulsars d'avoir des éphémérides valides sur la période d'observation $\gamma$. Au final, les astronomes de la radio et des $X$ nous ont fourni 767 éphémérides contemporaines. Une première partie des éphémérides est notre sélection de 224 pulsars candidats $\gamma$ avec $\dot{E}>10^{34}$\,erg\,s$^{-1}$ (voir le chapitre \ref{chap:candidat}), tandis que l'autre partie est un échantillon de 543 pulsars avec $\dot{E}<10^{34}$\,erg\,s$^{-1}$ permettant ainsi de réduire le biais possible dans notre recherche de pulsations. Plusieurs programmes d'analyse basés sur les tests de périodicité (section \ref{sec:signi}) tel que le \textit{H-test} ont été mis en place dans la collaboration \textit{Fermi} et notamment à Bordeaux pour suivre ces pulsars \citep{jrjc2007}. Les pulsars présentant une significativité supérieure à 5\,$\sigma$ pour les 6 premiers mois des données LAT ont été sélectionnés pour cette étude. Cette méthode a l'avantage d'être sensible à des flux plus faibles, et permet de comparer l'alignement en phase des profils aux différentes longueurs d'onde, ce qui apporte une excellente contrainte sur la géométrie de l'émission.

Le tableau \ref{tab:psrcatpar} liste les paramètres mesurés et intrinsèques des 46 pulsars $\gamma$ détectés. Les pulsars sont répertoriés selon 3 classes:~(r) identifie les pulsars normaux découverts en \textbf{r}adio, (g) les pulsars découverts en \textbf{g}amma, et (m) identifient les pulsars \textbf{m}illiseconde, (b) indiquent s'ils appartiennent à des systèmes binaires. La dérivée de la période $\dot{P}$ est corrigée de l'effet cinématique \textit{Shklovskii} \citep{shklovskii70}: $\dot{P} = \dot{P}_{obs} - \mu^2 P_{obs} d / c^2$, où $\mu$ est le mouvement propre du pulsar et $d$ la distance. Cette correction s'addresse essentiellement aux pulsars milliseconde \citep{abdo09msp}.

\fig[PSR~J1028$-$5819:~Courbes de lumière]{scale=0.65,angle=0.0}{c9_J1028-5820_catalog_lightcurve.pdf}{\textbf{Cadre du haut:}~Courbe de lumière de PSR~J1028$-$5819 au-dessus de 0.1 GeV. Deux rotations sont représentées avec 50 intervalles en phase par période ($P=91.4$\,ms). La ligne pointillée montre le niveau de fond estimé avec un anneau centré sur la position du pulsar (42 coups/intervalle). \textbf{Trois cadres suivants:}~Histogrammes du nombre de photons détectés en fonction de la phase pour quatre intervalles d'énergie. Le nombre de coups au-dessus de 3\,GeV est représenté en noir. \textbf{Cadre du bas:}~Profil radio pulsé enregistré par le télescope de Parkes à la fréquence de 1.4\,GHz (1024 intervalles de phase).}{fig:j1028}

\subsubsection{Courbes de lumière}

Un échantillon des courbes de lumière des 46 pulsars détectés par le LAT sont présentées dans l'annexe de cette thèse. La sélection des événements et la construction des phases pour chaque pulsar sont similaires à l'analyse décrite dans les chapitres précédents pour les pulsars PSRs~J0205+6449, J1048$-$5832, et J2229+6114. Pour un petit nombre de pulsars localisés près d'une source $\gamma$ brillante, la ROI a été redéfinie à 0.5$^{\circ}$ au lieu de 1$^{\circ}$. Soulignons que les éphémérides utilisées pour les 16 pulsars découverts en $\gamma$, ainsi que pour le pulsar Geminga et le pulsar radio PSR~J1124$-$5916\footnote{PSR~J1124$-$5916 étant très faible en radio (une mesure demande environ 4h d'observation au télescope Parkes), il est plus facile de déterminer des TOAs à partir des données LAT. L'éphéméride contient un TOA radio nécessaire pour avoir le décalage entre les pulsations radio et $\gamma$.}, dérivent de modèles de rotation construits à partir des données du LAT \citep{ray09}. Ces modèles sont basés sur des observations s'étendant du 25 juin 2008 au 1er mai 2009. Les éphémérides utilisées pour les pulsars émetteurs radio ont été fournies par les télescopes énumérés à la section \ref{sec:distribpsr}.

Pour illustrer ce travail, la figure \ref{fig:j1028} représente le résultat de l'analyse temporelle pour le pulsar radio PSR~J1028$-$5819 \citep{abdo09j1028}. Le profil gris en haut représente l'histogramme du nombre de coups en fonction de la phase au-dessus de 0.1\,GeV, tandis que les autres cadres montrent le profil $\gamma$ dans des intervalles d'énergies: 0.1\ --\ 0.3 GeV, 0.3\ --\ 1\,GeV, $>$1\,GeV (avec les photons supérieurs à 3\,GeV représentés en noir). Pour les pulsars émetteurs radio, le profil radio aligné en phase est montré dans le cadre le plus bas. Le nombre d'intervalles en phase pour chaque pulsar a été défini en fonction de la statistique et de la RMS des résidus de la solution temporelle, telle que la résolution d'un intervalle en phase $\sigma_t$ soit inférieure à $P/N \sqrt{12}$, avec $N =$ 25 ou 50 intervalles et $P$ la période de rotation du pulsar. La ligne pointillée sur le profil au-dessus de 0.1\,GeV représente le niveau de fond estimé avec un anneau entre 1$^{\circ}$ $< \theta <$ 2$^{\circ}$ entourant la source. Pour certains pulsars le niveau de fond ne correspond pas au minimum du phasogramme. La résolution spatiale du LAT étant large à basse énergie, dans le cas où le pulsar se trouve très proche d'une autre source, l'analyse est défaillante. Cet effet est observé pour les pulsars~PSR J1420-6048 et PSR~J1418-6058 situé à moins de 0.2\,$^{\circ}$ l'un de l'autre. Il peut également s'agir de la présence des PWNe \textit{K3} et du \textit{Rabbit}.

Si l'on regarde la dépendance en énergie des courbes de lumière à la fois pour les pulsars de type (r) et de type (g), le rapport P1/P2 décroît quand l'énergie augmente, ceci pour une grande partie des pulsars avec deux pics. Pour les pulsars avec un $\Delta \sim 0.5$, tels que PSRs J1028$-$5820, J2021+3651, J0633+0632, J1124$-$5916, J1813$-$1246, J1826$-$1256, J1836+5925, et J2238+59, cette tendance n'est pas observée, suggérant une symétrie du profil $\gamma$.

\subsection{Analyse Spectrale}

Pour étudier le spectre des pulsars et extraire le flux intégré au-dessus de 100\,MeV de la région pulsée des 46 pulsars détectés, une analyse spectrale par un maximum de vraisemblance a été effectuée en utilisant le catalogue des 6 mois de données. La source $\gamma$ Cygnus X-3 très proche du pulsar PSR~J2032+4127 a également été modélisée, bien qu'elle n'ait pas été détectée automatiquement dans la création du catalogue. L'analyse étant similaire à celle des pulsars étudiés aux chapitres 6 à 8, nous n'en rappellerons que les points essentiels. 

Tous les pulsars ont été modélisés par une loi de puissance avec une coupure exponentielle (équation \ref{eq:specj0205}). Le terme de normalisation $N_0$, l'indice spectral $\Gamma$, et l'énergie de coupure $E_{Cutoff}$ sont laissés libres dans l'ajustement. L'émission diffuse galactique a été modélisée en utilisant le modèle de diffus nommé 54\_77Xvarh7S basé sur GALPROP. L'émission extragalactique a été, quant à elle, modélisée par une simple loi de puissance. Pour chaque pulsar, nous avons sélectionné les événements au-dessus de 0.1\,GeV, à l'intérieur d'une région de 10$^{\circ}$ centrée sur la position de la source, et inclus dans notre modèle tous les objets jusqu'à 17$^{\circ}$. Rappellons que les sources à l'extérieur de la région analysée peuvent contribuer à l'émission de basse énergie. L'ajustement des paramètres spectraux a été effectué sur la région pulsée (ON) du pulsar, en tenant compte de la région non pulsée (OFF), modélisée par une simple loi de puissance (possible PWN). On trouvera la définition de la phase du OFF dans le tableau \ref{tab:pulseshapepar} pour chaque objet. Dans la plupart des cas, l'émission du OFF n'est pas significative. Un futur papier est dédié à l'étude de l'émission non pulsée (possible PWN) des pulsars détectés. \\

Le tableau \ref{tab:specpar} présente les résultats de l'ajustement des paramètres spectraux des 46 pulsars détectés par le LAT. Pour le calcul de la luminosité et de l'efficacité $\gamma$, nous avons supposé un facteur de correction $f_\Omega = 1$. Notons que les paramètres spectraux du pulsar du Crabe proviennent de \citet{abdo09crab}, qui prend en compte la modélisation de la PWN. Signalons que ce travail a été effectué par l'équipe du catalogue général $\textit{Fermi}$.

Pour un certain nombre de pulsars, le modèle spectral incluant une coupure exponentielle n'est pas plus significatif qu'une simple loi de puissance. Nous avons déterminé cela en calculant le $TS_{cutoff} = 2\Delta log(\mathrm{likelihood})$ (comparable à une distribution de $\chi^2$ pour un degré de liberté) entre les modèles avec et sans coupure. Les pulsars ayant un \textit{TS} $<$ 10 présente une mesure de l'énergie de coupure non significative. \\

Il faut signaler que la connaissance de l'émission diffuse galactique n'est pas parfaite, et peut avoir un impact sur le flux et les paramètres spectraux des sources étudiées. Ceci fait partie des erreurs systématiques qu'il est difficile d'estimer, étant donné que le modèle de diffus évolue en fonction de la statistique grandissante des données du LAT. Il est ainsi intéressant de comparer les résultats spectraux des trois pulsars individuels étudiés aux chapitres 6, 7, et 8, avec les résultats du catalogue, étant donné que ces études utilisent des modèles de diffus différents et des intervalles de temps différents. Aux incertitudes statistiques et systématiques près, l'accord entre les analyses individuelles des pulsars et le catalogue est excellent. Les conclusions théoriques à partir des observations ne sont donc pas érronées entre ces deux analyses. Cependant, notons que ces trois pulsars ont un flux relativement élevé. Des tests comparatifs ont été effectués pour des pulsars présentant un faible flux, tel que PSR~J0248+6021 ($F \sim 4 \times 10^{-8}$\,ph\,cm$^{-2}$\,s$^{-1}$). Dans ce cas, le changement de diffus peut affecter le flux du pulsar d'un facteur 1.5. Les observations spectrales apportées par les pulsars faibles sont donc à prendre avec précaution. 

%

\clearpage

\begin{landscape}
\begin{table}
\begin{tiny}
\begin{tabular}{llrrrrrrcc}
\hline
\hline
PSR & Type$^a$ & Flux de Photon ($F_{\rm 100}$)& Flux en Energie ($G_{\rm 100}$)& $\Gamma$ \ \ \ \ & $E_{\rm Cutoff}$ \ \ \ & TS & TS$_{\rm cutoff}$ & Luminosité & Efficacité \\
 &  & ($\rm 10^{-8} \,ph \ cm^{-2} \,s^{-1}$) & ($\rm 10^{-5} \,MeV \ cm^{-2}\, s^{-1}$) &  & ($\rm GeV$) \ \ & &  & ($\rm 10^{33}\, erg \,s^{-1}$) & ($f_{\Omega}=1$) \\
\hline
J0007$+$7303	&	g	&	30.7	$\pm$	1.3	&	38.2	$\pm$	1.3	&	1.38	$\pm$	0.05	&	4.6	$\pm$	0.4	&	7384 &	274.7	&	89	$\pm$	38	&	0.20	$\pm$	0.08 \\
J0030$+$0451	&	m	& 5.83	$\pm$	0.78	&	5.26	$\pm$	0.42	&	1.22	$\pm$	0.19	&	1.8	$\pm$	0.4	&	960	&	59.2	&	0.57	$\pm$ 0.35	&	0.17	$\pm$	0.10 \\
J0205$+$6449	&	r	&	13.2	$\pm$	2.0	&	6.64 $\pm$	0.65	&	2.09	$\pm$	0.17	&	3.5	$\pm$	1.4	&	346	&	12.5	&	54	--	81 & 0.002	--	0.003 \\
J0218$+$4232	&	m	&	6.2	$\pm$	1.7	&	3.62	$\pm$	0.64	&	2.02	$\pm$ 0.28	&	5.1	$\pm$	4.2	&	119	&	4.7	&	27	--	69	&	0.11	--	0.29	\\
J0248$+$6021	&	r	&	3.7	$\pm$	1.8	&	3.07	$\pm$	0.70	&	1.15	$\pm$	0.59	&	1.4	$\pm$	0.6	&	103	&	18.5	&	15	--	300	&	0.07	--	1.4	\\
J0357$+$32 	&	g	&	10.4	$\pm$	1.2	&	6.38	$\pm$	0.44	&	1.29	$\pm$	0.22	&	0.9	$\pm$	0.2	&	949	&	71.6	&	...		&	...					\\
J0437$-$4715	&	m	&	3.65	$\pm$	0.84	&	1.86	$\pm$	0.26	&	1.74	$\pm$	0.38	&	1.3	$\pm$	0.7	&	172	&	9.9	&	0.054	$\pm$	0.008	&	0.02	$\pm$	0.003		\\
J0534$+$2200$^c$	&	r	&	209	$\pm$	22	&	130.6	$\pm$	1.3	&	1.97	$\pm$	0.07	&	5.8	$\pm$	1.2	&	21507	&	80.2	&	620	$\pm$	310	
			&	0.001	$\pm$	0.001				\\J0613$-$0200	&	m	&	3.38	$\pm$	0.85	&	3.23	$\pm$
			0.42	&	1.38	$\pm$	0.29	&	2.7	$\pm$	1.0	&	285	&	18.5	&	0.89	$_{	-0.42	}^{+	0.71
			}$	&	0.07	$_{	-0.03	}^{+	0.06	}$	\\J0631$+$1036	&	r	&	2.8	$\pm$	1.2	&	3.04
			$\pm$	0.61	&	1.38	$\pm$	0.42	&	3.6	$\pm$	1.8	&	86	&	10	&	2.0	--	48	
					&	0.01	--	0.27				\\J0633$+$0632	&	g	&	8.4	$\pm$	1.4	&
					8.0	$\pm$	0.77	&	1.29	$\pm$	0.22	&	2.2	$\pm$	0.6	&	370	&	50.8	&	...
										&	...						\\J0633$+$1746	&	g
&	305.3	$\pm$	3.5	&	338.1	$\pm$	3.5	&	1.08	$\pm$	0.02	&	1.9	$\pm$	0.05	&	62307	&	5120.4	&	25
$_{	-12	}^{+	24	}$	&	0.78	$_{	-0.38	}^{+	0.74	}$	\\J0659$+$1414	&	r	&	10	$\pm$	1.4
&	3.17	$\pm$	0.36	&	2.37	$\pm$	0.50	&	0.7	$\pm$	0.5	&	206	&	6.9	&	0.31	$\pm$	0.08		
	&	0.01	$\pm$	0.002				\\J0742$-$2822	&	r	&	3.18	$\pm$	1.2	&	1.82	$\pm$	0.42	&
	1.76	$\pm$	0.48	&	2.0	$\pm$	1.4	&	47	&	4.2	&	9.0	$_{	-9	}^{+	12	}$	&	0.07	$_{
	-0.07	}^{+	0.09	}$	\\J0751$+$1807	&	m	&	1.35	$\pm$	0.66	&	1.09	$\pm$	0.38	&	1.56	$\pm$	0.70
	&	3.0	$\pm$	4.3	&	37	&	3.8	&	0.47	$_{	-0.35	}^{+	1	}$	&	0.08	$_{	-0.06	}^{+	0.17
	}$	\\J0835$-$4510	&	r	&	1061	$\pm$	7.0	&	879.4	$\pm$	5.4	&	1.57	$\pm$	0.01	&	3.2	$\pm$
	0.06	&	219585	&	5971	&	87	$\pm$	12				&	0.01	$\pm$	0.002			
	\\J1028$-$5819	&	r	&	19.6	$\pm$	3.1	&	17.7	$\pm$	1.4	&	1.25	$\pm$	0.20	&	1.9	$\pm$	0.5
	&	620	&	75.1	&	120	$\pm$	73				&	0.14	$\pm$	0.09				\\J1048$-$5832
	&	r	&	19.7	$\pm$	3.0	&	17.2	$\pm$	1.3	&	1.31	$\pm$	0.18	&	2.0	$\pm$	0.4	&	881	&
	81.8	&	150	$\pm$	90				&	0.08	$\pm$	0.05				\\J1057$-$5226	&	r	&
	30.45	$\pm$	1.7	&	27.2	$\pm$	0.98	&	1.06	$\pm$	0.10	&	1.3	$\pm$	0.1	&	4961	&	366.3	&	17
	$\pm$	9				&	0.56	$\pm$	0.31				\\J1124$-$5916	&	r	&	5.2	$\pm$	1.8
	&	3.79	$\pm$	0.70	&	1.43	$\pm$	0.40	&	1.7	$\pm$	0.7	&	111	&	16.7	&	100	$_{	-53	}^{+
	34	}$	&	0.01	$_{	-0.004	}^{+	0.003	}$	\\J1418$-$6058	&	g	&	27.7	$\pm$	8.3	&	23.5	$\pm$
	3.8	&	1.32	$\pm$	0.24	&	1.9	$\pm$	0.4	&	162	&	54.1	&	110	--	700				&
	0.02	--	0.14				\\J1420$-$6048	&	r	&	24.2	$\pm$	7.9	&	15.8	$\pm$	3.5	&	1.73
	$\pm$	0.24	&	2.7	$\pm$	1.0	&	63	&	21.4	&	590	$\pm$	380				&	0.06	$\pm$	0.04
				\\J1459$-$60	&	g	&	17.8	$\pm$	3.4	&	10.56	$\pm$	1.2	&	1.83	$\pm$	0.24	&
				2.7	$\pm$	1.1	&	337	&	21.1	&	...						&	...		
							\\J1509$-$5850	&	r	&	8.7	$\pm$	1.4	&	9.7	$\pm$	1.2	&	1.36
							$\pm$	0.28	&	3.5	$\pm$	1.1	&	262	&	26.3	&	78	$\pm$	49	
									&	0.15	$\pm$	0.10				\\J1614$-$2230	&	m	&
2.89	$\pm$	1.2	&	2.74	$\pm$	0.50	&	1.34	$\pm$	0.43	&	2.4	$\pm$	1.0	&	149	&	13.3	&	5.3	$\pm$
3.4				&	1.0	$\pm$	0.7				\\J1709$-$4429	&	r	&	149.8	$\pm$	4.1	&	124
$\pm$	2.6	&	1.70	$\pm$	0.04	&	4.9	$\pm$	0.4	&	16009	&	373.6	&	290	--	1900				&
0.09	--	0.57				\\J1718$-$3825	&	r	&	9.1	$\pm$	5.8	&	6.7	$\pm$	1.9	&	1.26	$\pm$
0.74	&	1.3	$\pm$	0.6	&	105	&	19.7	&	120	$\pm$	80				&	0.09	$\pm$	0.06		
	\\J1732$-$31	&	g	&	25.3	$\pm$	3.0	&	24.2	$\pm$	1.4	&	1.27	$\pm$	0.14	&	2.2	$\pm$	0.3
	&	1002	&	131.2	&	...						&	...						\\J1741$-$2054
	&	g	&	20.3	$\pm$	2.0	&	12.8	$\pm$	0.8	&	1.39	$\pm$	0.17	&	1.2	$\pm$	0.2	&	935	&
	92.6	&	2.2	$\pm$	1.3				&	0.24	$\pm$	0.14				\\J1744$-$1134	&	m	&
	4.3	$\pm$	1.6	&	2.8	$\pm$	0.6	&	1.02	$\pm$	0.71	&	0.7	$\pm$	0.4	&	78	&	20	&	0.43
	$\pm$	0.13				&	0.1	$\pm$	0.03				\\J1747$-$2958	&	r	&	18.2	$\pm$	4.2
	&	13.1	$\pm$	1.7	&	1.11	$\pm$	0.34	&	1.0	$\pm$	0.2	&	213	&	59.3	&	63	--	390	
			&	0.02	--	0.16				\\J1809$-$2332	&	g	&	49.5	$\pm$	3.0	&	41.3	$\pm$
			1.6	&	1.52	$\pm$	0.07	&	2.9	$\pm$	0.3	&	3451	&	201.9	&	140	$\pm$	140		
				&	0.33	$\pm$	0.33				\\J1813$-$1246	&	g	&	28.1	$\pm$	3.5	&	16.9
				$\pm$	1.3	&	1.83	$\pm$	0.14	&	2.9	$\pm$	0.8	&	482	&	39.7	&	...		
							&	...						\\J1826$-$1256	&	g	&	41.8	$\pm$
4.1	&	33.4	$\pm$	1.8	&	1.49	$\pm$	0.11	&	2.4	$\pm$	0.3	&	1152	&	138	&	...			
		&	...						\\J1833$-$1034	&	r	&	20.5	$\pm$	4.6	&	10.1	$\pm$	1.4
		&	2.24	$\pm$	0.18	&	7.7	$\pm$	4.8	&	110	&	4.9	&	270	$\pm$	60				&
		0.01	$\pm$	0.002				\\J1836$+$5925$^d$	&	g	&	65.6	$\pm$	1.8	&	59.9	$\pm$	1.3	&
		1.35	$\pm$	0.04	&	2.3	$\pm$	0.1	&	20982	&	674.6	&	$<$46						&
		$<$4.0						\\J1907$+$06	&	g	&	40.25	$\pm$	3.8	&	27.5	$\pm$	1.6	&
		1.84	$\pm$	0.10	&	4.6	$\pm$	1.0	&	1209	&	59.3	&	...						&	...
							\\J1952$+$3252	&	r	&	17.6	$\pm$	1.9	&	13.4	$\pm$	0.9	&	1.75
							$\pm$	0.12	&	4.5	$\pm$	1.2	&	1008	&	36.4	&	64	$\pm$	32	
									&	0.02	$\pm$	0.01				\\J1958$+$2846	&	g	&
7.65	$\pm$	1.6	&	8.45	$\pm$	0.83	&	0.77	$\pm$	0.31	&	1.2	$\pm$	0.2	&	491	&	89.2	&	...	
				&	...						\\J2021$+$3651	&	r	&	67.35	$\pm$	4.4	&	47
				$\pm$	1.8	&	1.65	$\pm$	0.07	&	2.6	$\pm$	0.3	&	3138	&	223.5	&	250	$_{	-240
				}^{+	500	}$	&	0.07	$_{	-0.07	}^{+	0.15	}$	\\J2021$+$4026$^d$	&	g	&	152.6	$\pm$
				4.9	&	97.6	$\pm$	2.2	&	1.79	$\pm$	0.04	&	3.0	$\pm$	0.2	&	10180	&	331.4	&
				260	$\pm$	150				&	2.2	$\pm$	1.3				\\J2032$+$4127	&	g
				&	6	$\pm$	2.3	&	11.1	$\pm$	1.4	&	0.68	$\pm$	0.46	&	2.1	$\pm$	0.6	&	487
				&	56.3	&	34	--	170				&	0.13	--	0.64			
				\\J2043$+$2740	&	r	&	2.41	$\pm$	0.90	&	1.55	$\pm$	0.32	&	1.07	$\pm$	0.66	&
				0.8	$\pm$	0.3	&	79	&	15.1	&	6.0	$\pm$	3.8				&	0.09	$\pm$	0.06
							\\J2124$-$3358	&	m	&	1.95	$\pm$	0.49	&	2.75	$\pm$	0.42	&	1.05
							$\pm$	0.34	&	2.7	$\pm$	1.0	&	226	&	22.9	&	0.21	$_{	-0.14	}^{+
							0.42	}$	&	0.05	$_{	-0.04	}^{+	0.11	}$	\\J2229$+$6114	&	r	&
							32.6	$\pm$	2.2	&	22.0	$\pm$	1.0	&	1.74	$\pm$	0.084	&	3.0	$\pm$	0.5
							&	1929	&	96	&	17	--	1100				&	0.001	--	0.05
			\\J2238$+$59	&	g	&	6.8	$\pm$	1.8	&	5.44	$\pm$	0.71	&	1.00	$\pm$	0.43	&	1.0	$\pm$	0.3	&	219	&	37.2	&	...						&	...						\\				
\hline
\hline

\end{tabular}
\caption{Résultats de l'ajustement des paramètres spectraux par un maximum de vraisemblance pour les 46 pulsars détectés par le LAT. \label{tab:specpar}}
\begin{scriptsize}
\textit{Les colonnes 3 et 4 listent respectivement le flux de photon ($F_{\rm 100}$) et le flux en énergie ($G_{\rm 100}$) intégrés au-dessus de 100\,MeV. Les colonnes 5 et 6 présentent l'indice spectral $\Gamma$ et l'énergie de coupure $E_{\rm C}$ de l'équation \ref{eq:specj0205}. Le test statistique (TS) renvoyé par l'ajustement est fourni en colonne 7, tandis que la significativité de la coupure est donnée en colonne 8. La luminosité et l'efficacité $\gamma$ sont listées en 9 et 10, et ont été calculées en supposant un facteur de correction $f_\Omega=1$. Rappelons que ces valeurs dépendent fortement de la distance.}
\end{scriptsize}

\end{tiny}
\end{table}
\end{landscape}


\section{Associations}

Le tableau \ref{tab:psrcat:association} fournit l'association spatiale des 46 pulsars détectés par le LAT avec des sources connues au GeV et au TeV; 22 des 46 pulsars sont associés à des sources EGRET, 11 sont des pulsars de type (g), 2 sont des pulsars milliseconde, et 9 sont des pulsars de type (r). Les pulsars PSR J1420$-$6048 et J1418$-$6058 sont associés tous les deux à la source EGRET 3EG~J1420$-$6038 (complexe de Kookaburra). Le LAT a permis de résoudre ces deux sources séparemment. Un certain nombre de sources EGRET non-identifiées avaient été au préalablement associées à des SNRs, PWNs, ou autres objets (e.g. \citealt{Walker2003,DeBecker2005}). Il n'est pas étonnant de voir que 19 des 38 pulsars normaux ont des associations avec une PWN ou/et une SNR \citep{roberts05,Green2009}. Finalement, 12 des 38 pulsars non-milliseconde sont associés avec des sources au TeV, dont la plupart (9 sur 12) sont associés avec une PWN. 

\tab[Associations spatiales des 46 pulsars détectés]{associations.tex}{
\scriptsize{
Associations spatiales des 46 pulsars détectés avec des sources observées au GeV et au TeV. \\
\textbf{Notes:}~$^a$Sources du catalogue LAT pour 3 mois de données \citep{abdo09bsl}. $^b$Sources des catalogues EGRET \citep{hartman99}, EGRET révisé \citep{casandjian08}, et GeV \citep{lamb97}. } \\
\tiny{
\textbf{Références:}~1. \citep{roberts05}, 2. \citep{Green2009}, 3. \citep{abdo09milagro}, 4. \citep{HESSCrab2006}, 5. \citep{HESSVela2006}, 6. \citep{HESSRabbit2006}, 7. \citep{HESS1709_2009}, 8. \citep{HESS1718_2007}, 9. \citep{HESS1833_2007}, 10. \citep{HESS1908_2009}, 11. \citep{ATEL2172}
}}{tab:psrcat:association}

\section{Résultats et Discussion}\label{sec:psrcatdiscussion}

\subsection{Luminosité Gamma}

Le potentiel au niveau des lignes de champs ouvertes, qui est proportionnel au courant de Goldreich et Julian, a été longtemps reconnu comme un paramètre important pour discrimer les modèles d'émission des pulsars $\gamma$ \citep{arons96}. Ce paramètre implique que la luminosité $\gamma$ est proportionnelle à la racine carrée de $\dot{E}$, c'est à dire, que l'efficacité $\gamma$ augmente quand le ralentissement $\dot{E}$ diminue. La figure \ref{fig:LvsEdot} présente la luminosité $\gamma$ en fonction du ralentissement $\dot{E}$ basée sur les flux en énergie mesurés pour chaque pulsar (voir le tableau \ref{tab:specpar}). Signalons deux points importants. Premièrement, la luminosité est très sensible à la distance estimée du pulsar (équation \ref{eq:luminosite}). Le tableau \ref{tab:psrcatdistance} présente la meilleure estimation de la distance pour 37 des 46 pulsars détectés par le LAT à partir de la mesure de dispersion, de la mesure de leur parallaxe, ou de l'observation de l'absorption de l'hydrogène. Les 9 pulsars restants sont des pulsars découverts en $\gamma$ (type g). Le second point est l'hypothèse d'un faisceau uniforme sur tout le ciel ($f_{\Omega}=1$), ce qui n'est pas réaliste pour certaines configurations géométriques des modèles d'émission $\gamma$.

Sur la figure \ref{fig:LvsEdot}, nous avons indiqué une ligne pointillée qui suppose une efficacité de conversion de 100\% pour transférer la perte d'énergie rotationnelle en photons $\gamma$ ($L_{\gamma} = \dot{E}$), ainsi qu'une ligne discontinue théorique qui suppose $L_{\gamma} = (10^{33}$\,erg\,s$^{-1}$\,$\dot{E}$)$^{1/2}$. Etant donné les larges incertitudes sur la luminosité, il est difficile pour l'instant d'étudier les détails de son évolution. Nous discuterons cependant de quelques tendances apparentes. 

Les pulsars avec un fort $\dot{E}$ semblent suivre relativement bien la tendance $\dot{E}^{1/2}$, malgré la large dispersion de la distance de PSR~J2021+3651 et PSR~J1709$-$4429. Dans l'intervalle 10$^{35}$\,erg\,s$^{-1}$ $< \dot{E} <$ 10$^{36.5}$\,erg\,s$^{-1}$, la luminosité semble plus ou moins constante, mais la plupart des pulsars de cette zone ont une distance basée sur la mesure de dispersion dont l'erreur peut dépasser 50\%. Il faut donc être prudent sur cette observation. De plus, l'une des deux distances (9\,kpc) du pulsar PSR~J0248+6021 demande une efficacité de 100\% pour produire la luminosité observée, ce qui est déraisonnable comparé à la tendance générale. Finalement, le pulsar PSR~J2021+4026 qui apparaît avec une grande efficacité est associé au reste de supernova $\gamma$ Cygni. L'âge caractéristique du pulsar est 10 $\times$ supérieur à l'âge de la SNR. La distance associée pour le pulsar doit donc être traité avec attention. Il est ainsi essentiel de maîtriser la distance des objets étudiés pour suivre l'évolution de la luminosité $\gamma$. 

\fig[Luminosité $L_{\gamma}$ en fonction de l'énergie de ralentissement $\dot{E}$]{scale=0.8,angle=90}{c9_LvsEdot.pdf}{Luminosité $\gamma$ en fonction du taux de perte d'énergie rotationnelle $\dot{E}$. La luminosité $\gamma$ est calculée à partir du flux en énergie intégré ($G_{100}$) et en utilisant un facteur de correction du faisceau $f_{\Omega}=1$ pour tous les pulsars. Nous avons également indiqué par une étoile la luminosité totale à haute énergie pour le pulsar du Crabe, $L_{Tot} = L_X + L_{\gamma}$. Carrés bleus: (g) pulsars découverts en $\gamma$. Triangles rouges: (m) pulsars $\gamma$ milliseconde. Cercles verts: (r) autres pulsars émetteurs radio et $\gamma$. Les points vides indiquent les pulsars pour lesquels la distance est uniquement connue par le DM. Les pulsars ayant deux distances sont indiqués avec deux points connectés par une ligne pointillée. Ligne pointillée : $L_{\gamma} = \dot{E}$. Ligne discontinue : $L_{\gamma}$ proportionnelle à $\sqrt{\dot{E}}$.}{fig:LvsEdot}

Pour $\dot{E}$ compris entre 10$^{34}$ et 10$^{35}$\,erg\,s$^{-1}$, les luminosités sont distribuées sur un large intervalle pour les pulsars normaux. Pour les modèles magnétosphériques externes OG et SG, cette zone correspond à un intervalle de saturation. Pour le modèle SG, l'interruption dans la production de photons $\gamma$ se produit à environ 10$^{35}$\,erg\,s$^{-1}$, tandis que pour le modèle OG la saturation est prédite pour $\dot{E} \sim 10^{34}$\,erg\,s$^{-1}$. Avec la statistique et les incertitudes actuelles, il n'est pas possible de discriminer ou de rejeter ces deux modèles. Signalons également le cas particulier du pulsar PSR~J0656+1414, qui a une luminosité 30 $\times$ inférieure à la valeur prédite pour une luminosité proportionnelle à $\dot{E}^{1/2}$. Une interprétation possible est l'observation de l'émission $\gamma$ de la calotte polaire du pulsar, qui est prévue pour $\zeta < 20^{\circ}$ \citep{everett01}. Il se peut que le profil pulsé constitué d'un seul pic et le spectre mou de ce pulsar, soit le résultat d'une autre zone d'émission non prédite par les modèles. Finalement, en dessous de 10$^{34}$\,erg\,s$^{-1}$, la majorité des objets sont des pulsars milliseconde dont la luminosité semble être proportionnelle à $\dot{E}$.

\fig[Diagramme $P-\dot{P}$]{scale=0.63}{c9_ppdot_distrib_v2.pdf}{Diagramme $P-\dot{P}$, distribution des pulsars en fonction de leur période de rotation et de leur dérivée. Les lignes vertes représentent l'âge caractéristique, tandis que les lignes rouges représentent le taux de perte d'énergie de rotation $\dot{E}$. Les points noirs indiquent les pulsars pour lesquels une recherche de pulsations a été effectuée en utilisant les éphémérides, tandis que les points gris indiquent les pulsars connus \citep{atnf}. Carrés bleus: (g) pulsars découverts en $\gamma$. Triangles rouges: (m) pulsars $\gamma$ milliseconde. Cercles verts: (r) autres pulsars émetteurs radio et $\gamma$. 
}{fig:ppdot2}

\subsection{Population}

La figure \ref{fig:ppdot2} présente le diagramme $P-\dot{P}$ mis à jour avec les 46 pulsars détectés avec le LAT, parmi lesquels 21 pulsars normaux radio et $\gamma$ (type r, cercle vert), 17 pulsars découverts en $\gamma$ (type g, carré bleu), et 8 pulsars $\gamma$ milliseconde (type m, triangle rouge). Les points noirs indiquent les pulsars pour lesquels une recherche de pulsations a été effectuée en utilisant les éphémérides, tandis que les points gris indiquent les pulsars connus \citep{atnf}, qui n'ont pas fait l'objet d'une recherche de signaux pulsés. Les 46 pulsars détectés sont concentrés en haut à gauche pour les pulsars normaux, avec une limite de détection inférieure à 10$^{34}$\,erg\,s$^{-1}$ pour le ralentissement $\dot{E}$ (PSR~J0357+32 et PSR~J1741-3054), et en bas à gauche pour cette nouvelle population d'émetteurs $\gamma$, les pulsars milliseconde \citep{abdo09msp}. Cependant, avec la statistique actuelle, il est encore difficile de délimiter une zone bien précise de production de photons $\gamma$ en fonction du ralentissement $\dot{E}$ et du champ magnétique au cylindre de lumière. Dans nos dernières analyses, un petit nombre de pulsars avec un $\dot{E} \sim$ 10$^{32-33}$\,erg\,s$^{-1}$ ont des significativités de 2 à 3\,$\sigma$ (notre limite de détection est définie à 5\,$\sigma$).

Rappelons que certains pulsars découverts en $\gamma$ (type g) ont été détectés en radio comme les pulsars \textit{radio faint} PSR J1741$-$2054 ($L_{1.4GHz} = 0.03$\,mJy) et PSR J2032+4127 \citep{camilo2009}. Des recherches de pulsations radio sur plusieurs autres objets comme Geminga, PSR~J0007+7303 (CTA1), PSR~J1836+5925, n'ont pas été conclues par une détection \citep{ka99,Halpern2004}. Ceci suggère que l'émission $\gamma$ a une extension plus large que le faisceau radio, comme prévu pour les modèles OG et SG/TPC. En d'autres termes, les modèles magnétosphériques externes prédisent un rapport des pulsars de type (r) sur les pulsars de type (g) beaucoup plus petit que les modèles \textit{Polar Cap} \citep{harding07}.

\subsection{Courbes de Lumière et \'Energie de Coupure}

La structure des courbes de lumière $\gamma$ ainsi que le décalage entre les pics à différentes longueurs d'onde peuvent nous aider à comprendre la géométrie et la physique de la région de l'émission. En se référant aux profils $\gamma$ des 46 pulsars \citep{abdo09psrcat}, on remarque que la majorité des pulsars montre deux pics relativement étroits, suggérant que l'on observe la caustique du pulsar à partir des bords. Quand un unique pic est observé, il a tendance à être large, suggérant une coupure tangentielle du cône d'émission. Cette description est réalisée par le modèle \textit{Outer Gap} et les modèles \textit{Slot Gap} et \textit{Two Pole Caustic} à haute altitude.

A partir des courbes de lumière des pulsars détectés, deux paramètres, dont le décalage en phase $\delta$ entre la position du pic radio principal et le premier pic $\gamma$, ainsi que l'écart $\Delta$ entre les pics $\gamma$, ont été extraits (voir la figure \ref{fig:psrcat:distribution} et le tableau \ref{tab:pulseshapepar}). Ces deux paramètres semblent corréler dans les modèles magnétosphériques externes \citep{romani95}. Cette distribution  $\delta - \Delta$ et en particulier la présence de valeurs de $\Delta \sim 0.2 - 0.3$ favorisent le modèle OG si l'on compare cela avec les prédictions des modèles OG et TPC dans \citep{watters09}, tandis que le nombre important de pulsars avec $\Delta \sim 0.4 - 0.5$ favorisent les modèles TPC.

\sfig[Distributions diverses]{pulsepar_figure.tex}{\textbf{Gauche:}~Séparation de la position en phase $\Delta$ des pics $\gamma$ en fonction du décalage en phase $\delta$ entre le principal pic radio et le premier pic $\gamma$. Les pulsars non détectés en radio sont distribués avec $\delta = 0$, tandis que les pulsars avec un simple pic sont distribués avec $\Delta = 0$. Carrés bleus: (g) pulsars découverts en $\gamma$. Triangles rouges: (m) pulsars $\gamma$ milliseconde. Cercles verts: (r) autres pulsars émetteurs radio et $\gamma$. \textbf{Droite:}~Distribution de l'énergie de coupure en fonction du champ magnétique au cylindre de lumière.}{fig:psrcat:distribution}

A présent, essayons de dégager une observation à partir des résultats spectraux. La coupure spectrale $E_{Cutoff}$ ne montre pas de corrélations avec le champ magnétique de surface (équation \ref{eq:magfieldsurface}), suggérant que l'émission $\gamma$ ne vient pas de la calotte polaire près du pôle magnétique (modèle \textit{Polar Cap}). Cependant, cette coupure spectrale semble légèrement corrélée au champ magnétique au niveau du cylindre de lumière (équation \ref{eq:magfieldcylinder}), c'est-à-dire, à une altitude haute de l'étoile (voir la figure \ref{fig:psrcat:distribution}). Les valeurs de $E_{Cutoff}$ s'étendent de 1\,GeV à 10\,GeV, et tous les types de pulsars semblent avoir la même corrélation, cela inclut les pulsars milliseconde. Les modèles magnétosphériques externes prédisent cette observation où l'émission $\gamma$ provient principalement du rayonnement de courbure des électrons. \\

\tab[Paramètres de forme des courbes de lumière]{pulsepar.tex}{Paramètres ($\delta$,$\Delta$) des courbes de lumière des 46 pulsars détectés par le LAT, estimés à partir du profil au-dessus de 100 MeV. La dernière colonne indique la définition de la région non pulsée du phasogramme. \\
$^{a}$ (r) pulsars découverts en \textbf{r}adio, (g) pulsars découverts en \textbf{g}amma, (m) pulsars \textbf{m}illiseconde \\
$^{b}$ Pour certains profils, les données actuelles ne permettent pas de discriminer entre un large pic et deux pics non-résolus (voir une discussion dans \citet{LAT6pulsars}).
}{tab:pulseshapepar}

Cette étude nous a permis d'établir une première avancé sur la population de pulsars. Nous reviendrons sur les conclusions de ce chapitre dans la conclusion générale de cette thèse.

\partie*{\textit{CONCLUSION}} 

\vspace*{40pt}

Le \textit{Large Area Telescope} à bord du satellite \textit{Fermi}, en orbite depuis le 11 juin 2008, est un instrument destiné à l'exploration du ciel des hautes énergies ($>$ 20\,MeV). Grâce à sa grande surface efficace, son pouvoir de localisation, et sa précision temporelle, le télescope \textit{Fermi} ouvre une nouvelle fenêtre pour l'étude passionnante de l'accélération de particules chargées à travers les cavités accélératrices de la magnétosphère des pulsars. Ces objets sont une étape dans l'évolution stellaire et sont d'excellents candidats pour localiser les nébuleuses à vent de pulsar et les restes de supernova, sources $\gamma$ possibles pour l'accélération des rayons cosmiques dans notre univers.  \\

Cette thèse constitue les premières observations des pulsars $\gamma$ avec le \textit{Large Area Telescope}. Nous avons mis en place une liste de candidats connus en radio ou en $X$, basée sur l'énergie de ralentissement et la distance de ces objets. Une analyse sur la recherche de signal pulsé à partir des données du LAT a été développée pour analyser les 767 éphémérides de pulsars fournies par les radiotélescopes et les télescopes $X$. Au total 22 pulsars radio relativement jeunes ont été détectés dans le domaine des hautes énergies pour les six premiers mois de la mission \textit{Fermi}. On ajoutera à cela la détection de nouvelles populations de pulsars, dont 8 pulsars milliseconde, et 16 peu ou pas émetteurs radio. Le nombre de pulsars $\gamma$ connus a donc évolué d'un ordre de grandeur depuis la fin de la mission EGRET, fournissant une nouvelle voie pour la démographie, la physique des pulsars, et la formation des étoiles à neutrons.  \\


La forme des courbes de lumières $\gamma$ des 46 pulsars détectés montre une intéressante uniformité. 75\% des profils $\gamma$ pulsés présentent deux pics séparés par plus de 0.2 en phase, tandis que les 25\% restants semblent montrer un seul et unique pic. Les spectres en énergie de l'émission pulsée peuvent être décrits par des lois de puissance avec une coupure exponentielle. Cette coupure en énergie s'étend de $\sim$ 1 à 6\,GeV, limitant actuellement l'observation de ces objets par les télescopes $\gamma$ au sol. De plus, leur luminosité à haute énergie semble occuper une grande fraction de l'énergie de rotation, et augmente à mesure que celle-ci diminue jusqu'à une limite de saturation $\dot{E} \sim 10^{33-34}$\,erg\,s$^{-1}$. Ces observations sur les profils et les spectres en énergie pulsés suggèrent que la zone d'émission $\gamma$ se trouve à haute altitude par rapport à l'étoile à neutrons et près du cylindre de lumière, même si pour certains cas particuliers, l'émission pourrait venir de la zone au-dessus des pôles de l'étoile. \\

Il s'avère qu'une large fraction de ces découvertes est corrélée à des sources telles que des PWN et/ou des SNRs observées en $X$ et/ou au TeV. Il sera ainsi essentiel de suivre les systèmes avec les instruments du TeV et des $X$ pour contraindre les différents modèles d'émission. De plus, la maîtrise du ciel $\gamma$ ouvre une nouvelle perspective pour explorer les chocs entre les vents stellaires des associations OB et des étoiles massives Wolf-Rayet, ou encore les jets des systèmes binaires $X$. Ces objets sont des candidats potentiels pour abriter un pulsar. Ils contribuent également au fond galactique pour la recherche de signaux plus subtils provenant d'autres types de sources telle que la matière noire. Enfin, cette nouvelle population de pulsars dévoile peu à peu l'origine d'une partie des sources EGRET non-identifiées à basse latitude galactique. Il semble en effet que les jeunes pulsars formés au sein des supernovae ou par accrétion de matière dominent la population des sources $\gamma$ de notre Galaxie. \\

Les résultats de cette thèse soutiennent l'importance de continuer l'étude de cette population de pulsars grandissante. Le rapport du nombre de pulsars normaux vus en radio et/ou en $\gamma$ permet de mettre des contraintes sur la géométrie des faisceaux, sur les sites d'accélération des particules, et éventuellement sur l'évolution de leurs propriétés d'émission au cours du temps. Il sera également essentiel de rechercher des pulsars millisecondes dans les données LAT. De plus, les spectres finement résolus en phase vont à présent nous permettre de définir plus en détails les modèles d'émission, tout en étudiant les profils pulsés des autres longueurs d'onde. Finalement, le débat sur l'association entre le pulsar PSR~J0205+6449 et la supernova historique nous rappelle qu'il est essentiel d'étudier les associations, d'une part pour l'évolution des systèmes stellaires, et d'autre part pour évaluer les distances de notre Galaxie.


\begin{appendix}               

\chapter{Première annexe}

Cette annexe présente un échantillon des courbes de lumière du premier catalogue de pulsars du télescope \textit{Fermi} \citep{abdo09psrcat}:
\begin{ablist} 
\item PSR J0007+7303 découvert en $\gamma$ dans le reste de supernova CTA1 \citep{abdo08}
\item le pulsar millisecond PSR J0030+0451 \citep{abdo09j0030}
\item le pulsar radio PSR J0659+1414 associé au reste de supernova \textit{Monogem ring}
\end{ablist} 

\begin{figure}
\centering
\includegraphics[scale=0.6,angle=-90]{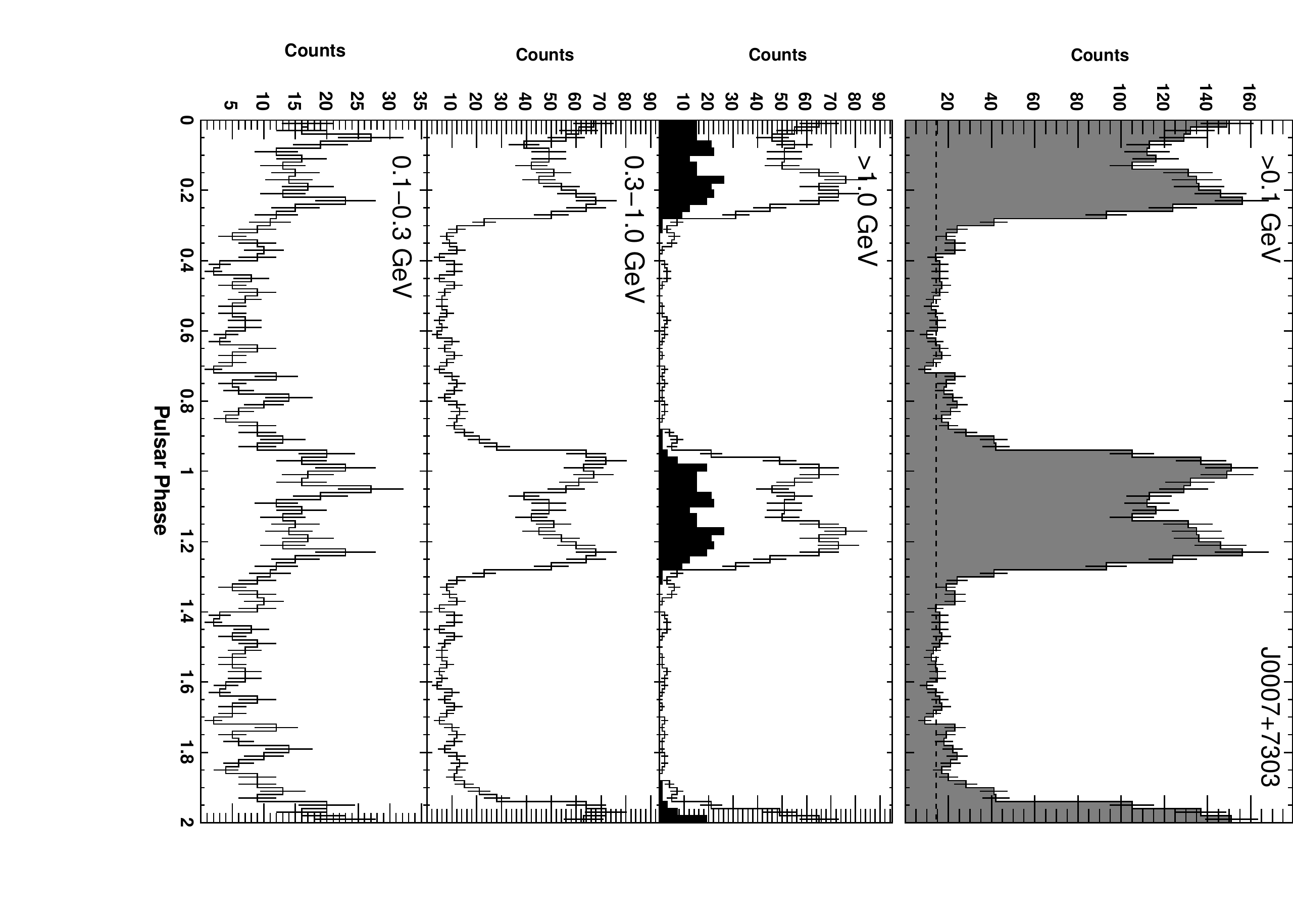}
\caption{Courbes de lumière de PSR J0007+7303 ($P=316$\,ms). 
\label{fig:J0007p7303_lightcurve}}
\end{figure}

\begin{sidewaysfigure}
\centering
\begin{minipage}[t]{0.45\linewidth}
\centering
\includegraphics[width=\linewidth]{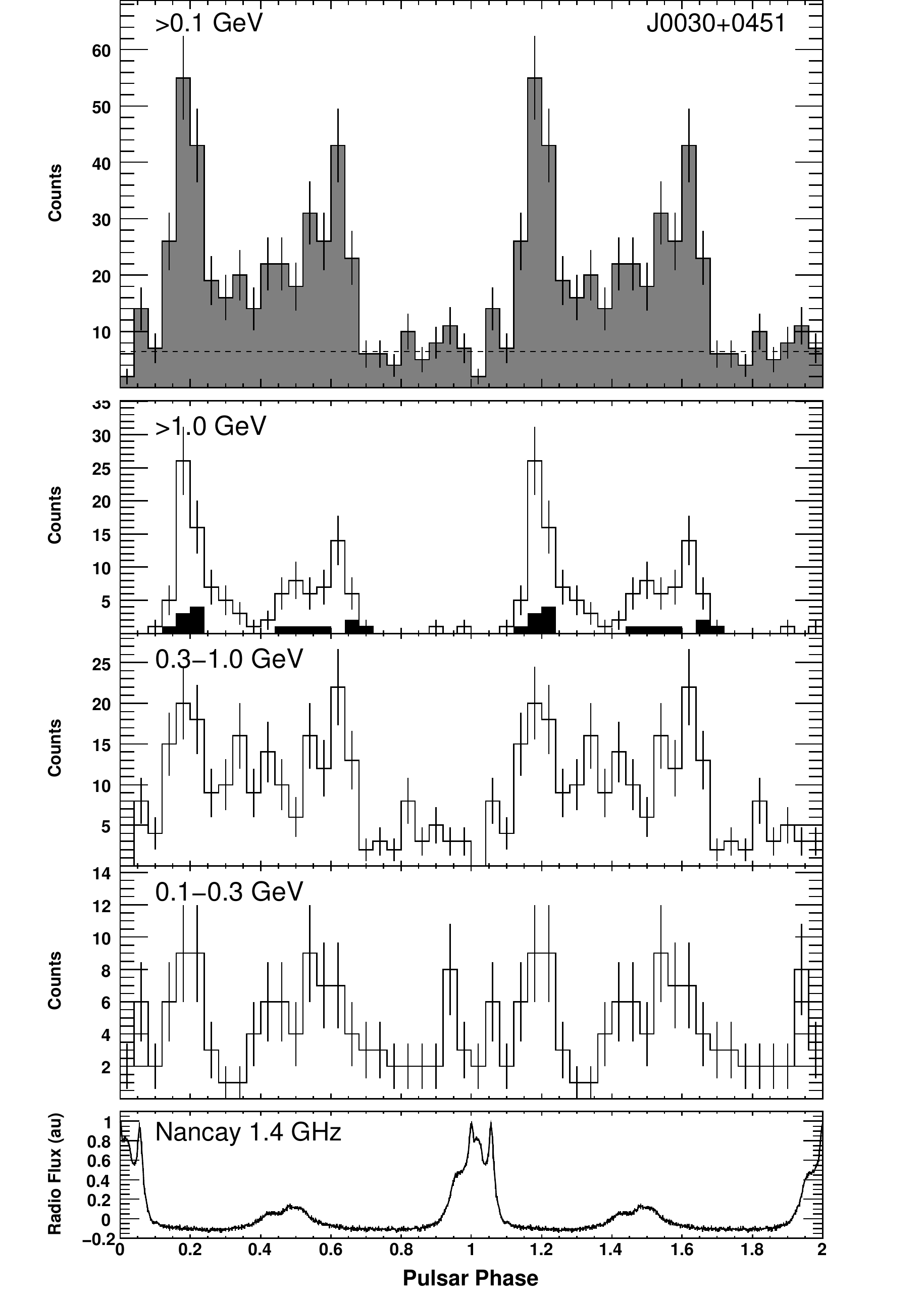}
\caption{Courbes de lumière de PSR J0030+0451 ($P=4.87$\,ms).
\label{fig:J0030p0451_lightcurve}}
\end{minipage}
\hspace{1cm}%
\begin{minipage}[t]{0.45\linewidth}
\centering
\includegraphics[width=\linewidth]{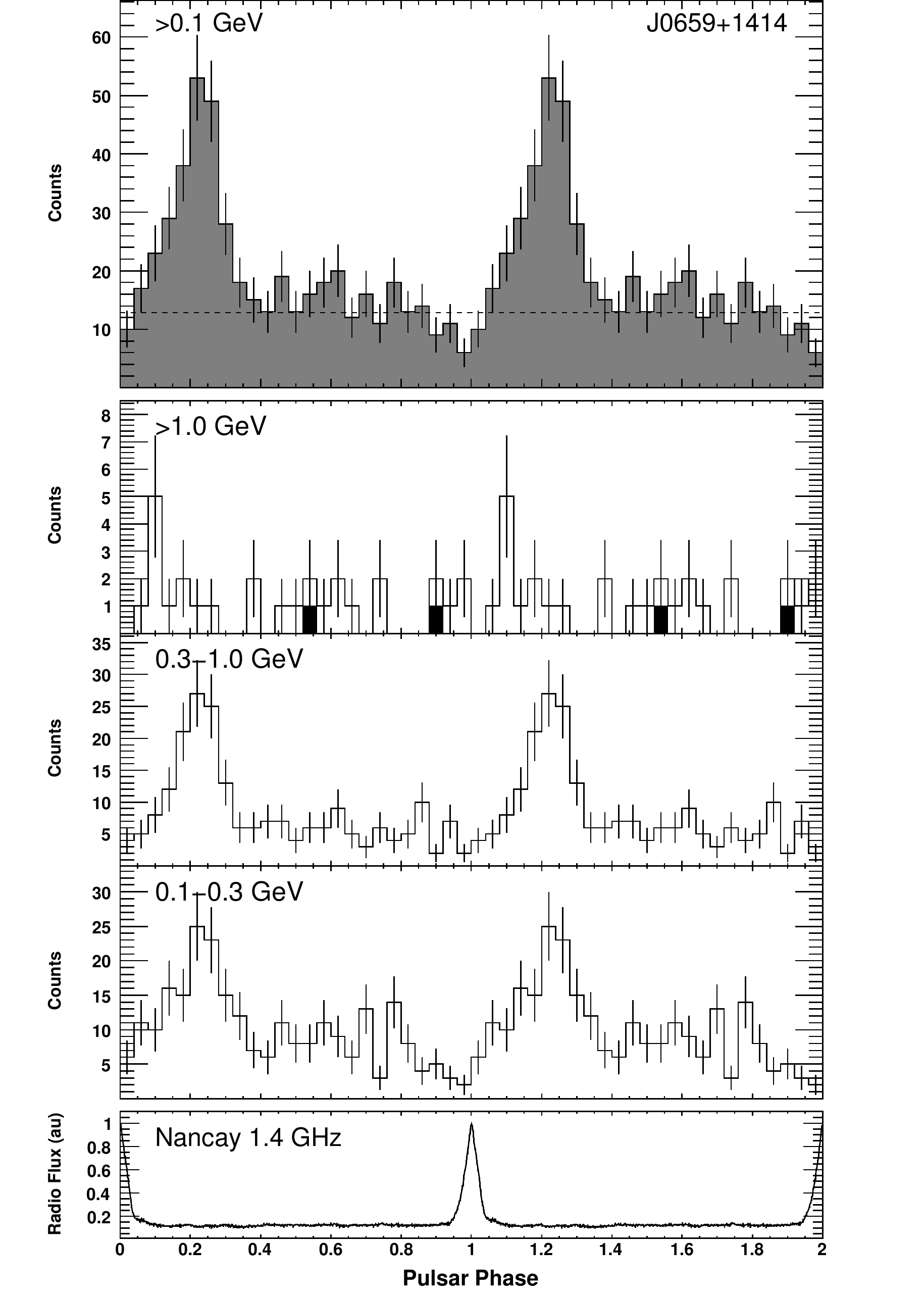}
\caption{Courbes de lumière de PSR J0659+1414 ($P=385$\,ms). 
\label{fig:J0659p1414_lightcurve}}
\end{minipage}
\end{sidewaysfigure}
\clearpage 

\end{appendix} 

\end{fmffile}         

\partie*{R\'ef\'erences}                    

\bibliographystyle{biblio/francais}   
\bibliography{main}            

\begin{thebibliography}{187}
\ProvideTextCommand{\guillemotleft}{OT1}{%
  \leavevmode\raise .27ex\hbox{$\scriptscriptstyle\ll$}}
\ProvideTextCommand{\guillemotright}{OT1}{%
  \leavevmode\raise .27ex\hbox{$\scriptscriptstyle\gg$}}
\newcommand{\enquote}[1]{\guillemotleft#1\guillemotright}
\providecommand{\natexlab}[1]{#1}
\providecommand{\bibnamefont}[1]{#1}
\providecommand{\url}[1]{\texttt{#1}}
\providecommand{\urlprefix}{URL }
\expandafter\ifx\csname urlstyle\endcsname\relax
  \providecommand{\doi}[1]{doi:\discretionary{}{}{}#1}\else
  \providecommand{\doi}{doi:\discretionary{}{}{}\begingroup
  \urlstyle{rm}\Url}\fi
\providecommand{\selectlanguage}[1]{\relax}
\input{babelbst.tex}
\newcommand{\Capitalize}[1]{\uppercase{#1}}
\newcommand{\capitalize}[1]{\expandafter\Capitalize#1}

\bibitem[{Abdo \bbletal{}(2009-calib)Abdo, Ackermann, Ajello, Ampe, Anderson,
  Atwood, Axelsson \bbland{} al.}]{abdo09calib}
\bibnamefont{Abdo}, A.~A., M.~\bibnamefont{Ackermann}, M.~\bibnamefont{Ajello},
  J.~\bibnamefont{Ampe}, B.~\bibnamefont{Anderson}, B.~\bibnamefont{Atwood},
  W., M.~\bibnamefont{Axelsson} \bbland{} \bibnamefont{al.} 2009-calib,
  \enquote{{The On-orbit Calibrations for the Fermi Large Area Telescope}},
  \emph{arXiv:0904.2226}.

\bibitem[{Abdo \bbletal{}(2009-BS)Abdo, Ackermann, Ajello, Anderson \bbland{}
  al.}]{abdo09blindsearch}
\bibnamefont{Abdo}, A.~A., M.~\bibnamefont{Ackermann}, M.~\bibnamefont{Ajello},
  B.~\bibnamefont{Anderson} \bbland{} \bibnamefont{al.} 2009-BS,
  \enquote{{Detection of 16 Gamma-Ray Pulsars Through Blind Frequency Searches
  Using the Fermi LAT}}, \emph{Science}, \bblvol{} 10.1126/science.1175558.

\bibitem[{{Abdo} \bbletal{}(2009-EGRET){Abdo}, {Ackermann}, {Ajello} \bbland{}
  {Atwood}}]{abdo09EGRETpsr}
\bibnamefont{{Abdo}}, A.~A., M.~\bibnamefont{{Ackermann}},
  M.~\bibnamefont{{Ajello}} \bbland{} B.~\bibnamefont{{Atwood}}, W. 2009-EGRET,
  \enquote{{The Fermi LAT View of Three EGRET Pulsars}}, \emph{ApJ}. - in prep.

\bibitem[{Abdo \bbletal{}(2009-geminga)Abdo, Ackermann, Ajello \bbland{}
  Atwood}]{abdo09geminga}
\bibnamefont{Abdo}, A.~A., M.~\bibnamefont{Ackermann}, M.~\bibnamefont{Ajello}
  \bbland{} B.~\bibnamefont{Atwood}, W. 2009-geminga, \enquote{{Fermi LAT
  Observations of the Geminga Pulsar}}, \emph{ApJ}.

\bibitem[{Abdo \bbletal{}(2009-J1907{\natexlab{a}})Abdo, Ackermann, Ajello
  \bbland{} Atwood}]{abdo09j2021p4026}
\bibnamefont{Abdo}, A.~A., M.~\bibnamefont{Ackermann}, M.~\bibnamefont{Ajello}
  \bbland{} B.~\bibnamefont{Atwood}, W. 2009-J1907{\natexlab{a}},
  \enquote{{Fermi LAT PSR~J2021+4026 near the gamma Cygni SNR}}, \emph{ApJ}. En
  préparation.

\bibitem[{Abdo \bbletal{}(2009-J1907{\natexlab{b}})Abdo, Ackermann, Ajello
  \bbland{} Atwood}]{abdo09j1907}
\bibnamefont{Abdo}, A.~A., M.~\bibnamefont{Ackermann}, M.~\bibnamefont{Ajello}
  \bbland{} B.~\bibnamefont{Atwood}, W. 2009-J1907{\natexlab{b}},
  \enquote{{PSR~J1907+0602 powering a bright TeV PWN}}, \emph{ApJ}. - in prep.

\bibitem[{Abdo \bbletal{}(2009-bsl)Abdo, Ackermann, Ajello, Atwood, Axelsson
  \bbland{} al.}]{abdo09bsl}
\bibnamefont{Abdo}, A.~A., M.~\bibnamefont{Ackermann}, M.~\bibnamefont{Ajello},
  B.~\bibnamefont{Atwood}, W., M.~\bibnamefont{Axelsson} \bbland{}
  \bibnamefont{al.} 2009-bsl, \enquote{{Fermi/Large Area Telescope Bright
  Gamma-Ray Source List}}, \emph{ApJS}, \bblvol{} 183, \bblp{}~46.

\bibitem[{Abdo \bbletal{}(2009-J0205)Abdo, Ackermann, Ajello, Atwood, Axelsson
  \bbland{} al.}]{abdo09j0205}
\bibnamefont{Abdo}, A.~A., M.~\bibnamefont{Ackermann}, M.~\bibnamefont{Ajello},
  B.~\bibnamefont{Atwood}, W., M.~\bibnamefont{Axelsson} \bbland{}
  \bibnamefont{al.} 2009-J0205, \enquote{{Discovery of pulsations from the
  pulsar J0205+6449 in SNR 3C~58 with the Fermi Gamma-Ray Space Telescope}},
  \emph{ApJL}, \bblvol{} 699, \bblp{} 102A.

\bibitem[{Abdo \bbletal{}(2009-LSI61)Abdo, Ackermann, Ajello, Atwood, Axelsson
  \bbland{} al.}]{abdo09lsi61}
\bibnamefont{Abdo}, A.~A., M.~\bibnamefont{Ackermann}, M.~\bibnamefont{Ajello},
  B.~\bibnamefont{Atwood}, W., M.~\bibnamefont{Axelsson} \bbland{}
  \bibnamefont{al.} 2009-LSI61, \enquote{{Fermi LAT Observations of
  LSI+61$^{\circ}$303: First Detection of an Orbital Modulation in GeV Gamma
  Rays}}, \emph{ApJL}, \bblvol{} 701, \bblp{} 123.

\bibitem[{Abdo \bbletal{}(2009-psrcat)Abdo, Ackermann, Ajello, Atwood, Axelsson
  \bbland{} al.}]{abdo09psrcat}
\bibnamefont{Abdo}, A.~A., M.~\bibnamefont{Ackermann}, M.~\bibnamefont{Ajello},
  B.~\bibnamefont{Atwood}, W., M.~\bibnamefont{Axelsson} \bbland{}
  \bibnamefont{al.} 2009-psrcat, \enquote{{The First Fermi Large Area Telescope
  Catalog of Gamma-Ray Pulsars}}, \emph{ApJ}.

\bibitem[{Abdo \bbletal{}(2009-J2021)Abdo, Ackermann, Ajello, Atwood, Baldini
  \bbland{} al.}]{abdo09j2021}
\bibnamefont{Abdo}, A.~A., M.~\bibnamefont{Ackermann}, M.~\bibnamefont{Ajello},
  B.~\bibnamefont{Atwood}, W., L.~\bibnamefont{Baldini} \bbland{}
  \bibnamefont{al.} 2009-J2021, \enquote{{Pulsed gamma-rays from PSR~J2021+3651
  with the Fermi Large Area Telescope}}, \emph{ApJ}, \bblvol{} 700, \bblp{}
  1059.

\bibitem[{Abdo \bbletal{}(2009-J2229)Abdo, Ackermann, Ajello, Atwood, Baldini
  \bbland{} al.}]{abdo09j2229}
\bibnamefont{Abdo}, A.~A., M.~\bibnamefont{Ackermann}, M.~\bibnamefont{Ajello},
  B.~\bibnamefont{Atwood}, W., L.~\bibnamefont{Baldini} \bbland{}
  \bibnamefont{al.} 2009-J2229, \enquote{{Fermi LAT detection of pulsed
  $\gamma$ rays from the Vela-like pulsars PSR~J1048-5832 and PSR~J2229+6114}},
  \emph{ApJ}, \bblvol{} 706, \bblp{} 1331.

\bibitem[{Abdo \bbletal{}(2009-Crab)Abdo, Ackermann, Atwood, Bagagli \bbland{}
  al.}]{abdo09crab}
\bibnamefont{Abdo}, A.~A., M.~\bibnamefont{Ackermann}, B.~\bibnamefont{Atwood},
  W., R.~\bibnamefont{Bagagli} \bbland{} \bibnamefont{al.} 2009-Crab,
  \enquote{{Fermi Large Area Telescope Observations of the Crab Pulsar and
  Nebula}}, \emph{ApJ}.

\bibitem[{Abdo \bbletal{}(2009-Vela)Abdo, Ackermann, Atwood, Bagagli \bbland{}
  al.}]{abdo09vela}
\bibnamefont{Abdo}, A.~A., M.~\bibnamefont{Ackermann}, B.~\bibnamefont{Atwood},
  W., R.~\bibnamefont{Bagagli} \bbland{} \bibnamefont{al.} 2009-Vela,
  \enquote{{Fermi LAT Observations of the Vela pulsar}}, \emph{ApJ}, \bblvol{}
  696, \bblp{} 1084.

\bibitem[{Abdo \bbletal{}(2008)Abdo, Ackermann, Atwood, Baldini \bbland{}
  al.}]{abdo08}
\bibnamefont{Abdo}, A.~A., M.~\bibnamefont{Ackermann}, B.~\bibnamefont{Atwood},
  W., L.~\bibnamefont{Baldini} \bbland{} \bibnamefont{al.} 2008, \enquote{{The
  Fermi Large Area Telescope discovers the pulsar in the young galactic
  supernova remnant}}, \emph{Science}, \bblvol{} 322, \bblp{} 1218.

\bibitem[{Abdo \bbletal{}(2009-J1028)Abdo, Ackermann, Atwood, Baldini \bbland{}
  Ballet}]{abdo09j1028}
\bibnamefont{Abdo}, A.~A., M.~\bibnamefont{Ackermann}, B.~\bibnamefont{Atwood},
  W., L.~\bibnamefont{Baldini} \bbland{} J.~a. \bibnamefont{Ballet}.
  2009-J1028, \enquote{{Discovery of Pulsed Gamma Rays from the Young Radio
  Pulsar PSR~J1028$-$5819 with the Fermi Large Area Telescope}}, \emph{ApJL},
  \bblvol{} 695, \bblp{}~72.

\bibitem[{{Abdo} \bbletal{}(2009{\natexlab{a}}){Abdo}, {Ackermann}, {Atwood},
  {Baldini}, {Ballet}, {Barbiellini}, {Baring}, {Bastieri} \bbland{}
  al.}]{abdo09msp}
\bibnamefont{{Abdo}}, A.~A., M.~\bibnamefont{{Ackermann}}, W.~B.
  \bibnamefont{{Atwood}}, L.~\bibnamefont{{Baldini}},
  J.~\bibnamefont{{Ballet}}, G.~\bibnamefont{{Barbiellini}}, M.~G.
  \bibnamefont{{Baring}}, D.~\bibnamefont{{Bastieri}} \bbland{}
  \bibnamefont{al.} 2009{\natexlab{a}}, \enquote{{A Population of Gamma-Ray
  Millisecond Pulsars Seen with the Fermi Large Area Telescope}},
  \emph{Science}, \bblvol{} 325, \bblp{} 848.

\bibitem[{{Abdo} \bbletal{}(2009{\natexlab{b}}){Abdo}, {Ackermann}, {Atwood},
  {Baldini}, {Ballet}, {Barbiellini}, {Baring}, {Bastieri} \bbland{}
  al.}]{abdo09j0030}
\bibnamefont{{Abdo}}, A.~A., M.~\bibnamefont{{Ackermann}}, W.~B.
  \bibnamefont{{Atwood}}, L.~\bibnamefont{{Baldini}},
  J.~\bibnamefont{{Ballet}}, G.~\bibnamefont{{Barbiellini}}, M.~G.
  \bibnamefont{{Baring}}, D.~\bibnamefont{{Bastieri}} \bbland{}
  \bibnamefont{al.} 2009{\natexlab{b}}, \enquote{{Pulsed Gamma Rays from the
  Millisecond Pulsar J0030+0451 with the Fermi Large Area Telescope}},
  \emph{ApJ}, \bblvol{} 699, \bblpp{} 1171--1177.

\bibitem[{Abdo \bbletal{}(2009-mil)Abdo, Allen, Aune, Berley, Chen,
  Christopher, DeYoung \bbland{} al.}]{abdo09milagro}
\bibnamefont{Abdo}, A.~A., B.~T. \bibnamefont{Allen}, T.~\bibnamefont{Aune},
  D.~\bibnamefont{Berley}, C.~\bibnamefont{Chen}, G.~E.
  \bibnamefont{Christopher}, T.~\bibnamefont{DeYoung} \bbland{}
  \bibnamefont{al.} 2009-mil, \enquote{{Milagro Observations of Multi-TeV
  Emission from Galactic Sources in the Fermi Bright Source List}}, \emph{ApJ},
  \bblvol{} 700, \bblp{} 127.

\bibitem[{Agostinelli \bbletal{}(2003)Agostinelli, Allison, Amako, Apostolakis
  \bbland{} al.}]{agostinelli03}
\bibnamefont{Agostinelli}, S., J.~\bibnamefont{Allison},
  K.~\bibnamefont{Amako}, J.~\bibnamefont{Apostolakis} \bbland{}
  \bibnamefont{al.} 2003, \enquote{{Geant 4 -- a Simulation Toolkit}},
  \emph{Nuclear Instruments and Methods in Physics Reasearch Section A},
  \bblvol{} 506, \bblp{} 250.

\bibitem[{{Aharonian} \bbletal{}(2009){Aharonian}, {Akhperjanian}, {Anton},
  {Barres de Almeida}, {Bazer-Bachi} \bbland{} al.}]{HESS1908_2009}
\bibnamefont{{Aharonian}}, F., A.~G. \bibnamefont{{Akhperjanian}},
  G.~\bibnamefont{{Anton}}, U.~\bibnamefont{{Barres de Almeida}}, A.~R.
  \bibnamefont{{Bazer-Bachi}} \bbland{} \bibnamefont{al.} 2009,
  \enquote{{Detection of very high energy radiation from HESS J1908+063
  confirms the Milagro unidentified source MGRO J1908+06}}, \emph{A\&A},
  \bblvol{} 499, \bblpp{} 723--728.

\bibitem[{{Aharonian} \bbletal{}(2007){Aharonian}, {Akhperjanian},
  {Bazer-Bachi}, {Behera}, {Beilicke}, {Benbow}, {Berge} \bbland{}
  al.}]{HESS1718_2007}
\bibnamefont{{Aharonian}}, F., A.~G. \bibnamefont{{Akhperjanian}}, A.~R.
  \bibnamefont{{Bazer-Bachi}}, B.~\bibnamefont{{Behera}},
  M.~\bibnamefont{{Beilicke}}, W.~\bibnamefont{{Benbow}},
  D.~\bibnamefont{{Berge}} \bbland{} \bibnamefont{al.} 2007,
  \enquote{{Discovery of two candidate pulsar wind nebulae in very-high-energy
  gamma rays}}, \emph{A\&A}, \bblvol{} 472, \bblpp{} 489--495.

\bibitem[{{Aharonian} \bbletal{}(2006{\natexlab{a}}){Aharonian},
  {Akhperjanian}, {Bazer-Bachi}, {Beilicke}, {Benbow}, {Berge} \bbland{}
  al.}]{HESSRabbit2006}
\bibnamefont{{Aharonian}}, F., A.~G. \bibnamefont{{Akhperjanian}}, A.~R.
  \bibnamefont{{Bazer-Bachi}}, M.~\bibnamefont{{Beilicke}},
  W.~\bibnamefont{{Benbow}}, D.~\bibnamefont{{Berge}} \bbland{}
  \bibnamefont{al.} 2006{\natexlab{a}}, \enquote{{Discovery of the two
  ''wings'' of the Kookaburra complex in VHE {$\gamma$}-rays with HESS}},
  \emph{A\&A}, \bblvol{} 456, \bblpp{} 245--251.

\bibitem[{{Aharonian} \bbletal{}(2006{\natexlab{b}}){Aharonian},
  {Akhperjanian}, {Bazer-Bachi}, {Beilicke}, {Benbow}, {Berge} \bbland{}
  al.}]{HESSVela2006}
\bibnamefont{{Aharonian}}, F., A.~G. \bibnamefont{{Akhperjanian}}, A.~R.
  \bibnamefont{{Bazer-Bachi}}, M.~\bibnamefont{{Beilicke}},
  W.~\bibnamefont{{Benbow}}, D.~\bibnamefont{{Berge}} \bbland{}
  \bibnamefont{al.} 2006{\natexlab{b}}, \enquote{{First detection of a VHE
  gamma-ray spectral maximum from a cosmic source: HESS discovery of the Vela X
  nebula}}, \emph{A\&A}, \bblvol{} 448, \bblpp{} L43--L47.

\bibitem[{{Aharonian} \bbletal{}(2006{\natexlab{c}}){Aharonian},
  {Akhperjanian}, {Bazer-Bachi}, {Beilicke}, {Benbow}, {Berge} \bbland{}
  al.}]{HESSCrab2006}
\bibnamefont{{Aharonian}}, F., A.~G. \bibnamefont{{Akhperjanian}}, A.~R.
  \bibnamefont{{Bazer-Bachi}}, M.~\bibnamefont{{Beilicke}},
  W.~\bibnamefont{{Benbow}}, D.~\bibnamefont{{Berge}} \bbland{}
  \bibnamefont{al.} 2006{\natexlab{c}}, \enquote{{Observations of the Crab
  nebula with HESS}}, \emph{A\&A}, \bblvol{} 457, \bblpp{} 899--915.

\bibitem[{Aliu(2008)}]{aliu08}
\bibnamefont{Aliu}, E. 2008, \enquote{{Search for VHE $\gamma$-ray emission in
  the vicinity of selected pulsars of the Northern Sky with VERITAS}},
  \emph{arXiv:0812.1415v1}.

\bibitem[{Anderson \bbland{} Itoh(1975)}]{anderson75}
\bibnamefont{Anderson}, P.~W. \bbland{} N.~\bibnamefont{Itoh}. 1975,
  \enquote{{Pulsar glitches and restlessness as a hard superfluidity
  phenomenon}}, \emph{Nature}, \bblvol{} 256, \bblp{}~25.

\bibitem[{Arons(1983)}]{arons83}
\bibnamefont{Arons}, J. 1983, \enquote{{Electron positron pairs in radio
  pulsars}}, \emph{AIPC}, \bblvol{} 101, \bblp{} 163.

\bibitem[{Arons(1996)}]{arons96}
\bibnamefont{Arons}, J. 1996, \enquote{{Pulsars as gamma ray sources}},
  \emph{A\&AS}, \bblvol{} 120, \bblp{}~49.

\bibitem[{Atwood \bbletal{}(2009)Atwood, Abdo, Ackermann, Althouse \bbland{}
  al.}]{atwood09}
\bibnamefont{Atwood}, W.~B., A.~A. \bibnamefont{Abdo},
  M.~\bibnamefont{Ackermann}, W.~\bibnamefont{Althouse} \bbland{}
  \bibnamefont{al.} 2009, \enquote{{The Large Area Telescope on the Fermi
  Gamma-Ray Space Telescope Mission}}, \emph{ApJ}, \bblvol{} 697, \bblp{}~6.

\bibitem[{{Atwood} \bbletal{}(2007){Atwood}, {Bagagli}, {Baldini},
  {Bellazzini}, {Barbiellini}, {Belli}, {Borden}, {Brez} \bbland{}
  al.}]{atwood07}
\bibnamefont{{Atwood}}, W.~B., R.~\bibnamefont{{Bagagli}},
  L.~\bibnamefont{{Baldini}}, R.~\bibnamefont{{Bellazzini}},
  G.~\bibnamefont{{Barbiellini}}, F.~\bibnamefont{{Belli}},
  T.~\bibnamefont{{Borden}}, A.~\bibnamefont{{Brez}} \bbland{}
  \bibnamefont{al.} 2007, \enquote{{Design and initial tests of the
  Tracker-converter of the Gamma-ray Large Area Space Telescope}},
  \emph{Astroparticle Physics}, \bblvol{}~28, \bblpp{} 422--434.

\bibitem[{{Atwood} \bbletal{}(2006){Atwood}, {Ziegler}, {Johnson} \bbland{}
  {Baughman}}]{atwood06}
\bibnamefont{{Atwood}}, W.~B., M.~\bibnamefont{{Ziegler}}, R.~P.
  \bibnamefont{{Johnson}} \bbland{} B.~M. \bibnamefont{{Baughman}}. 2006,
  \enquote{{A Time-differencing Technique for Detecting Radio-quiet Gamma-Ray
  Pulsars}}, \emph{ApJ}, \bblvol{} 652, \bblp{}~49.

\bibitem[{Baade \bbland{} Zwicky(1934)}]{baade34}
\bibnamefont{Baade}, W. \bbland{} F.~\bibnamefont{Zwicky}. 1934, \enquote{{On
  supernovae}}, \emph{PNAS}, \bblvol{}~20, \bblp{} 254.

\bibitem[{Backer \bbletal{}(1993)Backer, Hama, Van~Hook \bbland{}
  Foster}]{backer93}
\bibnamefont{Backer}, D.~C., S.~\bibnamefont{Hama}, S.~\bibnamefont{Van~Hook}
  \bbland{} R.~S. \bibnamefont{Foster}. 1993, \enquote{{Temporal Variations of
  Pulsar Dispersion Measures}}, \emph{ApJ}, \bblvol{} 404, \bblp{} 636.

\bibitem[{Backer \bbletal{}(1982)Backer, Kulkarni, Heiles, Davis \bbland{}
  Goss}]{backer82}
\bibnamefont{Backer}, D.~C., S.~R. \bibnamefont{Kulkarni},
  C.~\bibnamefont{Heiles}, M.~M. \bibnamefont{Davis} \bbland{} W.~M.
  \bibnamefont{Goss}. 1982, \enquote{{A millisecond pulsar}}, \emph{Nature},
  \bblvol{} 300, \bblp{} 615.

\bibitem[{Baldini(2007)}]{baldini07}
\bibnamefont{Baldini}, L. 2007, \enquote{{Preliminary results of the LAT
  Calibration Unit beam tests, The First Glast Symposium}}, \emph{AIP Conf.
  Proc.}, \bblvol{} 921.

\bibitem[{Becker \bbletal{}(1982)Becker, Helfand \bbland{}
  Szymkowiak}]{becker82}
\bibnamefont{Becker}, R.~H., D.~J. \bibnamefont{Helfand} \bbland{} A.~E.
  \bibnamefont{Szymkowiak}. 1982, \enquote{{An X-ray study of two Crablike
  supernova remnants - 3C 58 and CTB 80}}, \emph{ApJ}, \bblvol{} 255, \bblp{}
  557.

\bibitem[{Bignami \bbland{} Caraveo(1992)}]{bignami92}
\bibnamefont{Bignami}, G.~F. \bbland{} P.~A. \bibnamefont{Caraveo}. 1992,
  \enquote{{Geminga, New period old gamma-rays}}, \emph{Nature}, \bblvol{} 357,
  \bblp{} 287.

\bibitem[{Bisnovatyi-Kogan \bbland{} Komberg(1974)}]{bisnovatyi74}
\bibnamefont{Bisnovatyi-Kogan}, G.~S. \bbland{} B.~V. \bibnamefont{Komberg}.
  1974, \enquote{{Pulsars and close binary systems}}, \emph{Sov.Astron.},
  \bblvol{}~18, \bblp{} 217.

\bibitem[{Bradt \bbletal{}(1969)Bradt, Rappaport, Mayer, Nather, Warner,
  Macfarlane \bbland{} Kristian}]{bradt69}
\bibnamefont{Bradt}, H., S.~\bibnamefont{Rappaport}, W.~\bibnamefont{Mayer},
  R.~E. \bibnamefont{Nather}, B.~\bibnamefont{Warner},
  M.~\bibnamefont{Macfarlane} \bbland{} J.~\bibnamefont{Kristian}. 1969,
  \enquote{{X-Ray and Optical Observations of the Pulsar NP 0532 in the Crab
  Nebula}}, \emph{Nature}, \bblvol{} 222, \bblp{} 728.

\bibitem[{Bregeon(2005)}]{bregeon05}
\bibnamefont{Bregeon}, J. 2005, \emph{{Contribution à l'étalonnage en énergie
  du calorimètre du GLAST-LAT et qualification des modèles de cascades
  hadroniques disponibles sous GEANT4}}, Université de Bordeaux.

\bibitem[{Breiman \bbletal{}(1984)Breiman, Friedman, Stone \bbland{}
  Olshen}]{breiman84}
\bibnamefont{Breiman}, L., J.~\bibnamefont{Friedman}, C.~J. \bibnamefont{Stone}
  \bbland{} R.~A. \bibnamefont{Olshen}. 1984, \enquote{{Classification and
  Regression Trees}}, \emph{Wadsworth International Group, Belmont, CA}.

\bibitem[{Buccheri \bbletal{}(1983)Buccheri, Bennett, Bignami, Bloemen,
  Boriakoff \bbland{} al.}]{buccheri83}
\bibnamefont{Buccheri}, R., K.~\bibnamefont{Bennett}, G.~F.
  \bibnamefont{Bignami}, J.~B.~G.~M. \bibnamefont{Bloemen},
  V.~\bibnamefont{Boriakoff} \bbland{} \bibnamefont{al.} 1983, \enquote{{Search
  for pulsed gamma-ray emission from radio pulsars in the COS-B data}},
  \emph{A\&A}, \bblvol{} 128, \bblp{} 245.

\bibitem[{Camilo \bbletal{}(2009)Camilo, Abdo \bbland{} al.}]{camilo09j1833}
\bibnamefont{Camilo}, F., A.~A. \bibnamefont{Abdo} \bbland{} \bibnamefont{al.}
  2009, \enquote{{Pulsed Gamma-rays from the Young Pulsars PSR~J1833-1034 and
  PSR~J1747-2958}}, \emph{ApJ}. - in prep.

\bibitem[{Camilo \bbland{} Rasio(2005)}]{camilo05}
\bibnamefont{Camilo}, F. \bbland{} F.~A. \bibnamefont{Rasio}. 2005,
  \enquote{{Pulsars in Globular Clusters}}, \emph{ASP Conf. Ser.}, \bblvol{}
  328, \bblp{} 147.

\bibitem[{{Camilo} \bbletal{}(2009){Camilo}, {Ray}, {Ransom}, {Burgay},
  {Johnson}, {Kerr}, {Gotthelf}, {Halpern} \bbland{} al.}]{camilo2009}
\bibnamefont{{Camilo}}, F., P.~S. \bibnamefont{{Ray}}, S.~M.
  \bibnamefont{{Ransom}}, M.~\bibnamefont{{Burgay}}, T.~J.
  \bibnamefont{{Johnson}}, M.~\bibnamefont{{Kerr}}, E.~V.
  \bibnamefont{{Gotthelf}}, J.~P. \bibnamefont{{Halpern}} \bbland{}
  \bibnamefont{al.} 2009, \enquote{{Radio Detection of LAT PSRs J1741-2054 and
  J2032+4127: No Longer Just Gamma-Ray Pulsars}}, \emph{ApJ}. - submitted.

\bibitem[{Camilo \bbletal{}(2002)Camilo, Stairs, Lorimer, Backer, Ransom,
  Klein, Wielebinski, Kramer, McLaughlin, Arzoumanian \bbland{}
  Muller}]{camilo02}
\bibnamefont{Camilo}, F., I.~H. \bibnamefont{Stairs}, D.~R.
  \bibnamefont{Lorimer}, D.~C. \bibnamefont{Backer}, S.~M.
  \bibnamefont{Ransom}, B.~\bibnamefont{Klein}, R.~\bibnamefont{Wielebinski},
  M.~\bibnamefont{Kramer}, M.~A. \bibnamefont{McLaughlin},
  Z.~\bibnamefont{Arzoumanian} \bbland{} P.~\bibnamefont{Muller}. 2002,
  \enquote{{Discovery of Radio Pulsations from the X-Ray Pulsar J0205+6449 in
  Supernova Remnant 3C~58 with the Green Bank Telescope}}, \emph{ApJ},
  \bblvol{} 571, \bblp{} L41.

\bibitem[{Cappellaro \bbland{} Turatto(2000)}]{cappellaro00}
\bibnamefont{Cappellaro}, E. \bbland{} M.~\bibnamefont{Turatto}. 2000,
  \enquote{{Supernova types and rates}}, \emph{astro-ph/0012455}.

\bibitem[{Casandjian \bbland{} Grenier(2008)}]{casandjian08}
\bibnamefont{Casandjian}, J.-M. \bbland{} I.~A. \bibnamefont{Grenier}. 2008,
  \enquote{{A revised catalogue of EGRET $\gamma$-ray sources}}, \emph{A\&A},
  \bblvol{} 489, \bblp{} 849.

\bibitem[{Caswell(1970)}]{caswell70}
\bibnamefont{Caswell}, J.~L. 1970, \enquote{{The Frequency of Supernovae in our
  Galaxy, Estimated from Supernova Remnants Detected at 178 MHz}}, \emph{A\&A},
  \bblvol{}~7, \bblp{}~59.

\bibitem[{Chandrasekhar(1935)}]{chandrasekhar35}
\bibnamefont{Chandrasekhar}, S. 1935, \enquote{{The highly collapsed
  configurations of a stellar mass}}, \emph{MNRAS}, \bblvol{}~95, \bblp{} 207.

\bibitem[{Cheng \bbletal{}(1986)Cheng, Ho \bbland{} Ruderman}]{cheng86}
\bibnamefont{Cheng}, K.~S., C.~\bibnamefont{Ho} \bbland{}
  M.~\bibnamefont{Ruderman}. 1986, \enquote{{Energetic radiation from rapidly
  spinning pulsars. I - Outer magnetosphere gaps. II - VELA and Crab}},
  \emph{ApJ}, \bblvol{} 300, \bblp{} 500.

\bibitem[{Chevalier(2005)}]{chevalier05}
\bibnamefont{Chevalier}, R.~A. 2005, \enquote{{Young Core-Collapse Supernova
  Remnants and Their Supernovae}}, \emph{ApJ}, \bblvol{} 619, \bblp{} 839.

\bibitem[{Cocke \bbletal{}(1969)Cocke, Disney \bbland{} Taylor}]{cocke69}
\bibnamefont{Cocke}, W.~J., M.~J. \bibnamefont{Disney} \bbland{} D.~J.
  \bibnamefont{Taylor}. 1969, \enquote{{Discovery of optical signals from
  pulsar NP 0532}}, \emph{Nature}, \bblvol{} 221, \bblp{} 525.

\bibitem[{{Cognard} \bbletal{}(2009){Cognard}, {Theureau} \bbland{}
  al.}]{cognard09}
\bibnamefont{{Cognard}}, I., G.~\bibnamefont{{Theureau}} \bbland{}
  \bibnamefont{al.} 2009, \enquote{{PSR~J0248+6021, a young pulsar in the
  northern Galactic plane}}, \emph{A\&A}. - in prep.

\bibitem[{Cordes \bbland{} Lazio(2002)}]{cordes02}
\bibnamefont{Cordes}, J.~M. \bbland{} T.~J.~W. \bibnamefont{Lazio}. 2002,
  \enquote{{NE2001.I. A New Model for the Galactic Distribution of Free
  Electrons and its Fluctuations}}, \emph{arXiv:astro-ph/0207156v3}.

\bibitem[{Daugherty \bbland{} Harding(1986)}]{daugherty86}
\bibnamefont{Daugherty}, J.~K. \bbland{} A.~K. \bibnamefont{Harding}. 1986,
  \enquote{{Compton scattering in strong magnetic fields}}, \emph{ApJ},
  \bblvol{} 309, \bblp{} 362.

\bibitem[{D'Avezac(2006)}]{davezac06}
\bibnamefont{D'Avezac}, P. 2006, \emph{{Contribution à la calorimétrie du
  télescope spatial à rayon $\gamma$ GLAST et étude des cascades
  électron-photon sur le fond diffus extragalactique}}, Ecole Polytechnique.

\bibitem[{{De Becker} \bbletal{}(2005){De Becker}, {Rauw} \bbland{}
  {Swings}}]{DeBecker2005}
\bibnamefont{{De Becker}}, M., G.~\bibnamefont{{Rauw}} \bbland{} J.-P.
  \bibnamefont{{Swings}}. 2005, \enquote{{On the Multiplicity of the O-Star Cyg
  OB2 \#8a and its Contribution to the {$\gamma$}-ray Source 3EG J2033+4118}},
  \emph{{Ap\&SS}}, \bblvol{} 297, \bblpp{} 291--298.

\bibitem[{De~Jager \bbletal{}(1989)De~Jager, Raubenheimer \bbland{}
  Swanepoel}]{dejager89}
\bibnamefont{De~Jager}, O.~C., B.~C. \bibnamefont{Raubenheimer} \bbland{}
  J.~W.~H. \bibnamefont{Swanepoel}. 1989, \enquote{{A poweful test for weak
  periodic signals with unknown light curve shape in sparse data}},
  \emph{A\&A}, \bblvol{} 221, \bblp{} 180.

\bibitem[{{Djannati-Ata{\"i}} \bbletal{}(2007){Djannati-Ata{\"i}}, {De Jager},
  {Terrier}, {Gallant} \bbland{} {Hoppe}}]{HESS1833_2007}
\bibnamefont{{Djannati-Ata{\"i}}}, A., O.~C. \bibnamefont{{De Jager}},
  R.~\bibnamefont{{Terrier}}, Y.~A. \bibnamefont{{Gallant}} \bbland{}
  S.~\bibnamefont{{Hoppe}}. 2007, \enquote{{New Companions for the lonely Crab?
  VHE emission from young pulsar wind nebulae revealed by H.E.S.S}},
  \emph{arXiv:0710.2247}.

\bibitem[{Dowd \bbletal{}(2000)Dowd, Sisk \bbland{} Hagen}]{dowd00}
\bibnamefont{Dowd}, A., W.~\bibnamefont{Sisk} \bbland{} J.~\bibnamefont{Hagen}.
  2000, \enquote{{WAPP --- Wideband Arecibo Pulsar Processor}}, \emph{ASP Conf.
  Ser.}, \bblvol{} 202, \bblp{} 275.

\bibitem[{Dyks \bbland{} Rudak(2003)}]{dyks03}
\bibnamefont{Dyks}, J. \bbland{} B.~\bibnamefont{Rudak}. 2003,
  \enquote{{Two-Pole Caustic Model for High-Energy Light Curves of Pulsars}},
  \emph{ApJ}, \bblvol{} 598, \bblp{} 1201.

\bibitem[{Everett \bbland{} Weisberg(2001)}]{everett01}
\bibnamefont{Everett}, J.~E. \bbland{} J.~M. \bibnamefont{Weisberg}. 2001,
  \enquote{{Emission Beam Geometry of Selected Pulsars Derived from Average
  Pulse Polarization Data}}, \emph{ApJ}, \bblvol{} 553, \bblp{} 341.

\bibitem[{Fich \bbletal{}(1989)Fich, Blitz \bbland{} Stark}]{fich89}
\bibnamefont{Fich}, M., L.~\bibnamefont{Blitz} \bbland{} A.~A.
  \bibnamefont{Stark}. 1989, \enquote{{The rotation curve of the Milky Way to 2
  R(0)}}, \emph{ApJ}, \bblvol{} 342, \bblp{} 272.

\bibitem[{Fierro(1995)}]{fierro95}
\bibnamefont{Fierro}, J.~M. 1995, \enquote{{Observations of Spin-Powered
  Pulsars with the EGRET Gamma Ray Telescope}}, \emph{Thesis, Stanford
  University}.

\bibitem[{Fritz \bbletal{}(1969)Fritz, Henry, Meekins, Chubb \bbland{}
  Friedmann}]{fritz69}
\bibnamefont{Fritz}, G., R.~C. \bibnamefont{Henry}, J.~F.
  \bibnamefont{Meekins}, T.~A. \bibnamefont{Chubb} \bbland{}
  H.~\bibnamefont{Friedmann}. 1969, \enquote{{X-ray pulsar in the Crab
  Nebula}}, \emph{Science}, \bblvol{} 164, \bblp{} 709.

\bibitem[{Gold(1968)}]{gold68}
\bibnamefont{Gold}, T. 1968, \enquote{{Rotating Neutron Stars as the Origin of
  the Pulsating Radio Sources}}, \emph{Nature}, \bblvol{} 218, \bblp{} 731.

\bibitem[{Goldreich \bbland{} Julian(1969)}]{goldreich69}
\bibnamefont{Goldreich}, P. \bbland{} W.~H. \bibnamefont{Julian}. 1969,
  \enquote{{Pulsar Electrodynamics}}, \emph{ApJ}, \bblvol{} 157, \bblp{} 869.

\bibitem[{Gonzalez \bbletal{}(2006)Gonzalez, Kaspi, Pivovaroff \bbland{}
  Gaensler}]{gonzalez06}
\bibnamefont{Gonzalez}, M.~E., V.~M. \bibnamefont{Kaspi}, M.~J.
  \bibnamefont{Pivovaroff} \bbland{} B.~M. \bibnamefont{Gaensler}. 2006,
  \enquote{{Chandra and XMM-Newton Observations of the Vela-like Pulsar
  B1046$-$58}}, \emph{ApJ}, \bblvol{} 652, \bblp{} 569.

\bibitem[{{Goodman} \bbland{} {Sinnis}(2009)}]{ATEL2172}
\bibnamefont{{Goodman}}, J. \bbland{} G.~\bibnamefont{{Sinnis}}. 2009,
  \enquote{{New Multi-TeV Gamma-Ray Sources MGRO J0632+17 and MGRO J2228+61}},
  \emph{The Astronomer's Telegram}, \bblvol{} 2172, \bblpp{} 1--+.

\bibitem[{Gourgoulhon(2004\ --\ 2005)}]{gourgoulhon05}
\bibnamefont{Gourgoulhon}, E. 2004\ --\ 2005, \emph{{Objets Compacts}}, Master
  Sciences de l'Univers et Technologies Spatiales.

\bibitem[{{Green}(2009)}]{Green2009}
\bibnamefont{{Green}}, D.~A. 2009, \enquote{{A revised Galactic supernova
  remnant catalogue}}, \emph{Bulletin of the Astronomical Society of India},
  \bblvol{}~37, \bblp{}~45.

\bibitem[{Green \bbland{} Gull(1982)}]{green82}
\bibnamefont{Green}, D.~A. \bbland{} S.~F. \bibnamefont{Gull}. 1982,
  \enquote{{Distance to Crab-like supernova remnant 3C58}}, \emph{Nature},
  \bblvol{} 299, \bblp{} 606.

\bibitem[{Guillemot(2009)}]{guillemot09}
\bibnamefont{Guillemot}, L. 2009, \emph{{Détections de pulsars milliseconde
  avec le FERMI Large Area Telescope}}, Université de Bordeaux.

\bibitem[{Guillemot \bbland{} Parent(2007)}]{jrjc2007}
\bibnamefont{Guillemot}, L. \bbland{} D.~\bibnamefont{Parent}. 2007, \emph{{Les
  Pulsars Gamma avec GLAST}}, Journées de Rencontres Jeunes Chercheurs (JRJC).

\bibitem[{Halpern \bbletal{}(2008)Halpern, Camilo, Giuliani, Gotthelf,
  McLaughlin, Mukherjee, Pellizzoni, Ransom, Roberts \bbland{}
  Tavani}]{halpern08}
\bibnamefont{Halpern}, J.~P., F.~\bibnamefont{Camilo},
  A.~\bibnamefont{Giuliani}, E.~V. \bibnamefont{Gotthelf}, M.~A.
  \bibnamefont{McLaughlin}, R.~\bibnamefont{Mukherjee},
  A.~\bibnamefont{Pellizzoni}, S.~M. \bibnamefont{Ransom}, M.~S.~E.
  \bibnamefont{Roberts} \bbland{} M.~\bibnamefont{Tavani}. 2008,
  \enquote{{Discovery of High-Energy Gamma-Ray Pulsations from PSR J2021+3651
  with AGILE}}, \emph{ApJ}, \bblvol{} 688, \bblp{} L33.

\bibitem[{{Halpern} \bbletal{}(2007){Halpern}, {Camilo} \bbland{}
  {Gotthelf}}]{Halpern2007}
\bibnamefont{{Halpern}}, J.~P., F.~\bibnamefont{{Camilo}} \bbland{} E.~V.
  \bibnamefont{{Gotthelf}}. 2007, \enquote{{The Next Geminga: Search for Radio
  and X-Ray Pulsations from the Neutron Star Identified with 3EG J1835+5918}},
  \emph{ApJ}, \bblvol{} 668, \bblpp{} 1154--1157.

\bibitem[{Halpern \bbletal{}(2001b)Halpern, Camilo, Gotthelf, Helfand, Kramer,
  Lyne, Leighly \bbland{} Eracleous}]{halpern01b}
\bibnamefont{Halpern}, J.~P., F.~\bibnamefont{Camilo}, E.~V.
  \bibnamefont{Gotthelf}, D.~J. \bibnamefont{Helfand}, M.~\bibnamefont{Kramer},
  A.~G. \bibnamefont{Lyne}, K.~M. \bibnamefont{Leighly} \bbland{}
  \bibnamefont{Eracleous}. 2001b, \enquote{{PSR J2229+6114: Discovery of an
  Energetic Young Pulsar in the Error Box of the EGRET Source 3EG J2227+6122}},
  \emph{ApJL}, \bblvol{} 552, \bblp{} L125.

\bibitem[{{Halpern} \bbletal{}(2004){Halpern}, {Gotthelf}, {Camilo}, {Helfand}
  \bbland{} {Ransom}}]{Halpern2004}
\bibnamefont{{Halpern}}, J.~P., E.~V. \bibnamefont{{Gotthelf}},
  F.~\bibnamefont{{Camilo}}, D.~J. \bibnamefont{{Helfand}} \bbland{} S.~M.
  \bibnamefont{{Ransom}}. 2004, \enquote{{X-Ray, Radio, and Optical
  Observations of the Putative Pulsar in the Supernova Remnant CTA 1}},
  \emph{ApJ}, \bblvol{} 612, \bblpp{} 398--407.

\bibitem[{Halpern \bbletal{}(2001a)Halpern, Gotthelf, Leighly \bbland{}
  Helfand}]{halpern01a}
\bibnamefont{Halpern}, J.~P., E.~V. \bibnamefont{Gotthelf}, K.~M.
  \bibnamefont{Leighly} \bbland{} D.~J. \bibnamefont{Helfand}. 2001a,
  \enquote{{A Possible X-Ray and Radio Counterpart of the High-Energy Gamma-Ray
  Source 3EG J2227+6122}}, \emph{ApJ}, \bblvol{} 547, \bblp{} 323.

\bibitem[{Halpern \bbland{} Holt(1992)}]{halpern92}
\bibnamefont{Halpern}, J.~P. \bbland{} S.~S. \bibnamefont{Holt}. 1992,
  \enquote{{Discovery of soft X-ray pulsations from the gamma-ray source
  Geminga}}, \emph{Nature}, \bblvol{} 357, \bblp{} 222.

\bibitem[{Han(2009)}]{han09}
\bibnamefont{Han}, J.~L. 2009, \enquote{{Pulsars as Fantastic Objects and
  Probes}}, \emph{arXiv:0901.1593v1}.

\bibitem[{Harding(2006)}]{harding06}
\bibnamefont{Harding}, A.~K. 2006, \enquote{{High-Energy emission from the
  pulsar polar cap and slot gap}}, \emph{363rd Heraeus Seminar: Neutron Stars
  and Pulsars Bad Honnef}, \bblp{}~14.

\bibitem[{Harding(2007b)}]{harding07b}
\bibnamefont{Harding}, A.~K. 2007b, \enquote{{Pulsar Physics and GLAST}},
  \emph{AIPC}, \bblvol{} 921, \bblp{}~49.

\bibitem[{Harding \bbletal{}(2007)Harding, Grenier \bbland{}
  Gonthier}]{harding07}
\bibnamefont{Harding}, A.~K., I.~A. \bibnamefont{Grenier} \bbland{} P.~L.
  \bibnamefont{Gonthier}. 2007, \enquote{{The Geminga Fraction}},
  \emph{Ap\&SS}, \bblvol{} 309, \bblp{} 221.

\bibitem[{Hartman \bbletal{}(1999)Hartman, Bertsch, Bloom, Chen \bbland{}
  al.}]{hartman99}
\bibnamefont{Hartman}, R.~C., D.~L. \bibnamefont{Bertsch}, S.~D.
  \bibnamefont{Bloom}, A.~W. \bibnamefont{Chen} \bbland{} \bibnamefont{al.}
  1999, \enquote{{Third EGRET catalog (3EG)}}, \emph{ApJ}, \bblvol{} 123,
  \bblp{}~79.

\bibitem[{Hermsen \bbletal{}(1977)Hermsen, Swanenburg, Bignami, Boella
  \bbland{} al.}]{hermsen77}
\bibnamefont{Hermsen}, W., B.~N. \bibnamefont{Swanenburg}, G.~F.
  \bibnamefont{Bignami}, G.~\bibnamefont{Boella} \bbland{} \bibnamefont{al.}
  1977, \enquote{{New high energy gamma-ray sources observed by COS B}},
  \emph{Nature}, \bblvol{} 269, \bblp{} 494.

\bibitem[{Hewish \bbletal{}(1968)Hewish, Bell, Pilkington, Scott \bbland{}
  Collins}]{hewish68}
\bibnamefont{Hewish}, A., S.~J. \bibnamefont{Bell}, J.~D.~H.
  \bibnamefont{Pilkington}, P.~F. \bibnamefont{Scott} \bbland{} R.~A.
  \bibnamefont{Collins}. 1968, \enquote{{Observation of a rapidly pulsating
  radio source}}, \emph{Nature}, \bblvol{} 217, \bblp{} 709.

\bibitem[{Hobbs \bbletal{}(2006)Hobbs, Edwards \bbland{} Manchester}]{hobbs06}
\bibnamefont{Hobbs}, G.~B., R.~T. \bibnamefont{Edwards} \bbland{} R.~N.
  \bibnamefont{Manchester}. 2006, \enquote{{TEMPO2, a new pulsar-timing package
  - I. An overview}}, \emph{Mon. Not. R. Astron. Soc.}, \bblvol{} 369, \bblp{}
  655.

\bibitem[{Hobbs \bbletal{}(2006b)Hobbs, Lorimer, Lyne \bbland{}
  Kramer}]{hobbs06b}
\bibnamefont{Hobbs}, G.~B., D.~R. \bibnamefont{Lorimer}, A.~G.
  \bibnamefont{Lyne} \bbland{} M.~\bibnamefont{Kramer}. 2006b, \enquote{{Proper
  motions of pulsars}}, \emph{Mon. Not. R. Astron. Soc.}, \bblvol{} 360,
  \bblp{} 974.

\bibitem[{Hobbs \bbletal{}(2004)Hobbs, Lyne, Kramer, Martin \bbland{}
  Jordan}]{hobbs04}
\bibnamefont{Hobbs}, G.~B., A.~G. \bibnamefont{Lyne}, M.~\bibnamefont{Kramer},
  C.~E. \bibnamefont{Martin} \bbland{} C.~\bibnamefont{Jordan}. 2004,
  \enquote{{Long-term timing observations of 374 pulsars}}, \emph{Mon. Not. R.
  Astron. Soc.}, \bblvol{} 353, \bblp{} 1311.

\bibitem[{Holloway(1973)}]{holloway73}
\bibnamefont{Holloway}, N.~J. 1973, \enquote{{Pulsars-p-n junctions in pulsar
  magnetospheres}}, \emph{Nature}, \bblvol{} 246, \bblp{}~6.

\bibitem[{{Hoppe} \bbletal{}(2009){Hoppe}, {de O{\~n}a-Wilhemi}, {Kh{\'e}lifi},
  {Chaves}, {de Jager}, {Stegmann}, {Terrier} \bbland{} {for the
  H.~E.~S.~S.~Collaboration}}]{HESS1709_2009}
\bibnamefont{{Hoppe}}, S., E.~\bibnamefont{{de O{\~n}a-Wilhemi}},
  B.~\bibnamefont{{Kh{\'e}lifi}}, R.~C.~G. \bibnamefont{{Chaves}}, O.~C.
  \bibnamefont{{de Jager}}, C.~\bibnamefont{{Stegmann}},
  R.~\bibnamefont{{Terrier}} \bbland{} \bibnamefont{{for the
  H.~E.~S.~S.~Collaboration}}. 2009, \enquote{{Detection of very-high-energy
  gamma-ray emission from the vicinity of PSR B1706-44 with H.E.S.S}},
  \emph{arXiv:0906.5574}.

\bibitem[{Ivanov \bbletal{}(2004)Ivanov, Rakhimov, Smolentsev, Stankevitch
  \bbland{} Finkelstein}]{ivanov04}
\bibnamefont{Ivanov}, V.~P., I.~A. \bibnamefont{Rakhimov}, S.~G.
  \bibnamefont{Smolentsev}, K.~S. \bibnamefont{Stankevitch} \bbland{} A.~M.
  \bibnamefont{Finkelstein}. 2004, \enquote{{Nonstationary Radio Luminosity of
  the Remnant of Supernova 1181 (3C 58)}}, \emph{Astron. Lett.}, \bblvol{}~30,
  \bblp{} 240.

\bibitem[{Janka \bbletal{}(2005)Janka, Scheck, Kifonidis, Müller \bbland{}
  Plewa}]{janka05}
\bibnamefont{Janka}, H.~T., L.~\bibnamefont{Scheck},
  K.~\bibnamefont{Kifonidis}, E.~\bibnamefont{Müller} \bbland{}
  T.~\bibnamefont{Plewa}. 2005, \enquote{{Supernova Asymmetries and Pulsar
  Kicks --- Views on Controversial Issues}}, \emph{ASP Conf. Series.},
  \bblvol{} 332, \bblp{} 363.

\bibitem[{Janssen \bbland{} Stappers(2007)}]{janssen07}
\bibnamefont{Janssen}, G.~H. \bbland{} B.~W. \bibnamefont{Stappers}. 2007,
  \enquote{{30 glitches in slow pulsars}}, \emph{A\&A}, \bblvol{} 457, \bblp{}
  611.

\bibitem[{Jiang \bbland{} Zhang(2006)}]{jiang06}
\bibnamefont{Jiang}, Z.~J. \bbland{} L.~\bibnamefont{Zhang}. 2006,
  \enquote{{Statistical Properties of High-Energy Radiation from Young
  Pulsars}}, \emph{ApJ}, \bblvol{} 643, \bblp{} 1130.

\bibitem[{Johnson(2002)}]{johnson02}
\bibnamefont{Johnson}, T. 2002, \enquote{{Fabrication and Assembly Procedure
  for the Anti-coincidence Detector (ACD) Tile Detector Assembly (TDA)}},
  \emph{ACD-PROC-000059}.

\bibitem[{Johnston \bbletal{}(2006)Johnston, Karastergiou \bbland{}
  Willett}]{johnston06}
\bibnamefont{Johnston}, S., A.~\bibnamefont{Karastergiou} \bbland{}
  K.~\bibnamefont{Willett}. 2006, \enquote{{High-frequency observations of
  southern pulsars}}, \emph{MNRAS}, \bblvol{} 369, \bblp{} 1916.

\bibitem[{Johnston \bbletal{}(1996)Johnston, Koribalski, Weisberg \bbland{}
  Wilson}]{johnston96}
\bibnamefont{Johnston}, S., B.~\bibnamefont{Koribalski}, J.~M.
  \bibnamefont{Weisberg} \bbland{} W.~\bibnamefont{Wilson}. 1996, \enquote{{HI
  line measurements of pulsars towards the GUM nebula and the Carina arm}},
  \emph{MNRAS}, \bblvol{} 279, \bblp{} 661.

\bibitem[{Johnston \bbletal{}(1992)Johnston, Lyne, Manchester, Kniffen,
  D'Amico, Lim \bbland{} Ashworth}]{johnston92}
\bibnamefont{Johnston}, S., A.~G. \bibnamefont{Lyne}, R.~N.
  \bibnamefont{Manchester}, D.~A. \bibnamefont{Kniffen},
  N.~\bibnamefont{D'Amico}, J.~\bibnamefont{Lim} \bbland{}
  M.~\bibnamefont{Ashworth}. 1992, \enquote{{A high-frequency survey of the
  southern Galactic plane for pulsars}}, \emph{MNRAS}, \bblvol{} 255, \bblp{}
  401.

\bibitem[{Joncas \bbland{} Higgs(1990)}]{joncas90}
\bibnamefont{Joncas}, G. \bbland{} L.~A. \bibnamefont{Higgs}. 1990,
  \enquote{{The DRAO galactic-plane survey. II - Field at L = 105 deg}},
  \emph{A\&AS}, \bblvol{}~82, \bblp{} 113.

\bibitem[{Jones(1998)}]{jones98}
\bibnamefont{Jones}, B. 1998, \emph{{A Search for Gamma-Ray Bursts, and the
  Application of Kalman Filters to Gamma-Ray Reconstruction}}, Phd, Stanford
  University.

\bibitem[{Kaplan \bbletal{}(2005)Kaplan, Escoffier, Lacasse, O'Neil, Ford
  \bbland{} Ransom}]{kaplan05}
\bibnamefont{Kaplan}, D.~L., R.~P. \bibnamefont{Escoffier}, R.~J.
  \bibnamefont{Lacasse}, K.~\bibnamefont{O'Neil}, J.~M. \bibnamefont{Ford}
  \bbland{} S.~M. \bibnamefont{Ransom}. 2005, \enquote{{The Green Bank
  Telescope Pulsar Spigot}}, \emph{PASP}, \bblvol{} 117, \bblp{} 643.

\bibitem[{Karastergiou \bbletal{}(2005{\natexlab{a}})Karastergiou, Johnston
  \bbland{} Manchester}]{karastergiou05}
\bibnamefont{Karastergiou}, A., S.~\bibnamefont{Johnston} \bbland{} R.~N.
  \bibnamefont{Manchester}. 2005{\natexlab{a}}, \enquote{{Polarization profiles
  of southern pulsars at 3.1 GHz}}, \emph{MNRAS}, \bblvol{} 359, \bblp{} 481.

\bibitem[{Karastergiou \bbletal{}(2005{\natexlab{b}})Karastergiou, Johnston
  \bbland{} Manchester}]{kara05}
\bibnamefont{Karastergiou}, A., S.~\bibnamefont{Johnston} \bbland{} R.~N.
  \bibnamefont{Manchester}. 2005{\natexlab{b}}, \enquote{{Polarization profiles
  of southern pulsars at 3.1 GHz}}, \emph{MNRAS}, \bblvol{} 359, \bblp{} 481.

\bibitem[{Kargaltsev \bbland{} Pavlov(2009)}]{kargaltsev08}
\bibnamefont{Kargaltsev}, O. \bbland{} G.~G. \bibnamefont{Pavlov}. 2009,
  \enquote{{Pulsar Wind Nebulae in the Chandra Era}}, \emph{arXiv:0801.2602v2}.

\bibitem[{Kaspi(1998)}]{kaspi98}
\bibnamefont{Kaspi}, V.~M. 1998, \enquote{{Radio pulsar/supernova remnant
  associations}}, \emph{Advances in Space Sciences}, \bblvol{}~21, \bblp{} 167.

\bibitem[{Kaspi \bbletal{}(2000)Kaspi, Lackey, Mattox, Manchester, Bailes
  \bbland{} Pace}]{kaspi00}
\bibnamefont{Kaspi}, V.~M., J.~R. \bibnamefont{Lackey},
  J.~\bibnamefont{Mattox}, R.~N. \bibnamefont{Manchester},
  M.~\bibnamefont{Bailes} \bbland{} R.~\bibnamefont{Pace}. 2000,
  \enquote{{High-Energy Gamma-Ray Observations of Two Young, Energetic Radio
  Pulsars}}, \emph{ApJ}, \bblvol{} 528, \bblp{} 445.

\bibitem[{{Kassim} \bbland{} {Lazio}(1999)}]{ka99}
\bibnamefont{{Kassim}}, N.~E. \bbland{} T.~J.~W. \bibnamefont{{Lazio}}. 1999,
  \enquote{{Upper Limits on the Continuum Emission from Geminga at 74 and 326
  MHZ}}, \emph{ApJL}, \bblvol{} 527, \bblp{} L101.

\bibitem[{Kniffen \bbletal{}(1974)Kniffen, Hartman, Thompson, Bignami, Fichtel,
  Tümer \bbland{} Ögelman}]{kniffen74}
\bibnamefont{Kniffen}, D.~A., R.~C. \bibnamefont{Hartman}, D.~J.
  \bibnamefont{Thompson}, G.~F. \bibnamefont{Bignami}, C.~E.
  \bibnamefont{Fichtel}, T.~\bibnamefont{Tümer} \bbland{}
  H.~\bibnamefont{Ögelman}. 1974, \enquote{{Gamma radiation from the Crab
  Nebula above 35 MeV}}, \emph{Nature}, \bblvol{} 251, \bblp{} 397.

\bibitem[{Kothes \bbletal{}(2001)Kothes, Uyaniker \bbland{}
  Pineault}]{kothes01}
\bibnamefont{Kothes}, R., B.~\bibnamefont{Uyaniker} \bbland{}
  S.~\bibnamefont{Pineault}. 2001, \enquote{{The Supernova Remnant G106.3+2.7
  and Its Pulsar-Wind Nebula: Relics of Triggered Star Formation in a Complex
  Environment}}, \emph{ApJ}, \bblvol{} 560, \bblp{} 236.

\bibitem[{Kramer \bbletal{}(2003)Kramer, Bell, Manchester, Lyne, Camilo, Stairs
  \bbland{} al.}]{kramer03}
\bibnamefont{Kramer}, M., J.~F. \bibnamefont{Bell}, R.~N.
  \bibnamefont{Manchester}, A.~G. \bibnamefont{Lyne}, F.~\bibnamefont{Camilo},
  I.~H. \bibnamefont{Stairs} \bbland{} \bibnamefont{al.} 2003, \enquote{{The
  Parkes Multibeam Pulsar Survey - III. Young pulsars and the discovery and
  timing of 200 pulsars}}, \emph{MNRAS}, \bblvol{} 342, \bblp{} 1299.

\bibitem[{{Kuiper} \bbletal{}(2000){Kuiper}, {Hermsen}, {Verbunt}, {Thompson},
  {Stairs}, {Lyne}, {Strickman} \bbland{} {Cusumano}}]{kuiper00}
\bibnamefont{{Kuiper}}, L., W.~\bibnamefont{{Hermsen}},
  F.~\bibnamefont{{Verbunt}}, D.~J. \bibnamefont{{Thompson}}, I.~H.
  \bibnamefont{{Stairs}}, A.~G. \bibnamefont{{Lyne}}, M.~S.
  \bibnamefont{{Strickman}} \bbland{} G.~\bibnamefont{{Cusumano}}. 2000,
  \enquote{{The likely detection of pulsed high-energy gamma-ray emission from
  millisecond pulsar PSR J0218+4232}}, \emph{A\&A}, \bblvol{} 359, \bblp{} 615.

\bibitem[{Lamb \bbland{} Macomb(1997)}]{lamb97}
\bibnamefont{Lamb}, R.~C. \bbland{} D.~J. \bibnamefont{Macomb}. 1997,
  \enquote{{Point Sources of GeV Gamma Rays}}, \emph{ApJ}, \bblvol{} 488,
  \bblp{} 872.

\bibitem[{Large \bbletal{}(1968)Large, Vaughan \bbland{} Mills}]{large68}
\bibnamefont{Large}, M.~I., A.~E. \bibnamefont{Vaughan} \bbland{} B.~Y.
  \bibnamefont{Mills}. 1968, \enquote{{A Pulsar Supernova Association?}},
  \emph{Nature}, \bblvol{} 220, \bblp{} 340.

\bibitem[{Lattimer \bbland{} Prakash(2007)}]{lattimer07}
\bibnamefont{Lattimer}, J.~M. \bbland{} M.~\bibnamefont{Prakash}. 2007,
  \enquote{{Neutron star observations: Prognosis for equation of state
  constraints}}, \emph{PhR}, \bblvol{} 442, \bblp{} 109.

\bibitem[{Leahy \bbletal{}(1983)Leahy, Darbro, Elsner, Weisskopf, Kahn,
  Sutherland \bbland{} Grindlay}]{leahy83}
\bibnamefont{Leahy}, D.~A., W.~\bibnamefont{Darbro}, R.~F.
  \bibnamefont{Elsner}, M.~C. \bibnamefont{Weisskopf}, S.~\bibnamefont{Kahn},
  P.~G. \bibnamefont{Sutherland} \bbland{} J.~E. \bibnamefont{Grindlay}. 1983,
  \enquote{{On searches for pulsed emission with application to four globular
  cluster X-ray sources - NGC 1851, 6441, 6624, and 6712}}, \emph{ApJ},
  \bblvol{} 266, \bblp{} 160.

\bibitem[{Livingstone \bbletal{}(2008)Livingstone, Ransom, Camilo, Kaspi, Lyne,
  Kramer \bbland{} Stairs}]{livingstone08}
\bibnamefont{Livingstone}, M.~A., S.~M. \bibnamefont{Ransom},
  F.~\bibnamefont{Camilo}, V.~M. \bibnamefont{Kaspi}, A.~G. \bibnamefont{Lyne},
  M.~\bibnamefont{Kramer} \bbland{} I.~H. \bibnamefont{Stairs}. 2008,
  \enquote{{Timing the Young Pulsar at the Centre of SNR 3C~58}}, \emph{AIP
  Conf. Proc.}, \bblvol{} 983, \bblp{} 160.

\bibitem[{Livingstone \bbletal{}(2009)Livingstone, Ransom, Camilo, Kaspi, Lyne,
  Kramer \bbland{} Stairs}]{livingstone09}
\bibnamefont{Livingstone}, M.~A., S.~M. \bibnamefont{Ransom},
  F.~\bibnamefont{Camilo}, V.~M. \bibnamefont{Kaspi}, A.~G. \bibnamefont{Lyne},
  M.~\bibnamefont{Kramer} \bbland{} I.~H. \bibnamefont{Stairs}. 2009,
  \enquote{{X-ray and Radio Timing of the Pulsar in 3C~58}},
  \emph{arXiv:0901.2119}.

\bibitem[{Lorimer \bbland{} Kramer(2005)}]{handbook05}
\bibnamefont{Lorimer}, D. \bbland{} M.~\bibnamefont{Kramer}. 2005,
  \emph{{Handbook of Pulsar Astronomy}}, Cambridge University Press.

\bibitem[{Lott \bbletal{}(2006)Lott, Piron, Blank, Bogaert, Bregeon, Canchel
  \bbland{} al.}]{lott06}
\bibnamefont{Lott}, B., F.~\bibnamefont{Piron}, B.~\bibnamefont{Blank},
  G.~\bibnamefont{Bogaert}, J.~\bibnamefont{Bregeon}, G.~\bibnamefont{Canchel}
  \bbland{} \bibnamefont{al.} 2006, \enquote{{Response of the GLAST LAT
  calorimeter to relativistic heavy ions}}, \emph{NIM}, \bblvol{} 560, \bblp{}
  395.

\bibitem[{Lyne \bbland{} Graham-Smith(1998)}]{psrastro}
\bibnamefont{Lyne}, A.~G. \bbland{} F.~\bibnamefont{Graham-Smith}. 1998,
  \emph{{Pulsar Astronomy}}, Cambridge Astrophysics Series, Second Edition.

\bibitem[{MAGIC(2008)}]{magic08}
\bibnamefont{MAGIC}, C. 2008, \enquote{{Observation of Pulsed $\gamma$-Rays
  Above 25 GeV from the Crab Pulsar with MAGIC}}, \emph{Science}, \bblvol{}
  322, \bblp{} 1221.

\bibitem[{Manchester(2008)}]{manchester08}
\bibnamefont{Manchester}, R.~N. 2008, \enquote{{40 Years of Pulsars:
  Millisecond Pulsars, Magnetars and More}}, \emph{AIP Conf. Ser.}, \bblvol{}
  983, \bblp{} 584.

\bibitem[{Manchester \bbletal{}(2005)Manchester, Hobbs, Teoh \bbland{}
  Hobbs}]{atnf}
\bibnamefont{Manchester}, R.~N., G.~B. \bibnamefont{Hobbs},
  A.~\bibnamefont{Teoh} \bbland{} M.~\bibnamefont{Hobbs}. 2005, \enquote{{The
  Australia Telescope National Facility Pulsar Catalogue}}, \emph{AJ},
  \bblvol{} 129, \bblp{} 1993.

\bibitem[{Manchester \bbletal{}(2001)Manchester, Lyne, Camilo, Bell \bbland{}
  al.}]{manchester01}
\bibnamefont{Manchester}, R.~N., A.~G. \bibnamefont{Lyne},
  F.~\bibnamefont{Camilo}, J.~F. \bibnamefont{Bell} \bbland{} \bibnamefont{al.}
  2001, \enquote{{The Parkes multi-beam pulsar survey - I. Observing and data
  analysis systems, discovery and timing of 100 pulsars}}, \emph{MNRAS},
  \bblvol{} 328, \bblp{}~17.

\bibitem[{Mattox \bbletal{}(1996)Mattox, Bertsch, Chiang, Dingus, Digel,
  Esposito \bbland{} al.}]{mattox96}
\bibnamefont{Mattox}, J.~R., D.~L. \bibnamefont{Bertsch},
  J.~\bibnamefont{Chiang}, B.~L. \bibnamefont{Dingus}, S.~W.
  \bibnamefont{Digel}, J.~A. \bibnamefont{Esposito} \bbland{} \bibnamefont{al.}
  1996, \enquote{{The Likelihood Analysis of EGRET Data}}, \emph{ApJ},
  \bblvol{} 461, \bblp{} 396.

\bibitem[{Matz \bbletal{}(1994)Matz, Ulmer, Grabelsky, Purcell, Grove \bbland{}
  al.}]{matz94}
\bibnamefont{Matz}, S., M.~P. \bibnamefont{Ulmer}, D.~A.
  \bibnamefont{Grabelsky}, W.~R. \bibnamefont{Purcell}, J.~E.
  \bibnamefont{Grove} \bbland{} \bibnamefont{al.} 1994, \enquote{{The pulsed
  hard X-ray spectrum of PSR~B1509$-$58}}, \emph{ApJ}, \bblvol{} 434, \bblp{}
  288.

\bibitem[{McEnery \bbletal{}(2004)McEnery, Moskalenko \bbland{}
  Ormes}]{mcenery04}
\bibnamefont{McEnery}, J.~E., I.~V. \bibnamefont{Moskalenko} \bbland{} F.~J.
  \bibnamefont{Ormes}. 2004, \enquote{{GLAST: Understanding the High Energy
  Gamma-Ray Sky}}, \emph{astro-ph/0406250}.

\bibitem[{Meegan \bbletal{}(2009)Meegan, Lichti, Bhat, Bissaldi, Briggs
  \bbland{} al.}]{meegan09}
\bibnamefont{Meegan}, C., G.~\bibnamefont{Lichti}, P.~N. \bibnamefont{Bhat},
  E.~\bibnamefont{Bissaldi}, M.~S. \bibnamefont{Briggs} \bbland{}
  \bibnamefont{al.} 2009, \enquote{{The Fermi Gamma-ray Burst Monitor}},
  \emph{ApJ}, \bblvol{} 702, \bblp{} 791.

\bibitem[{Melzer \bbland{} Thorne(1966)}]{melzer66}
\bibnamefont{Melzer}, D.~W. \bbland{} K.~S. \bibnamefont{Thorne}. 1966,
  \enquote{{}}, \emph{ApJ}, \bblvol{} 145, \bblp{} 514.

\bibitem[{Morini(1983)}]{morini83}
\bibnamefont{Morini}, M. 1983, \enquote{{Inverse Compton gamma-ray from
  pulsars. I - The VELA pulsar}}, \emph{MNRAS}, \bblvol{} 202, \bblp{} 495.

\bibitem[{Murray \bbletal{}(2002)Murray, Slane, Seward \bbland{}
  Ransom}]{murray02}
\bibnamefont{Murray}, S.~S., P.~O. \bibnamefont{Slane}, F.~D.
  \bibnamefont{Seward} \bbland{} S.~M. \bibnamefont{Ransom}. 2002,
  \enquote{{Discovery of X-Ray Pulsations from the Compact Central Source in
  the Supernova Remnant 3C~58}}, \emph{ApJ}, \bblvol{} 568, \bblp{} 226.

\bibitem[{de~Naurois \bbletal{}(2002)de~Naurois, Holder, Bazer-Bachi, Bergeret,
  Bruel \bbland{} al.}]{denaurois02}
\bibnamefont{de~Naurois}, M., J.~\bibnamefont{Holder},
  R.~\bibnamefont{Bazer-Bachi}, H.~\bibnamefont{Bergeret},
  P.~\bibnamefont{Bruel} \bbland{} \bibnamefont{al.} 2002,
  \enquote{{Measurement of the Crab Flux above 60 GeV with the CELESTE Cerenkov
  Telescope}}, \emph{ApJ}, \bblvol{} 566, \bblp{} 343.

\bibitem[{Nel \bbland{} De~Jager(1995)}]{nel95}
\bibnamefont{Nel}, H.~I. \bbland{} O.~C. \bibnamefont{De~Jager}. 1995,
  \enquote{{Gamma-Ray Pulsars: Polar Cap or Outer Gap Emission ?}}, \emph{ASS},
  \bblvol{} 230, \bblp{} 299.

\bibitem[{Ng \bbland{} Romani(2004)}]{ng04}
\bibnamefont{Ng}, C.-Y. \bbland{} R.~W. \bibnamefont{Romani}. 2004,
  \enquote{{Fitting Pulsar Wind Tori}}, \emph{ApJ}, \bblvol{} 601, \bblp{} 479.

\bibitem[{Ng \bbland{} Romani(2008)}]{ng08}
\bibnamefont{Ng}, C.-Y. \bbland{} R.~W. \bibnamefont{Romani}. 2008,
  \enquote{{Fitting Pulsar Wind Tori. II. Error Analysis and Applications}},
  \emph{ApJ}, \bblvol{} 673, \bblp{} 411.

\bibitem[{Nolan \bbletal{}(2003)Nolan, Tompkins, Grenier \bbland{}
  Michelson}]{nolan03}
\bibnamefont{Nolan}, P.~L., W.~F. \bibnamefont{Tompkins}, I.~A.
  \bibnamefont{Grenier} \bbland{} P.~F. \bibnamefont{Michelson}. 2003,
  \enquote{{Variability of EGRET Gamma-Ray Sources}}, \emph{ApJ}, \bblvol{}
  597, \bblp{} 615.

\bibitem[{Noutsos \bbletal{}(2009)Noutsos, A. \bbland{} al.}]{noutsos09}
\bibnamefont{Noutsos}, A., A.~A. \bibnamefont{A.} \bbland{} \bibnamefont{al.}
  2009, \enquote{{Fermi LAT Observations of the pulsar J2043+2740}},
  \emph{ApJ}. - in prep.

\bibitem[{O'Brien \bbletal{}(2008)O'Brien, Johnston, ..., Parent \bbland{}
  ...}]{obrien08}
\bibnamefont{O'Brien}, J.~T., S.~\bibnamefont{Johnston}, \bibnamefont{...},
  D.~\bibnamefont{Parent} \bbland{} \bibnamefont{...} 2008, \enquote{{PSR
  J1410$-$6132: a young, energetic pulsar associated with the EGRET source 3EG
  J1410$-$6147}}, \emph{MNRAS}, \bblvol{} 388, \bblp{}~L1.

\bibitem[{Ostriker \bbland{} Gunn(1969)}]{ostriker69}
\bibnamefont{Ostriker}, J.~P. \bbland{} J.~E. \bibnamefont{Gunn}. 1969,
  \enquote{{On the Nature of Pulsars. I. Theory}}, \emph{ApJ}, \bblvol{} 157,
  \bblp{} 1395.

\bibitem[{Pacini(1968)}]{pacini68}
\bibnamefont{Pacini}, F. 1968, \enquote{{Rotating Neutron Stars, Pulsars and
  Supernova Remnants}}, \emph{Nature}, \bblvol{} 219, \bblp{} 145.

\bibitem[{Pellizzoni \bbletal{}(2009)Pellizzoni, Pilia, Possenti, Chen
  \bbland{} al.}]{pellizzoni09}
\bibnamefont{Pellizzoni}, A., M.~\bibnamefont{Pilia},
  A.~\bibnamefont{Possenti}, A.~\bibnamefont{Chen} \bbland{} \bibnamefont{al.}
  2009, \enquote{{Discovery of New Gamma-Ray Pulsars with AGILE}}, \emph{ApJ},
  \bblvol{} 695, \bblp{} 115.

\bibitem[{Pivovaroff \bbletal{}(2000)Pivovaroff, Kaspi \bbland{}
  Gotthelf}]{pivovaroff00}
\bibnamefont{Pivovaroff}, M.~J., V.~M. \bibnamefont{Kaspi} \bbland{} E.~V.
  \bibnamefont{Gotthelf}. 2000, \enquote{{ASCA Observations of the Young
  Rotation-powered Pulsars PSR B1046$-$58 and PSR B1610$-$50}}, \emph{ApJ},
  \bblvol{} 528, \bblp{} 436.

\bibitem[{Porter \bbletal{}(2008)Porter, Moskalenko, Strong, Orlando \bbland{}
  Bouchet}]{porter08}
\bibnamefont{Porter}, T.~A., I.~V. \bibnamefont{Moskalenko}, A.~W.
  \bibnamefont{Strong}, E.~\bibnamefont{Orlando} \bbland{}
  L.~\bibnamefont{Bouchet}. 2008, \enquote{{Inverse Compton Origin of the Hard
  X-Ray and Soft Gamma-Ray Emission from the Galactic Ridge}}, \emph{ApJ},
  \bblvol{} 682, \bblp{} 400.

\bibitem[{Radhakrishnan \bbland{} Cooke(1969)}]{radhna69}
\bibnamefont{Radhakrishnan}, V. \bbland{} D.~J. \bibnamefont{Cooke}. 1969,
  \enquote{{Magnetic Poles and the Polarization Structure of Pulsar
  Radiation}}, \emph{Astrophys.Lett.}, \bblvol{}~3, \bblp{} 225.

\bibitem[{Ransom \bbletal{}(2004)Ransom, Camilo, Kaspi, Slane, Gaensler,
  Gotthelf \bbland{} Murray}]{ransom04}
\bibnamefont{Ransom}, S., F.~\bibnamefont{Camilo}, V.~\bibnamefont{Kaspi},
  P.~\bibnamefont{Slane}, B.~\bibnamefont{Gaensler}, E.~\bibnamefont{Gotthelf}
  \bbland{} S.~\bibnamefont{Murray}. 2004, \enquote{{X-Ray Timing of the Young
  Pulsar in 3C 58}}, \emph{AIPC}, \bblvol{} 714, \bblp{} 350.

\bibitem[{{Ray} \bbland{} al.(2009)}]{ray09}
\bibnamefont{{Ray}}, P.~S. \bbland{} \bibnamefont{al.} 2009, \enquote{{Precise
  Timing of Fermi Gamma-Ray Pulsars}}, \emph{ApJ}. In prep.

\bibitem[{Richards \bbland{} Comella(1969)}]{richards69}
\bibnamefont{Richards}, D.~W. \bbland{} J.~M. \bibnamefont{Comella}. 1969,
  \enquote{{The period of pulsar NP 0532}}, \emph{Nature}, \bblvol{} 222,
  \bblp{} 551.

\bibitem[{Roberts \bbletal{}(1993)Roberts, Goss, Kalberla, Herbstmeier
  \bbland{} Schwarz}]{roberts93}
\bibnamefont{Roberts}, D.~A., W.~M. \bibnamefont{Goss}, P.~M.~W.
  \bibnamefont{Kalberla}, U.~\bibnamefont{Herbstmeier} \bbland{} U.~J.
  \bibnamefont{Schwarz}. 1993, \enquote{{High Resolution HI Observations of
  3C~58}}, \emph{A\&A}, \bblvol{} 274, \bblp{} 427.

\bibitem[{Roberts \bbletal{}(2005)Roberts, Brogan, Gaensler, Hessels, Ng
  \bbland{} Romani}]{roberts05}
\bibnamefont{Roberts}, M.~S.~E., C.~L. \bibnamefont{Brogan}, B.~M.
  \bibnamefont{Gaensler}, J.~W.~T. \bibnamefont{Hessels}, C.-Y.
  \bibnamefont{Ng} \bbland{} R.~W. \bibnamefont{Romani}. 2005, \enquote{{Pulsar
  Wind Nebulae in Egret Error Boxes}}, \emph{Ap\&SS}, \bblvol{} 297,
  \bblp{}~93.

\bibitem[{{Roberts} \bbletal{}(2002){Roberts}, {Hessels}, {Ransom}, {Kaspi},
  {Freire}, {Crawford} \bbland{} {Lorimer}}]{Roberts2002}
\bibnamefont{{Roberts}}, M.~S.~E., J.~W.~T. \bibnamefont{{Hessels}}, S.~M.
  \bibnamefont{{Ransom}}, V.~M. \bibnamefont{{Kaspi}}, P.~C.~C.
  \bibnamefont{{Freire}}, F.~\bibnamefont{{Crawford}} \bbland{} D.~R.
  \bibnamefont{{Lorimer}}. 2002, \enquote{{PSR J2021+3651: A Young Radio Pulsar
  Coincident with an Unidentified EGRET {$\gamma$}-Ray Source}}, \emph{ApJL},
  \bblvol{} 577, \bblpp{} L19--L22.

\bibitem[{Romani \bbland{} Yadigaroglu(1995)}]{romani95}
\bibnamefont{Romani}, R.~W. \bbland{} I.~A. \bibnamefont{Yadigaroglu}. 1995,
  \enquote{{Gamma-ray pulsars: Emission zones and viewing geometries}},
  \emph{ApJ}, \bblvol{} 438, \bblp{} 314.

\bibitem[{Rudak \bbletal{}(2002)Rudak, Dyks \bbland{} Bulik}]{rudak02}
\bibnamefont{Rudak}, B., J.~\bibnamefont{Dyks} \bbland{}
  T.~\bibnamefont{Bulik}. 2002, \enquote{{High-energy radiation from pulsars: a
  challenge to polar-cap models}}, \emph{MPE Report}, \bblvol{} 278, \bblp{}
  142.

\bibitem[{Ruderman \bbland{} Sutherland(1975)}]{ruderman75}
\bibnamefont{Ruderman}, M.~A. \bbland{} P.~G. \bibnamefont{Sutherland}. 1975,
  \enquote{{Theory of pulsars - Polar caps, sparks, and coherent microwave
  radiation}}, \emph{ApJ}, \bblvol{} 196, \bblp{}~51.

\bibitem[{Russell(1921)}]{russell21}
\bibnamefont{Russell}, H.~N. 1921, \enquote{{Stellar Evolution}}, \emph{PA},
  \bblvol{}~29, \bblp{} 541.

\bibitem[{Schmidt(1963)}]{schmidt63}
\bibnamefont{Schmidt}, M. 1963, \enquote{{3C 273 : A Star-Like Object with
  Large Red-Shift}}, \emph{Nature}, \bblvol{} 197, \bblp{} 1040.

\bibitem[{Shearer \bbland{} Neustroev(2008)}]{shearer08}
\bibnamefont{Shearer}, A. \bbland{} V.~V. \bibnamefont{Neustroev}. 2008,
  \enquote{{Optical observations of PSR J0205+6449}}, \emph{MNRAS}, \bblvol{}
  390, \bblp{} 235.

\bibitem[{{Shklovskii}(1970)}]{shklovskii70}
\bibnamefont{{Shklovskii}}, I.~S. 1970, \enquote{{Possible Causes of the
  Secular Increase in Pulsar Periods}}, \emph{Soviet Astronomy}, \bblvol{}~13,
  \bblpp{} 562--+.

\bibitem[{Smith \bbletal{}(2008)Smith, Guillemot, Camilo, Cognard, Dumora
  \bbland{} al.}]{smith08}
\bibnamefont{Smith}, D.~A., L.~\bibnamefont{Guillemot},
  F.~\bibnamefont{Camilo}, I.~\bibnamefont{Cognard}, D.~\bibnamefont{Dumora}
  \bbland{} \bibnamefont{al.} 2008, \enquote{{Pulsar Timing for the Fermi
  Gamma-ray Space Telescope}}, \emph{A\&A}, \bblvol{} 492, \bblp{} 923.

\bibitem[{Staelin \bbland{} Reifenstein~III(1968)}]{staelin68}
\bibnamefont{Staelin}, D.~H. \bbland{} E.~C. \bibnamefont{Reifenstein~III}.
  1968, \enquote{{Pulsating radio sources near the Crab Nebula}},
  \emph{Science}, \bblvol{} 162, \bblp{} 1481.

\bibitem[{Standish(1998)}]{standish98}
\bibnamefont{Standish}, E.~M. 1998, \emph{{JPL Planetary and Lunar Ephemerides,
  DE405/LE405}}, Memo IOM 312.F-98-048.

\bibitem[{Stephenson(1971)}]{stephenson71}
\bibnamefont{Stephenson}, F.~R. 1971, \enquote{{Suspected Supernova in A.D.
  1181}}, \emph{QJRAS}, \bblvol{}~12, \bblp{}~10.

\bibitem[{Stephenson \bbland{} Green(2002)}]{stephenson02}
\bibnamefont{Stephenson}, F.~R. \bbland{} D.~A. \bibnamefont{Green}. 2002,
  \emph{{Historical Supernovae and their Remnants}}, Oxford: Clarendon.

\bibitem[{Strong \bbletal{}(2004a)Strong, Moskalenko \bbland{}
  Reimer}]{Strong04a}
\bibnamefont{Strong}, A.~W., I.~V. \bibnamefont{Moskalenko} \bbland{}
  O.~\bibnamefont{Reimer}. 2004a, \enquote{{Diffuse Galactic Continuum Gamma
  Rays: A Model Compatible with EGRET Data and Cosmic-Ray Measurements}},
  \emph{ApJ}, \bblvol{} 613, \bblp{} 962.

\bibitem[{Strong \bbletal{}(2004b)Strong, Moskalenko, Reimer, Digel \bbland{}
  Diehl}]{Strong04b}
\bibnamefont{Strong}, A.~W., I.~V. \bibnamefont{Moskalenko},
  O.~\bibnamefont{Reimer}, S.~\bibnamefont{Digel} \bbland{}
  R.~\bibnamefont{Diehl}. 2004b, \enquote{{The distribution of cosmic-ray
  sources in the Galaxy, $\gamma$-rays and the gradient in the CO-to-H$_2$
  relation}}, \emph{A\&A}, \bblvol{} 422, \bblp{}~47.

\bibitem[{Sturrock(1971)}]{sturrock71}
\bibnamefont{Sturrock}, P.~A. 1971, \enquote{{A Model of Pulsars}}, \emph{ApJ},
  \bblvol{} 164, \bblp{} 529.

\bibitem[{Swanenburg \bbletal{}(1981)Swanenburg, Bennett, Bignami \bbland{}
  al.}]{swan81}
\bibnamefont{Swanenburg}, B.~N., K.~\bibnamefont{Bennett}, G.~F.
  \bibnamefont{Bignami} \bbland{} \bibnamefont{al.} 1981, \enquote{{Second COS
  B catalog of high-energy gamma-ray sources}}, \emph{ApJ}, \bblvol{} 243,
  \bblp{}~69.

\bibitem[{Taylor \bbland{} Cordes(1993)}]{taylor93}
\bibnamefont{Taylor}, J.~H. \bbland{} J.~M. \bibnamefont{Cordes}. 1993,
  \enquote{{Pulsar distances and the galactic distribution of free electrons}},
  \emph{ApJ}, \bblvol{} 411, \bblp{} 674.

\bibitem[{Theureau \bbletal{}(2005)Theureau, Coudreau, Hallet, Hanski, L.
  \bbland{} al.}]{theureau05}
\bibnamefont{Theureau}, G., N.~\bibnamefont{Coudreau}, N.~\bibnamefont{Hallet},
  M.~\bibnamefont{Hanski}, A.~\bibnamefont{L.} \bbland{} \bibnamefont{al.}
  2005, \enquote{{Kinematics of the local universe . XII. 21-cm line
  measurements of 586 galaxies with the new Nançay receiver}}, \emph{A\&A},
  \bblvol{} 430, \bblp{} 373.

\bibitem[{Thompson(2001)}]{thompson01}
\bibnamefont{Thompson}, D.~J. 2001, \enquote{{High Energy Gamma Ray Astronomy,
  ed. F.~A. Aharonian, H.~J. Volk}}, \emph{AIP Conf. Proc.}, \bblvol{} 157,
  \bblp{} 103.

\bibitem[{Thompson(2008)}]{thompson08}
\bibnamefont{Thompson}, D.~J. 2008, \enquote{{Gamma ray astrophysics: the EGRET
  results}}, \emph{RPPh}, \bblvol{}~71, \bblp{} 6901.

\bibitem[{Thompson \bbletal{}(1999)Thompson, Bailes, Bertsch, Cordes, D'Amico
  \bbland{} al.}]{thompson99}
\bibnamefont{Thompson}, D.~J., M.~\bibnamefont{Bailes}, D.~L.
  \bibnamefont{Bertsch}, J.~\bibnamefont{Cordes}, N.~\bibnamefont{D'Amico}
  \bbland{} \bibnamefont{al.} 1999, \enquote{{Gamma Radiation from PSR
  B1055-52}}, \emph{ApJ}, \bblvol{} 516, \bblp{} 297.

\bibitem[{Thompson \bbletal{}(2005)Thompson, Bertsch \bbland{}
  O'Neal}]{thompson05}
\bibnamefont{Thompson}, D.~J., D.~L. \bibnamefont{Bertsch} \bbland{} R.~H.
  \bibnamefont{O'Neal}. 2005, \enquote{{The Highest-Energy Photons Seen by the
  Energetic Gamma Ray Experiment Telescope (EGRET) on the Compton Gamma Ray
  Observatory}}, \emph{ApJS}, \bblvol{} 157, \bblp{} 324.

\bibitem[{Thompson \bbletal{}(2002)Thompson, Digel, Nolan \bbland{}
  Reimer}]{thompson02}
\bibnamefont{Thompson}, D.~J., S.~W. \bibnamefont{Digel}, P.~L.
  \bibnamefont{Nolan} \bbland{} O.~\bibnamefont{Reimer}. 2002,
  \enquote{{High-Energy Gamma Rays from Neutron Stars in Supernova Remnants:
  From EGRET to GLAST}}, \emph{ASPC}, \bblvol{} 271, \bblp{}~65.

\bibitem[{Thompson \bbletal{}(1975)Thompson, Fichtel, Kniffen \bbland{}
  Ögelman}]{thompson75}
\bibnamefont{Thompson}, D.~J., C.~E. \bibnamefont{Fichtel}, D.~A.
  \bibnamefont{Kniffen} \bbland{} H.~B. \bibnamefont{Ögelman}. 1975,
  \enquote{{SAS-2 high-energy gamma-ray observations of the VELA pulsar}},
  \emph{ApJ}, \bblvol{} 200, \bblp{} L79.

\bibitem[{Torii \bbletal{}(2000)Torii, Slane, Kinugasa, Hashimotodani \bbland{}
  Tsunemi}]{torii00}
\bibnamefont{Torii}, K., P.~O. \bibnamefont{Slane}, K.~\bibnamefont{Kinugasa},
  K.~\bibnamefont{Hashimotodani} \bbland{} H.~\bibnamefont{Tsunemi}. 2000,
  \enquote{{ASCA Observations of the Crab-Like Supernova Remnant 3C 58}},
  \emph{PASJ}, \bblvol{}~52, \bblp{} 875.

\bibitem[{Van~Etten \bbletal{}(2008)Van~Etten, Romani \bbland{}
  Ng}]{vanetten08}
\bibnamefont{Van~Etten}, A., R.~W. \bibnamefont{Romani} \bbland{} C.-Y.
  \bibnamefont{Ng}. 2008, \enquote{{Rings and Jets around PSR J2021+3651: The
  ``Dragonfly Nebula''}}, \emph{ApJ}, \bblvol{} 680, \bblp{} 1417.

\bibitem[{{Walker} \bbletal{}(2003){Walker}, {Mori} \bbland{}
  {Ohishi}}]{Walker2003}
\bibnamefont{{Walker}}, M., M.~\bibnamefont{{Mori}} \bbland{}
  M.~\bibnamefont{{Ohishi}}. 2003, \enquote{{Dense Gas Clouds and the
  Unidentified EGRET Sources}}, \emph{{ApJ}}, \bblvol{} 589, \bblpp{} 810--817.

\bibitem[{Watters \bbletal{}(2009)Watters, Romani, Weltevrede \bbland{}
  Johnston}]{watters09}
\bibnamefont{Watters}, K.~P., R.~W. \bibnamefont{Romani},
  P.~\bibnamefont{Weltevrede} \bbland{} S.~\bibnamefont{Johnston}. 2009,
  \enquote{{An Atlas for Interpreting Gamma-Ray Pulsar Light Curves}},
  \emph{ApJ}, \bblvol{} 695, \bblp{} 1289.

\bibitem[{Weiler \bbland{} Panagia(1978)}]{weiler78}
\bibnamefont{Weiler}, K.~W. \bbland{} N.~\bibnamefont{Panagia}. 1978,
  \enquote{{Are Crab-type Supernova Remnants (Plerions) Short-lived?}},
  \emph{A\&A}, \bblvol{}~70, \bblp{} 419.

\bibitem[{Weltevrede(2009)}]{weltevrede09}
\bibnamefont{Weltevrede}, P. 2009, \enquote{soumis}, \emph{PASA}.

\bibitem[{{Weltevrede} \bbletal{}(2009){Weltevrede}, {Abdo}, {Ackermann},
  {Atwood} \bbland{} al.}]{LAT6pulsars}
\bibnamefont{{Weltevrede}}, P., A.~A. \bibnamefont{{Abdo}},
  M.~\bibnamefont{{Ackermann}}, W.~B. \bibnamefont{{Atwood}} \bbland{}
  \bibnamefont{al.} 2009, \enquote{{Gamma-ray and Radio Properties of Six
  Pulsars Detetcted by the \textit{Fermi} LAT}}, \emph{ApJ}. Submitted.

\bibitem[{Weltevrede \bbland{} Johnston(2008a)}]{welte08a}
\bibnamefont{Weltevrede}, P. \bbland{} S.~\bibnamefont{Johnston}. 2008a,
  \enquote{{The population of pulsars with interpulses and the implications for
  beam evolution}}, \emph{MNRAS}, \bblvol{} 387, \bblp{} 1755.

\bibitem[{Weltevrede \bbland{} Wright(2009)}]{Weltevrede09b}
\bibnamefont{Weltevrede}, P. \bbland{} G.~\bibnamefont{Wright}. 2009,
  \enquote{{Mapping the magnetosphere of PSR B1055-52}}, \emph{MNRAS},
  \bblvol{} 395, \bblp{} 2117.

\end{thebibliography}

\listoffigures          
\listoftables           

\printindex                   

\phantomsection

\addstarredchapter{Résumé - Abstract} 

\thispagestyle{empty2}

\markboth{}{}

\paragraph{Résumé:} Le Large Area Telescope à bord du satellite \textit{Fermi}, lancé le 11 juin 2008, est un télescope spatial observant l'univers des hautes énergies. L'instrument couvre l'intervalle en énergie de 20\,MeV à 300\,GeV avec une sensibilité nettement améliorée et la capacité de localiser des sources ponctuelles. Il détecte les photons $\gamma$ par leur conversion en paire électron-positron, et mesure leur direction et leur énergie grâce à un trajectographe et un calorimètre.

Cette thèse présente les courbes de lumières et les mesures spectrales résolues en phase des pulsars radio et gamma détectés par le LAT. La mesure des paramètres spectraux (flux, indice spectral, et énergie de coupure) dépend des fonctions de réponse de l'instrument (IRFs). Une méthode développée pour la validation en orbite de la surface efficace est présentée en utilisant le pulsar de Vela. Les efficacités des coupures entre les données du LAT et les données simulées sont comparées à chaque niveau de la rejection du fond. Les résultats de cette analyse sont propagés vers les IRFs pour évaluer les systématiques des mesures spectrales.

La dernière partie de cette thèse présente les découvertes de nouveaux pulsars $\gamma$ individuels tels que PSR J0205+6449, J2229+6114, et J1048$-$5832 à partir des données du LAT et des éphémérides radio et $X$. Des analyses temporelles et spectrales sont investies dans le but de contraindre les modèles d'émission gamma. Finalement, nous discutons les propriétés d'une large population de pulsars gamma détectés par le LAT, incluant les pulsars normaux et les pulsars milliseconde.

\paragraph{Mots-clés:}{\it Pulsars - Fermi Large Area Telescope - Surface Efficace}


\paragraph{Abstract:} The Large Area Telescope (LAT) on \textit{Fermi}, launched on 2008 June 11, is a space telescope to explore the high energy $\gamma$-ray universe. The instrument covers the energy range from 20\,MeV to 300\,GeV with greatly improved sensitivity and ability to localize $\gamma$-ray point sources. It detects $\gamma$-rays through conversion to electron-positron pairs and measurement of their direction in a tracker and their energy in a calorimeter. 

This thesis presents the $\gamma$-ray light curves and the phase-resolved spectral measurements of radio-loud gamma-ray pulsars detected by the LAT. The measurement of pulsar spectral parameters (i.e. integrated flux, spectral index, and energy cut-off) depends on the instrument response functions (IRFs). A method developed for the on-orbit validation of the effective area is presented using the Vela pulsar. The cut efficiencies between the real data and the simulated data are compared at each stage of the background rejection. The results are then propagated to the IRFs, allowing the systematic uncertainties of the spectral parameters to be estimated. 

The last part of this thesis presents the discoveries, using both the LAT observations and the radio and $X$ ephemeredes, of new individual $\gamma$-ray pulsars such as PSR~J0205+6449, and the Vela-like pulsars J2229+6114 and J1048$-$5832. Timing and spectral analysis are investigated in order to constrain the $\gamma$-ray emission model. In addition, we discuss the properties of a large population of $\gamma$-ray pulsars detected by the LAT, including normal pulsars, and millisecond pulsars.

\paragraph{Keywords:}{\it Pulsars - Fermi Large Area Telescope - Effective Area}  

\end{document}